\documentclass[]{aastex631}
\usepackage{multirow}

\usepackage{xspace}
\usepackage{wrapfig}

\begin{document}

\title{The protostars in Orion: Characterizing the properties of their magnetized envelopes}

\author[0000-0001-7393-8583]{Bo Huang}
\email{huang@ice.csic.es}
\affiliation{Institut de Ciències de l'Espai (ICE-CSIC), Campus UAB, Can Magrans S/N, E-08193 Cerdanyola del Vallès, Catalonia, Spain}

\author[0000-0002-3829-5591]{Josep M. Girart}
\affiliation{Institut de Ciències de l'Espai (ICE-CSIC), Campus UAB, Can Magrans S/N, E-08193 Cerdanyola del Vallès, Catalonia, Spain}
\affiliation{Institut d'Estudis Espacials de Catalunya (IEEC), c/Gran Capita, 2-4, E-08034 Barcelona, Catalonia, Spain}

\author[0000-0003-3017-4418]{Ian W. Stephens}
\affiliation{Department of Earth, Environment, and Physics, Worcester State University, Worcester, MA 01602, USA}

\author[0000-0001-5811-0454]{Manuel Fern\'andez L\'opez}
\affiliation{Instituto Argentino de Radioastronomía (CCT-La Plata, CONICET; CICPBA), C.C. No. 5, 1894, Villa Elisa, Buenos Aires, Argentina}

\author[0000-0002-2885-1806]{Philip C. Myers}
\affiliation{Center for Astrophysics $\mid$ Harvard $\&$ Smithsonian, 60 Garden Street, Cambridge, MA 02138, USA}

\author[0000-0003-2384-6589]{Qizhou Zhang}
\affiliation{Center for Astrophysics $\mid$ Harvard $\&$ Smithsonian, 60 Garden Street, Cambridge, MA 02138, USA}

\author[0000-0002-6195-0152]{John J. Tobin}
\affiliation{National Radio Astronomy Observatory, 520 Edgemont Rd., Charlottesville, VA 22093, USA}

\author[0000-0002-3583-780X]{Paulo Cortes}
\affiliation{National Radio Astronomy Observatory, 520 Edgemont Rd., Charlottesville, VA 22093, USA}
\affiliation{Joint ALMA Observatory, Alonso de Córdova 3107, Vitacura, Santiago, Chile}

\author{Nadia M. Murillo}
\affiliation{Instituto de Astronom\'ia, Universidad Nacional Aut\'onoma de M\'exico, AP106, Ensenada CP 22830, B. C., M\'exico}
\affiliation{Star and Planet Formation Laboratory, RIKEN Cluster for Pioneering Research, Wako, Saitama 351-0198, Japan}

\author[0000-0001-7474-6874]{Sarah Sadavoy}
\affiliation{Department of Physics, Engineering and Astronomy, Queen’s University, 64 Bader Lane, Kingston, ON, K7L 3N6, Canada}

\author[0000-0001-5653-7817]{Hector G. Arce}
\affiliation{Department of Astronomy, Yale University, New Haven, CT 06511, USA}

\author[0000-0003-2251-0602]{John M. Carpenter}
\affiliation{Joint ALMA Observatory, Av. Alonso de Córdova 3107, Vitacura, Santiago, Chile}

\author[0000-0003-4022-4132]{Woojin Kwon}
\affiliation{Department of Earth Science Education, Seoul National University, 1 Gwanak-ro, Gwanak-gu, Seoul 08826, Republic of Korea}
\affiliation{SNU Astronomy Research Center, Seoul National University, 1 Gwanak-ro, Gwanak-gu, Seoul 08826, Republic of Korea}
\affiliation{The Center for Educational Research, Seoul National University, 1 Gwanak-ro, Gwanak-gu, Seoul 08826, Republic of Korea}

\author[0000-0002-5714-799X]{Valentin J. M. Le Gouellec}
\affiliation{NASA Ames Research Center, Space Science and Astrobiology Division M.S. 245-6 Moffett Field, CA 94035, USA}
\affiliation{NASA Postdoctoral Program Fellow}

\author[0000-0002-7402-6487]{Zhi-Yun Li}
\affiliation{Astronomy Department, University of Virginia, Charlottesville, VA 22904, USA}

\author[0000-0002-4540-6587]{Leslie W. Looney}
\affiliation{Department of Astronomy, University of Illinois, 1002 West Green Street, Urbana, IL 61801, USA}

\author[0000-0001-7629-3573]{Tom Megeath}
\affiliation{Department of Astronomy, University of Toledo, Toledo, OH 43606, USA}

\author[0000-0002-5216-8062]{Erin G. Cox}
\affiliation{Center for Interdisciplinary Exploration and Research in Astrophysics (CIERA), 1800 Sherman Avenue, Evanston, IL 60201, USA}

\author[0000-0003-3682-854X]{Nicole Karnath}
\affiliation{SOFIA Science Center, Universities Space Research Association, NASA Ames Research Center, Moffett Field, California 94035, USA}
\affiliation{Space Science Institute, 4765 Walnut St, Suite B Boulder, CO 80301, USA}
\affiliation{Center for Astrophysics $\mid$ Harvard $\&$ Smithsonian, 60 Garden Street, Cambridge, MA 02138, USA}

\author[0000-0003-3172-6763]{Dominique Segura-Cox}
\affiliation{Center for Astrochemical Studies, Max Planck Institute for Extraterrestrial Physics, D-85748 Garching, Germany}
\affiliation{Department of Astronomy, University of Texas, 2515 Speedway, Stop C1400, Austin, TX 78712, USA}

\begin{abstract}

We present a study connecting the physical properties of protostellar envelopes to the morphology of the envelope-scale magnetic field.  
We used the ALMA polarization observations of 61 young prtostars at 0.87 mm on $\sim400-3000$ au scales from the {\em B}-field Orion Protostellar Survey to infer the envelope-scale magnetic field, and used the dust emission to measure the envelope properties on comparable scales. 
We find that protostars showing standard-hourglass-field morphology tend to have larger masses and lower velocity dispersions in their envelopes, whereas systems with spiral-field morphologies have higher velocity dispersion.
Combining with the disk properties taken from the Orion VLA/ALMA Nascent Disk and Multiplicity survey, we connect envelope properties to fragmentation.
Our results show that the fragmentation level is positively correlated with the angle dispersion of the magnetic field, suggesting that the envelope fragmentation tends to be suppressed by the magnetic field.
We also find that protostars exhibiting standard hourglass magnetic field structure tend to have a smaller disk and smaller angle dispersion of the magnetic field than other field configurations, specially the rotated hourglass, but also the spiral and others, suggesting a more effective magnetic braking in the standard hourglass morphology of magnetic fields.
Nevertheless, significant misalignment between the magnetic field and outflow axes tends to reduce magnetic braking, leading to the formation of larger disks.
\end{abstract}


\section{Introduction} \label{sec:intro}

Fragmentation is believed to be a critical process in the formation of star and stellar systems \citep[e.g.,][]{matsumoto2003fragmentation}.
Initially, turbulence induces chaotic motions and density fluctuations across various scales, compressing the gas in certain regions and leading to localized density enhancements \citep[e.g.,][]{mac2004tur}.
Within these dense regions, gravitational forces overcome thermal pressure if the mass exceeds the Jeans mass.
As the cloud collapses, the large structures fragment into smaller, dense cores, each of which forms individual stars or multiple star systems \citep[e.g.,][]{mckee2007star}.
This hierarchical fragmentation continues until individual cores reach densities high enough to resist further fragmentation \citep[e.g.,][]{beuther2015frag}.
However, the specific mechanisms by which a cloud core fragments to ultimately form a multiple system or small cluster remain uncertain.

Theoretical and numerical studies \citep[e.g.,][]{maury2022, offner2023ppvii} suggest that cores with some angular momentum are prone to fragmentation into multiple objects, indicating the need for additional mechanisms to reduce the fragmentation to the observed levels.
Since non-ideal MHD effects do not inhibit protostellar multiplicity \citep[e.g.,][]{wurster2019nonmhd}, further research points to turbulence and magnetic fields (hereafter, \textit{B}-fields) as playing a pivotal role in the fragmentation process \citep[e.g.,][]{padoan2001turbulence, price2009magnetic}. 
For instance, compressive turbulence promotes fragmentation and the formation of structures with high density contrasts, whereas solenoidal turbulence \citep[e.g.,][]{federrath2010frag, girichidis2011frag} suppresses fragmentation and maintains a smoother density structure within the cloud. 
On the other hand, the ratio of rotational to magnetic energy has been found to influence core fragmentation \citep[e.g.,][]{machida2008formation}.
Moreover, some studies suggest that {\em B}-field tends to suppress fragmentation by providing additional support against gravitational collapse, and/or through magnetic braking, which carries away the angular momentum from the inner regions of the core \citep[e.g.,][]{commercon2011collapse, guszejnov2021starforge, mignon2021collapse}.

Dust polarization is the most commonly used tracer of {\em B}-fields in star-forming regions, as ``radiative torques'' \citep[e.g.,][]{hoang2009grain, andersson2015interstellar} tend to align spinning and elongated dust grains with their long axes perpendicular to the ambient {\em B}-field direction.
Over the last decades, dust polarization observations carried out with (sub)millimeter interferometers have increasingly proven effective in mapping {\em B}-fields at the scales of cores ($\sim 10^{4}$ au) and envelopes ($\sim10^{2}$ to $10^{4}$ au) \citep[e.g.][]{girart1999detection, girart2013dr, cox2018alma, galametz2018sma, hull2019interferometric, le2020IMS, cortes2021magnetic}.
The theory predicts that the {\em B}-field regulates the collapse of clouds, whereas ambipolar diffusion allows the formation of ``supercritical'' dense cores, where the gravitational force dominates over the magnetic support.
As cloud cores collapse, {\em B}-field lines are dragged inward to be drawn into pinched morphology.
This has been observed in various studies \citep[e.g.,][]{girart2006magnetic, girart2009magnetic, qiu2014submillimeter, beltran2019alma, le2019characterizing, hull2020understanding}.
In the case of flux freezing, the pinched geometry in the pseudo-disk strengthens the {\em B}-field and carries away angular momentum, leading to catastrophic magnetic braking and potentially preventing the disk formation \citep[e.g.,][]{galli2006mag}.
However, disk may form if the magnetic and rotational axes are misaligned \citep[e.g.,][]{joos2012protostellar, li2013misalignment, machida2020misalignment}.
To statistically test these scenarios related to fragmentation and star/disk formation, it is crucial to understand the connection between envelope properties and the resulting fragmentation.

Recently, the first results from the \textit{B}-fields Orion protostellar survey \cite[BOPS, with a resolution of $\sim 1^{\prime\prime}$, corresponding to $\sim$ 400 au, ][]{huang2024magnetic} presented the largest polarization study of \textit{B}-fields in protostellar envelopes using dust polarization observations of 61 protostars in Orion.
We have successfully mapped the \textit{B}-fields morphology at scales ranging from several hundred to a few thousand au, and classified the protostars using two criteria: polarization/{\em B}-field structure and the alignment between the outflow directions and velocity gradients.
On one hand, the morphological categories of the {\em B}-field for protostars are classified as: ``standard hourglass" (with the main axis coincident with the outflow axis), ``rotated hourglass" (similar to the ``standard hourglass" but rotated by $\sim90^{\circ}$, with its axis perpendicular to the outflow), ``spiral" configuration, ``complex" field morphology, ``not enough data" (insufficient polarized data to identify the {\em B}-field structure), and ``self-scattering" (compact polarized emission in the innermost regions, with the polarization direction perpendicular to the disk major axis).
On the other hand, BOPS protostars are classified as ``Perp-Type" (velocity gradient is perpendicular to the outflow), ``Nonperp-Type" (velocity gradient is not perpendicular to the outflow), and ``Unres-Type" (velocity gradient is unresolved).
In addition, observations from a major Orion interferometric study, known as the Orion VLA/ALMA Nascent Disk and Multiplicity \citep[VANDAM, ][]{tobin2020vla}, have identified protostellar disks in the Orion molecular clouds with a resolution of $\sim 0\farcs1$ (corresponding to $\sim$ 40 au).
This study provided data on the radius and mass of these disks.
Since all BOPS protostars were originally observed at high resolution in the VANDAM survey, the combination of these datasets enables a comprehensive test of the magnetic braking model and allows for a more effective linkage of magnetic and turbulent properties to the characteristics of protostellar disks and multiplicity.

This paper builds upon the first results of BOPS \citep{huang2024magnetic} and is organized as follows: Section \ref{sec:obs} introduces the observations and data reduction processes.
Section \ref{sec:analysis} presents the methods for the calculation and estimation of physical parameters, the velocity dispersion, the angle dispersion of the {\em B}-field, and the fragmentation level.
A detailed discussion of the correlation between these parameters is provided in Section \ref{sec:Dis}.
Finally, the conclusions are summarized in Section \ref{sec:Con}.

\section{Observations} \label{sec:obs}

The 870 $\mu$m (Band 7) dust polarization observations were taken with ALMA, and the spatial resolution and the Largest Angular Scale (LAS) are $\sim$1\arcsec and $\sim$8\arcsec, corresponding to a linear resolution of $\sim$400 and $\sim$3200 au at a distance of $\sim$400 pc (see \citealt{huang2024magnetic} for further details on the observation setup).

The dust continuum images were produced using the \texttt{CASA} \citep{casa2022} task \texttt{tclean} with a Briggs weighting parameter of robust = 0.5.
The images were iteratively improved through three rounds of phase-only self-calibration, using the Stokes \textit{I} image as a model.
The solution intervals for the first, second, and third iterations were set to 600, 30, and 10 seconds, respectively.
The final Stokes \textit{I}, \textit{Q}, and \textit{U} continuum maps were generated independently using line-free and self-calibrated data of all spectral windows.
The noise level in the final Stokes \textit{I} dust map is $\sigma_{I}\sim0.1$~mJy~beam$^{-1}$, while the rms noise level in the Stokes \textit{Q} and \textit{U} dust maps is $\sigma_{QU}\sim0.07$~mJy~beam$^{-1}$.
This difference arises from the total intensity emission (Stokes \textit{I}) image being more dynamic-range limited than the polarized emission (Stokes \textit{Q} and \textit{U}).
The debiased polarized intensity $P$, the fractional polarization $P_{\rm frac}$, and the polarization position angle $\theta$ were calculated using $P=(Q^{2}+U^{2}-\sigma_{QU}^{2})^{1/2}$ \citep{vaillancourt2006placing}, $P_{\text{frac}} = P/I$, and $\theta=0.5\arctan(U/Q)$, respectively.

For the C$^{17}$O (3--2) line data, we used the \texttt{CASA} \texttt{uvcontsub} task to perform continuum fitting and subtraction in the $u$-$v$ plane.
We then utilized the \texttt{tclean} task to produce Stokes \textit{I} image cubes using self-calibrated, continuum-subtracted data for each spectral window.
The C$^{17}$O (3--2) line, optically thin in dense regions \citep[with spectral resolution of $\sim$ 0.87 km s$^{-1}$, ][]{huang2024magnetic}, was employed to trace the velocity field of the envelope.

\section{Analysis} \label{sec:analysis}

The first results of the BOPS have revealed the dust polarized continuum emission for 61 protostars \citep{huang2024magnetic}, and some parameters are listed in Appendix \ref{app:A}.
In this work, we use the classification criteria based on the {\em B}-field/polarization structure because the velocity gradients in a large fraction of the sample are unresolved.
Then the 61 protostars are classified into 5 types: ``standard hourglass", ``rotated hourglass", ``spiral", ``others" (including ``complex" and ``not enough data"), and ``self-scattering", as shown in Appendix \ref{app:B}.
The following sections presents the estimation of physical envelope parameters, velocity dispersion, the angle dispersion of \textit{B}-field, and fragmentation level for all BOPS protostars.
These properties will then be compared to the disk properties to understand which factors could potentially influence the envelope fragmentation and whether the initial misalignment between the \textit{B}-field and outflows could result in reduced magnetic braking during disk formation.

\subsection{Physical Envelope Properties}\label{subsec:envelope}

We use the 0.87 mm continuum emission to derive the dust mass $M_{\rm dust}$ of the envelope within the inner 6$^{\prime\prime}$ (corresponding to a scale of 2400 au, with a radius of 1200 au).
This region covers at least $\sim$ 5$\sigma$ of the intensity detection in Stokes \textit{I} map for all protostars. 
The dust envelope mass is estimated under the assumption of optically thin dust emission, using the following expression:
\begin{eqnarray}\label{eq1}
M_{\rm dust} = \frac{S_{\nu}d^{2}}{\kappa_{\nu} B_{\nu}(T_{\rm dust})}
\end{eqnarray}
where $d$ is the distance, $\nu$ is the frequency corresponding to the observed wavelength, $B_{\nu}(T_{\rm dust})$ is the Planck function at the dust temperature of $T_{\rm dust}$, and $\kappa_{\nu}$ is the dust opacity.
We adopt $\kappa_{870 \mu{\rm m}}\approx1.84\ {\rm cm}^{2}\ {\rm g}^{-1}$, which is appropriate for dusty particles with ice mantles at densities of approximately $10^{6}$ cm$^{-3}$, as reported by \cite{ossenkopf1994dust}.

\begin{longtable}{lrcccccccccrcc}
\caption{\label{Tab:paras} Envelope and disk parameters for the BOPS sample\tablenotemark{$\dagger$}}\\
\hline
\hline
\multirow{2}*{Name} & $L_{\rm bol}$ & $S_{\nu}$ & $\overline{T}_{\rm env}$ & $T_{\rm c}$ & $M_{\rm env}$ & $N_{\rm gas}$ & $n_{\rm gas}$ & $\sigma_{\rm nth}$ & $\delta \phi$ & $N_{F}$  & $N_{B}$ \\
~ & (L$_{\odot}$) & (Jy) & (K) & (K) & (M$_{\odot}$) & (cm$^{-2}$) & (cm$^{-3}$) & (km s$^{-1}$) & ($\arcdeg$) & ~ & ~  \\
\hline
\endfirsthead
\caption{Envelope and disk parameters for the BOPS sample\tablenotemark{$\dagger$}}\\
\hline\hline
\multirow{2}*{Name} & $L_{\rm bol}$ & $S_{\nu}$ & $\overline{T}_{\rm env}$ & $T_{\rm c}$ & $M_{\rm env}$ & $N_{\rm gas}$ & $n_{\rm gas}$ & $\sigma_{\rm nth}$ & $\delta \phi$ & $N_{F}$  & $N_{B}$ \\
~ & (L$_{\odot}$) & (Jy) & (K) & (K) & (M$_{\odot}$) & (cm$^{-2}$) & (cm$^{-3}$) & (km s$^{-1}$) & ($\arcdeg$) & ~ & ~  \\
\hline
\endhead
\hline
\endfoot
\textbf{Standard}&&&&&&&&&&& \\
\hline
HOPS-11 & 9.00 & 0.40 & 26.1 & 17.7 $\pm$ 0.9 & 0.26 & 1.3E+23 & 5.4E+06 & 0.67 $\pm$ 0.20 & / & 1.00 & 6 \\
HOPS-87 & 36.49 & 3.89 & 37.0 & 21.2 $\pm$ 1.9 & 1.46 & 7.3E+23 & 3.0E+07 & 0.43 $\pm$ 0.13 & 17.6 $\pm$ 1.4 & 1.00 & 86  \\
HOPS-359 & 10.00 & 0.91 & 26.8 & 17.4 $\pm$ 2.0 & 0.62 & 3.1E+23 & 1.3E+07 & 0.45 $\pm$ 0.11  & 29.2 $\pm$ 2.3 & 1.00 & 80 \\
HOPS-395 & 0.50 & 0.58 & 20.2 & 14.3 $\pm$ 1.1 & 0.49 & 2.4E+23 & 1.0E+07 & 0.40 $\pm$ 0.12 & 15.0 $\pm$ 2.0 & 2.00 & 29 \\
HOPS-400 & 2.94  & 1.18 & 21.5 & 15.8 $\pm$ 0.7 & 1.01 & 5.0E+23 & 2.1E+07 & 0.45 $\pm$ 0.22 & 18.2 $\pm$ 1.9 & 2.00 & 48 \\
HOPS-407 & 0.71  & 0.61 & 20.2 & 17.8 $\pm$ 2.0 & 0.59 & 2.9E+23 & 1.2E+07 & 0.44 $\pm$ 0.12 & 10.5 $\pm$ 1.0 & 1.50 & 53 \\
OMC1N-8-N & / & 0.63 & / & 37.5 $\pm$ 10.5 & 0.31 & 1.6E+23 & 6.5E+06 & 0.43 $\pm$ 0.31 & 20.6 $\pm$ 2.1 & 2.00 & 51 \\
\hline
\textbf{Rotated} &&&&&&&&&&&\\
\hline
HH270IRS & 7.70 & 0.49 & 25.1 & 16.7 $\pm$ 0.1 & 0.35 & 1.7E+23 & 7.2E+06 & 0.45 $\pm$ 0.15 & 33.8 $\pm$ 4.1 & 2.00 & 35 \\
HOPS-78  & 8.93 & 0.81 & 26.0 & 22.6 $\pm$ 2.1 & 0.48 & 2.4E+23 & 9.9E+06 & 0.59 $\pm$ 0.27 & 40.2 $\pm$ 4.3 & 2.25 & 45 \\
HOPS-168 & 48.07 & 0.51 & 39.6 & 16.5 $\pm$ 1.4 & 0.17 & 8.4E+22 & 3.5E+06 & 0.64 $\pm$ 0.26 & 28.4 $\pm$ 3.9 & 2.25 & 28 \\
HOPS-169 & 3.91 & 0.67 & 22.2 & 19.2 $\pm$ 1.4 & 0.47 & 2.3E+23 & 9.7E+06 & 0.30 $\pm$ 0.18 & 32.9 $\pm$ 3.9 & 1.00 & 37 \\
HOPS-288 & 135.47 & 1.60 & 51.4 & 13.7 $\pm$ 1.0 & 0.43 & 2.1E+23 & 8.9E+06 & 1.34 $\pm$ 0.51 & 24.5 $\pm$ 2.9 & 3.00 & 37 \\
HOPS-310 & 13.83 & 0.58 & 29.1 & 17.0 $\pm$ 1.2 & 0.33 & 1.6E+23 & 6.8E+06 & 0.75 $\pm$ 0.15 & / & 1.00 & 9 \\
HOPS-317S & 4.90 & 2.33 & 22.9 & 14.7 $\pm$ 0.9 & 1.95 & 9.7E+23 & 4.0E+07 & 0.45 $\pm$ 0.16 & / & 1.25 & 15 \\
HOPS-370 & 360.86 & 1.08 & 65.6 & 25.1 $\pm$ 1.9 & 0.21 & 1.0E+23 & 4.3E+06 & 0.80 $\pm$ 0.32 & 33.8 $\pm$ 3.1 & 1.00 & 62 \\
HOPS-401 & 0.61 & 0.37 & 20.2 & 14.8 $\pm$ 1.0 & 0.37 & 1.9E+23 & 7.7E+06 & 0.69 $\pm$ 0.12 & / & 1.00 & 13 \\
HOPS-409 & 8.18 & 0.38 & 25.4 & 30.8 $\pm$ 7.9 & 0.23 & 1.2E+23 & 4.8E+06 & 0.83 $\pm$ 0.34 & 17.0 $\pm$ 2.5 & 1.25 & 25 \\
OMC1N-4-5-ES & / & 0.72 & / & 36.1 $\pm$ 8.9 & 0.28 & 1.4E+23 & 5.8E+06 & 0.63 $\pm$ 0.21 & 34.6 $\pm$ 4.1 & 2.25 & 66 \\
\hline
\textbf{Spiral} &&&&&&&&&&&\\
\hline
HOPS-96 & 6.19 & 0.44 & 23.8 & 25.8 $\pm$ 4.3 & 0.29 & 1.4E+23 & 6.0E+06 & 0.86 $\pm$ 0.15 & / & 1.00 & 11 \\
HOPS-124 & 58.29 & 1.43 & 41.6 & 20.3 $\pm$ 0.6 & 0.48 & 2.4E+23 & 9.9E+06 & 1.20 $\pm$ 0.43 & / & 2.25 & 8 \\
HOPS-182 & 71.12 & 1.43 & 43.7 & 16.9 $\pm$ 1.2 & 0.42 & 2.1E+23 & 8.7E+06 & 1.44 $\pm$ 0.54 & 46.4 $\pm$ 4.6 & 3.50 & 51 & \\
HOPS-303 & 1.49 & 0.42 & 20.6 & 21.4 $\pm$ 1.0 & 0.37 & 1.9E+23 & 7.8E+06 & 0.56 $\pm$ 0.21 & / & 1.00 & 13 \\
HOPS-361N & 478.99 & 1.70 & 70.4 & 18.6 $\pm$ 3.1 & 0.36 & 1.8E+23 & 7.5E+06 & 0.69 $\pm$ 0.33 & 44.2 $\pm$ 3.4 & 3.25 & 86 \\
HOPS-361S & 478.99 & 1.10 & 70.4 & 18.8 $\pm$ 3.2 & 0.23 & 1.2E+23 & 4.8E+06 & 0.66 $\pm$ 0.23 & 43.1 $\pm$ 3.8 & 3.50 & 65 \\
HOPS-384 & 1477.95 & 1.98 & 93.3 & 51.2 $\pm$ 15.6 & 0.28 & 1.4E+23 & 5.8E+06 & 0.50 $\pm$ 0.33 & 30.8 $\pm$ 2.2 & 2.50 & 98 \\
HOPS-403 & 4.14 & 0.89 & 22.3 & 19.0 $\pm$ 3.0 & 0.77 & 3.8E+23 & 1.6E+07 & 0.80 $\pm$ 0.03 & / & 2.00 & 10 \\
HOPS-404 & 0.95 & 0.53 & 20.4 & 16.0 $\pm$ 0.7 & 0.53 & 2.6E+23 & 1.1E+07 & 0.37 $\pm$ 0.23 & / & 1.00 & 11 \\
\hline
\textbf{Others} &&&&&&&&&&&\\
\hline
HOPS-12W & 7.31 & 0.48 & 24.8 & 18.2 $\pm$ 1.4 & 0.30 & 1.5E+23 & 6.1E+06 & 0.57 $\pm$ 0.18 & 39.7 $\pm$ 5.0 & 2.00 & 23 \\
HOPS-87S & / & 0.07 & / & 21.2 $\pm$ 1.8 & 0.06 & 2.8E+22 & 1.2E+06 & 0.76 $\pm$ 0.18 & / & 1.00 & 0 \\
HOPS-88 & 15.81 & 0.70 & 30.0 & 21.3 $\pm$ 2.1 & 0.34 & 1.7E+23 & 7.1E+06 & 0.83 $\pm$ 0.25 & 35.2 $\pm$ 4.3 & 1.00 & 35 \\
HOPS-164 & 0.58 & 0.24 & 20.2 & 15.9 $\pm$ 1.2 & 0.19 & 9.7E+22 & 4.0E+06 & 0.63 $\pm$ 0.25 & / & 1.00 & 7 \\
HOPS-224 & 2.99 & 0.36 & 21.5 & 13.6 $\pm$ 0.5 & 0.34 & 1.7E+23 & 7.1E+06 & 0.59 $\pm$ 0.26 & / & 1.00 & 6 \\
HOPS-250 & 6.79 & 0.20 & 24.3 & 14.4 $\pm$ 0.9 & 0.16 & 7.8E+22 & 3.3E+06 & 0.70 $\pm$ 0.24 & / & 1.00 & 5 \\
HOPS-317N & 4.76 & 0.11 & 22.9 & 14.7 $\pm$ 0.8 & 0.09 & 4.5E+22 & 1.9E+06 & 0.37 $\pm$ 0.19 & / & 1.25 & 5 \\
HOPS-325 & 6.20 & 0.28 & 23.9 & 24.6 $\pm$ 5.0 & 0.22 & 1.1E+23 & 4.6E+06 & 0.44 $\pm$ 0.04 & / & 1.00 & 12 \\
HOPS-341 & 2.07 & 0.24 & 21.0 & 16.0 $\pm$ 0.8 & 0.23 & 1.1E+23 & 4.8E+06 & 0.19 $\pm$ 0.12 & / & 1.25 & 12 \\
HOPS-373E & 5.32 & 0.53 & 23.2 & 17.8 $\pm$ 1.5 & 0.43 & 2.2E+23 & 9.0E+06 & 0.28 $\pm$ 0.18 & 28.9 $\pm$ 3.6 & 1.50 & 33 \\
HOPS-373W & 5.32 & 0.64 & 23.2 & 17.8 $\pm$ 1.5 & 0.52 & 2.6E+23 & 1.1E+07 & 0.50 $\pm$ 0.16 & / & 1.50 & 15 \\
HOPS-398 & 1.01 & 0.62 & 20.4 & 19.8 $\pm$ 2.2 & 0.55 & 2.8E+23 & 1.2E+07 & 0.50 $\pm$ 0.19 & 32.0 $\pm$ 5.0 & 1.00 & 20 \\
HOPS-399 & 6.34 & 1.79 & 24.0 & 20.8 $\pm$ 2.3 & 1.28 & 6.3E+23 & 2.6E+07 & 0.42 $\pm$ 0.15 & 46.4 $\pm$ 3.2 & 1.00 & 103 \\
HOPS-402 & 0.55 & 0.51 & 20.2 & 14.9 $\pm$ 1.1 & 0.51 & 2.5E+23 & 1.1E+07 & 0.66 $\pm$ 0.17 & / & 1.00 & 10 \\
HOPS-408 & 0.52 & 0.31 & 20.2 & 13.9 $\pm$ 0.6 & 0.27 & 1.3E+23 & 5.6E+06 & / & / & 1.00 & 8 \\
OMC1N-4-5-EN & / & 0.47 & / & 36.0 $\pm$ 8.7 & 0.18 & 9.0E+22 & 3.8E+06 & 0.65 $\pm$ 0.26 & 16.4 $\pm$ 1.4 & 1.50 & 66 \\
OMC1N-6-7 & / & 0.95 & / & 35.5 $\pm$ 8.8 & 0.38 & 1.9E+23 & 7.8E+06 & 0.80 $\pm$ 0.58 & 39.2 $\pm$ 3.8 & 1.25 & 53 \\
OMC1N-8-S & / & 0.16 & / & 37.0 $\pm$ 10.3 & 0.06 & 3.0E+22 & 1.3E+06 & 0.48 $\pm$ 0.21 & / & 1.50 & 0 \\
\hline
\textbf{Scattering} &&&&&&&&&&&\\
\hline
HH212M & 14.00 & 0.52 & 29.1 & 17.2 $\pm$ 0.7 & 0.32 & 1.6E+23 & 6.6E+06 & 0.39 $\pm$ 0.15 & / & 1.00 & / \\
HOPS-10 & 3.33 & 0.17 & 21.8 & 17.6 $\pm$ 0.6 & 0.12 & 6.2E+22 & 2.6E+06 & 0.49 $\pm$ 0.12 & / & 1.00 & / \\
HOPS-12E & 7.31 & 0.33 & 24.8 & 18.2 $\pm$ 1.4 & 0.21 & 1.0E+23 & 4.3E+06 & 0.51 $\pm$ 0.23 & / & 2.50 & / \\
HOPS-50 & 4.20 & 0.32 & 22.4 & 23.8 $\pm$ 1.9 & 0.23 & 1.2E+23 & 4.8E+06 & 0.64 $\pm$ 0.29 & / & 1.00 & / \\
HOPS-53 & 26.42 & 0.37 & 34.1 & 36.8 $\pm$ 4.0 & 0.15 & 7.6E+22 & 3.2E+06 & 0.36 $\pm$ 0.15 & / & 1.00 & / \\
HOPS-60 & 21.93 & 0.55 & 32.6 & 27.5 $\pm$ 4.7 & 0.24 & 1.2E+23 & 5.0E+06 & 0.92 $\pm$ 0.30 & / & 1.00 & / \\
HOPS-81 & 1.24 & 0.14 & 20.5 & 22.1 $\pm$ 1.6 & 0.11 & 5.6E+22 & 2.4E+06 & 0.56 $\pm$ 0.20 & / & 1.00 & / \\
HOPS-84 & 49.11 & 0.59 & 39.8 & 21.7 $\pm$ 1.7 & 0.20 & 1.1E+23 & 4.2E+06 & 0.39 $\pm$ 0.23 & / & 2.00 & / \\
HOPS-153 & 4.43 & 0.22 & 22.5 & 15.1 $\pm$ 1.1 & 0.16 & 7.8E+22 & 3.3E+06 & 0.87 $\pm$ 0.27 & / & 1.00 & / \\
HOPS-203N & 20.44 & 0.42 & 32.0 & 16.0 $\pm$ 1.1 & 0.18 & 9.0E+22 & 3.8E+06 & 0.50 $\pm$ 0.19 & / & 2.50 & / \\
HOPS-203S & 20.44 & 0.38 & 32.0 & 16.0 $\pm$ 1.1 & 0.16 & 8.1E+22 & 3.4E+06 & 0.43 $\pm$ 0.20 & / & 2.00 & / \\
HOPS-247 & 3.09 & 0.37 & 21.6 & 13.4 $\pm$ 0.7 & 0.34 & 1.7E+23 & 7.1E+06 & 0.74 $\pm$ 0.30 & / & 1.00 & / \\
HOPS-340 & 1.85 & 0.21 & 20.8 & 16.0 $\pm$ 0.8 & 0.20 & 1.0E+23 & 4.2E+06 & 0.36 $\pm$ 0.22 & / & 1.25 & / \\
HOPS-354 & 6.57 & 0.34 & 24.2 & 16.2 $\pm$ 0.1 & 0.18 & 9.1E+22 & 3.8E+06 & 0.54 $\pm$ 0.23 & / & 1.00 & / \\
HOPS-358 & 24.96 & 0.44 & 33.7 & 14.7 $\pm$ 1.0 & 0.22 & 1.1E+23 & 4.6E+06 & 1.34 $\pm$ 0.56 & / & 1.00 & / \\
HOPS-383 & 7.83 & 0.22 & 25.2 & 23.4 $\pm$ 3.0 & 0.14 & 6.9E+22 & 2.9E+06 & 0.70 $\pm$ 0.34 & / & 1.00 & / \\
\hline 
\multicolumn{14}{p{10cm}}{\footnotesize \parbox{17.4cm}{%
Column 2 to 12 present the bolometric luminosity ($L_{\rm{bol}}$), total continuum flux density at 345 GHz ($S_{\nu}$), averaged envelope dust temperature ($\overline{T}_{\rm env}$), cloud dust temperature ($T_{\rm c}$), envelope gas mass ($M_{\rm env}$), gas column density ($N_{\rm gas}$), gas number density ($n_{\rm gas}$), non-thermal velocity dispersion ($\sigma_{\rm nth}$), {\em B}-field angle dispersion ($\delta \phi$), fragmentation level ($N_{F}$), and number of Nyquist-sampling {\em B}-field segments ($N_{B}$), respectively. 
``Standard'', ``Rotated'', and ``Spiral'' indicate protostars with different types of {\em B}-field structure of standard hourglass, rotated hourglass, and spiral configuration, respectively.
``Others'' means the {\em B}-field pattern is complex or  insufficient to identify its morphology.
``Scattering'' represents protostars with compact polarized emission, which is thought to be produced by self-scattering \citep{huang2024magnetic}.

\tablenotemark{$\dagger$} Typical uncertainties for the different parameters are: 10\% for $S_\nu$; 20--30\% for $\overline{T}_{\rm env}$; 30--40\% for $M_{\rm env}$, $N_{\rm gas}$, and $n_{\rm gas}$.
}
}
\end{longtable}

In the case of a dusty cloud heated by internal stellar radiation, the radial temperature profile can be expressed as a power law: $T_{\rm dust} \propto r^{-2/(4+\beta)}$ \citep{kenyon1993model}, where $\beta$ is the dust opacity spectral index.
\cite{girart2009ism} adopted a value of $\beta=1.5$ when studying low-mass protostars at 1000 au scales.
\cite{kwon2009opacity} suggested that $\beta$ should be around or less than 1 for Class 0 protostars.
\cite{tobin2020vla} fitted a temperature power-law index of -0.46 for hundreds-au-size disks within the Orion star-forming protostellar system, corresponding to $\beta \sim 0.35$, which is scaled by luminosity.
However, this profile shows a slight decline in the outer 100 au regions due to backheating from the surrounding envelope \citep{tobin2020vla}, it is subsequently recommended to refine the index to -0.50 \citep{sheehan2022vandam}.
For this study, we use the temperature model proposed by \cite{tobin2020vla} for Orion star-forming regions, with $\beta$ set to 1 (corresponding to the index of $\sim-0.40$) for BOPS protostellar envelopes (most of them are in Class 0 stage).
Therefore, the dust temperature $T_{\rm dust}$ can be expressed as:
\begin{eqnarray}\label{eq2}
T_{\rm dust} = T_{0}\biggl(\frac{L_{\rm bol}}{L_{\odot}}\biggr)^{0.25}\biggl(\frac{r}{50~{\rm au}}\biggr)^{-0.40}
\end{eqnarray}
where $T_{0}=43$ K is the average temperature for a $\sim$ 1 $L_{\odot}$ protostar at a radius of $\sim$50 au \citep{whitney2003radiative, tobin2013resolved, tobin2020vla},  $r$ is the distance from the protostar, and $L_{\rm bol}$ is the bolometric luminosity taken from \cite{furlan2016herschel}.
Given the clear detection of C$^{17}$O emission in our sample, it is inferred that the dust temperature cannot fall below 20 K, as CO would otherwise be frozen onto the dust grains, making it undetectable \citep[e.g.,][]{jorgensen2015molecule}. 
The Equation \ref{eq2} may yield very low dust temperatures (below 20 K) at the edge of the selected regions, especially when the bolometric luminosity $L_{\rm bol}$ is very low.
In such cases, we use 20 K as the minimum dust temperature for pixels where the calculated temperature falls below this value.
Excluding the extremely high temperature in the central pixel, we calculate the average envelope dust temperature $\overline{T}_{\rm env}$ within inner regions with a radius of $R = 1200$ au.

\begin{figure*}
\centering
\includegraphics[clip=true,trim=0.2cm 0.28cm 0.2cm 0.2cm,width=0.5 \textwidth]{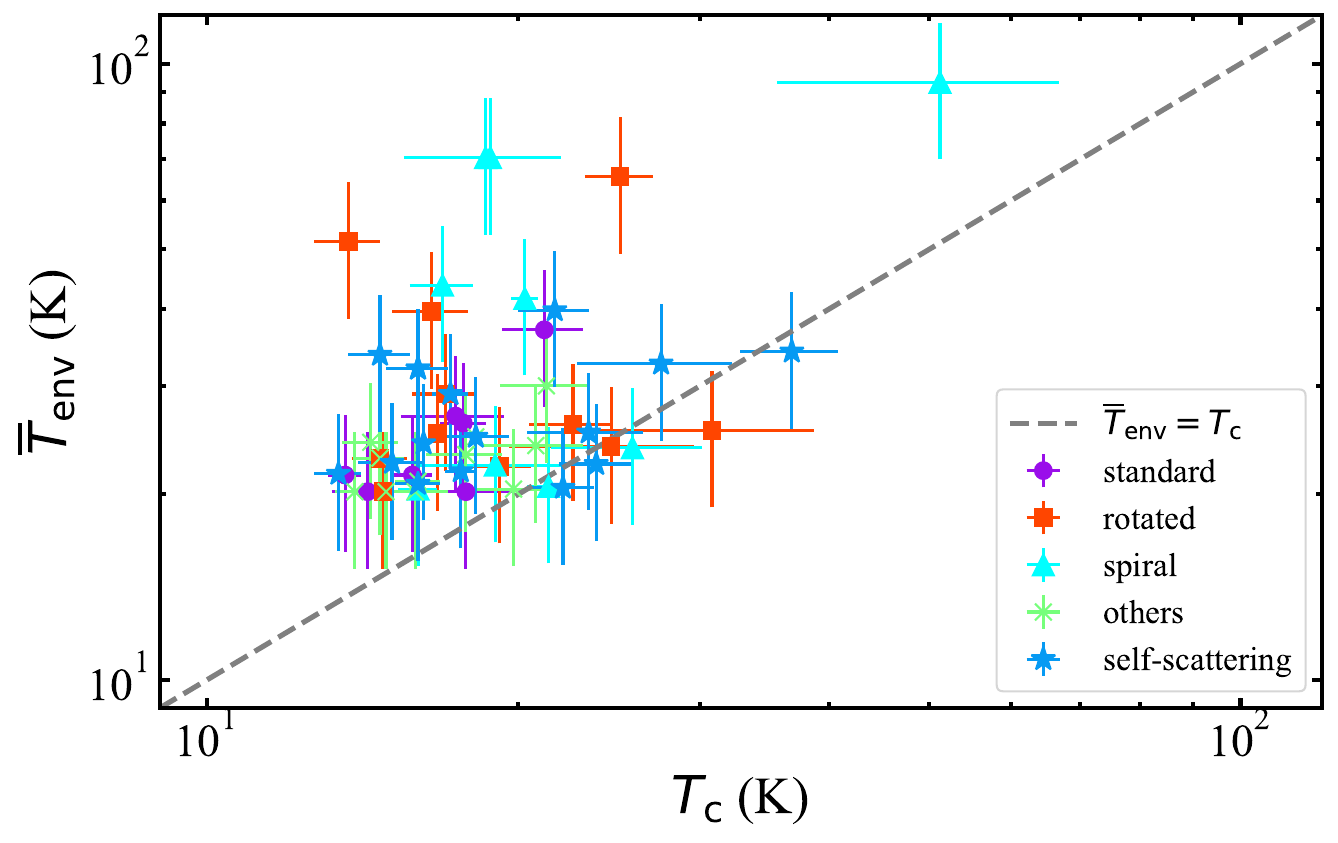}
\caption{The difference between derived dust temperature within envelope $\overline{T}_{\rm env}$ and cloud dust temperature $T_{\rm c}$ obtained from \cite{lombardi2014temperature}.
Grey dashed line indicates when $\overline{T}_{\rm env} = T_{\rm c}$.} 
\label{fig:tem}
\end{figure*}

\cite{lombardi2014temperature} used Planck and Herschel observations of Orion to construct and fit SEDs, and further to obtain a map of dust temperature.
As shown in Figure \ref{fig:tem}, comparing the cloud dust temperature $T_{\rm c}$ \citep{lombardi2014temperature} with the envelope dust temperature $\overline{T}_{\rm env}$ (this work, derived using our method described above), we find $\overline{T}_{\rm env}$ in most BOPS protostars is larger than $T_{\rm c}$, while 10 protostars have values of $\overline{T}_{\rm env}$ comparable to the dust temperature of the surrounding molecular cloud ($\vert\overline{T}_{\rm env}-T_{\rm c}\vert \lesssim 3$, the value of 3 is the half of the mean error for $\vert\overline{T}_{\rm env}-T_{\rm c}\vert$, see the notemark of Table \ref{Tab:paras} or Section \ref{sec:Dis} for error discussions).
Given that the scales of molecular clouds in question are on the parsec level, which is much larger than the envelope scales we focus on, it is expected that $\overline{T}_{\rm env}$ should be comparable to or higher than $T_{\rm c}$.
However, HOPS-409 shows that the value of $\overline{T}_{\rm env}$ is lower than that of $T_{\rm c}$, probably due to the cloud providing shielding from the interstellar radiation field.
We note that 6 protostars lack information of $\overline{T}_{\rm env}$, in this case, $T_{\rm c}$ is employed to estimate the physical properties of the envelope instead.

The gas column density $N_{\rm gas}$ is estimated using the relation:
\begin{eqnarray}\label{eq3}
N_{\rm gas} = \frac{M_{\rm env}}{\pi R^{2}\mu_{\rm gas}m_{\rm H}}
\end{eqnarray}
where $R=1200$ au is the radius, $M_{\rm env} = \eta M_{\rm dust}$ is the envelope gas mass, $\eta=100$ is the gas-to-dust mass ratio, $\mu_{\rm gas} = 2.37$ is the mean molecular weight per free particle \citep{kauffmann2008mambo}, and $m_{\rm H}$ is the mass of the hydrogen atom.
For a spherical region, the average gas number density can be calculated as
\begin{eqnarray}\label{eq4}
n_{\rm gas} = \frac{3N_{\rm gas}}{4R}
\end{eqnarray}

\begin{figure*}
\centering
\includegraphics[clip=true,trim=0.2cm 0cm 0.2cm 0.2cm,width=0.4 \textwidth]{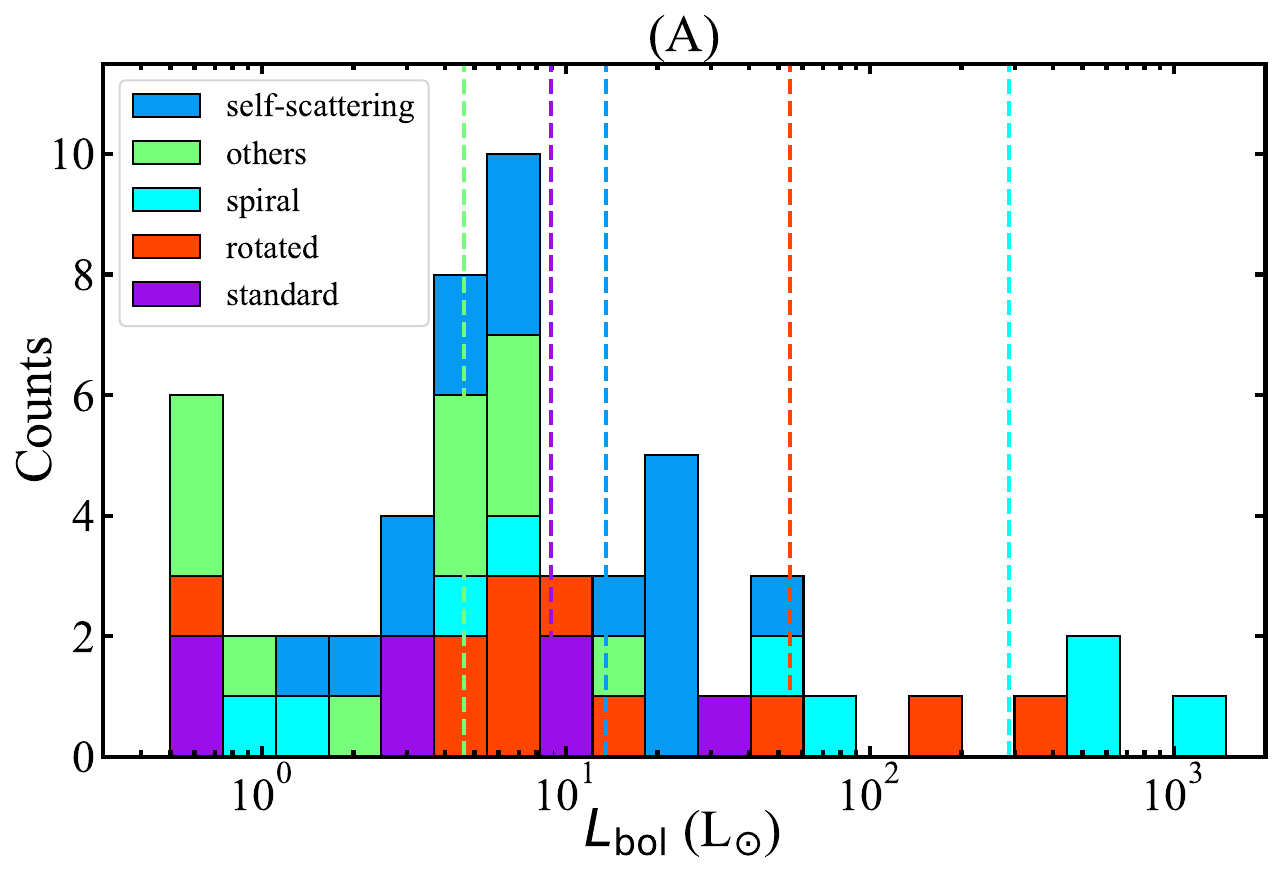}
~~~~~~~~~~
\includegraphics[clip=true,trim=0.2cm 0.2cm 0.2cm 0.2cm,width=0.405 \textwidth]{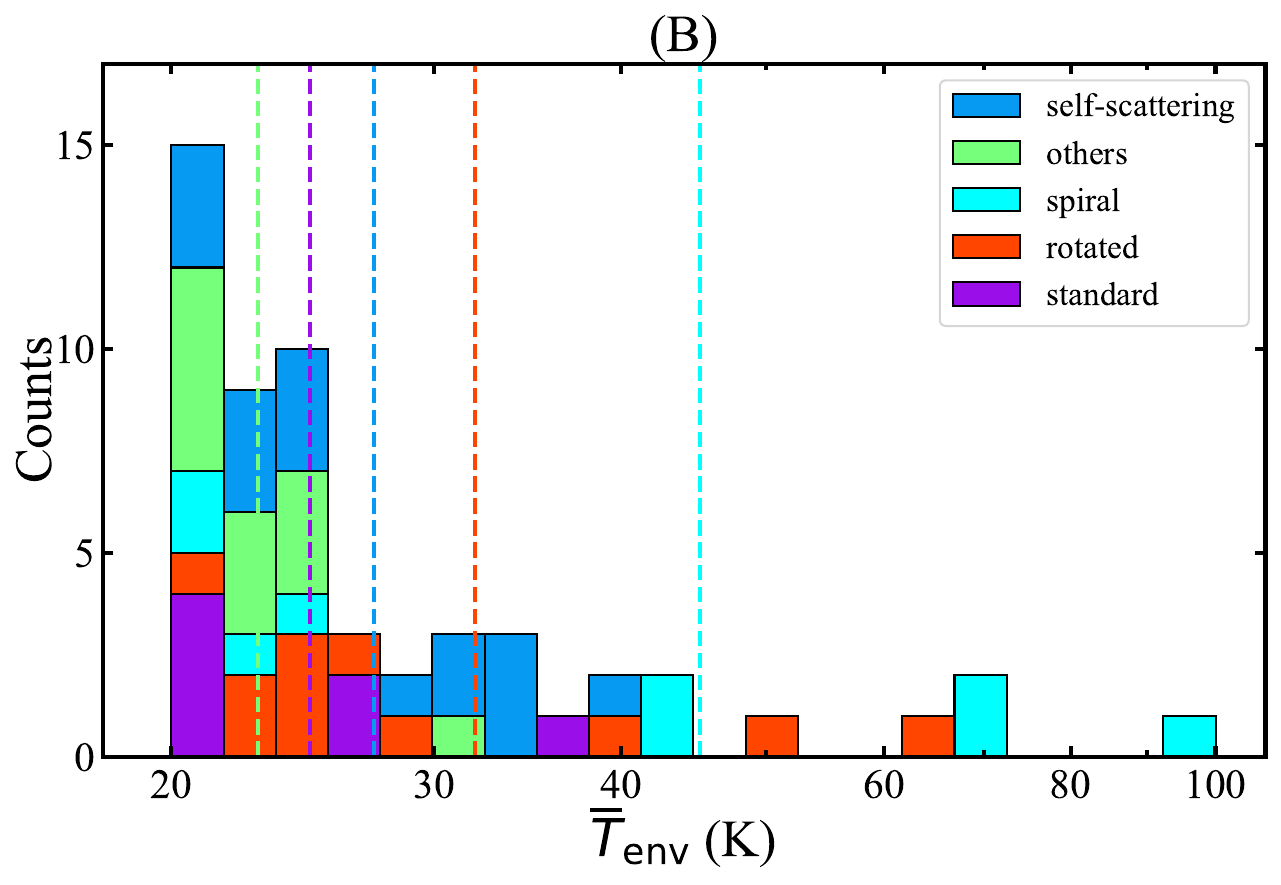}
~\\
~\\
\includegraphics[clip=true,trim=0.2cm 0.2cm 0.2cm 0.2cm,width=0.4 \textwidth]{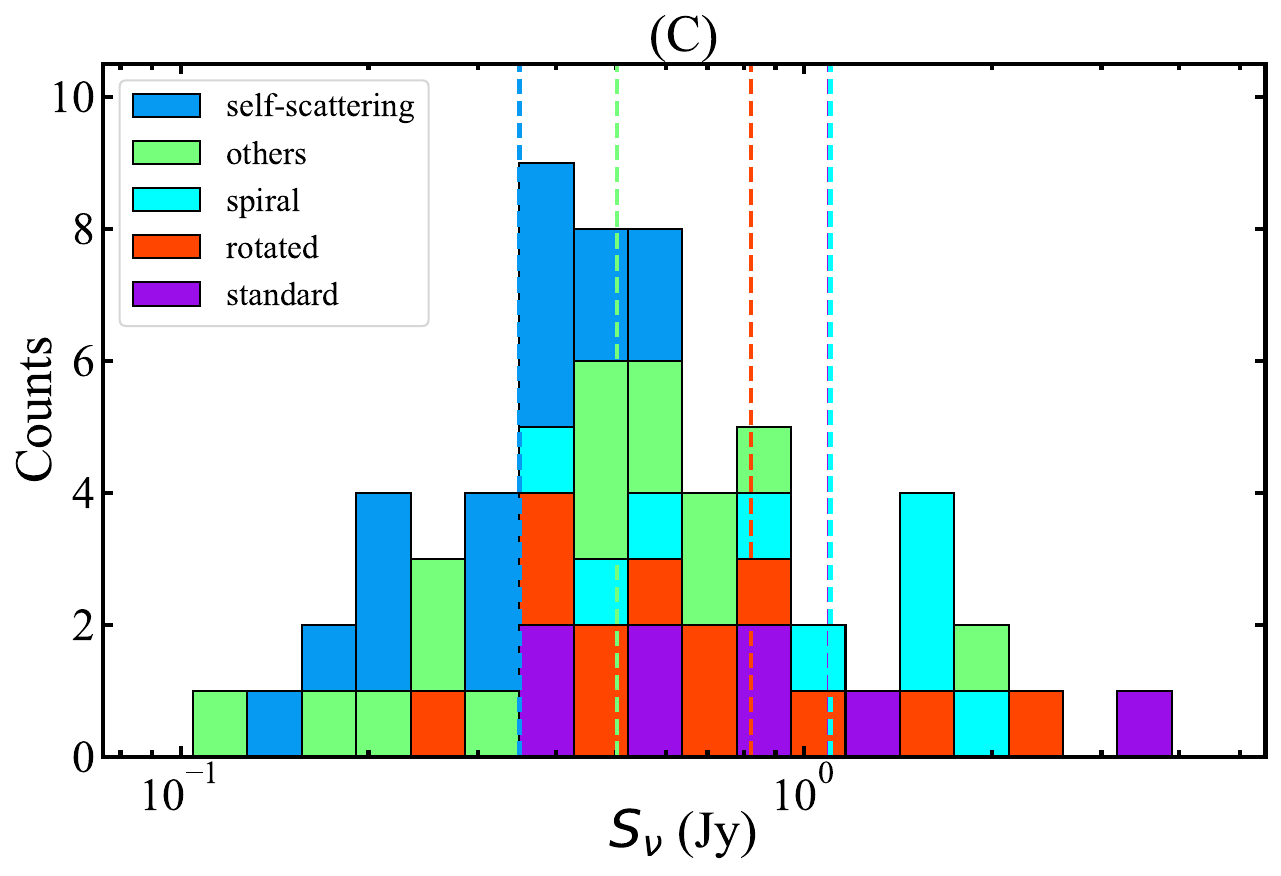}
~~~~~~~~~~
\includegraphics[clip=true,trim=0.2cm 0.2cm 0.2cm 0.2cm,width=0.4 \textwidth]{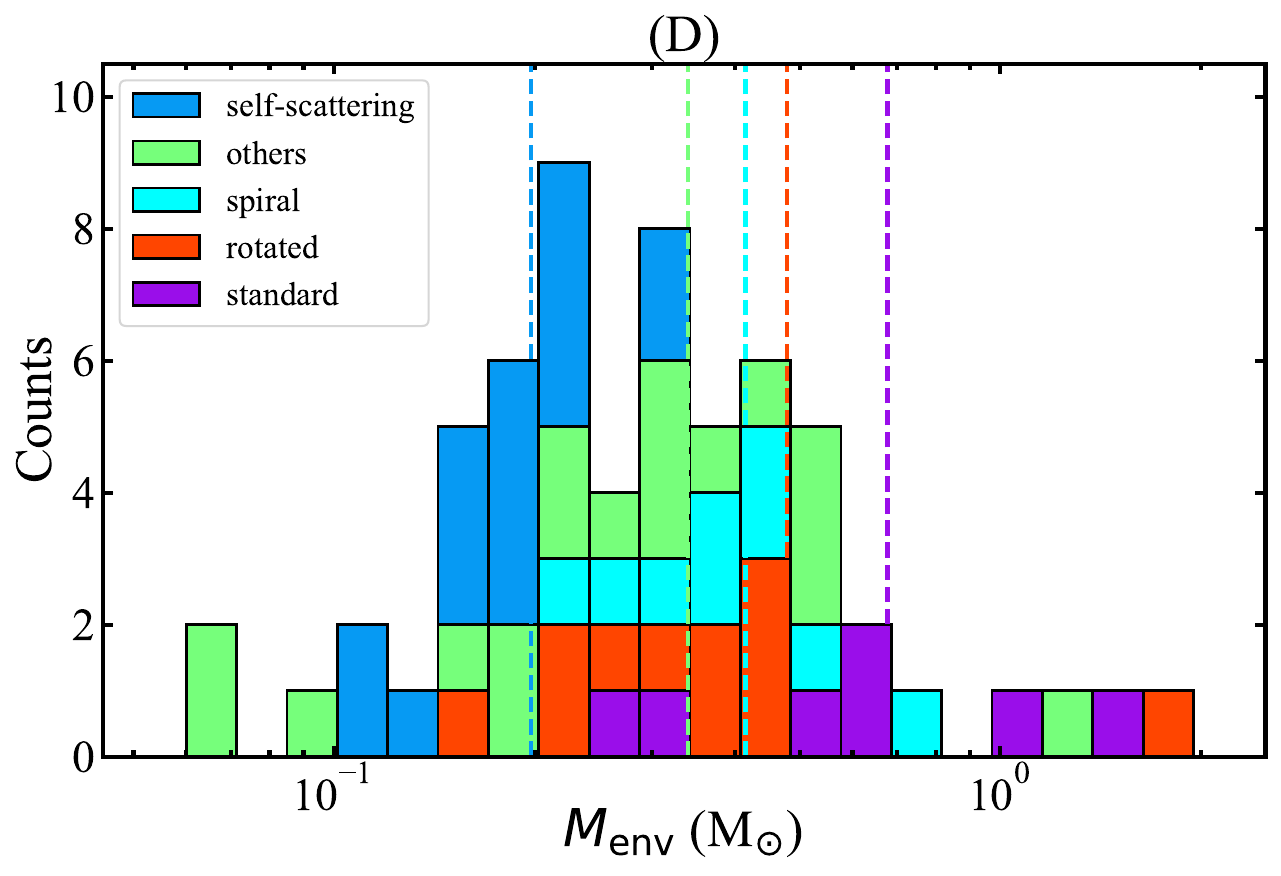}
~\\
~\\
\includegraphics[clip=true,trim=0.2cm 0.2cm 0.2cm 0.2cm,width=0.4 \textwidth]{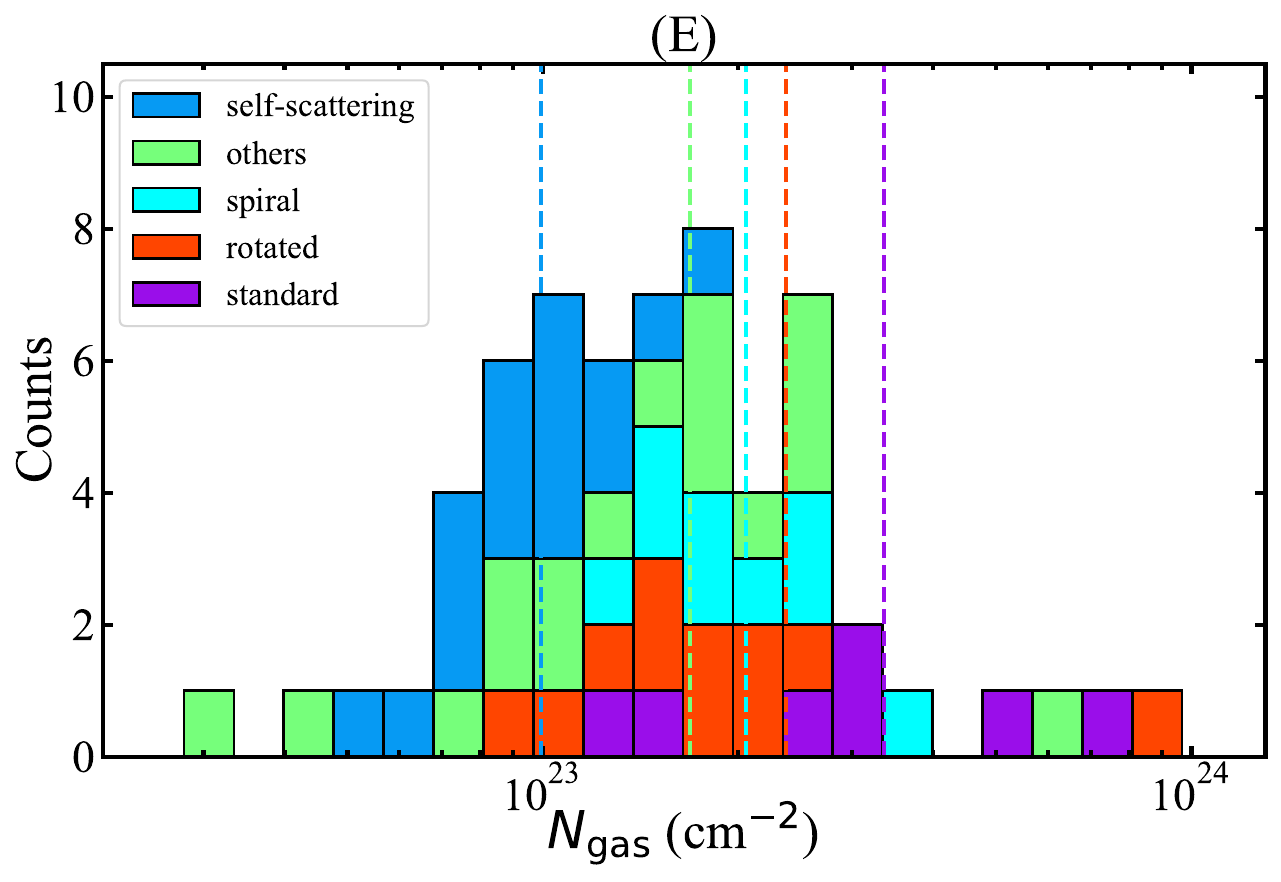}
~~~~~~~~~~
\includegraphics[clip=true,trim=0.2cm 0.2cm 0.2cm 0.2cm,width=0.4 \textwidth]{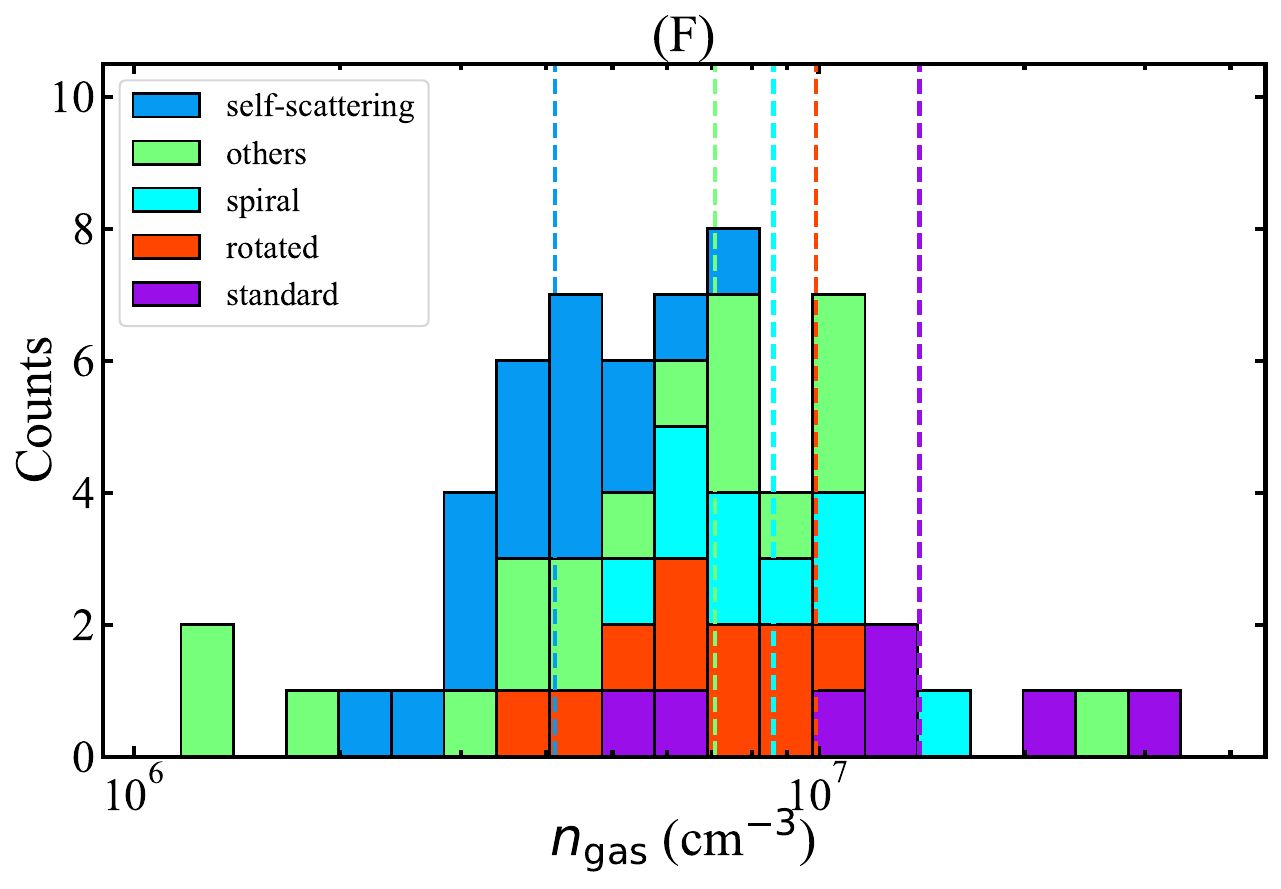}
~\\
~\\
\includegraphics[clip=true,trim=0.2cm 0.25cm 0.2cm 0.2cm,width=0.4 \textwidth]{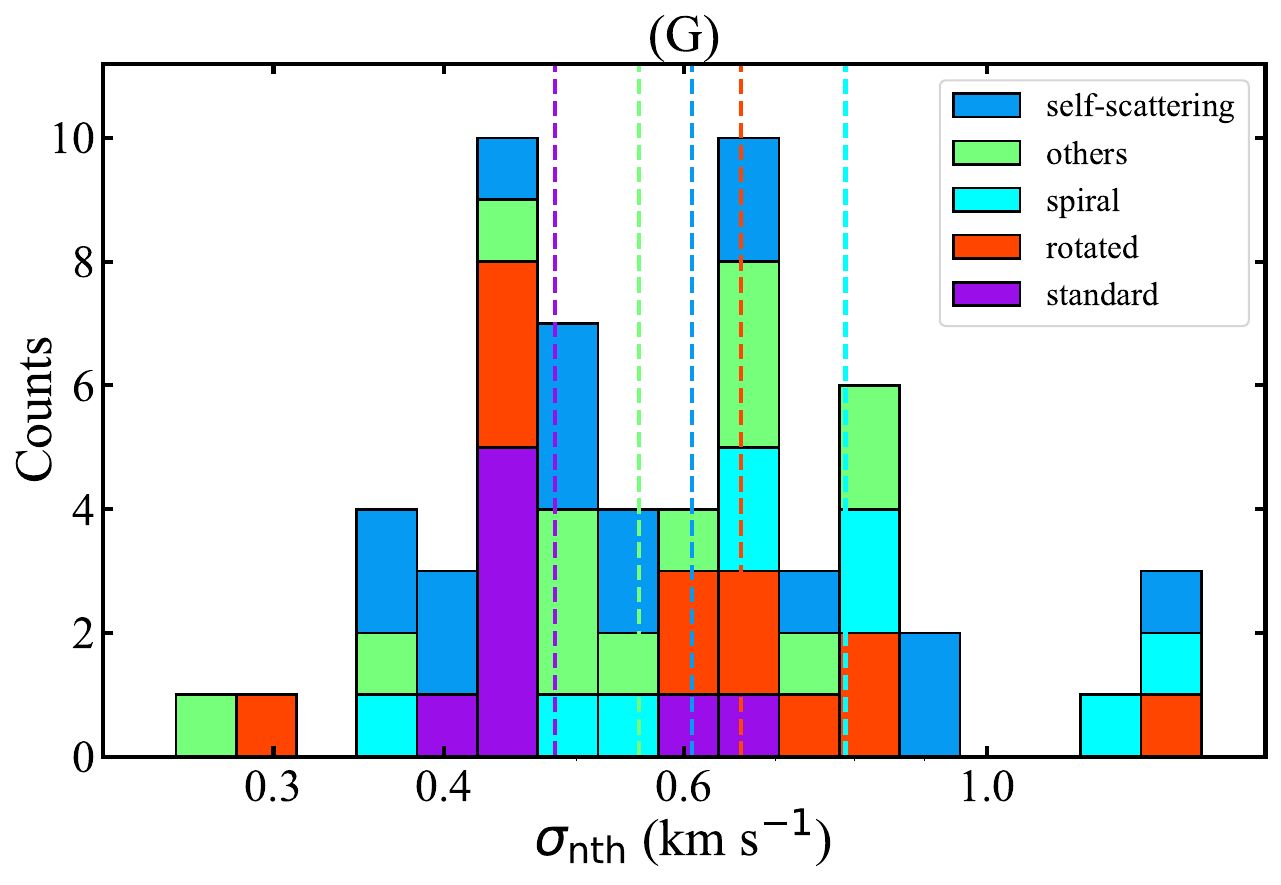}
~~~~~~~~~~
\includegraphics[clip=true,trim=0.2cm 0cm 0.2cm 0.2cm,width=0.4 \textwidth]{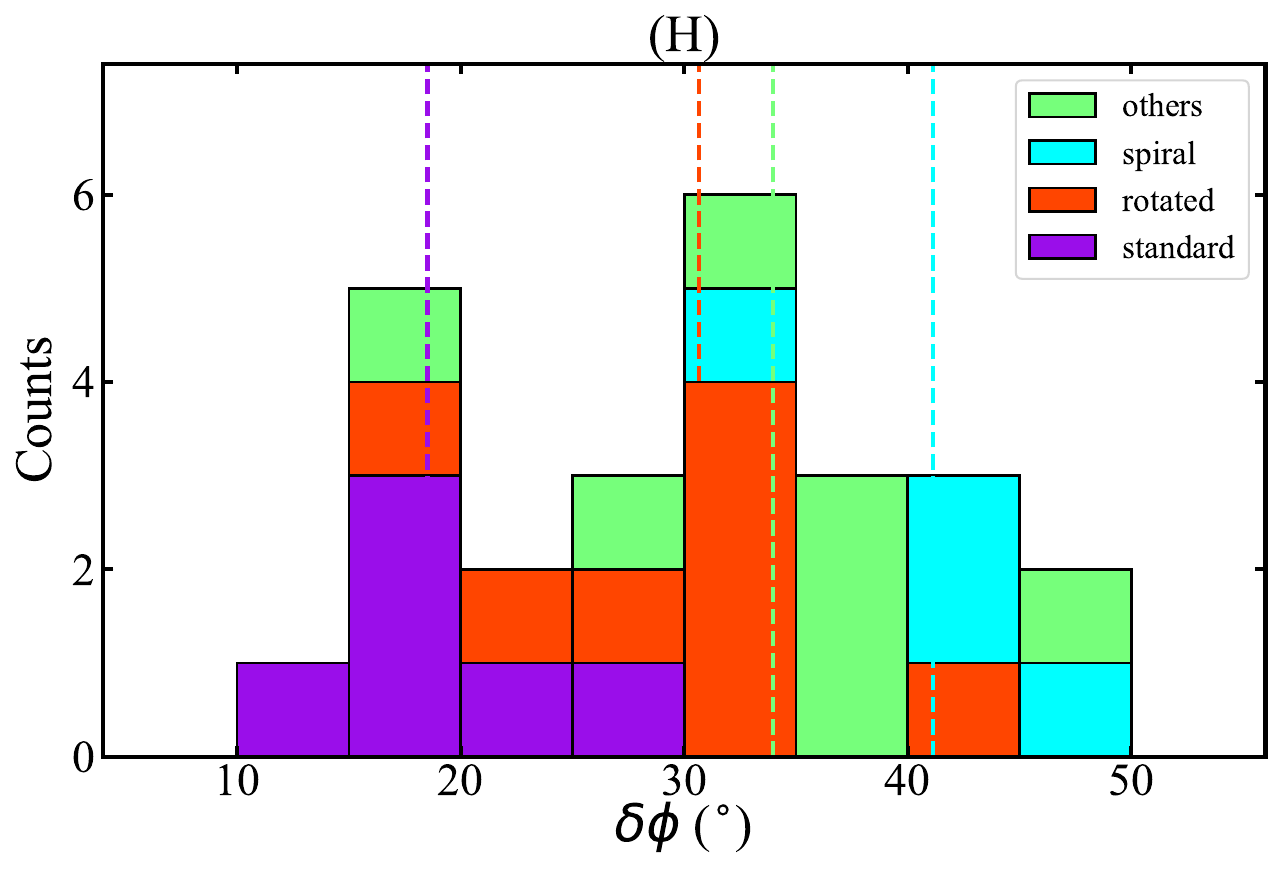}
\caption{The stacked histogram of physical envelope parameters for protostars with different {\em B}-field structure.
Panel A to H: bolometric luminosity $L_{\rm bol}$, mean dust envelope temperature $\overline{T}_{\rm env}$, total flux density $S_{\nu}$, envelope gas mass $M_{\rm env}$, gas column density $N_{\rm gas}$, gas number density $n_{\rm gas}$, non-thermal velocity dispersion $\sigma_{\rm nth}$ and angle dispersion of {\em B}-field $\delta\phi$, respectively.
Dashed lines indicate the mean values of the distribution for protostars with different types of {\em B}-field structure.} 
\label{fig:distribution}
\end{figure*}

Table \ref{Tab:paras} summarizes the derived parameters for each protostar, including information from the literature. 
Column 2 to 8 of Table \ref{Tab:paras} provide the following data: bolometric luminosity $L_{\rm bol}$, total flux density $S_{\nu}$, mean envelope dust temperature $\overline{T}_{\rm env}$, cloud dust temperature $T_{\rm c}$, envelope gas mass $M_{\rm env}$, gas column density $N_{\rm gas}$, and gas number density $n_{\rm gas}$.
Figure \ref{fig:distribution}(A--B) illustrate the distributions of bolometric luminosity and envelope dust temperature for 55 protostars (the remaining 6 lack luminosity information), while Figure \ref{fig:distribution}(C--F) present the distributions for parameters of flux intensity, envelope mass, gas column density and number density for 61 targets.

\subsection{Line Width}

Assuming that non-thermal motions on envelope scales arise from turbulence, we use the line width of the C$^{17}$O (3--2) to estimate the turbulence.
C$^{17}$O has hyperfine lines, but they can be ignored because the hyperfine splitting is minimal ($\sim0.0012$ km s$^{-1}$) compared to the spectral resolution ($\sim0.87$ km s$^{-1}$).
To determine the velocity dispersion, we correct the observed velocity dispersion by deconvolving it from the minor contribution from the velocity resolution \citep[e.g.,][]{huang2023clump}:
\begin{eqnarray}\label{eq5}
\sigma_{\rm v}^{2} = \sigma_{\rm obs}^{2}-\frac{\Delta v_{\rm r}^{2}}{8\rm ln2}
\end{eqnarray}
where $\sigma_{\rm v}$ is the velocity dispersion along the line of sight, and $\Delta v_{\rm r}$ is the spectral resolution of the observations $\sim 0.87\ \rm km\ s^{-1}$.
The observed velocity dispersion $\sigma_{\rm obs}=\Delta V_{\rm FWHM}/(2\sqrt{2\rm ln 2})$, where $\Delta V_{\rm FWHM}$ is estimated using the moment one map of C$^{17}$O within the same regions as that for the estimation of the physical parameters.
In light of the potential impact of the gravity and outflows on observed dispersion, we exculded the innermost regions with a radius of 200 au around the dust peak, and the region where the CO outflow emission is larger than 5-$\sigma$.
Then the non-thermal component of the velocity dispersion can be expressed as: 
\begin{eqnarray}\label{eq6}
\sigma_{\rm nth}^{2}=\sigma_{\rm v}^{2}-\sigma_{\rm th}^{2}=\sigma_{\rm v}^{2}-\frac{k_{\rm B}T_{\rm kin}}{m_{\rm obs}}
\end{eqnarray}
where $\sigma_{\rm nth}$ and $\sigma_{\rm th}$ are the non-thermal and the thermal velocity dispersion, respectively. 
$k_{\rm B}$ is the Boltzmann constant, and $m_{\rm obs}$ is the mass of the observed molecule (for C$^{17}$O, $m_{\rm obs}=29\ m_{\rm H}$). 
$T_{\rm kin}$ is the kinetic temperature of the gas, which we assume to be equal to $\overline{T}_{\rm env}$.
The non-thermal velocity dispersions are listed in column 9 of Table \ref{Tab:paras}.
Figure \ref{fig:distribution}(G) shows the distribution of $\sigma_{\rm nth}$ for 60 protostars (HOPS-408 has no detection of velocity dispersion within the selected region).

\subsection{Angle Dispersion of B-field} \label{subsec:disp}

The conventional approach for estimating the angle dispersion of {\em B}-fields ($\delta\phi$) is by measuring its standard deviation.
However, this approach is limited by the 180$^{\circ}$ ambiguity in polarization angle, which complicates the accurate estimation of the angle dispersion.
To address this challenge, further methods have been proposed to quantify $\delta\phi$.
\cite{heitsch2001magnetic} proposed a variation in which the restriction for large angles was removed.
Later \cite{hildebrand2009dispersion} and \cite{houde2009dispersion} applied a more sophisticated technique which uses an analytical derivation of the dispersion in polarization angles within a turbulent molecular cloud.
In a recent study, \cite{cortes2024magmar} proposed an alternative approach to derive the angle dispersion by assuming that the local turbulence perturbing the field follows a Gaussian distribution, and they suggested that 16 Nyquist-sampling segments of {\em B}-field are required for the Gaussian distribution under the assumption of 5\arcdeg ~for the margin of error and 10\arcdeg ~for the mean standard deviation of polarization angles.
In this paper, we follow the typical method to estimate the angle dispersion $\delta\phi$ by calculating the standard deviation of the distribution of {\em B}-field segments.
In order to reduce the ambiguity associated with the 180° orientation of the {\em B}-field, the distribution of the position angles is shifted when the angles are centred near 0° or 180°.
The standard deviation of the shifted distribution is regarded as $\delta\phi$.
To avoid the contamination of self-scattering \citep{huang2024magnetic}, the polarization segments within the innermost 400 au region ($\sim 1\arcsec$, corresponding to the beam size) have been masked.
We also mask the polarization data for regions beyond a radius of 1200 au, consistent with the scales used to derive the physical parameters in Section \ref{subsec:envelope}.
Protostellar envelopes with less than 16 Nyquist-sampling segments of {\em B}-field within 1200 au regions have large uncertainties ($>5^{\circ}$) according to \cite{cortes2024magmar}, therefore we only estimate the angle dispersion of {\em B}-field for 25 protostars with sufficient statastics ($>$ 16 Nyquist-sampling segments), which is listed in column 10 of Table \ref{Tab:paras}.
Among these 25 protostellar envelopes, 6 exhibit {\em B}-field morphology that is standard-hourglass, 8 are rotated-hourglass, 4 are spiral, and 7 are complex or lack polarized emission to identify their {\em B}-field morphology.
Figure \ref{fig:distribution}(H) shows the distribution of $\delta\phi$ for each type of {\em B}-field morphology.

\subsection{Fragmentation Level and Disk Properties}

\begin{figure*}
\centering
\includegraphics[clip=true,trim=2.4cm 6.5cm 4.8cm 7.5cm,width=0.45 \textwidth]{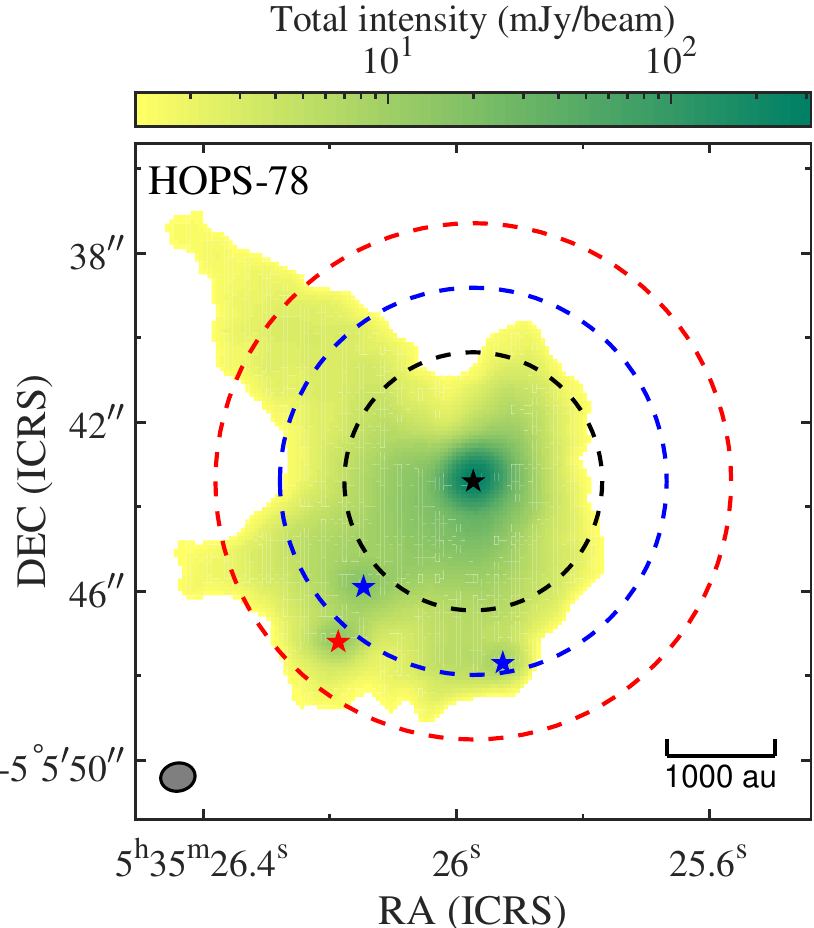}
\caption{Dust emission map at 0.87 mm toward an example of HOPS-78.
The color scale shows the dust emission greater than 10$\sigma$.
The black, blue, and red dashed circles represent regions with radii of 1200, 1800, and 2400 au, respectively.
The asterisks indicate the presence of disks identified by \cite{tobin2020vla}.} 
\label{fig:fra_example}
\end{figure*}

\begin{figure*}
\centering
\includegraphics[clip=true,trim=0cm 0.cm 0.22cm 0.2cm,width=0.325 \textwidth]{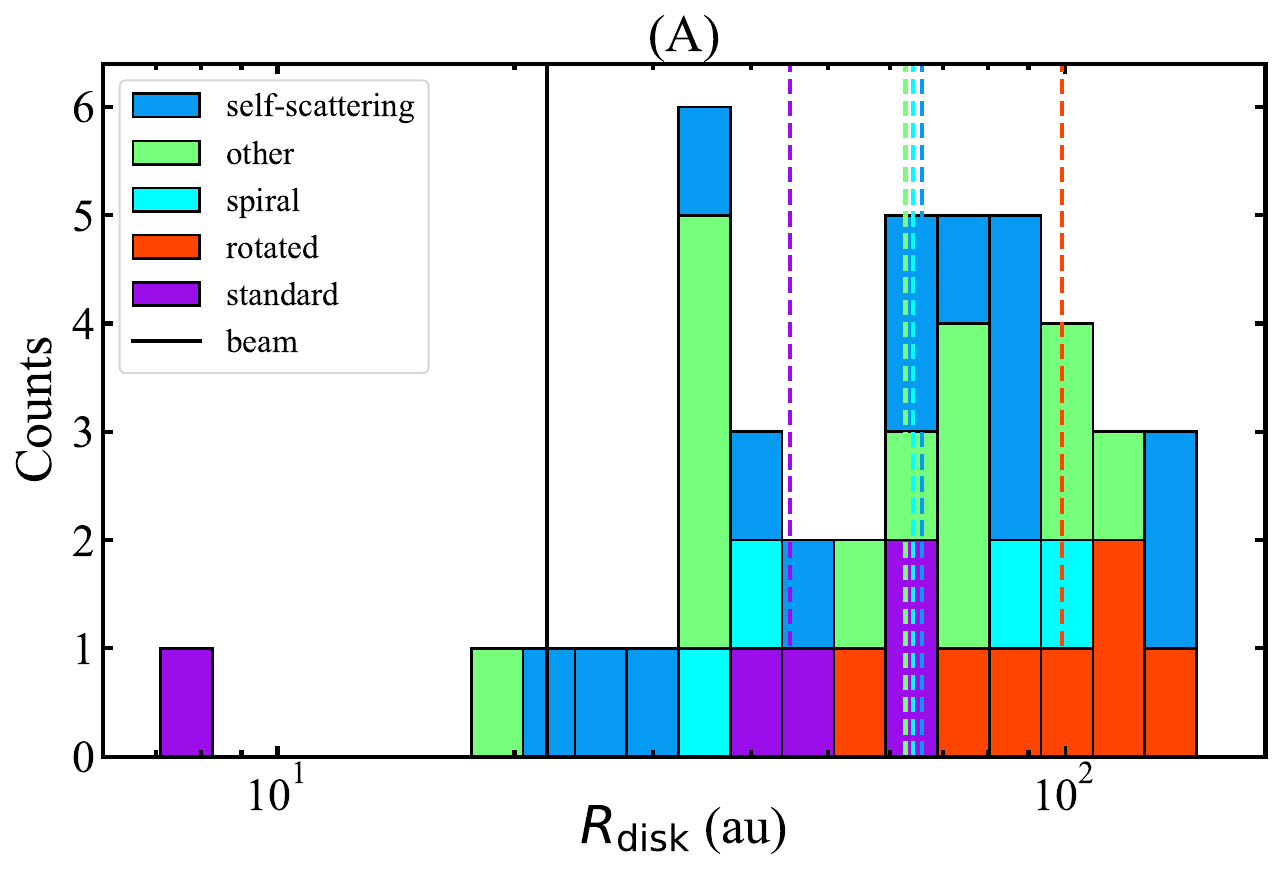}
\includegraphics[clip=true,trim=0cm 0.2cm 0.2cm 0.2cm,width=0.325 \textwidth]{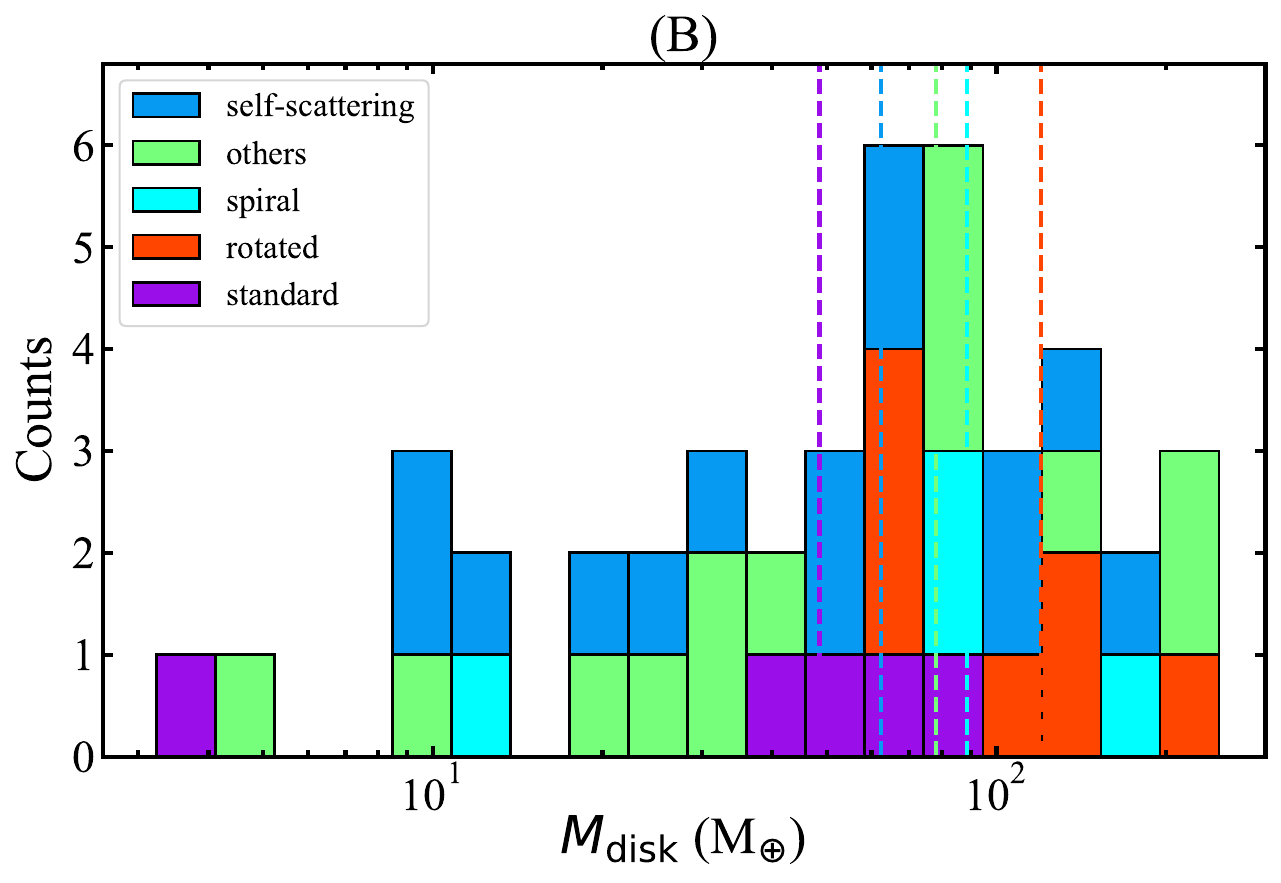}
\includegraphics[clip=true,trim=0.28cm 0.2cm 0.2cm 0.2cm,width=0.325 \textwidth]{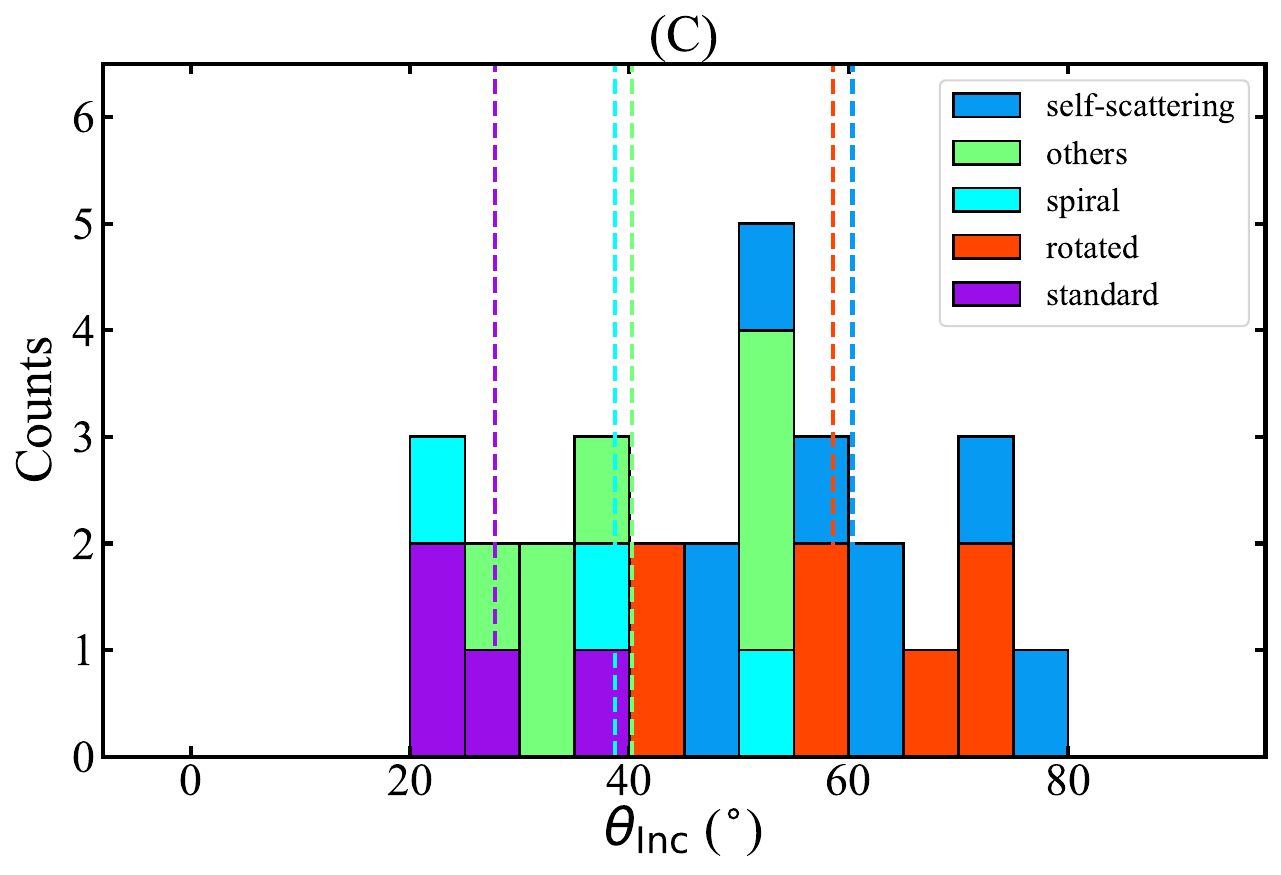}
\caption{The stacked histogram of disk radius $R_{\rm disk}$ (panel A), disk mass $M_{\rm disk}$ (panel B), and disk inclination angle $\theta_{\rm Inc}$ (panel C) for protostars with different {\em B}-field structure.
Black solid line in panel A indicates the synthesized beam radius.
Dashed lines in A, B and C panels indicate the mean values of the distribution for different {\em B}-field structures excluding sources with unresolved disk.
The disk masses obtained from \cite{tobin2020vla} are refined using Equation \ref{eq7} from \cite{sheehan2022vandam}, the disk radii and disk inclination angles are obtained from \cite{tobin2020vla}.} 
\label{fig:disk}
\end{figure*}

\cite{tobin2020vla} identified protostellar disks in the Orion molecular clouds using both VLA and ALMA observations, with further refinements presented in their subsequent study of \cite{tobin2022vla}.
All objects in our sample are included within the VANDAM survey, which enables us to estimate the multiplicity of these systems.
\cite{palau2013frag} defined the fragmentation level as the number of millimeter sources inside a region with a radius of 0.05 pc for massive dense cores, whereas \cite{busquet2016frag} used a different radius of 0.15 pc.
We follow the above methods, but define a different radius of 1200 au, which is the scale for estimating the physical parameters above.
Then the number of fragments (including small-scale fragmentation, that was only resolved in the survey with higher resolution of $\sim$40--400 au-scale companions) identified from the VANDAM and the BOPS projects within the innermost envelope region is indicative of the fragmentation level $N_{\rm F}$.
It is possible that the outer ($>1200$ au) regions may influence the fragmentation of the envelopes, given that the typical scale for the star-forming envelopes is up to several thousand au \citep[e.g.,][]{myers2000obs}.
We thus define larger radii of 1800 au and 2400 au for estimating extra contribution of $N_{\rm F}$, but with less weighting values of 0.5 and 0.25, respectively.
For example, in Figure \ref{fig:fra_example}, which shows the 0.87 mm dust emission map of an example HOPS-78, a protostar (black asterisk) is located in the innermost region with a radius of 1200 au (black dashed circle), then $N_{\rm F}$ in this region is counted as 1.
While there are 2 (blue asterisks) and 1 (red asterisk) protostars in annular regions with inner and outer radii of 1200 and 1800 au (blue dashed circle), and 1800 and 2400 au (red dashed circle), then the $N_{\rm F}$ for each region is counted as 1 and 0.25, respectively.
Note that these protostars are identified from continuum maps of the BOPS and VANDAM survey.
The total fragmentation level $N_{\rm F}$ for HOPS-78 is therefore assigned a value of 2.25.
Column 11 of Table \ref{Tab:paras} lists the fragmentation level for 61 targets.

The mass, radius, size of the disks have been reported in \cite{tobin2020vla}.
\cite{sheehan2022vandam} found that the disk masses derived from \cite{tobin2020vla} are overestimated compared to radiative transfer model, with the discrepancy becoming more pronounced for lower-mass disks.
To address this, they proposed a log-linear relation to reconcile the disk masses derived from \cite{tobin2020vla} with those obtained through radiative transfer modelling, expressed as:
\begin{eqnarray}\label{eq7}
\log(M_{\rm disk}^{\rm Sheehan}) = \frac{\log(M_{\rm disk})-(0.83\pm0.05)}{(0.67\pm0.03)}
\end{eqnarray}
Figure \ref{fig:disk}(A--C) show the distributions of disk mass refined by Equation \ref{eq7}, disk radius, and disk inclination angle for 43 protostars, respectively.
The rest 18 protostars lack information of the disk, or the disk dust emission is potentially contaminated by the envelope or other disk emission (see Appendix \ref{app:B} for figures).
The disk inclination angle is with respect to the line of sight, which is estimated by $\theta_{\rm inc} = \arccos(\Theta_{\rm min}/\Theta_{\rm maj})$ if the disk is assumed to be geometrically thin and circular, where $\Theta_{\rm maj}$ and $\Theta_{\rm min}$ are the deconvolved major and minor axes of the disk.
It is important to note that there are 14 out of 43 protostars, the disk dust emission is compact and only partially resolved, as the disk size, particularly the disk deconvolved minor axis is less than the VANDAM's spatial resolution of $0.1^{\prime\prime}$ \footnote{The 14 sources with compact and partially resolved disk emission are: HOPS-10, HOPS-12W, HOPS-53, HOPS-81, HOPS-88, HOPS-182, HOPS-203S, HOPS-317N, HOPS-325, HOPS-341, HOPS-340, HOPS-354, HOPS-373W and HOPS-395. See also Table \ref{Tab:parameters}.}. 
The trend of the distribution for disk properties, as well as other parameters, will be discussed in the following.

\section{Discussion} \label{sec:Dis}

In Section \ref{sec:analysis}, we have derived physical parameters of envelope gas mass, gas column density, gas number density for entire sample of BOPS (61 protostars) based on the available bolometric luminosity \citep{furlan2016herschel} and dust temperature \citep{lombardi2014temperature}, and non-thermal velocity dispersion for 60 protostars within the inner region of 6\arcsec ~(corresponding to a scale of 2400 au, with a radius of 1200 au).
Angle dispersion of {\em B}-field is also estimated within the same regions for 25 BOPS targets with sufficient Nyquist-sampling statistics.
Combining with the VANDAM survey, the fragmentation level has been identified for 61 BOPS targets.
In addition, disk properties have been presented for 43 BOPS targets from \cite{tobin2020vla}, without any contamination of envelope emission and other components of disk emission.

The uncertainties of the derived physical parameters are due to unknowns and errors in factors such as dust opacities, dust temperatures, and observed flux.
The uncertainty in physical parameters resulting from dust opacity, an internally consistent systematic error, does not significantly impede the identification of correlations between physical parameters, as these are calculated in a uniform manner, which translates the distributions up or down.
Whereas the dust temperature and the flux calibration are considered random.
For the temperature, based on the model that we used, we consider typical uncertainties of 20--30\%, and we expect flux calibration to be accurate within 10\%. 
All combined, we derive uncertainties of about 30--40\% for the masses and densities.

The distributions of the these parameters are shown in Figure \ref{fig:distribution}.
Combining both the distribution and the mean values (dashed lines) for each type of {\em B}-field morphologies, we find that protostars with high bolometric luminosity ($>$ 50 L$_{\odot}$) and/or envelope dust temperature ($>$ 35 K) tend to exhibit spiral or rotated-hourglass {\em B}-field morphology (Figure \ref{fig:distribution}(A--B)).
In Figure \ref{fig:distribution}(C--F), envelopes with compact polarized emission (self-scattering in blue) tend to have lower envelope masses, gas column densities, and gas number densities.
Whereas protostars exhibiting a standard-hourglass {\em B}-field pattern tend to exhibit relatively larger values in these parameters.
However, these standard-hourglass protostars have relatively low non-thermal velocity dispersion, which is potentially indicative of the magnetic braking effect, whereby the strong field exerts a braking influence on the rotational rate (as shown in panel G of Figure \ref{fig:distribution}).
In Figure \ref{fig:distribution}(H), we find standard-hourglass-{\em B}-field protostars have relatively low angle dispersion, in contrast to rotated-hourglass-, spiral-, and complex-field sources, which display a relatively larger angle dispersion.

In Figure \ref{fig:disk}, which shows the distribution of disk properties, we find from mean values of disk mass/size in the first row (dashed lines in panel A and B) that protostars exhibiting standard hourglass {\em B}-field structure have smaller disk mass/size than those exhibiting other field configurations,  with these sources lying in the left-side within the distribution after excluding the compact disks.
However, more statistics are needed to confirm this scenario.
In panel C, the mean values for protostars with different type of {\em B}-field structure are ranging from 25$^{\circ}$ to 60$^{\circ}$. 
Note that the finite resolution and vertical thickness of the disks \citep{sheehan2022vandam, lin2023disk, van2023disk} will bias this simple inclination estimate against 90$^{\circ}$, which would influence the location of the mean values.
\cite{frau2011model} found from the magnetohydrodynamics collapse model that sources with hourglass (including standard- and rotated-hourglass) {\em B}-field structure can be easier seen when $\theta_{\rm Inc}\gtrsim30^{\circ}$.
In our results, the disk inclination angles of hourglass-field protostars can be found within a range of $\theta_{\rm Inc}\gtrsim20^{\circ}$.
However, protostars with standard-hourglass {\em B}-field morphology have small disk size closed to the VAMDAN's spatial resolution of $\sim0.1^{\prime\prime}$ (see Table \ref{Tab:parameters} in Appendix \ref{app:A}), which would introduce a significant degree of uncertainty in determining the disk inclination
angle.
Higher angular resolution observations are needed to better determine the disk inclination for the compact disks.

Below we discuss the potential correlation between these parameters and disk properties.
To statistically examine the relations between two pairs of parameters, we performed both the Pearson linear correlation coefficient ($r_{p}$) and the Spearman's rank correlation coefficient ($r_{s}$).
The values of $r_{p}$ and $r_{s}$ are each bounded from -1 to 1, for negative to positive slope correlations, respectively. 
In each case, the closer the absolute value is to 1, the stronger the correlation.
The p--value is used for the purpose of testing the null hypothesis that the two variables are uncorrelated.
In general, statistically significant correlations are identified by p--value $<$ 0.05.
Here we analyze the following figures, and based on these details, we will motivate a more detailed analysis on fragmentation and the possible effect of magnetic braking on disk formation.

\subsection{Physical Envelope Properties}

\begin{figure*}
\centering
\includegraphics[clip=true,trim=0.3cm 0.2cm 0.2cm 0.25cm,width=0.415 \textwidth]{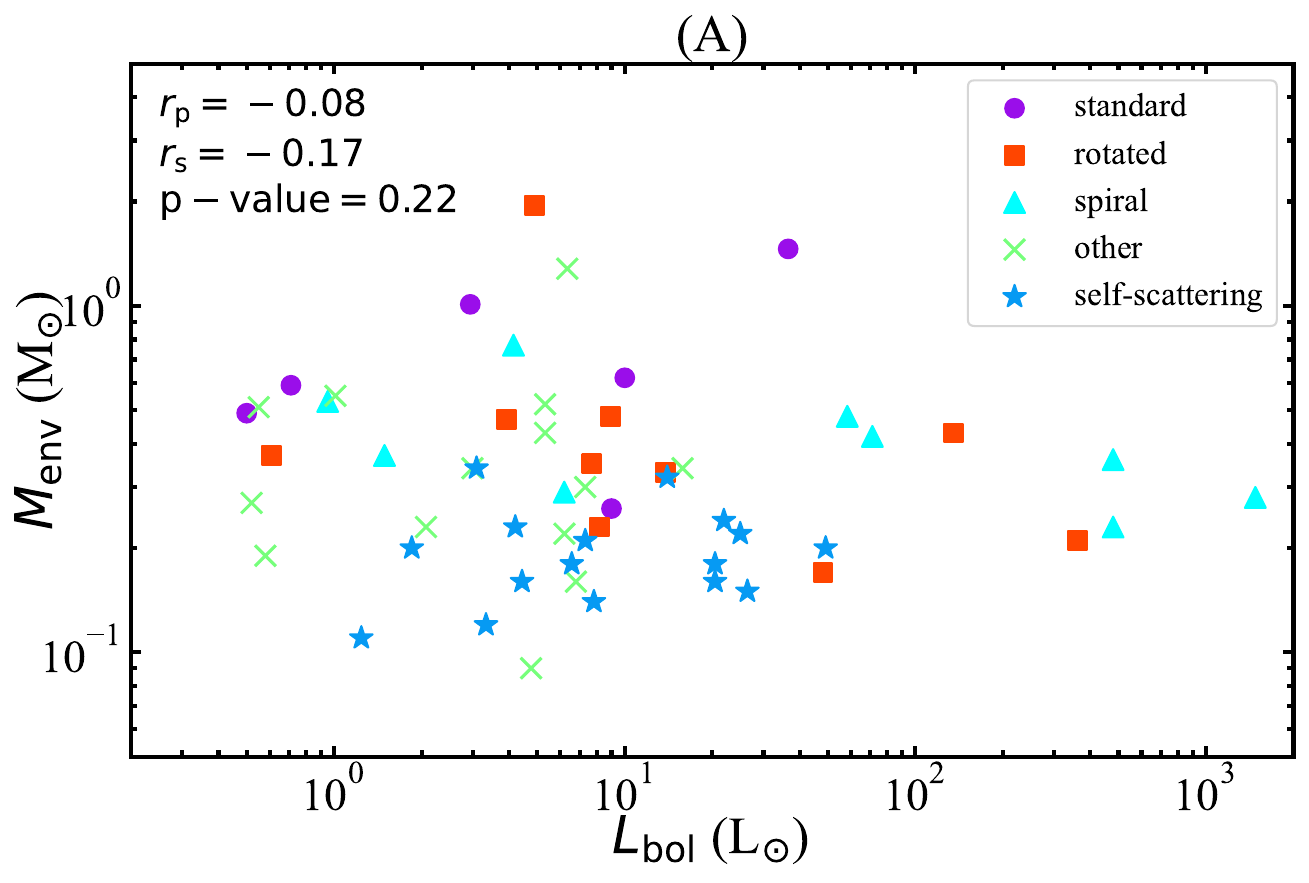}
~~~~~~~~~~
\includegraphics[clip=true,trim=0.3cm 0.3cm 0.2cm 0.25cm,width=0.415 \textwidth]{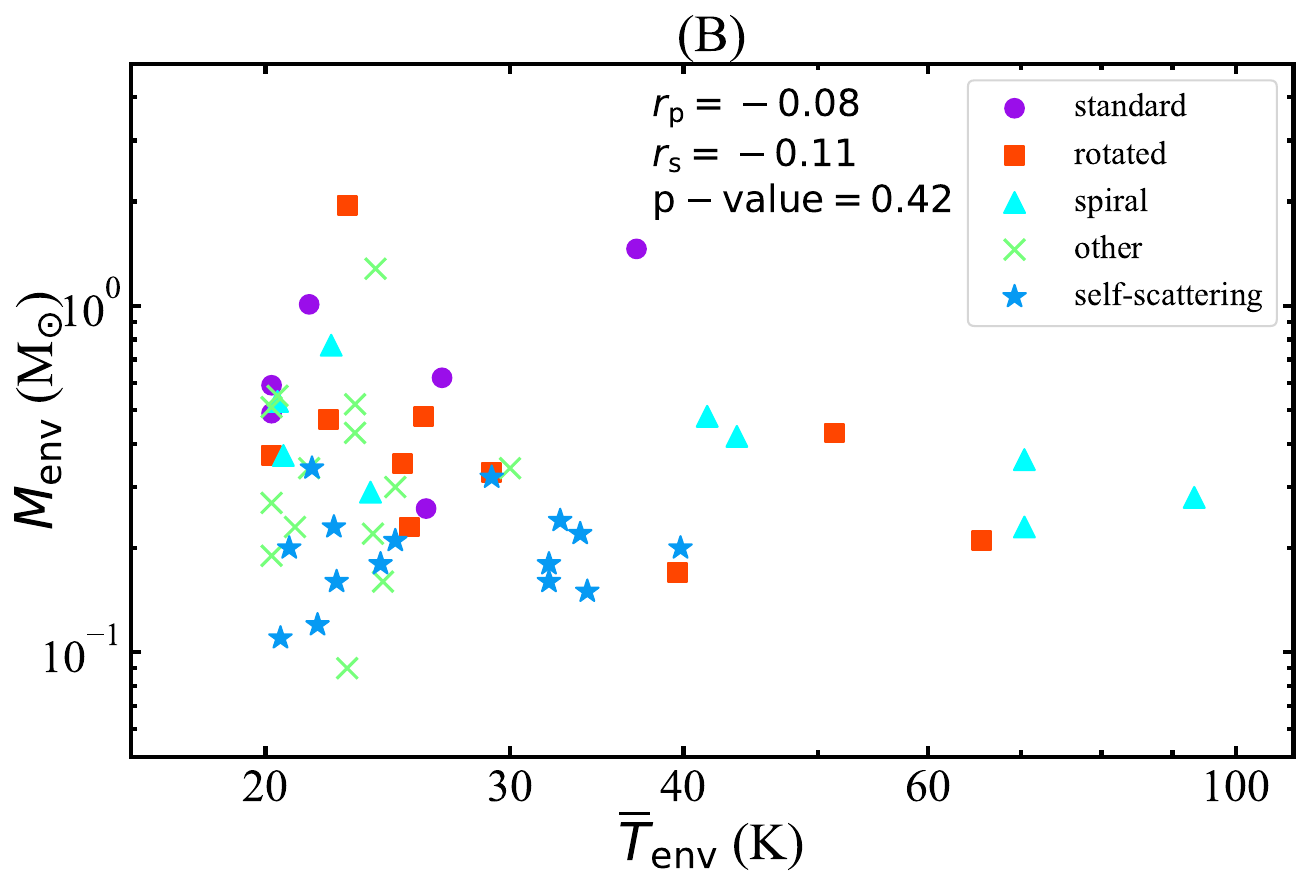}
~\\
~\\
\includegraphics[clip=true,trim=0.25cm 0.3cm 0.2cm 0.25cm,width=0.42 \textwidth]{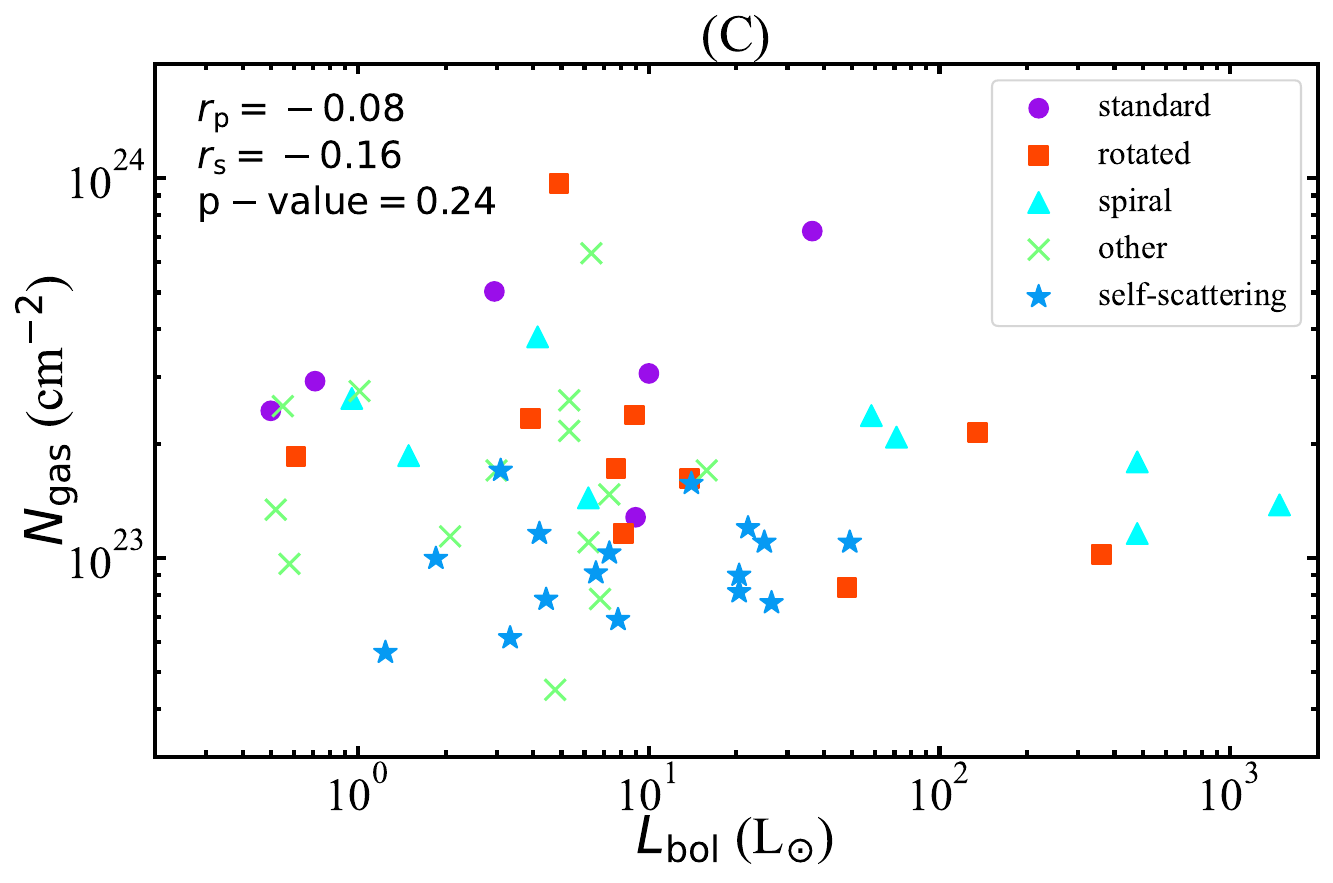}
~~~~~~~~~~
\includegraphics[clip=true,trim=0.25cm 0.3cm 0.2cm 0.25cm,width=0.42 \textwidth]{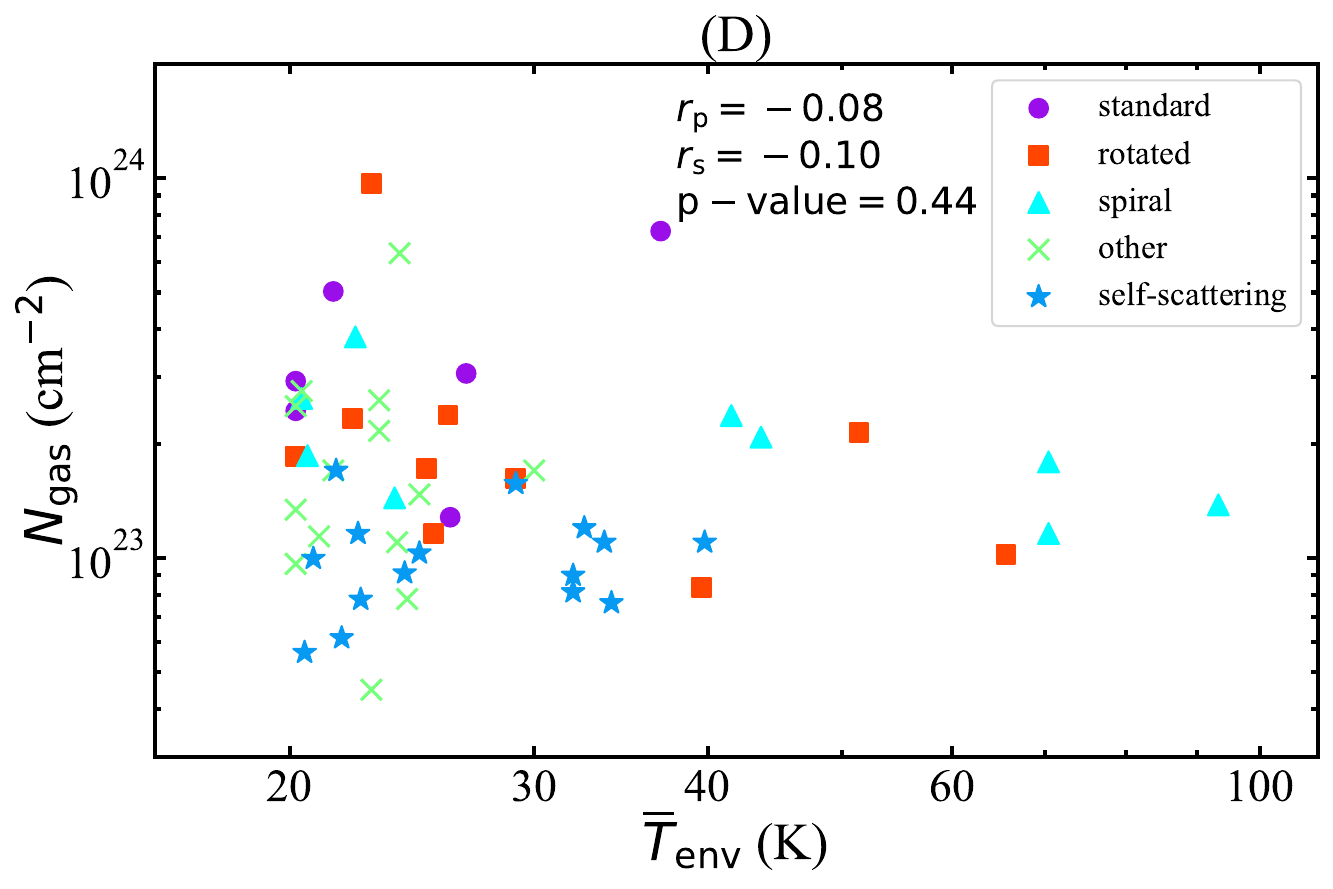}
~\\
~\\
\includegraphics[clip=true,trim=0.25cm 0.2cm 0.2cm 0.25cm,width=0.42 \textwidth]{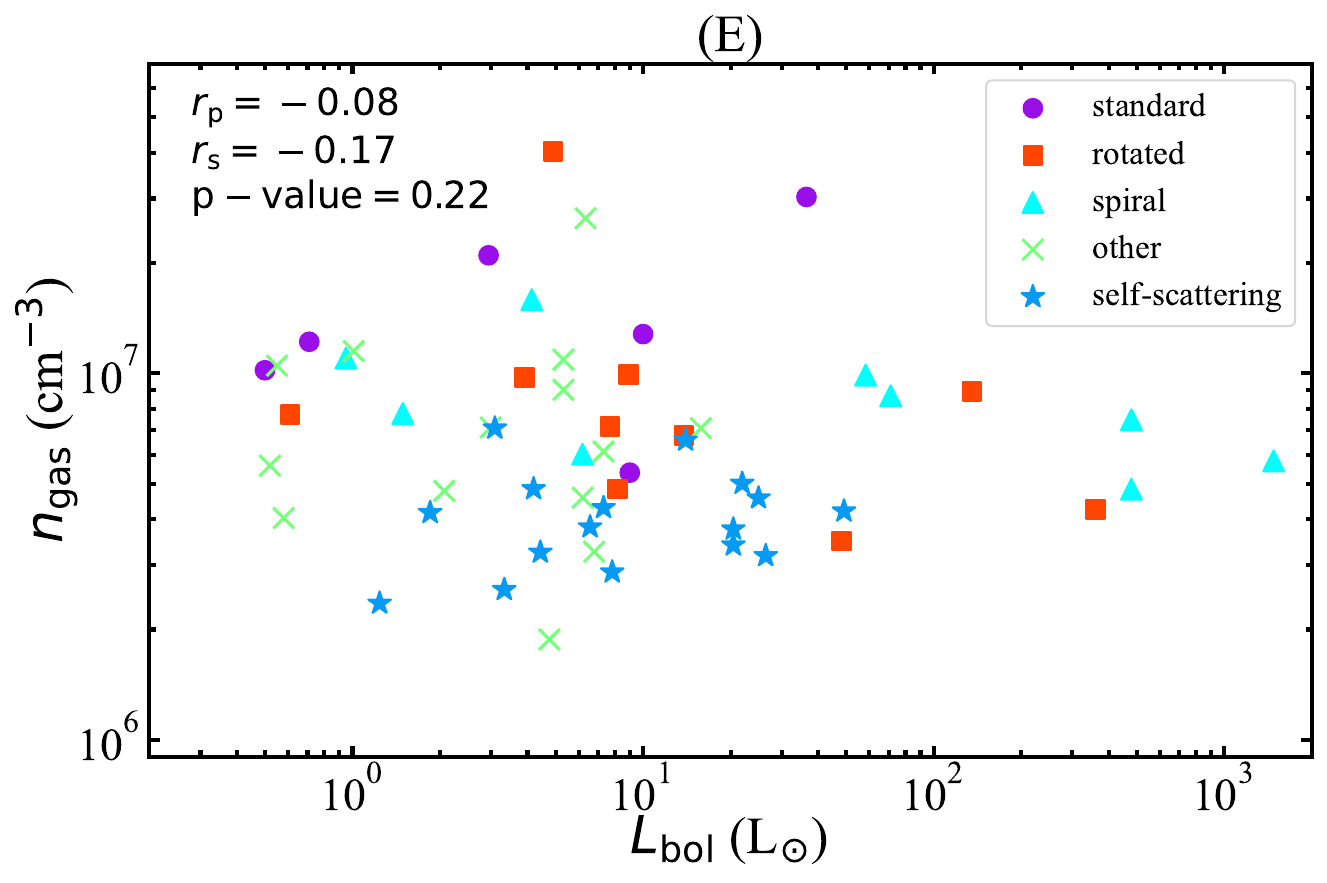}
~~~~~~~~~~
\includegraphics[clip=true,trim=0.25cm 0.3cm 0.2cm 0.25cm,width=0.42 \textwidth]{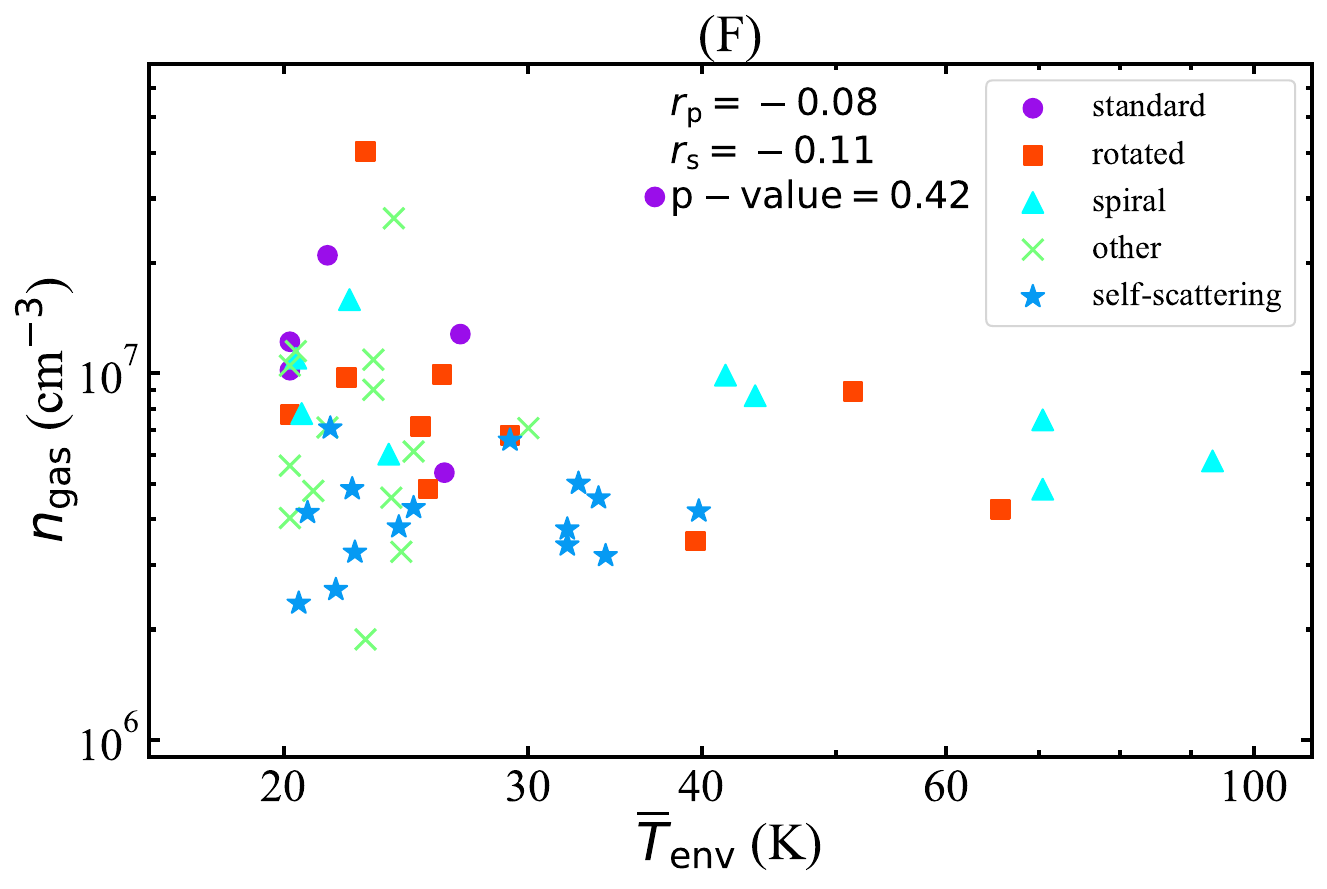}
~\\
~\\
\includegraphics[clip=true,trim=0.25cm 0.25cm 0.2cm 0.25cm,width=0.42 \textwidth]{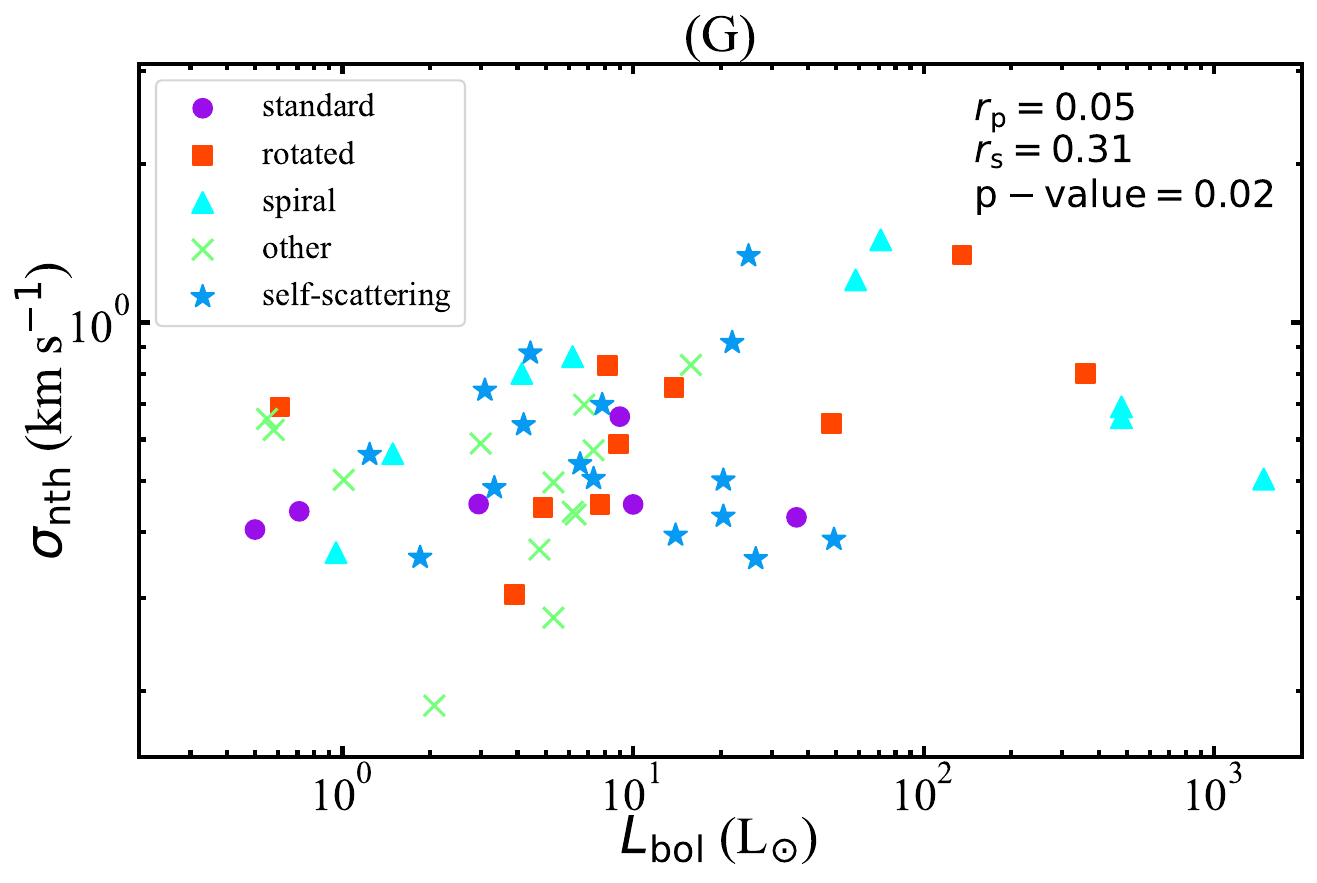}
~~~~~~~~~~
\includegraphics[clip=true,trim=0.25cm 0.3cm 0.2cm 0.25cm,width=0.42 \textwidth]{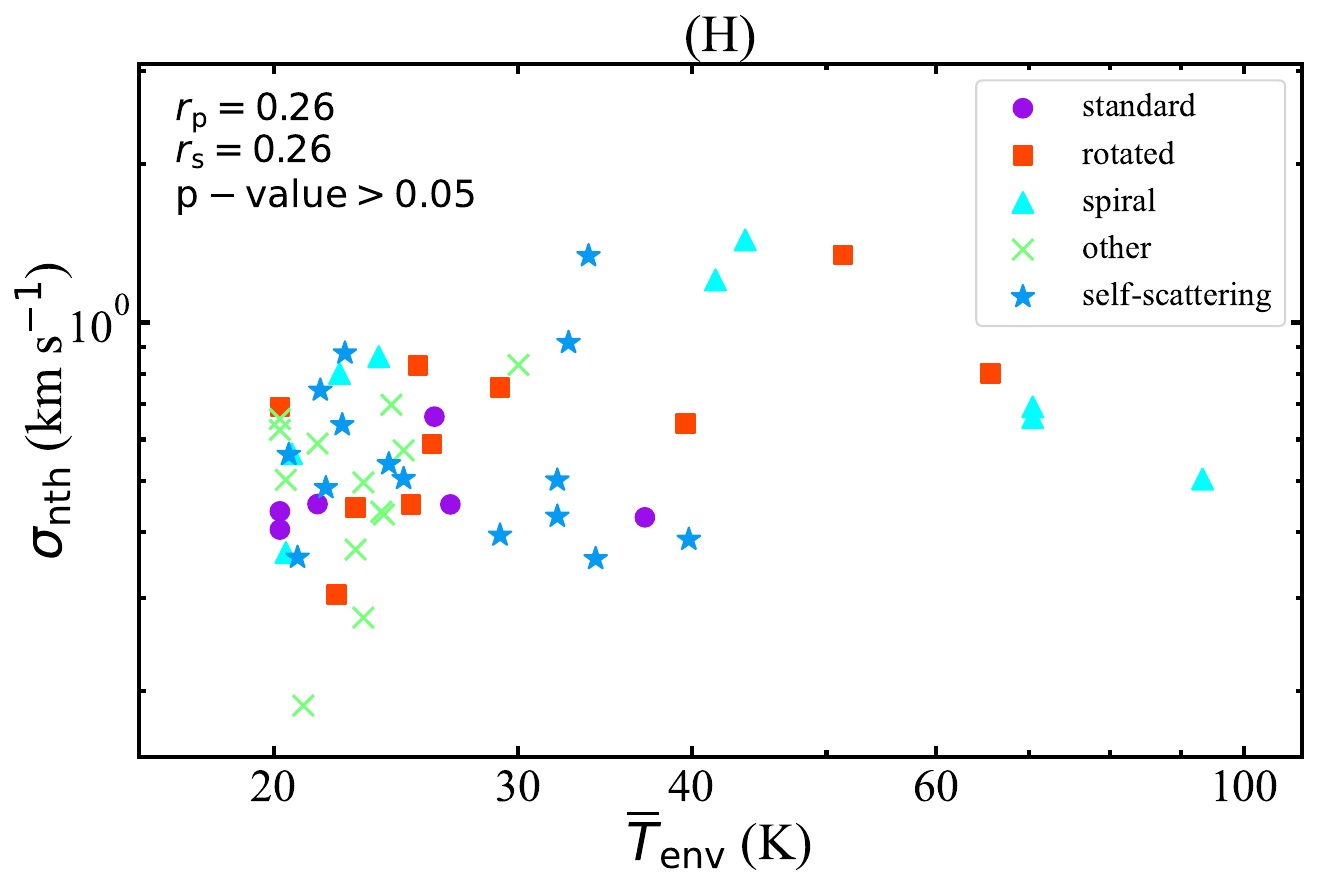}
\caption{Left panels: Distributions between the bolometric luminosity $L_{\rm bol}$ and the envelope gas mass $M_{\rm env}$ (panel A), the gas column density $N_{\rm gas}$ (C), the gas volumn density $n_{\rm gas}$ (E), and the non-thermal velocity dispersion $\sigma_{\rm nth}$ (G). 
Right panels: Distribution between the mean dust temperature $\overline{T}_{\rm env}$ and the envelope gas mass $M_{\rm env}$ (B), the gas column density $N_{\rm gas}$ (D), the gas volume density $n_{\rm gas}$ (F), and the non-thermal velocity dispersion $\sigma_{\rm nth}$ (H).} 
\label{fig:disp}
\end{figure*}

The dust temperature is expected to be correlated with the bolometric luminosity if the dust is primarily heated by the stellar radiation (see Section \ref{subsec:envelope}).
During the early stages of star formation (Class 0 and early Class I), the protostar is deeply embedded within a dense envelope.
In this phase, the luminosity is predominantly due to the accretion of material from the envelope through the disk onto the protostar, which releases gravitational energy.
High-density regions are often associated with intense accretion activity as the protostar accumulates material.
The inflow of material from the envelope onto the protostar leads to an increase in $L_{\rm bol}$.
Additionally, a more massive envelope absorbs and re-radiates more energy and it is more efficient at cooling due to its higher optical depth.
\ref{fig:disp}(A--F) shows the distribution between these physical parameters for a sample of 55 envelopes (the remaining 6 targets have no bolometric luminosity information).
However, our results reveal no significant correlations between $L_{\rm bol}$ and $M_{\rm env}$, $N_{\rm gas}$, $n_{\rm gas}$, and between $\overline{T}_{\rm env}$ and $M_{\rm env}$, $N_{\rm gas}$, $n_{\rm gas}$, as their correlation coefficients are close to 0, with p--values much higher than 0.05.
The lack of direct correlations is likely because protostellar envelope masses are influenced by both the final stellar mass of the protostar and the evolutionary stage, with envelope masses decreasing rapidly over time \citep{fischer2017hops}.
Luminosity, on the other hand, is influenced by episodic accretion, which may vary independently of the envelope mass \citep[e.g.,][]{zakri2022orion, narang2024lumi}. 

Non-thermal motions, often driven by turbulence, are of significant importance in star-forming regions and affect the overall dynamics and evolution of the protostellar system. 
Therefore, $\sigma_{\rm nth}$ serves as an indicator of turbulence level, which enhance the accretion process by increasing the rate at which material is fed onto the protostar.
Protostars with higher value of $L_{\rm bol}$ typically drive powerful outflows and winds, injecting energy into the surrounding medium and thereby increasing the non-thermal velocity dispersion. 
In high turbulent regions, the non-thermal velocity dispersion provide support against gravitational collapse. 
Once, however, collapse is triggered, the turbulent motions channel material onto the forming protostar, influencing its luminosity.
Theoretical models also suggest that higher levels of turbulence ($\sigma_{\rm nth}$) lead to more efficient accretion processes, leading to increased protostellar luminosities \citep[e.g.,][]{krumholz2005ism, padoan2011ism, federrath2015ism}.
Additionally, in regions with high $\sigma_{\rm nth}$, which are characterised by intense star formation activity (e.g., jets and winds), one might expect that higher dust temperatures may result from the combined effects of shock heating and enhanced radiative feedback from very young protostars.
To test these senarios, we plot the distribution between $\sigma_{\rm nth}$ and $L_{\rm bol}$, and between $\sigma_{\rm nth}$ and $T_{\rm env}$ for 54 protostars (5 protostars lack luminosity information, while one protostar of HOPS-408 has no detection of velocity dispersion), as shown in Figure \ref{fig:disp}(G--H).
A slight positive correlation is observed between $\sigma_{\rm nth}$ and $L_{\rm bol}$ from $r_{s}$ and p-value, as well as between $\overline{T}_{\rm env}$ and $L_{\rm bol}$ from $r_{p}$ and $r_{s}$.
However, $r_{p} \sim 0$ in panel G and the p-value $>$ 0.05 in panel H suggest insignificant correlations between these parameters.
The interplay between localized feedback, external turbulence, {\em B}-fields, and accretion variability makes it difficult to isolate a direct relationship, especially when dealing with deeply embedded young protostars in Orion, a highly turbulent environment.
In addition, the limited spectral resolution would also affect these results.
More focused studies on individual cores or higher-resolution observations might reveal more localized correlations that are otherwise masked by these factors.

We also note that these results will be influenced by the assumption of a typical temperature of 20 K.
However, we find that 7 protostars exhibit relatively high values $\overline{T}_{\rm env}$ ($>$ 40 K). 
Of these, 2 display a rotated hourglass field shape (HOPS-288 and HOPS-370), 5 of the protostars are spiral (HOPS-124, HOPS-182, HOPS-361N, HOPS-361S, and HOPS-384).

\subsection{Fragmentation}

\begin{figure*}
\centering
\includegraphics[clip=true,trim=0.3cm 0.25cm 0.2cm 0.2cm,width=0.46 \textwidth]{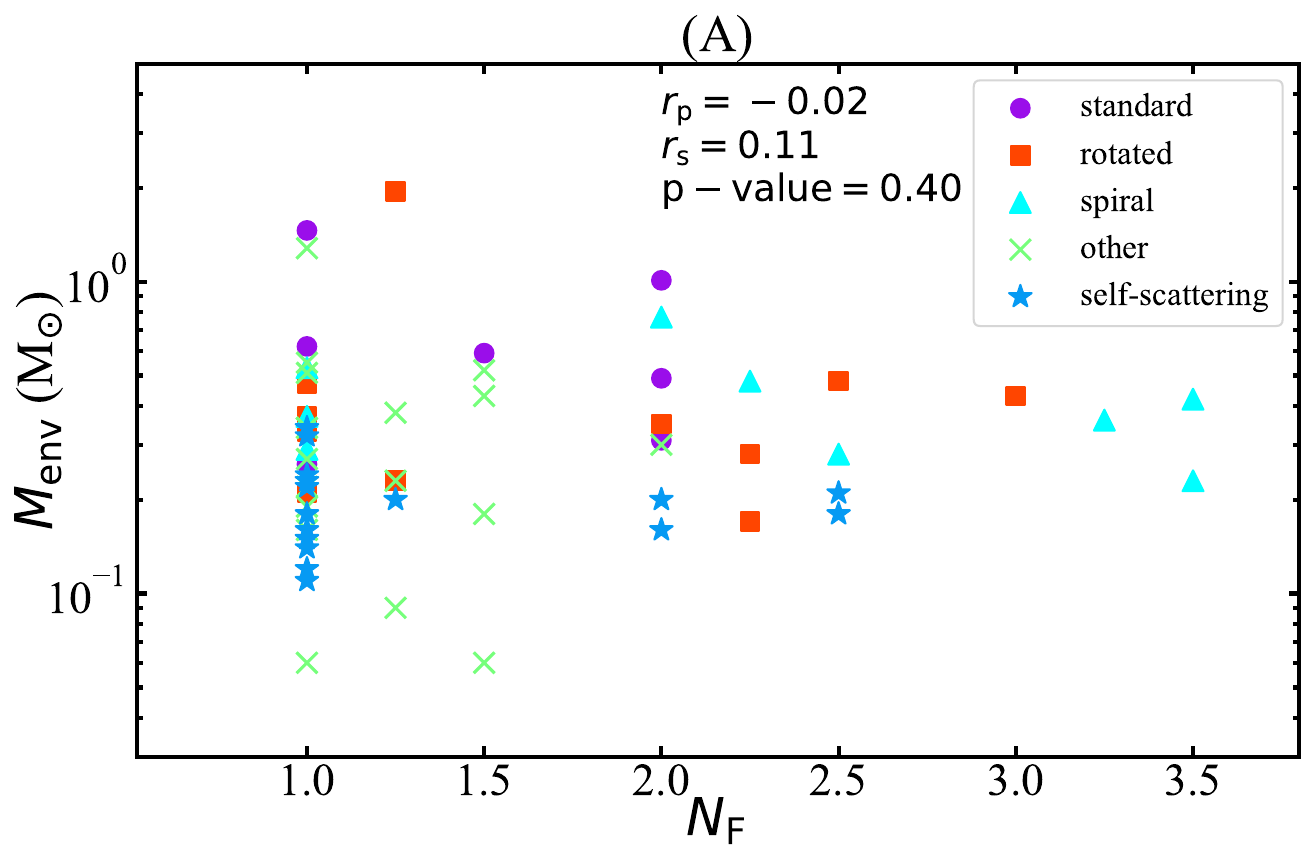}
~~~~~~~~~~~
\includegraphics[clip=true,trim=0.25cm 0.25cm 0.2cm 0cm,width=0.465 \textwidth]{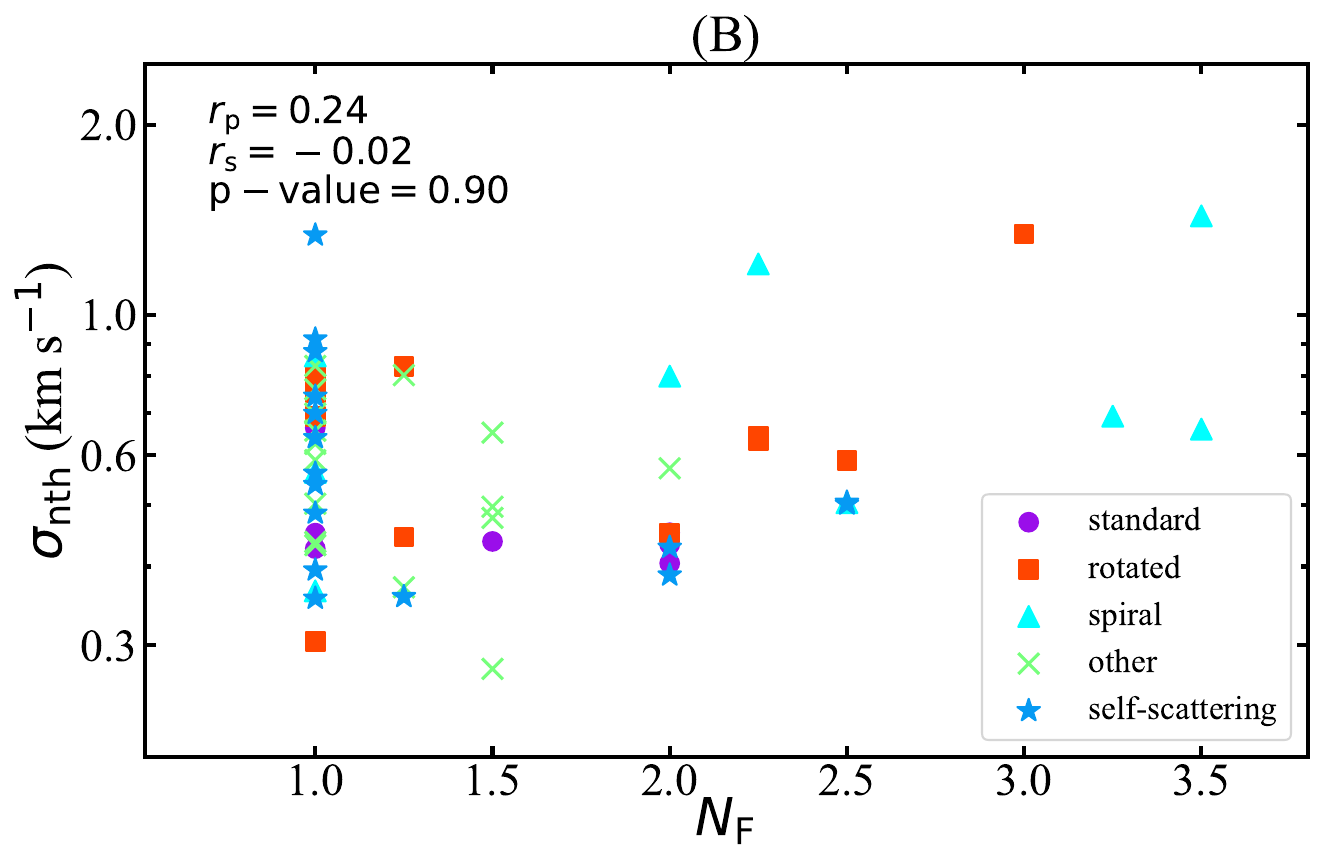}
~\\
~\\
\includegraphics[clip=true,trim=0.2cm 0.2cm 0.2cm 0.25cm,width=0.46 \textwidth]{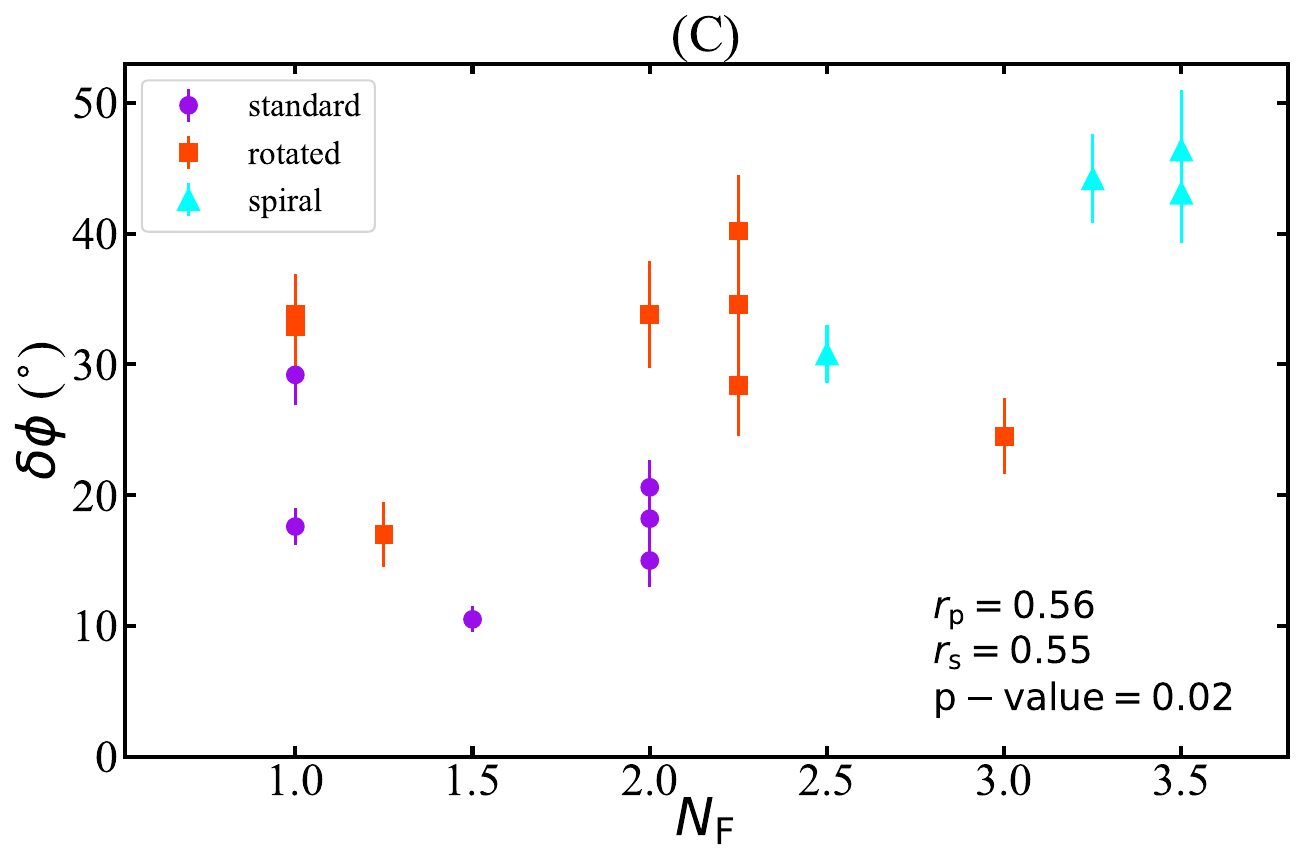}
~~~~~~~~~~~
\includegraphics[clip=true,trim=0.2cm 0.2cm 0.2cm 0.25cm,width=0.46 \textwidth]{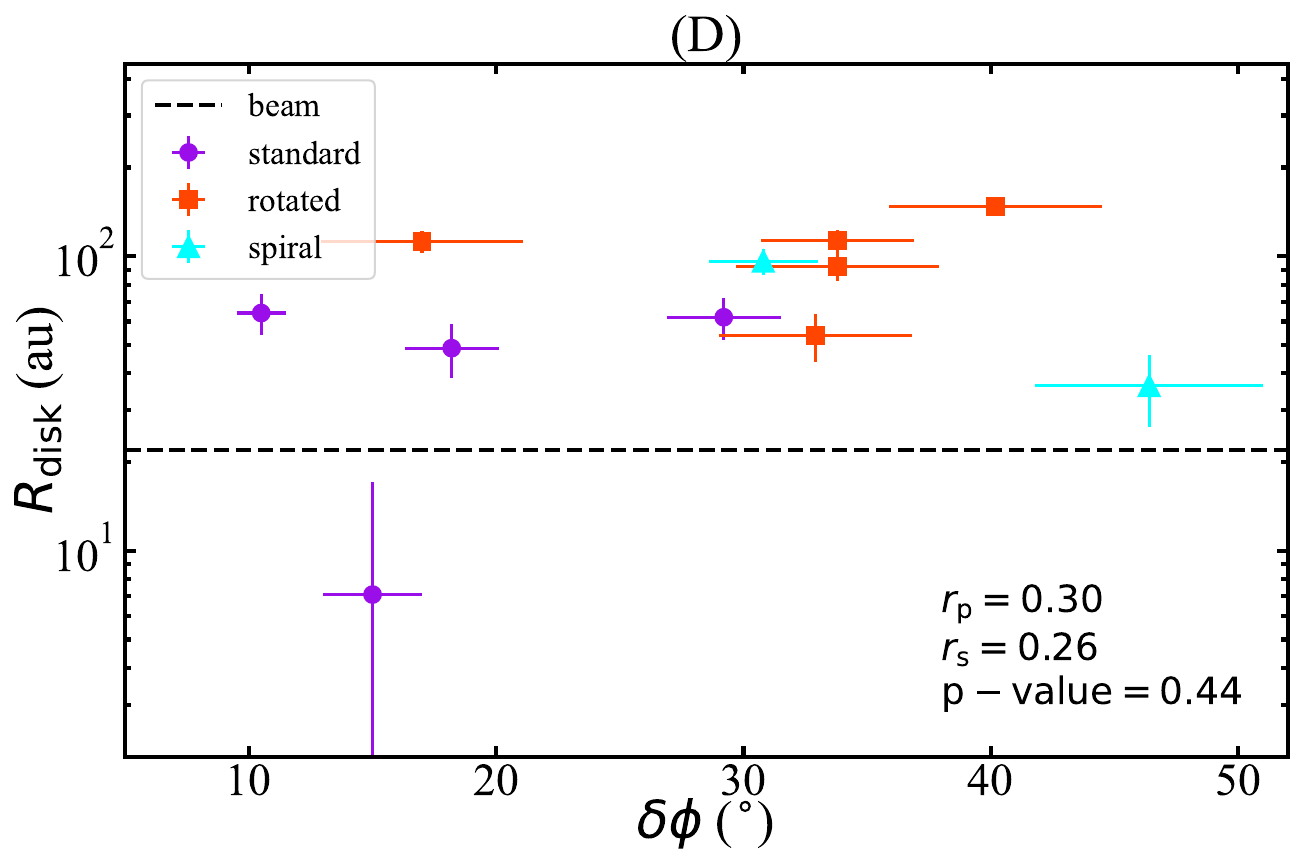}
~\\
~\\
\includegraphics[clip=true,trim=0.2cm 0.2cm 0.2cm 0.2cm,width=0.46 \textwidth]{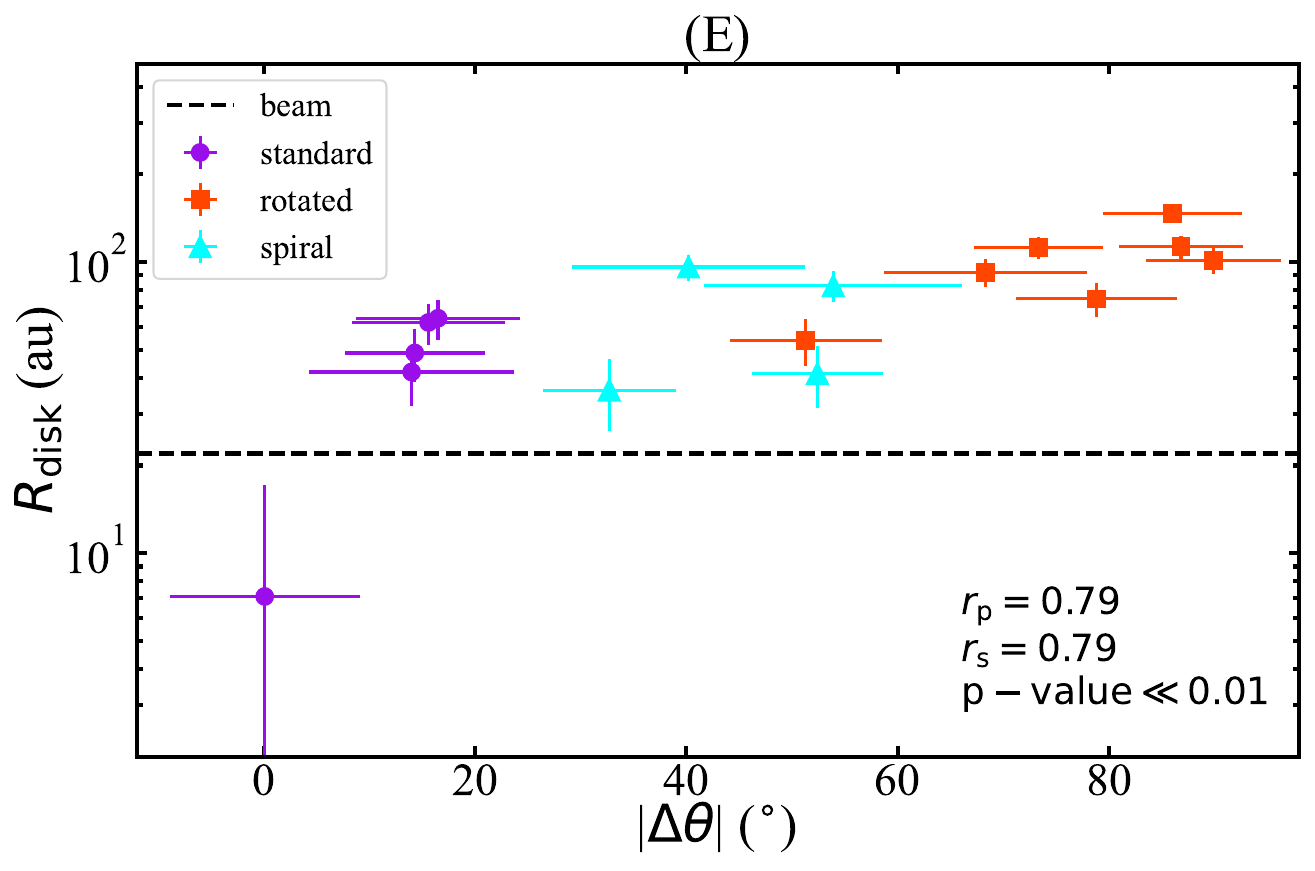}
~~~~~~~~~~~
\includegraphics[clip=true,trim=0.2cm 0.2cm 0.2cm 0.2cm,width=0.46 \textwidth]{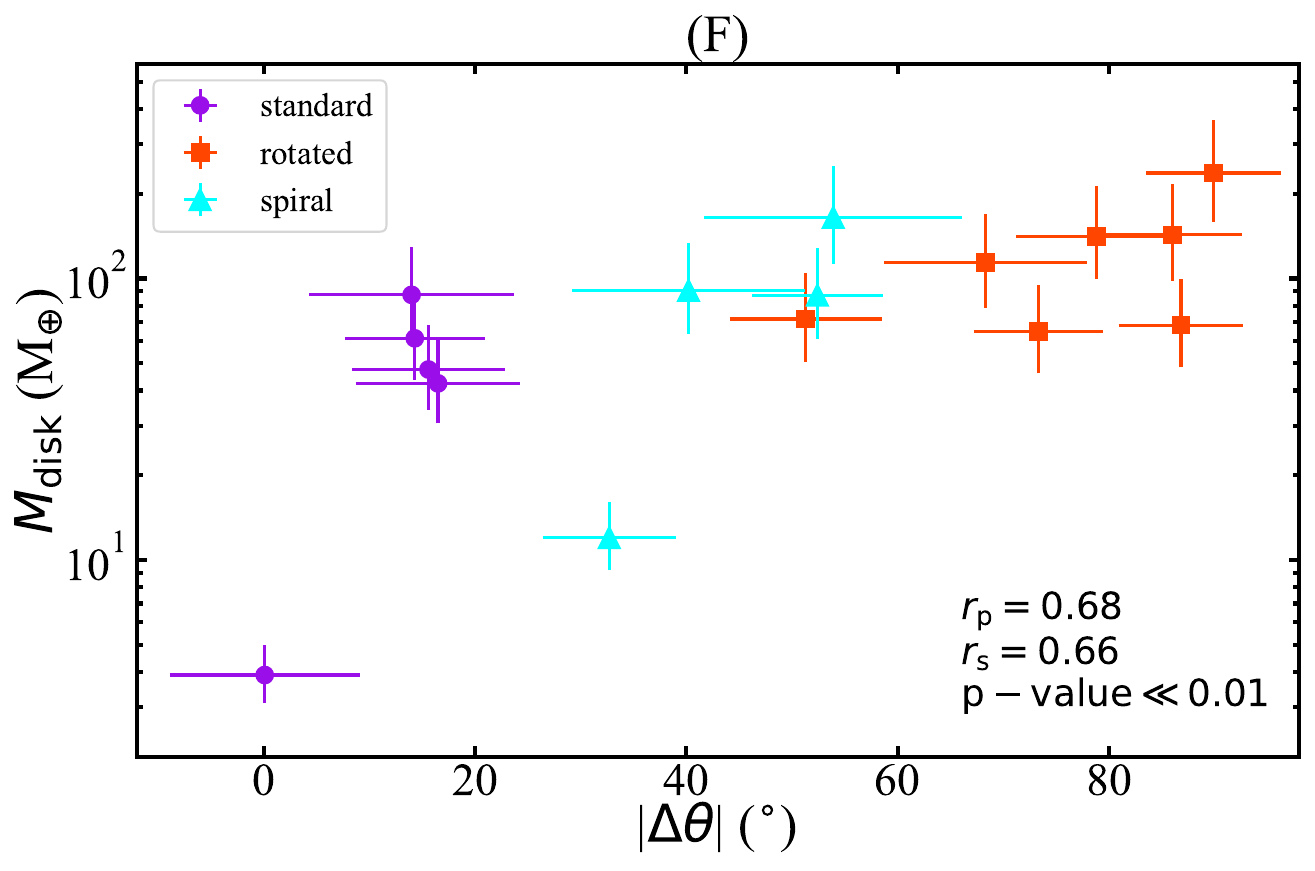}
\caption{Panel A--C: Distributions between the fragmentation level $N_{\rm F}$ and the envelope mass $M_{\rm env}$ (panel A),  the non-thermal velocity dispersion $\sigma_{\rm nth}$ (panel B), and the angle dispersion of {\em B}-field $\delta\phi$ (panel C).
Panel D--F: Distributions between the disk radius $R_{\rm disk}$ and the alignment of the {\em B}-field and outflow axes $\vert \Delta\theta \vert$ (panel D), and between the disk mass $M_{\rm disk}$ and $\vert \Delta\theta \vert$ (panel E and F).
In panel F, disk masses with their disk radii less than 100 au are derived from \cite{sheehan2022vandam}.} 
\label{fig:fr}
\end{figure*}

Figure \ref{fig:fr}(A--C) show the relation between $N_{\rm F}$ and $M_{\rm env}$ for 61 protostars, between $N_{\rm F}$ and $\sigma_{\rm nth}$ for 60 protostars (HOPS-408 has no C$^{17}$O detection, so its $\sigma_{\rm nth}$ cannot be resolved), and between $N_{\rm F}$ and $\delta\phi$ for 18 protostars out of 25 with enough polarzation statistics (the other 7 have a complex large-scale {\em B}-field morphology), respectively.
It is important to note that the values of $N_{\rm F}$ are dominated by multiples within the inner 2400 au region, a scale that is consistent with the one used to estimate the physical parameters as discussed in Section \ref{sec:analysis}. 
Combined with the statistical test methods, we discuss the potential influence of specific factors on envelope fragmentation at the 2400 au scale in the following.

Optical and near infrared studies suggest that the multiplicity rate drops with decreasing mass \citep[e.g.,][]{chini2012multi, chini2013multi}, we thus investigated the correlation between $N_{\rm F}$ and $M_{\rm env}$.
Based on the correlation coefficients shown in Figure \ref{fig:fr}(A), there is no clear correlation between $N_{\rm F}$ and $M_{\rm env}$.
This result may suggest that the correlation observed in the optical/infrared is not yet established at the very early stages of star and cluster formation. 
It is possible, therefore, that the final number of members in a multiple system is determined after multiple star formation episodes, as suggested by both theoretical \citep[e.g.,][]{myers2011ism} and observational\citep[e.g.,][]{bik2012fra} studies. 
Furthermore, the process of fragmentation and the formation of multiple objects are typically governed by a complex interplay of factors beyond just the envelope mass.
These interactions may lead to significant variability in outcomes, masking any simple correlation between envelope mass and multiplicity.
Previous observational studies of massive dense cores also suggest that there is no correlation between mass and multiplicity within a radius of 0.05 pc of the dust peak \citep[e.g.,][]{palau2013frag}.

In recent decades, several studies on the turbulence scenario \citep[e.g.,][]{padoan2001turbulence, vazquez2007frag} have demonstrated that turbulence is capable of generating density structures through the action of supersonic shocks, which compress and fragment the gas efficiently, particularly in the presence of significant turbulence.
Subsequent studies have indicated that different types of turbulence, i.e. compressive and solenoidal, exert divergent effects on fragmentation \citep[e.g.,][]{federrath2008frag}.
Compressive turbulence is conducive to fragmentation, whereas solenoidal turbulence has the effect on slowing down fragmentation \citep[e.g.,][]{federrath2010frag, girichidis2011frag}.
As shown in Figure \ref{fig:fr}(B), the fragmentation level $N_{\rm F}$ is uncorrelated with $\sigma_{\rm nth}$.
However, it is important to note that $\sigma_{\rm nth}$ may include contributions from rotational and infalling motions on envelope scales, which cannot be distinguished with the limited spectral resolution of C$^{17}$O.
Therefore, deeper studies with sufficient spectral resolution should be carried out to reach more robust conclusions.

On the other hand, the fragmentation level in massive dense cores could be explained by a {\em B}-field/turbulence balance \citep{palau2013frag} or changes in the mass-to-flux ratio \citep{busquet2016frag, nacho2020frag, palau2021frag} on scales of several thousands au, suggesting a potential correlation between {\em B}-fields and fragmentation on envelope scales.
One of the most longstanding methods for estimating the {\em B}-field strength is the Davis-Chandrasekhar-Fermi \cite[DCF,][]{davis1951dcf, chandrasekhar1953dcf} method, which relies on the principle that turbulent motions should visibly affect the distribution of the {\em B}-field in the plane of the sky.
The basic form of the DCF formula is
$B_{\rm{pos}} = \sqrt{4\pi\rho} \sigma_{\rm nth}/\delta\phi$, where $\rho$ is the gas density, and $\delta\phi$ is the measured angle dispersion of the {\em B}-field (as discussed in Section \ref{subsec:disp}).
Since $\sigma_{\rm nth}$ is barely resolved for most sources in our sample, which significantly affect the estimation of {\em B}-field strength.
However, the above equation indicates that the field strength is inversely proportional to the dispersion in the polarization position angles, which provides a potential way to indirectly study the correlation between field strength and other physical parameters on envelope scales.
In order to statistically test this correlation, we plot the distribution between the angle dispersion of the {\em B}-field $\delta\phi$ and the fragmentation level $N_{\rm F}$, as shown in Figure \ref{fig:fr}(C).
Combing correlation coefficients, we find that $N_{\rm F}$ is positively correlated with $\delta\phi$.
This suggests that higher angle dispersions in the {\em B}-field is associated with higher multiplicities, and thus envelope fragmentation tends to be suppressed by the {\em B}-field.
Previous studies have indicated that strong {\em B}-fields suppress fragmentation, as they constitute an additional form of support against gravitational contraction and also via the magnetic effect (inward pinching and gathering of the field lines dragged during collapse makes the magnetic strength larger at inner radius, and acts against rotation), which reduces the amount of angular momentum in the inner part of cores \citep[e.g.,][]{galli2006mag, mellon2008mag, hennebelle2008mag, price2009magnetic, myers2014star, cunningham2018mag, kwon2019mag, yen2021mag}.

In addition, multiplicity may also decrease with the evolution from Class 0 to Class I \citep{tobin2020vla, tobin2022vla}.
However, since the BOPS sample predominantly consists of objects at the Class 0 stage, the sample size in our study is insufficient to differentiate evolutionary factors over time.

\subsection{\textit{B}-field and Disk Formation}

Observations indicate that cloud cores in star-forming environments are magnetized \citep[e.g.,][]{crutcher1999ism}.
Theoretical models suggest that {\em B}-field tension forces deflect infalling gas towards the equatorial plane, resulting in a dense, flattened pseudo-disk \citep[e.g.,][]{allen2003collapse}.
During the collapse phase, the field lines are drawn inward into a pinched geometry \citep[e.g.,][]{girart2006magnetic, girart2009magnetic, stephens2013hourglass, kwon2019mag}, which strengthens the {\em B}-field and carries the angular momentum outward.
This process is known as magnetic braking, which reduce angular momentum by one order of magnitude, potentially suppressing the formation of a centrifugally supported disk \citep[e.g.,][]{galli2006mag}.
Figure \ref{fig:fr}(D) presents the distribution between the disk size $R_{\rm disk}$ and the {\em B}-field angle dispersion ($\delta \phi$, which is inversely proportional to the {\em B}-field strength) for 11 protostars with good detection in the disk radius. 
We find that protostellar envelopes with a standard hourglass {\em B}-field structure tend to have a smaller disk radius and angle dispersion of the {\em B}-field than those with a rotated hourglass field structure, suggesting that magnetic braking is more effective for the standard hourglass field shape.
Some studies propose that an initial misalignment between the {\em B}-field and the rotation axis would lead to weaker magnetic braking of the collapsing core, enabling earlier disk formation \citep[e.g.,][]{joos2012protostellar, li2013misalignment, hirano2019Bmisalign, machida2020misalignment}.

\cite{huang2024magnetic} have revealed the dust polarized continuum emission and outflow signatures in protostars, finding that $\sim$ 40\% of the protostars exhibit a mean {\em B}-field direction that is approximately perpendicular to the outflow on envelope scales.
However, the relative orientation in the remaining sample appears to be random.
We combine the disk radius/mass and envelope misalignment between the {\em B}-field and outflow for BOPS protostars, as shown in Figure \ref{fig:fr}(E--F).
Protostars with complex and unresolved polarized emission, or disks significantly affected by envelope emission and other component of disk emission (see Table \ref{Tab:parameters} in Appendix \ref{app:A} and figures in Appendix \ref{app:B}) are excluded in our analysis.
This exclusion is due to the substantial uncertainty in the averaged {\em B}-field direction or disk radius/mass for these sources.
Then the number of sources used in Figure \ref{fig:fr}(E) and Figure \ref{fig:fr}(F) is 16.
In panel E, protostars with standard hourglass, rotated hourglass, and spiral {\em B}-field structures exhibit a strong positive correlation between disk radius $R_{\rm disk}$ and the misalignment angle between magnetic and outflow axes $\vert\Delta\theta\vert$.
A similar discernible trend can be found from Figure \ref{fig:fr}(F) that $M_{\rm disk}$ is positively correlated with $\vert\Delta\theta\vert$, but the correlation is weaker compared to the $R_{\rm disk}$-$\vert\Delta\theta\vert$ relation.
If the bottom-left point with compact disk emission in Figure \ref{fig:fr}(E--F) is excluded, the updated correlation coefficients are $r_{p}\sim0.74$, $r_{s}\sim0.75$, p--value $<0.01$ for panel E, and $r_{p}\sim0.62$, $r_{s}\sim0.59$, p--value $\sim0.02$ for panel F, respectively.
These results indicate that the correlations between the disk radius/mass and the misalignment angle remain consistent, suggesting that the correlation is less affected by this point.
The p-values, which are $<$ 0.05, suggest that these correlations are statistically significant.
In addition, it is important to note from the distribution of both panel E and panel F that protostars exhibiting standard hourglass {\em B}-field tend to have lower disk size, compared to those exhibiting spiral and rotated hourglass field structure.

\cite{huang2024magnetic} applied both the intensity-weighted method (which gives more weight to the inner protostellar envelope) and the error-weighted method (which provides an averaged weight across the selected envelope) to estimate the mean direction of the (\textit{B}-field). 
The mean direction distributions of \textit{B}-field obtained from both weighting methods were found to be comparable throughout the sample, suggesting that the measured \textit{B}-field is minimally influenced by gravitational collapse and outflows.
In this case, the measured \textit{B}-field is likely representative of the original pre-stellar core \textit{B}-field. 
Furthermore, the outflow direction aligns with the angular momentum of the original core. 
Our findings suggest that a greater misalignment between these two directions lead to weaker magnetic braking of the collapsing core and result in a more prominent disk, which is consistent with the misalignment-induced disk formation model \citep[e.g.,][]{joos2012protostellar}.

\section{Conclusions} \label{sec:Con}

We have calculated and analyzed the physical parameters for 61 protostelar envelopes within the BOPS sample. 
The combination of the envelope properties derived from the BOPS \citep{huang2024magnetic} and the disk properties derived from the VANDAM survey \citep{tobin2020vla} allows us to link the turbulence and {\em B}-field in the envelope with the resulting fragmentation and disk properties.
The main results are as follows:

\begin{enumerate}
\item Protostellar envelopes exhibiting standard-hourglass field shape show higher values of envelope mass, gas column density, and gas number density than those with other field configurations, lower non-thermal velocity dispersion have also be observed in this type. 
In contrast, envelopes with a spiral \textit{B}-field structure tend to have higher non-thermal velocity dispersion, indicating a more turbulent environment.

\item The angle dispersion of the {\em B}-field, which is inversely proportional to the {\em B}-field strength, is positively correlated with the fragmentation level, suggesting that the pinched field suppresses the fragmentation, as it provides additional support against gravitational collapse, and/or brakes the rotation efficiently.
However, the lack of correlations between the fragmentation level and the envelope mass, and between the fragmentation level and the non-thermal velocity dispersion, may be due to the complex interplay of various factors influencing fragmentation beyond just the envelope mass, or to the limited spectral resolution.

\item Protostars with standard hourglass {\em B}-field structure tend to have smaller disk size and angle dispersion of the magnetic field than those with other field configurations, specially rotated hourglasses, but also spirals and others.
This likely suggests that in the case of the standard hourglass field structure, the magnetic field strength is stronger, ultimately leading to a smaller disk through the more effective magnetic braking.
However, due to the limited sample size of protostars with standard hourglass {\em B}-field structure, further study with more statistics is required to substantiate this scenario.

\item A more significant misalignment between the magnetic and outflow axes is associated with larger disk radii and disk masses.
This suggests that increased misalignment results in weaker magnetic braking during core collapse, which aligns with the misalignment-induced disk formation model.

\item The dust temperature and bolometric luminosity are not correlated with envelope masses and densities, probably because the cloud with low bolometric luminosity provides shielding from the interstellar radiation field.
In addition, no clear correlation can be found between non-thermal velocity dispersion and luminosity on envelope scales, contrast with the simulation models predicted.
Deeper studies on individual cores or higher-resolution observations might reveal more localized correlations that are otherwise masked by the larger-scale dynamics.

\end{enumerate}

Insignificant correlations found for some pairs of parameters, especially in the analysis of the non-thermal velocity dispersion, may be due to the limited sensitivity and resolution of the observations.
It is important to note that the non-thermal velocity dispersion may include contributions from disk rotation, envelope rotation, and/or infalling motions, further observations with higher sensitivity and resolution should be carried out to distinguish them and further refine our conclusions about the non-thermal velocity dispersion or the turbulence.

\begin{acknowledgments}
B.H. and J.M.G. acknowledge support by the grant PID2020-117710GB-I00 and PID2023-146675NB-I0 (MCI-AEI-FEDER, UE). 
B.H. also acknowledges financial support from the China Scholarship Council (CSC) under grant No. 202006660008.
This work is also partially supported by the program Unidad de Excelencia María de Maeztu CEX2020-001058-M.
M.F.L. acknowledges support from the European Research Executive Agency HORIZON-MSCA-2021-SE-01 Research and Innovation programme under the Marie Skłodowska-Curie grant agreement number 101086388 (LACEGAL). M.F.L. also acknowledges the warmth and hospitality of the ICE-UB group of star formation.
L.W.L. acknowledges support by NSF AST-1910364 and NSF AST-2307844.
Z.Y.L. is supported in part by NASA 80NSSC20K0533 and NSF AST-2307199.
W.K. is supported by the National Research Foundation of Korea (NRF) grant funded by the Korea government (MSIT) (RS-2024-00342488).
The authors acknowledge Anaëlle Maury and Chat Hull for helpful discussions, Jacob Labonte for early analysis of the BOPS data.
This paper makes use of the following ALMA data: ADS/JAO.ALMA\#2019.1.00086.
ALMA is a partnership of ESO (representing its member states), NSF (USA) and NINS (Japan), together with NRC (Canada), MOST and ASIAA (Taiwan), and KASI (Republic of Korea), in cooperation with the Republic of Chile. The Joint ALMA Observatory is operated by ESO, AUI/NRAO and NAOJ.
\end{acknowledgments}

\appendix

\section{Table} \label{app:A}

\begin{longtable}{lccccccccccccc}
\caption{\label{Tab:parameters} Some parameters for the BOPS sample}\\
\hline
\hline
\multirow{2}*{Name} & R.A. & Decl. & $D$ & $M_{\rm disk}$ & $R_{\rm disk}$ & $\langle \Theta_{\rm max} \times \Theta_{\rm min} \rangle_{\rm Decon.}$ & $\theta_{\rm Inc}$ & $\Delta\theta$ \\
~ & (h:m:s) & (d:m:s) & (pc) & (M$_{\oplus}$) & (au) & ($^{\prime\prime}\times^{\prime\prime}$) & ($^{\circ}$) & ($^{\circ}$) \\
\hline
\endfirsthead
\caption{Envelope and disk parameters for the BOPS sample\tablenotemark{$\dagger$}}\\
\hline\hline
\multirow{2}*{Name} & R.A. & Decl. & $D$ & $M_{\rm disk}$ & $R_{\rm disk}$ & $\langle \Theta_{\rm max} \times \Theta_{\rm min} \rangle_{\rm Decon.}$ & $\theta_{\rm Inc}$ & $\Delta\theta$ \\
~ & (h:m:s) & (d:m:s) & (pc) & (M$_{\oplus}$) & (au) & ($^{\prime\prime}\times^{\prime\prime}$) & ($^{\circ}$) & ($^{\circ}$) \\
\hline
\endhead
\hline
\endfoot
\textbf{Standard} \\
\hline
HOPS-11 & 05:35:13:43 & -05:57:57.91 & 388.3 & 87.5$^{+41.7}_{-26.2}$ & 41.9 $\pm$ 10.0 & 0.13 $\times$ 0.12 & 22.6 & 14.0 $\pm$ 9.7  \\
HOPS-87N\tablenotemark{a} & 05:35:23.42 & -05:01:30.57 & 392.7 & / & / & / & / & / \\
HOPS-359 & 05:47:24.84 & ~00:20:59.39 & 429.4 & 47.5$^{+20.6}_{-13.3}$ & 62.0 $\pm$ 10.0 & 0.17 $\times$ 0.15 & 28.1 & 15.6 $\pm$ 7.2 \\
HOPS-395\tablenotemark{d} & 05:39:17.09 & -07:24:24.59 & 397.2 & 3.9$^{+1.1}_{-0.8}$ & 7.1 $\pm$ 10.0 & 0.02 $\times$ 0.02 & / & 0.1 $\pm$ 9.0 \\
HOPS-400 & 05:42:45.26 & -01:16:13.89 & 415.4 & 61.3$^{+27.7}_{-17.7}$ & 48.7 $\pm$ 10.0 & 0.14 $\times$ 0.13 & 21.3 & 14.3 $\pm$ 6.6 \\
HOPS-407 & 05:46:28.25 & ~00:19:27.95 & 419.1 & 42.4$^{+18.1}_{-11.8}$ & 64.1 $\pm$ 10.0 & 0.18 $\times$ 0.14 & 38.9 & 16.5 $\pm$ 7.8 \\
OMC1N-8-N\tablenotemark{c} & 05:35:18.20 & -05:20:48.57 & 392.8 & / & / & / & / & / \\
\hline
\textbf{Rotated} \\
\hline
HH270IRS & 05:51:22.72 & ~02:56:05:00 & 405.7 & 113.8$^{+56.3}_{-35.0}$ & 92.3 $\pm$ 10.0 & 0.27 $\times$ 0.14 & 58.8 & 68.3 $\pm$ 9.6 \\
HOPS-78  & 05:35:25.96 & -05:05:43.38 & 392.8 & 142.9$^{+73.0}_{-44.9}$ & 147.1 $\pm$ 10.0 & 0.44 $\times$ 0.14 & 71.4 & 86.0 $\pm$ 6.6 \\
HOPS-168\tablenotemark{b} & 05:36:18.94 & -06:45:23.58 & 383.3 & / & / & / & / & / \\
HOPS-169 & 05:36:36.16 & -06:38:54.41 & 384.0 & 71.8$^{+33.2}_{-21.1}$ & 53.8 $\pm$ 10.0 & 0.17 $\times$ 0.13 & 40.1 & 51.3 $\pm$ 7.2 \\
HOPS-288\tablenotemark{b} & 05:39:56.00 & -07:30:27.62 & 405.5 & / & / & / & / & / \\
HOPS-310 & 05:42:27.68 & -01:20:01.35 & 414.3 & 237.2$^{+129.7}_{-78.0}$ & 101.3 $\pm$ 10.0 & 0.29 $\times$ 0.15 & 58.9 & 89.9 $\pm$ 6.4 \\
HOPS-317S\tablenotemark{a} & 05:46:08.38 & -00:10:43.59 & 427.1 & / & / & / & / & / \\
HOPS-370 & 05:35:27.64 & -05:09:34.40 & 392.8 & 68.3$^{+31.4}_{-20.0}$ & 112.8 $\pm$ 10.0 & 0.34 $\times$ 0.11 & 71.1 & 86.8 $\pm$ 5.9 \\
HOPS-401 & 05:46:07.73 & -00:12:21.31 & 426.9 & 140.7$^{+71.7}_{-41.1}$ & 74.7 $\pm$ 10.0 & 0.21 $\times$ 0.16 & 40.4 & 78.8 $\pm$ 7.6 \\
HOPS-409 & 05:35:21.37 & -05:13:17.88 & 392.8 & 64.9$^{+29.6}_{-18.9}$ & 112.1 $\pm$ 10.0 & 0.34 $\times$ 0.12 & 69.3 & 73.3 $\pm$ 6.1 \\
OMC1N-4-5-ES\tablenotemark{c} & 05:35:15.97 & -05:20:14.23 & 392.8 & / & / & / & / & / \\
\hline
\textbf{Spiral} \\
\hline
HOPS-96 & 05:35:29.72 & -04:58:48.63 & 392.7 & 87.3$^{+41.5}_{-26.1}$ & 41.4 $\pm$ 10.0 & 0.12 $\times$ 0.11 & 23.6 & 52.4$\pm$ 6.2 \\
HOPS-124\tablenotemark{ab} & 05:39:19.91 & -07:26:11.22 & 398.0 & / & / & / & / & / \\
HOPS-182\tablenotemark{d} & 05:36:18.79 & -06:22:10.24 & 385.1 & 12.0$^{+4.1}_{-2.8}$ & 36.3 $\pm$ 10.0 & 0.11 $\times$ 0.06 & / & 32.7 $\pm$ 6.3 \\
HOPS-303 & 05:42:02.65 & -02:07:45.94 & 410.0 & 164.9$^{+85.9}_{-52.5}$ & 82.9 $\pm$ 10.0 & 0.24 $\times$ 0.19 & 37.7 & 53.9 $\pm$ 12.2 \\
HOPS-361N\tablenotemark{b} & 05:47:04.63 & ~00:21:47.82 & 430.4 & / & / & / & / & / \\
HOPS-361S & 05:47:04.78 & ~00:21:42.79 & 430.4 & 90.8$^{+43.4}_{-27.3}$ & 96.1 $\pm$ 10.0 & 0.26 $\times$ 0.15 & 54.8 & 40.2 $\pm$ 11.0 \\
HOPS-384\tablenotemark{b} & 05:41:44.14 & -01:54:46.00 & 409.5 & / & / & / & / & / \\
HOPS-403\tablenotemark{a} & 05:46:27.91 & -00:00:52.15 & 428.2 & / & / & / & / & / \\
HOPS-404\tablenotemark{a} & 05:48:07.72 & ~00:33:51.81 & 430.1 & / & / & / & / & / \\
\hline
\textbf{Others} \\
\hline
HOPS-12W\tablenotemark{d} & 05:35:08.63 & -05:55:54.65 & 388.6 & 5.1$^{+1.5}_{-1.1}$ & 20.5 $\pm$ 10.0 & 0.06 $\times$ 0.05 & / & / \\
HOPS-87S\tablenotemark{c} & 05:35:23.67 & -05:01:40.27 & 392.7 & / & / & / & / & / \\
HOPS-88\tablenotemark{d} & 05:35:22.47 & -05:01:14.33 & 392.7 & 33.1$^{+13.6}_{-8.9}$ & 34.0 $\pm$ 10.0 & 0.10 $\times$ 0.09 & / & / \\
HOPS-164 & 05:37:00.43 & -06:37:10.91 & 385.0 & 85.8$^{+40.8}_{-25.7}$ & 70.6 $\pm$ 10.0 & 0.22 $\times$ 0.14 & 50.5 & / \\
HOPS-224 & 05:41:32.07 & -08:40:09.82 & 440.3 & 201.6$^{+107.9}_{-65.4}$ & 107.7 $\pm$ 10.0 & 0.29 $\times$ 0.18 & 51.6 & / \\
HOPS-250 & 05:40:48.85 & -08:06:57.11 & 428.5 & 86.2$^{+41.0}_{-25.8}$ & 117.2 $\pm$ 10.0 & 0.32 $\times$ 0.19 & 53.6 & / \\
HOPS-317N\tablenotemark{d} & 05:46:08.60 & -00:10:38.49 & 427.1 & 20.8$^{+7.9}_{-5.3}$ & 34.5 $\pm$ 10.0 & 0.10 $\times$ 0.09 & / & / \\
HOPS-325\tablenotemark{d} & 05:46:39.20 & ~00:01:12.30 & 428.5 & 25.1$^{+9.9}_{-6.6}$ & 34.9 $\pm$ 10.0 & 0.10 $\times$ 0.09 & / & / \\
HOPS-341\tablenotemark{d} & 05:47:00.92 & ~00:26:21.50 & 430.9 & 8.5$^{+2.7}_{-1.9}$ & 54.5 $\pm$ 10.0 & 0.15 $\times$ 0.09 & / & / \\
HOPS-373E & 05:46:31.10 & -00:02:33.05 & 428.1 & 39.4$^{+16.6}_{-10.8}$ & 69.8 $\pm$ 10.0 & 0.19 $\times$ 0.17 & 26.5 & / \\
HOPS-373W\tablenotemark{d} & 05:46:30.91 & -00:02:35.16 & 428.1 & 32.9$^{+13.5}_{-8.9}$ & 36.0 $\pm$ 10.0 & 0.10 $\times$ 0.09 & / & / \\
HOPS-398 & 05:41:29.41 & -02:21:16.39 & 408.0 & 248.0$^{+136.4}_{-81.9}$ & 76.9 $\pm$ 10.0 & 0.22 $\times$ 0.19 & 30.3 & / \\
HOPS-399 & 05:41:24.93 & -02:18:06.66 & 407.9 & 153.0$^{+78.9}_{-48.4}$ & 66.2 $\pm$ 10.0 & 0.19 $\times$ 0.15 & 37.9 & / \\
HOPS-402\tablenotemark{a} & 05:46:10.03 & -00:12:16.92 & 426.9 & / & / & / & / & / \\
HOPS-408 & 05:39:30.90 & -07:23:59.75 & 398.9 & 77.3$^{+36.2}_{-22.9}$ & 93.5 $\pm$ 10.0 & 0.28 $\times$ 0.24 & 31.0 & / \\
OMC1N-4-5-EN\tablenotemark{c} & 05:35:16.05 & -05:20:05.75 & 392.8 & / & / & / & / & / \\
OMC1N-6-7\tablenotemark{c} & 05:35:15.70 & -05:20:39.31 & 392.8 & / & / & / & / & / \\
OMC1N-8-S\tablenotemark{c} & 05:35:18.00 & -05:20:55.77 & 392.8 & / & / & / & / & / \\
\hline
\textbf{Scattering} \\
\hline
HH212M & 05:43:51.41 & -01:02:53.20 & 413.2 & 58.2$^{+26.1}_{-16.7}$ & 78.6 $\pm$ 10.0 & 0.22 $\times$ 0.10 & 63.0 & / \\
HOPS-10\tablenotemark{d} & 05:35:09.05 & -05:58:26.89 & 388.2 & 21.3$^{+8.1}_{-5.5}$  & 46.8 $\pm$ 10.0 & 0.14 $\times$ 0.07 & / & / \\
HOPS-12E\tablenotemark{b} & 05:35:08.95 & -05:55:54.99 & 388.6 & / & / & / & / & / \\
HOPS-50 & 05:34:40.92 & -05:31:44.74 & 391.5 & 126.1$^{+63.3}_{-39.2}$ & 82.8 $\pm$ 10.0 & 0.25 $\times$ 0.13 & 58.7 & / \\
HOPS-53\tablenotemark{d} & 05:33:57.40 & -05:23:30.00 & 390.5 & 10.9$^{+3.7}_{-2.5}$ & 21.6 $\pm$ 10.0 & 0.07 $\times$ 0.05 & / & / \\
HOPS-60 & 05:35:23.29 & -05:12:03.45 & 392.8 & 104.9$^{+51.3}_{-32.0}$ & 62.7 $\pm$ 10.0 & 0.19 $\times$ 0.11 & 54.6 & / \\
HOPS-81\tablenotemark{d} & 05:35:28.01 & -05:04:57.36 & 392.8 & 30.3$^{+12.2}_{-8.1}$ & 32.4 $\pm$ 10.0 & 0.10 $\times$ 0.06 & / & / \\
HOPS-84 & 05:35:26.56 & -05:03:55.15 & 392.8 & 111.6$^{+55.0}_{-34.2}$ & 87.4 $\pm$ 10.0 & 0.26 $\times$ 0.12 & 62.5 & / \\
HOPS-153 & 05:37:57.03 & -07:06:56.27 & 387.9 & 58.6$^{+26.3}_{-16.8}$ & 129.1 $\pm$ 10.0 & 0.39 $\times$ 0.10 & 75.1 & / \\
HOPS-203N\tablenotemark{b} & 05:36:22.87 & -06:46:06.63 & 383.5 & / & / & / & / & / \\
HOPS-203S\tablenotemark{d} & 05:36:22.90 & -06:46:09.54 & 383.5 & 10.7$^{+3.6}_{-2.5}$ & 30.9 $\pm$ 10.0 & 0.10 $\times$ 0.06 & / & / \\
HOPS-247 & 05:41:26.19 & -07:56:51.90 & 430.9 & 189.5$^{+100.6}_{-61.1}$ & 85.3 $\pm$ 10.0 & 0.23 $\times$ 0.16 & 45.9 & / \\
HOPS-340\tablenotemark{d} & 05:47:01.31 & ~00:26:23.05 & 430.9 & 25.9$^{+10.2}_{-6.8}$ & 41.7 $\pm$ 10.0 & 0.11 $\times$ 0.07 & / & / \\
HOPS-354\tablenotemark{d} & 05:54:24.26 & ~01:44:19.87 & 355.4 & 9.4$^{+3.1}_{-2.2}$ & 24.4 $\pm$ 10.0 & 0.08 $\times$ 0.07 & / & / \\
HOPS-358 & 05:46:07.26 & -00:13:30.25 & 426.8 & 64.9$^{+29.6}_{-18.9}$ & 136.3 $\pm$ 10.0 & 0.38 $\times$ 0.10 & 74.7 & / \\
HOPS-383 & 05:35:29.79 & -04:59:50.38 & 392.8 & 52.5$^{+23.1}_{-14.9}$ & 61.7 $\pm$ 10.0 & 0.18 $\times$ 0.12 & 48.2 & / \\
\hline 
\multicolumn{14}{p{10cm}}{\footnotesize \parbox{17.4cm}{%
Column 2 to 9 present the R.A., Decl., distance ($D$), disk mass ($M_{\rm disk}$), disk radius ($R_{\rm disk}$), deconvolved disk size ($\langle r_{\rm max} \rangle \times \langle r_{\rm min} \rangle$), disk inclined angle with respect to the line of sight ($\theta_{\rm Inc} = \arccos(\langle \Theta_{\rm min} \rangle / \langle \Theta_{\rm max} \rangle)$), and the difference between outflow and {\em B}-field direction ($\Delta\theta$), respectively.
The source distance is obtained from \cite{furlan2016herschel}; the disk radius and disk size and their uncertainty are obtained from \cite{tobin2020vla}; the disk mass is refined by \cite{sheehan2022vandam}.
The difference between outflow and {\em B}-field is obtained from \cite{huang2024magnetic}, which is the alignment between the mean outflow direction and the uncertainty-weighted mean {\em B}-field direction.

\tablenotemark{a} The disk emission is contaminated by the envelope dust emission.

\tablenotemark{b} The disk emission is contaminated by other disk emission.

\tablenotemark{c} No information on disk properties presented.

\tablenotemark{d} The disk is compact and partially resolved.
}
}
\end{longtable}

\section{Figures} \label{app:B}
\subsection{Self-scattering dominated protostars}
Left panels: the velocity field in color scale (obtained from the C$^{17}$O (3--2) line) overlaid with the 870~$\mu$m dust polarization segments and Stokes {\em I} contours.
Blue/red contours indicate the blue-/red-shifted outflow, with counter levels set at 5 times the outflow {\em rms} × (1, 2, 4, 8, 16, 32).
The black contour levels for the Stokes {\textit I} image are 10 times the {\em rms} × (1, 2, 4, 8, 16, 32, 64, 128, 256, 512).
For the velocity field, regions with an S/N less than 4 have been flagged.
The purple segments represent the polarization.
The magenta dashed circle corresponds to scales with a radius of 1200 au.
Right panels: 9 mm dust maps obtained from \cite{tobin2020vla}.

\begin{figure*}[!ht]
\centering
\includegraphics[clip=true,trim=0cm 0cm 0cm 0cm,width=0.32 \textwidth]{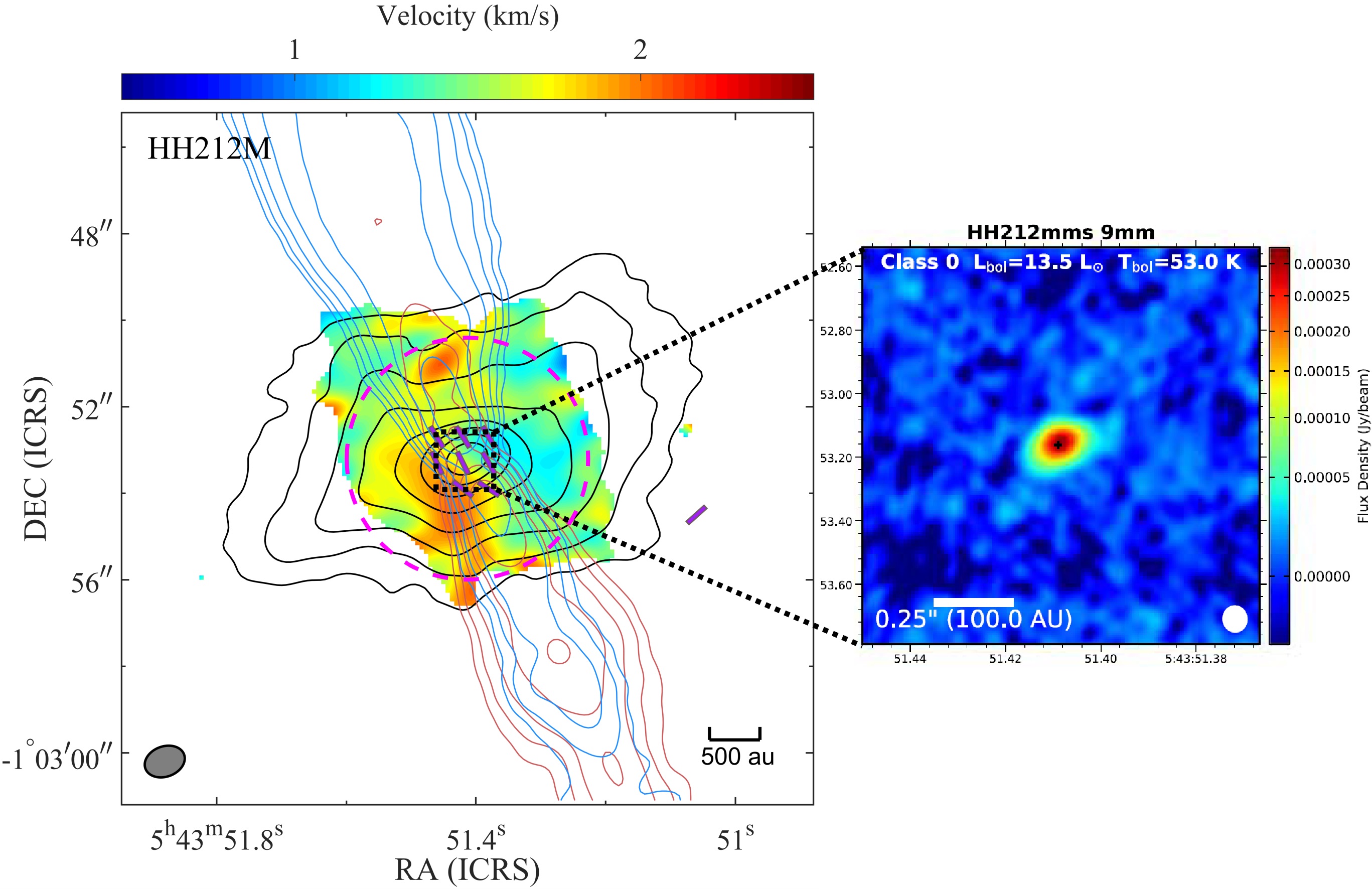}~~~~~
\includegraphics[clip=true,trim=0cm 0cm 0cm 0cm,width=0.32 \textwidth]{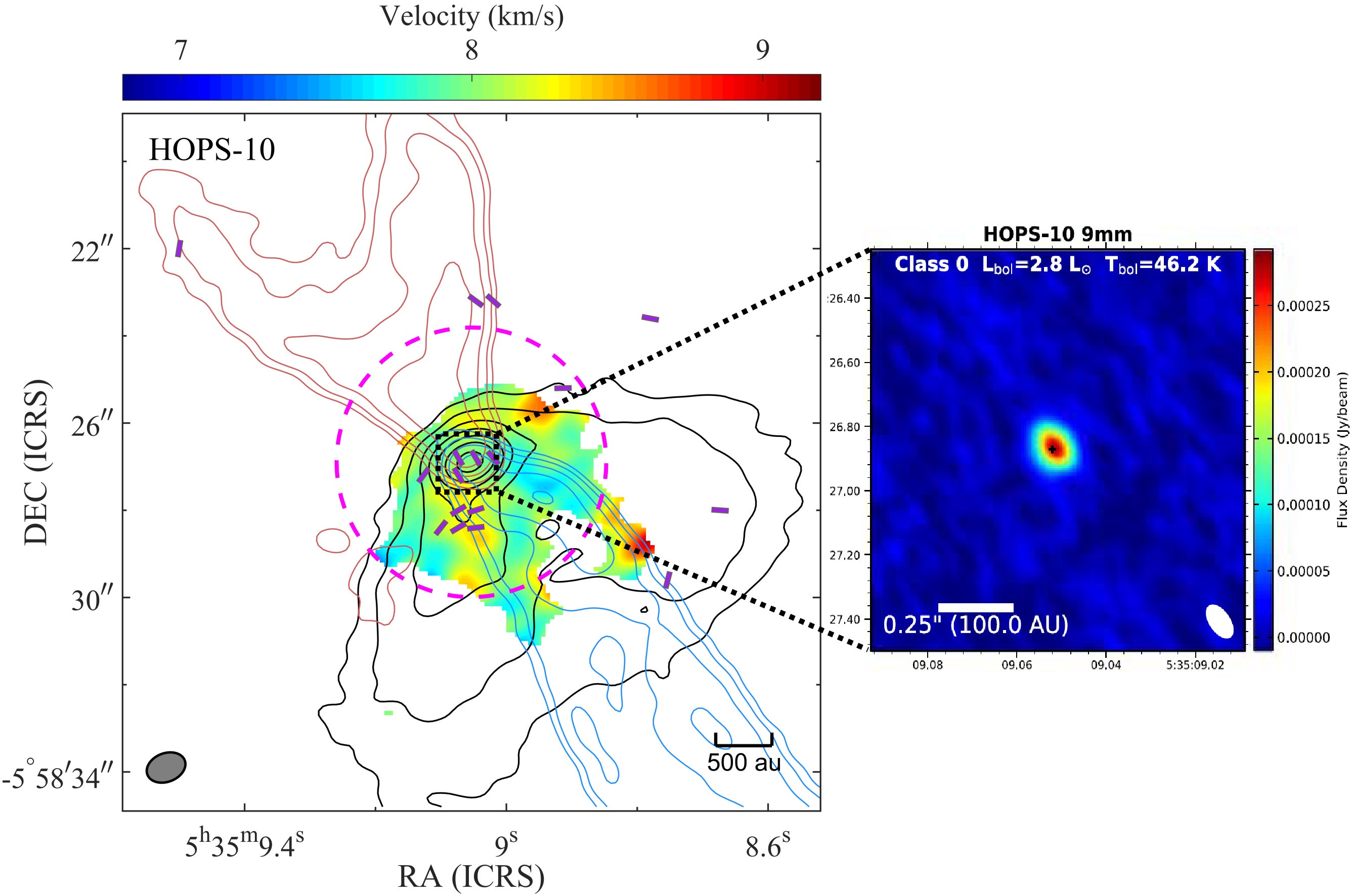}~~~~~
\includegraphics[clip=true,trim=0cm 0cm 0cm 0cm,width=0.327 \textwidth]{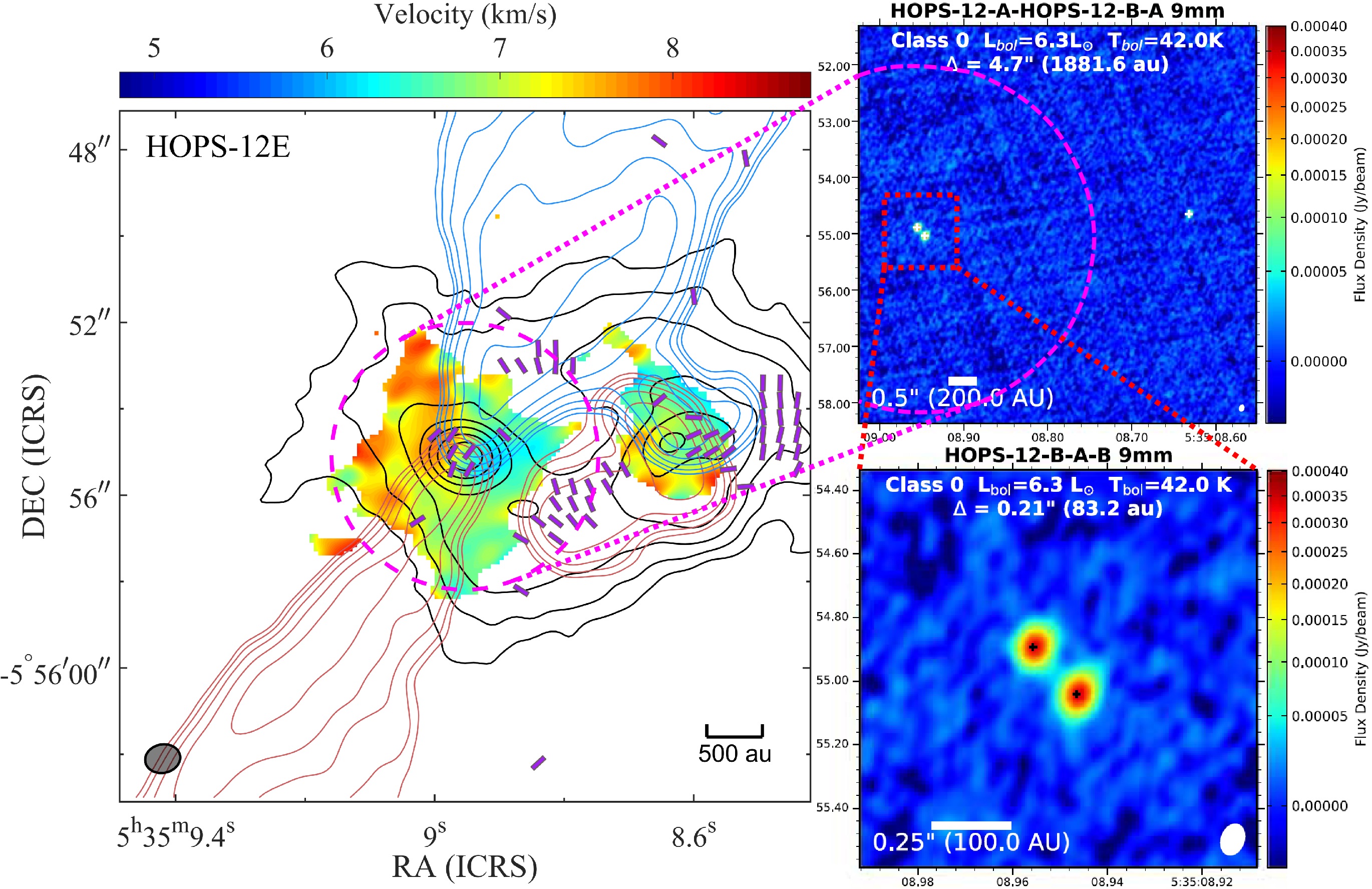}
~\\
~\\
\includegraphics[clip=true,trim=0cm 0cm 0cm 0cm,width=0.32 \textwidth]{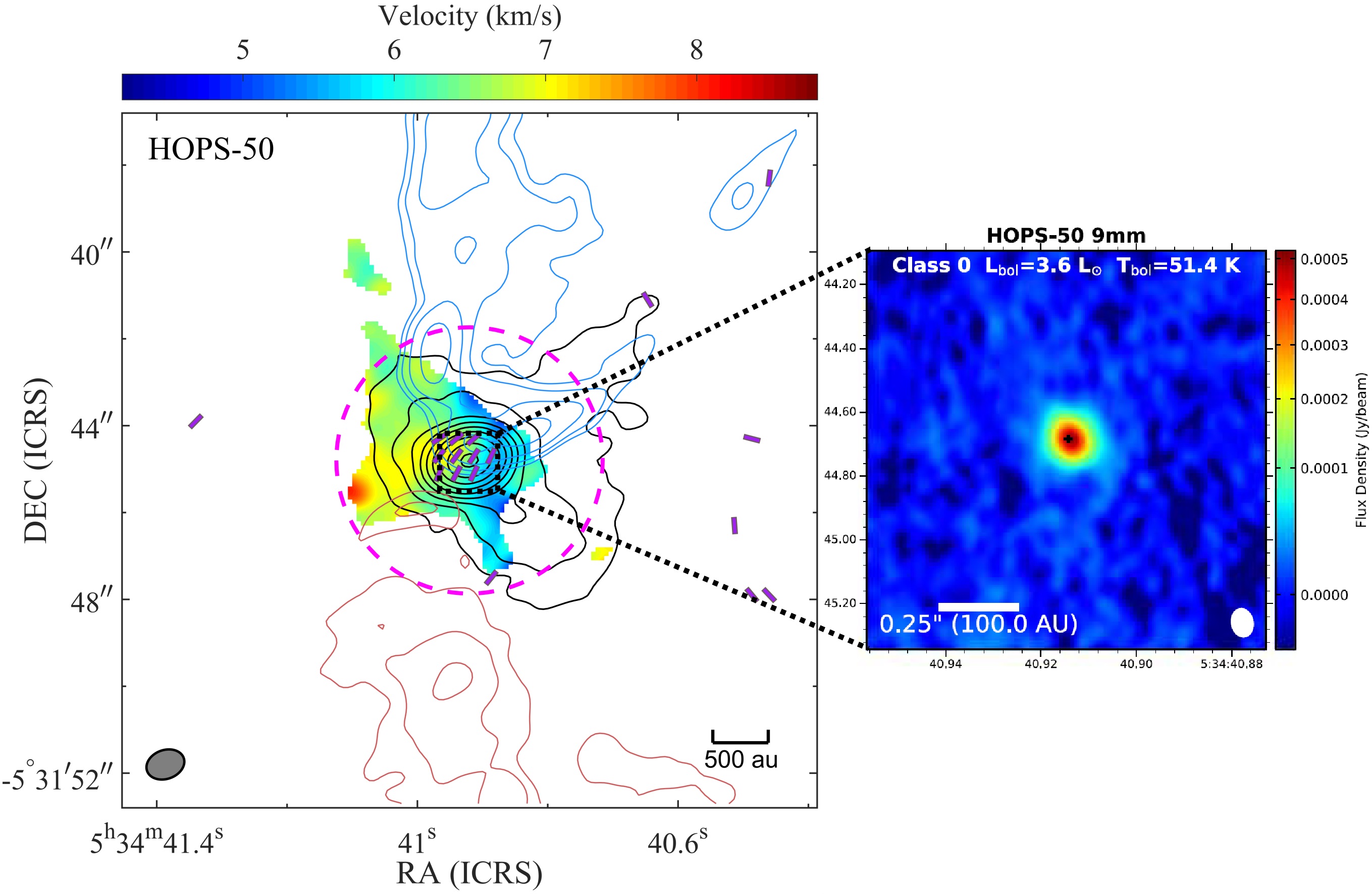}~~~~~
\includegraphics[clip=true,trim=0cm 0cm 0cm 0cm,width=0.32 \textwidth]{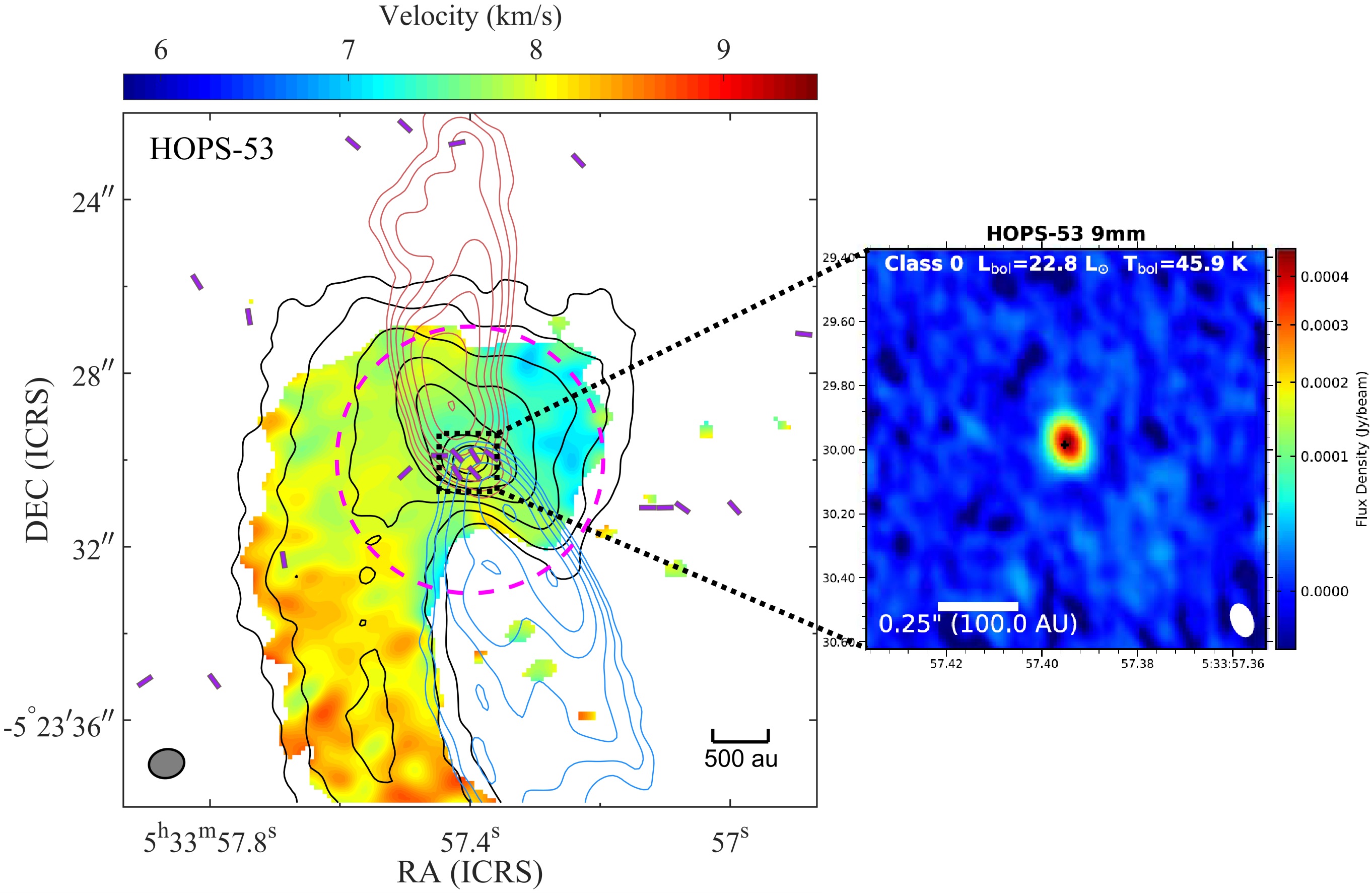}~~~~~
\includegraphics[clip=true,trim=0cm 0cm 0cm 0cm,width=0.32 \textwidth]{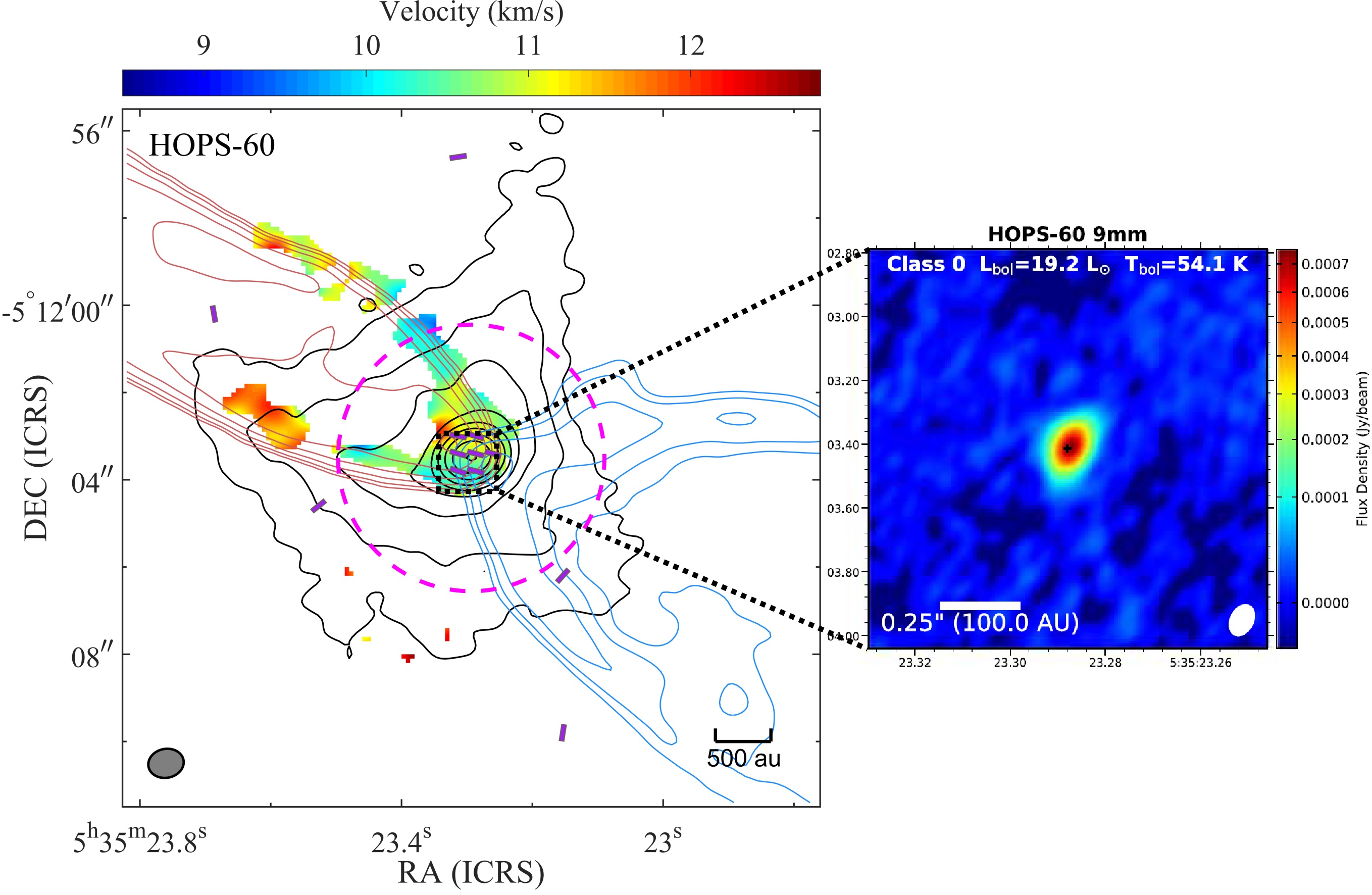}
~\\
~\\
\includegraphics[clip=true,trim=0cm 0cm 0cm 0cm,width=0.32 \textwidth]{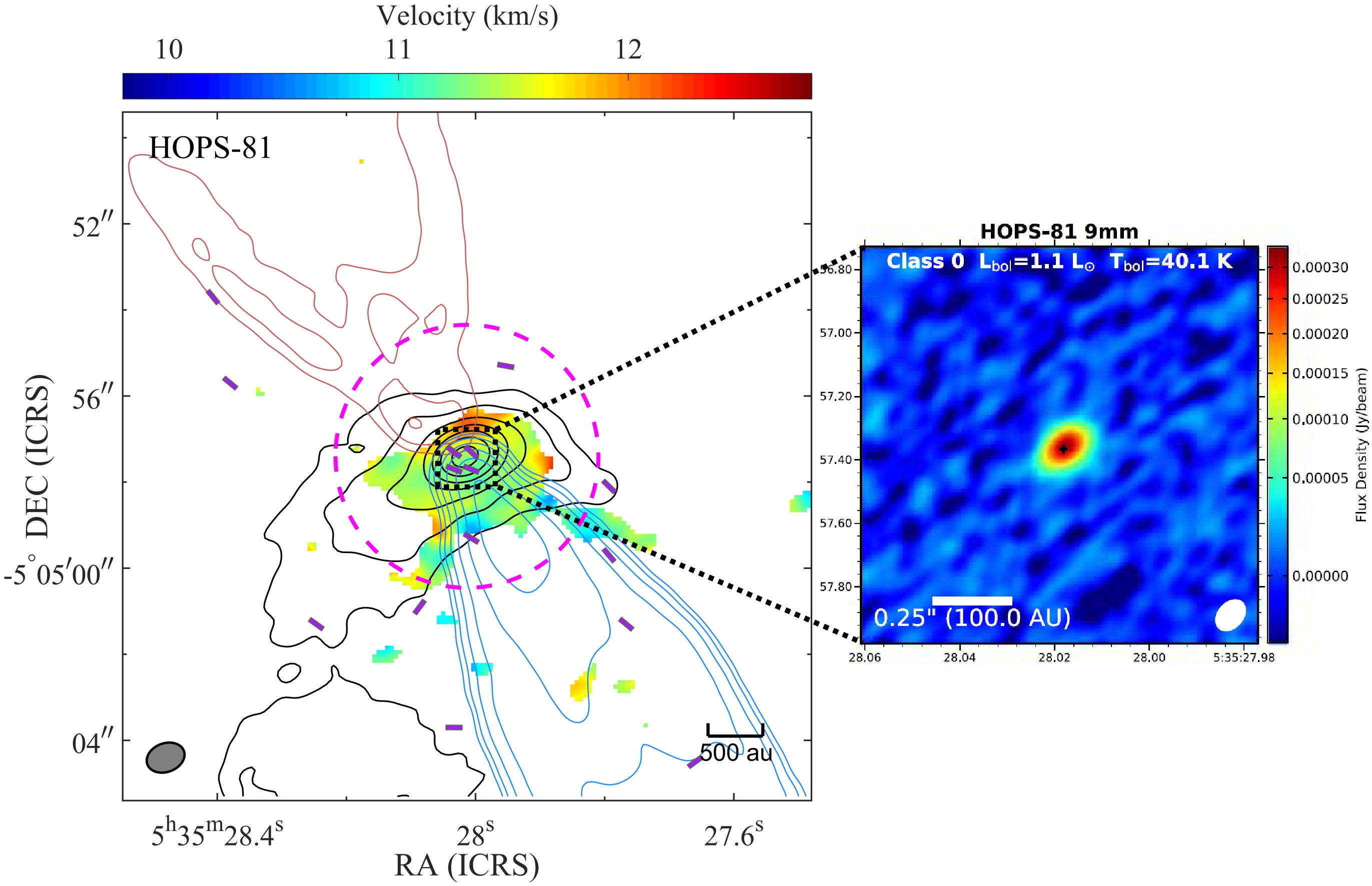}~~~~~
\includegraphics[clip=true,trim=0cm 0cm 0cm 0cm,width=0.32 \textwidth]{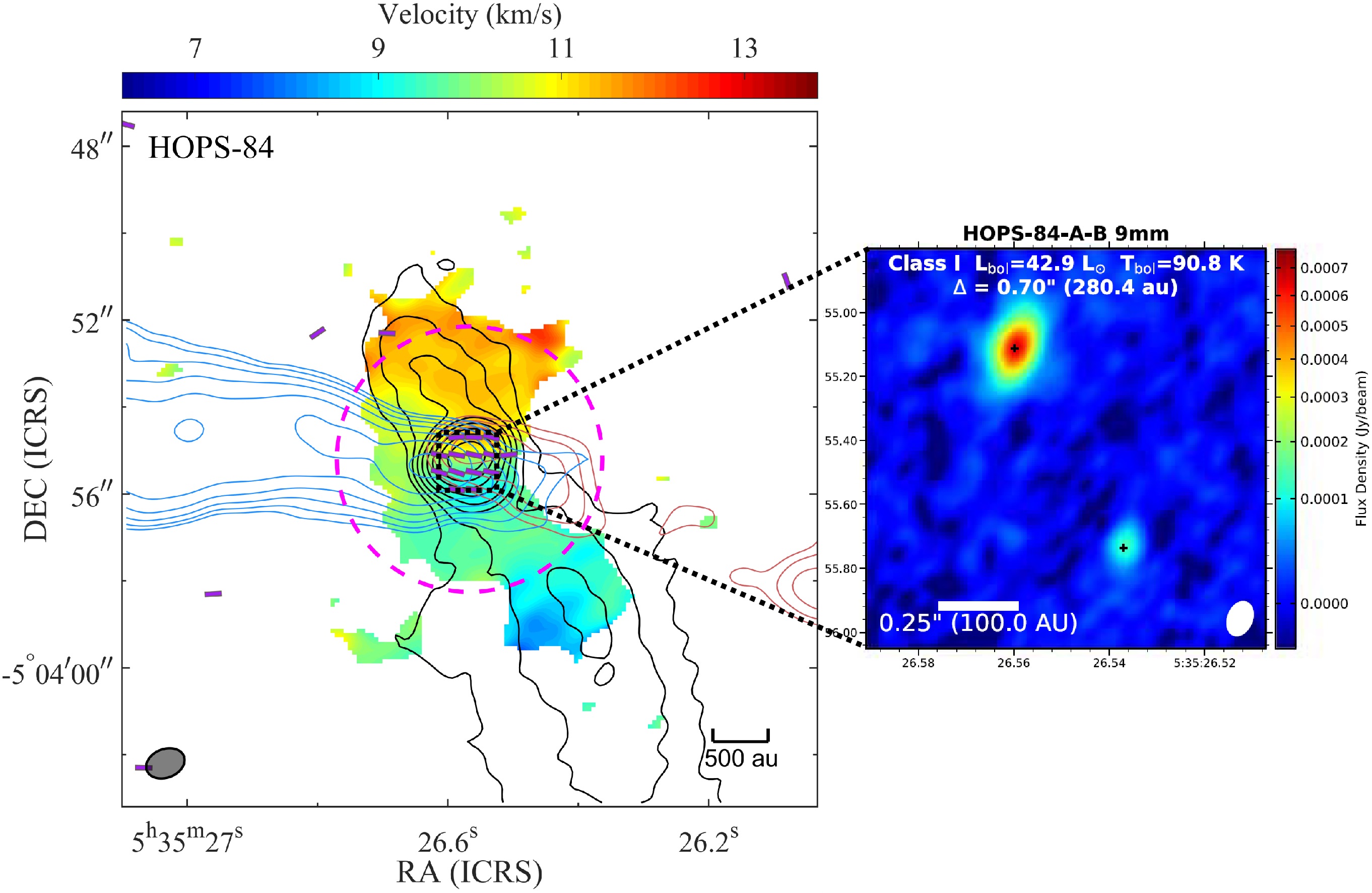}~~~~~
\includegraphics[clip=true,trim=0cm 0cm 0cm 0cm,width=0.32 \textwidth]{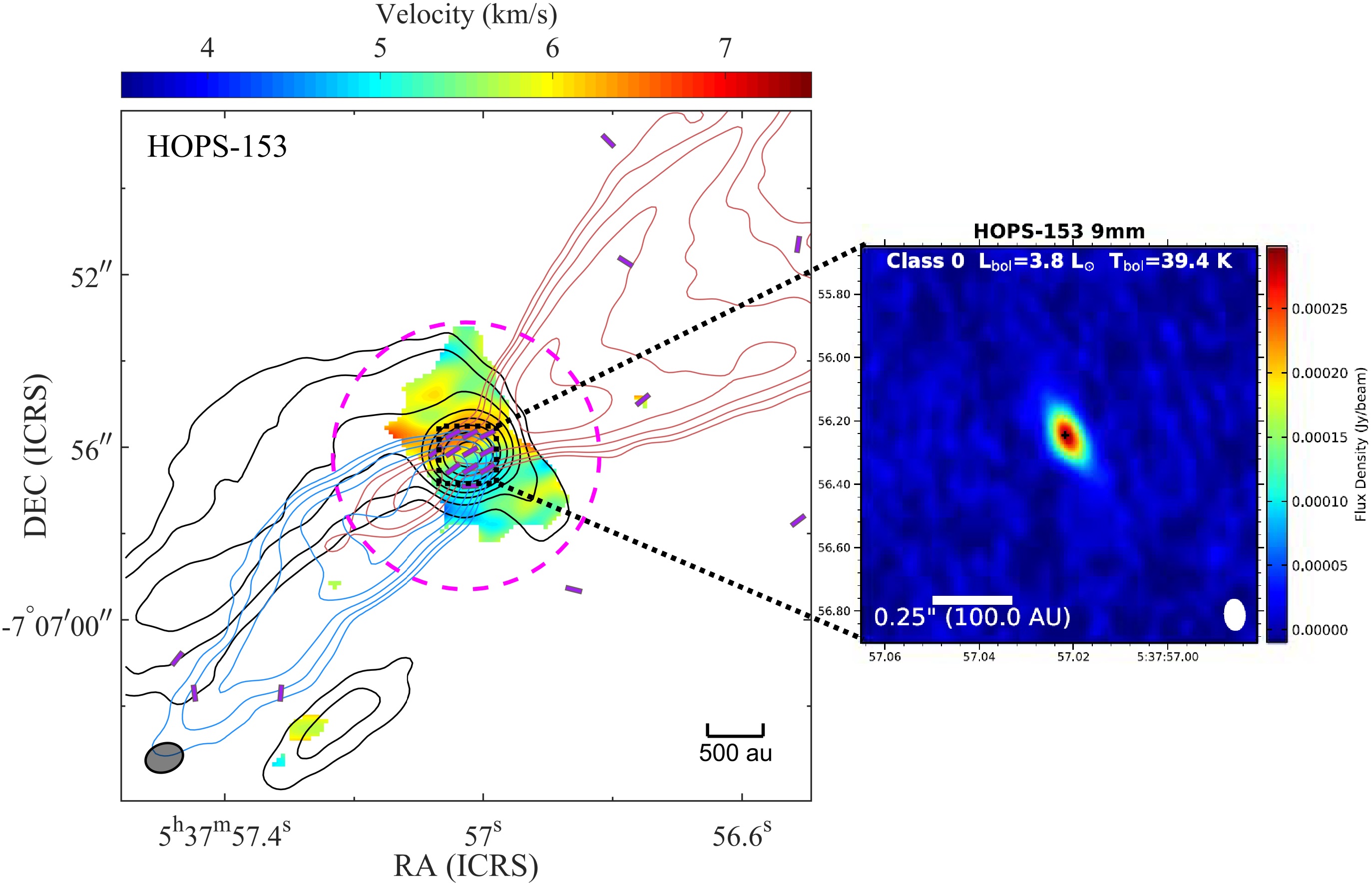}
~\\
~\\
\includegraphics[clip=true,trim=0cm 0cm 0cm 0cm,width=0.32 \textwidth]{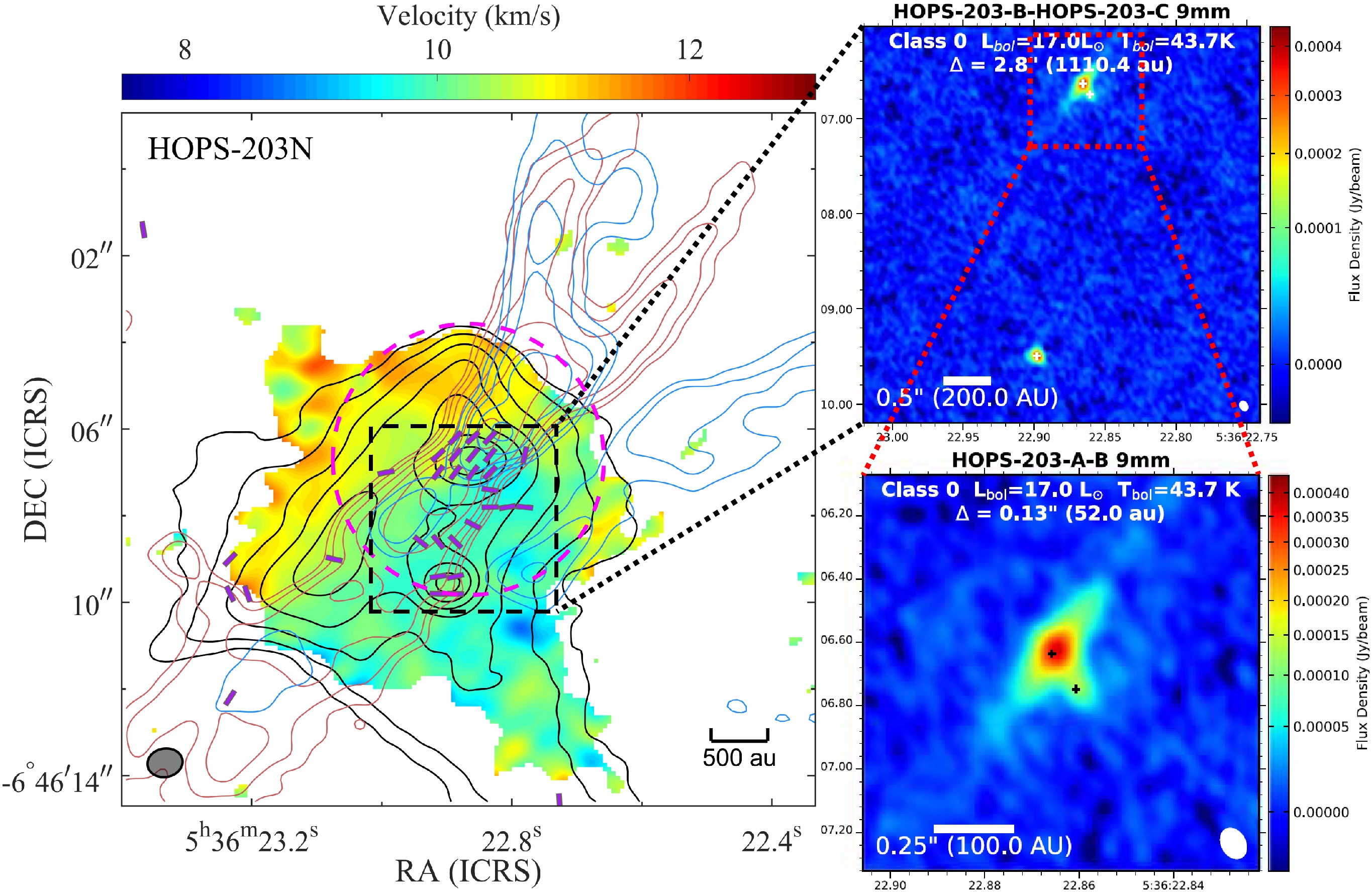}~~~~~
\includegraphics[clip=true,trim=0cm 0cm 0cm 0cm,width=0.32 \textwidth]{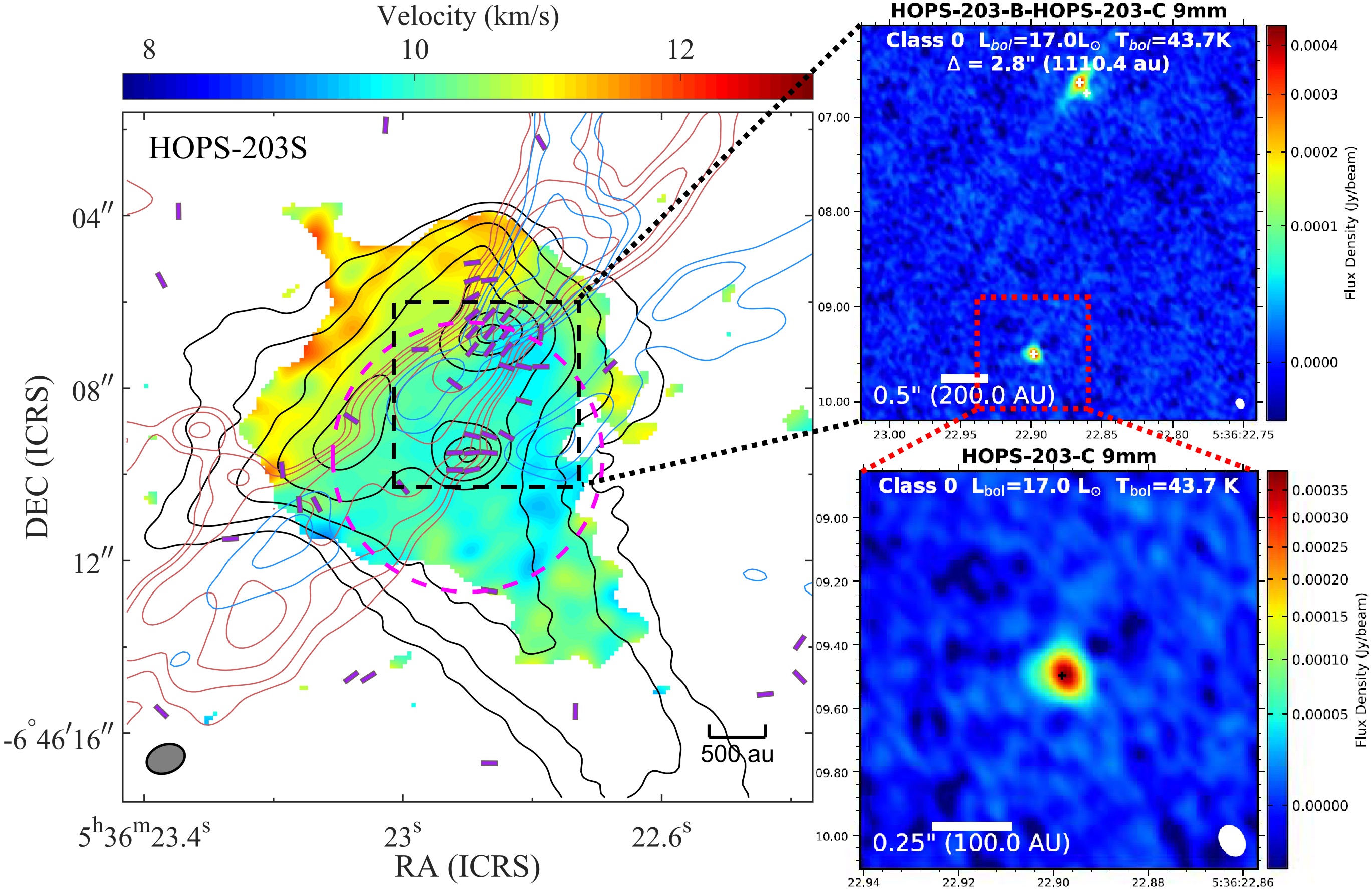}~~~~~
\includegraphics[clip=true,trim=0cm 0cm 0cm 0cm,width=0.32 \textwidth]{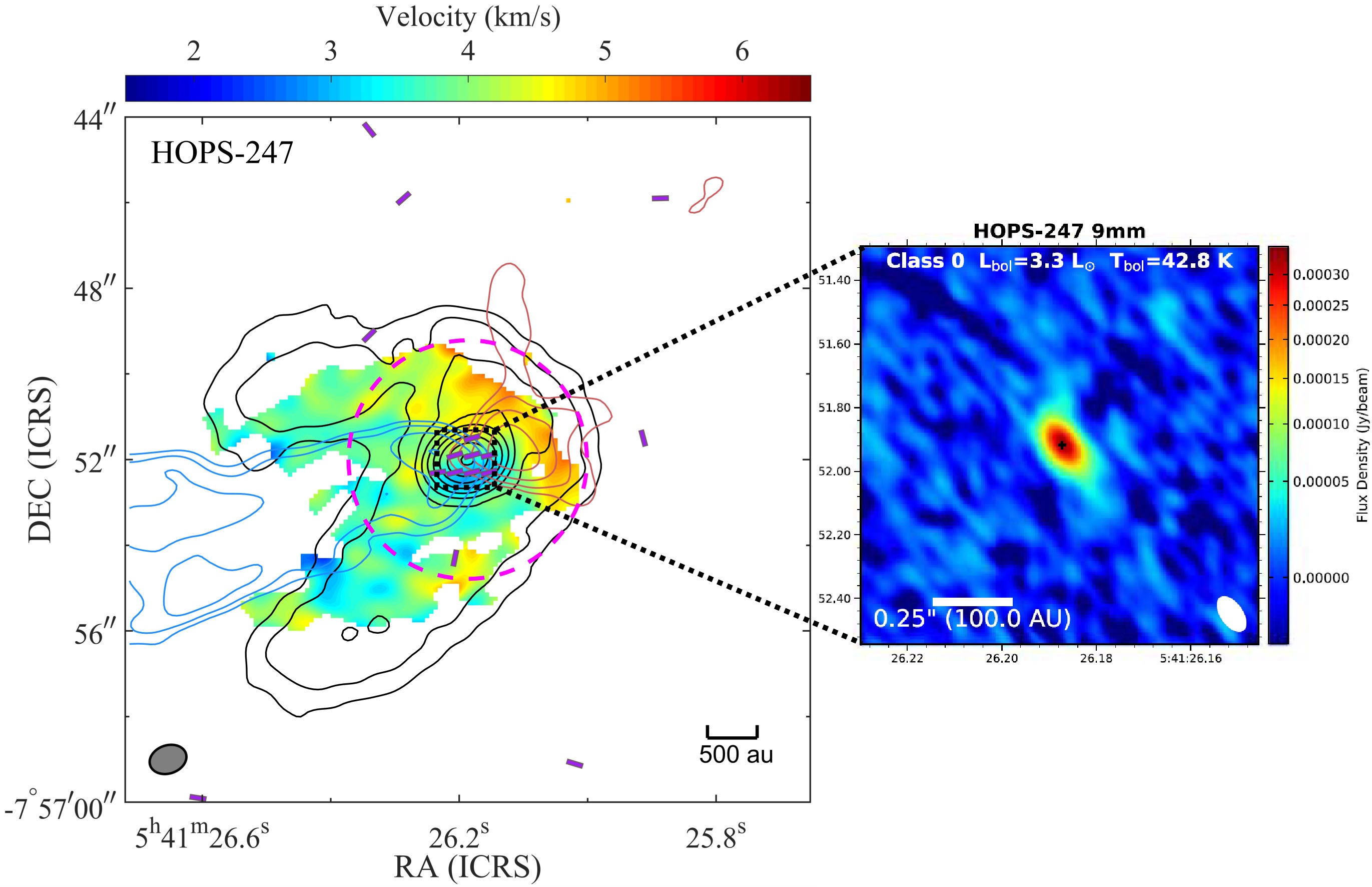}
~\\
~\\
\includegraphics[clip=true,trim=0cm 0cm 0cm 0cm,width=0.32 \textwidth]{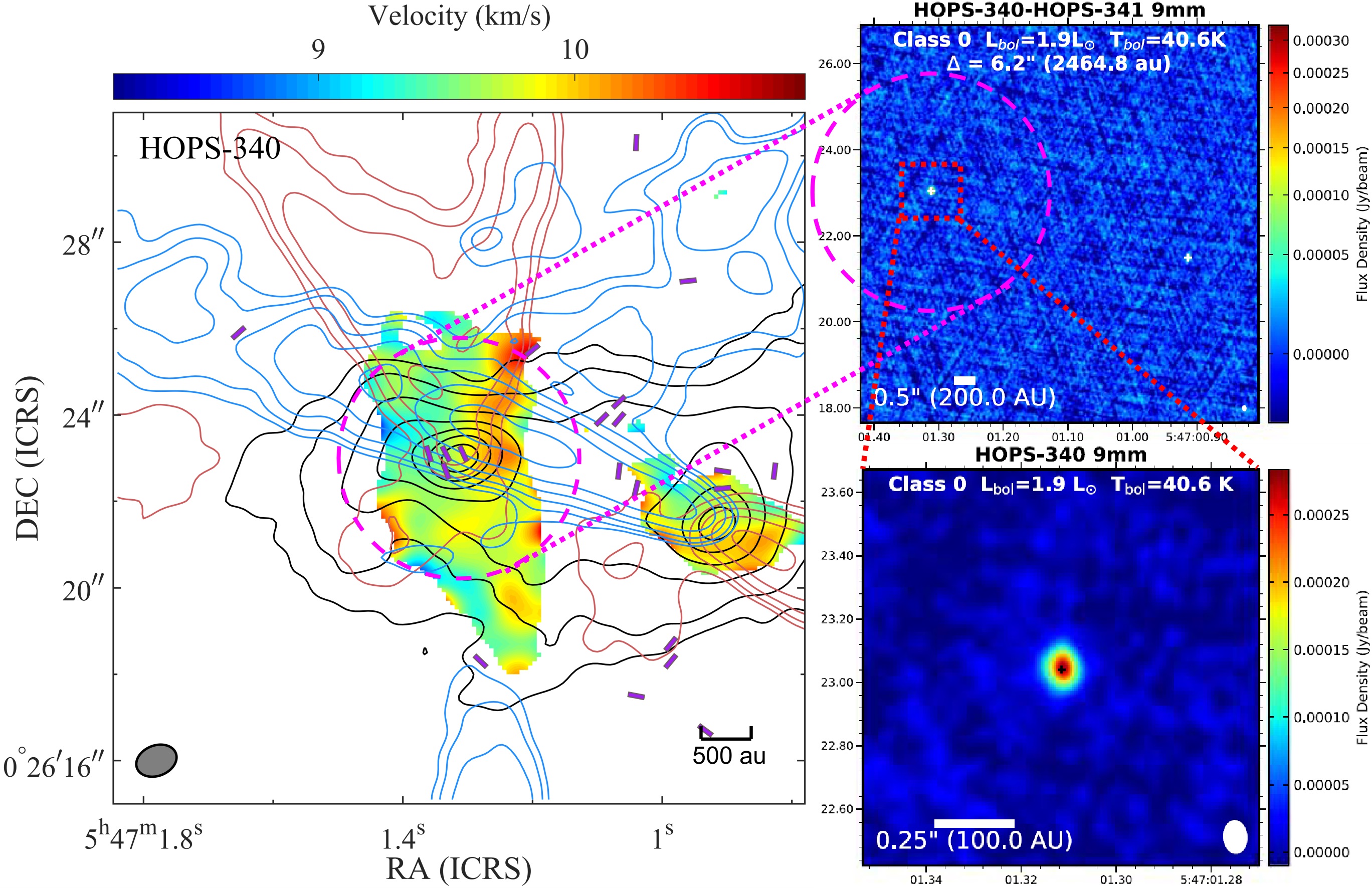}~~~~~
\includegraphics[clip=true,trim=0cm 0cm 0cm 0cm,width=0.32 \textwidth]{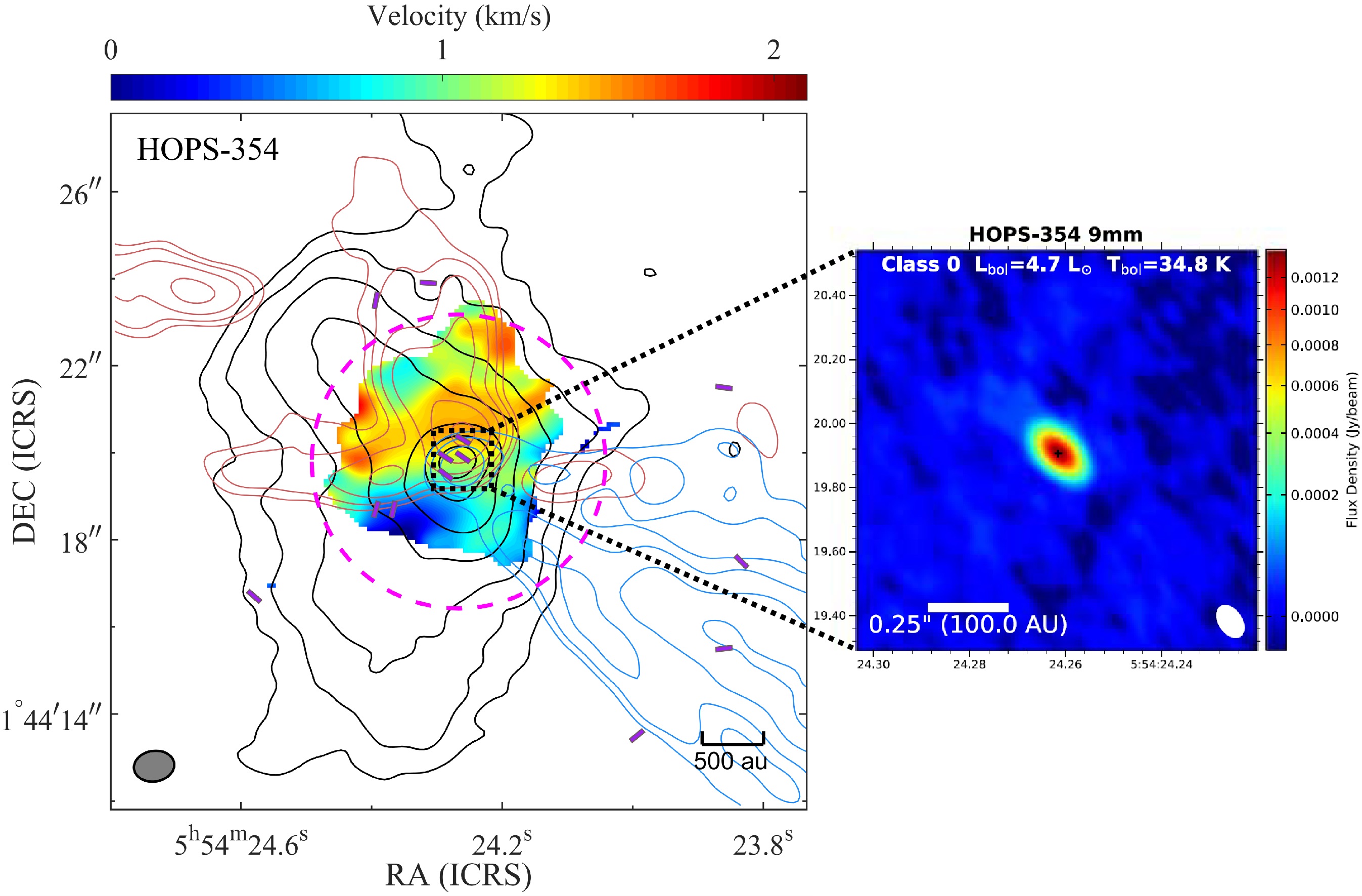}~~~~~
\includegraphics[clip=true,trim=0cm 0cm 0cm 0cm,width=0.32 \textwidth]{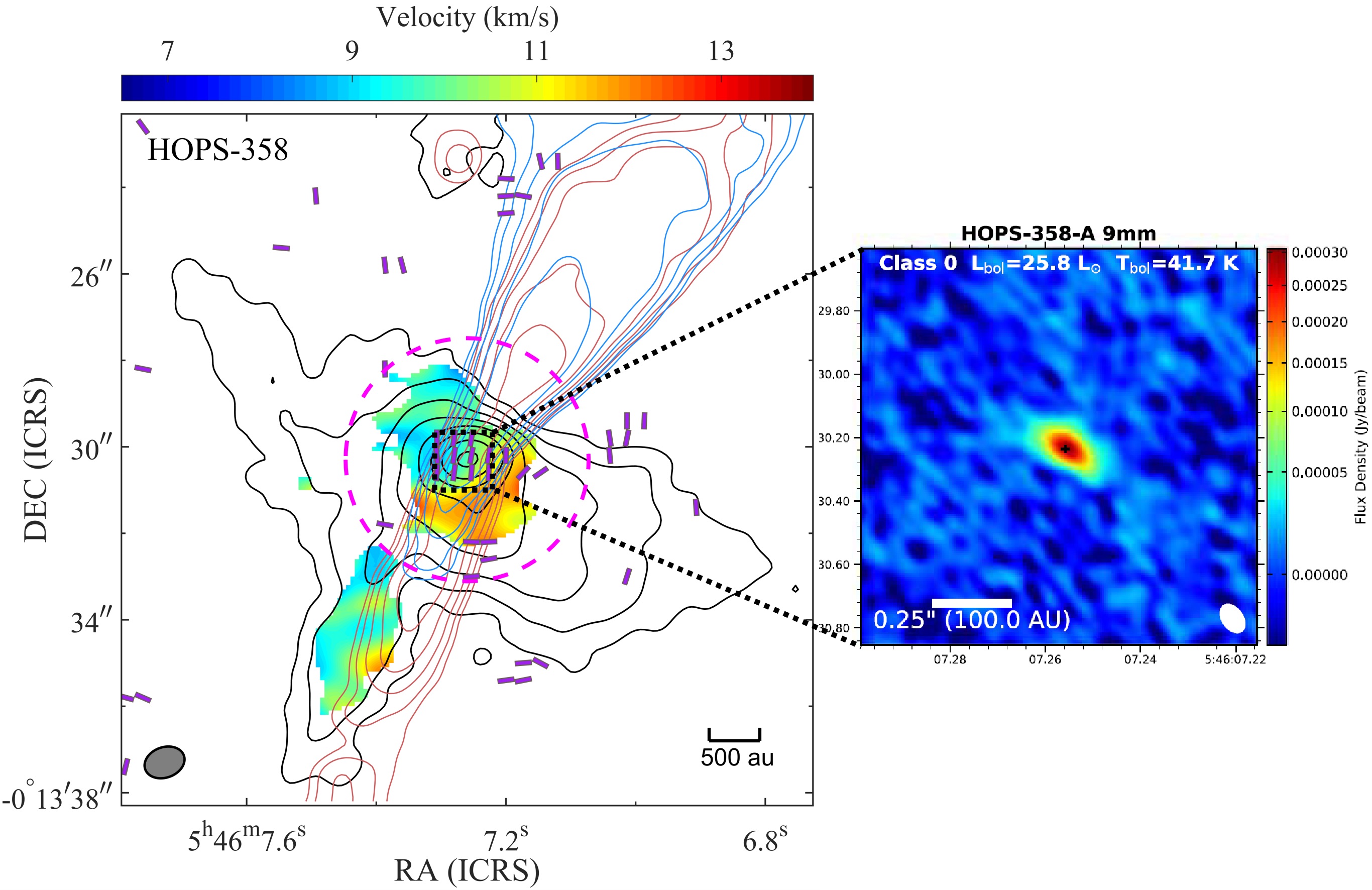}
\end{figure*}

\begin{figure*}[!ht]
\centering
\includegraphics[clip=true,trim=0cm 0cm 0cm 0cm,width=0.327 \textwidth]{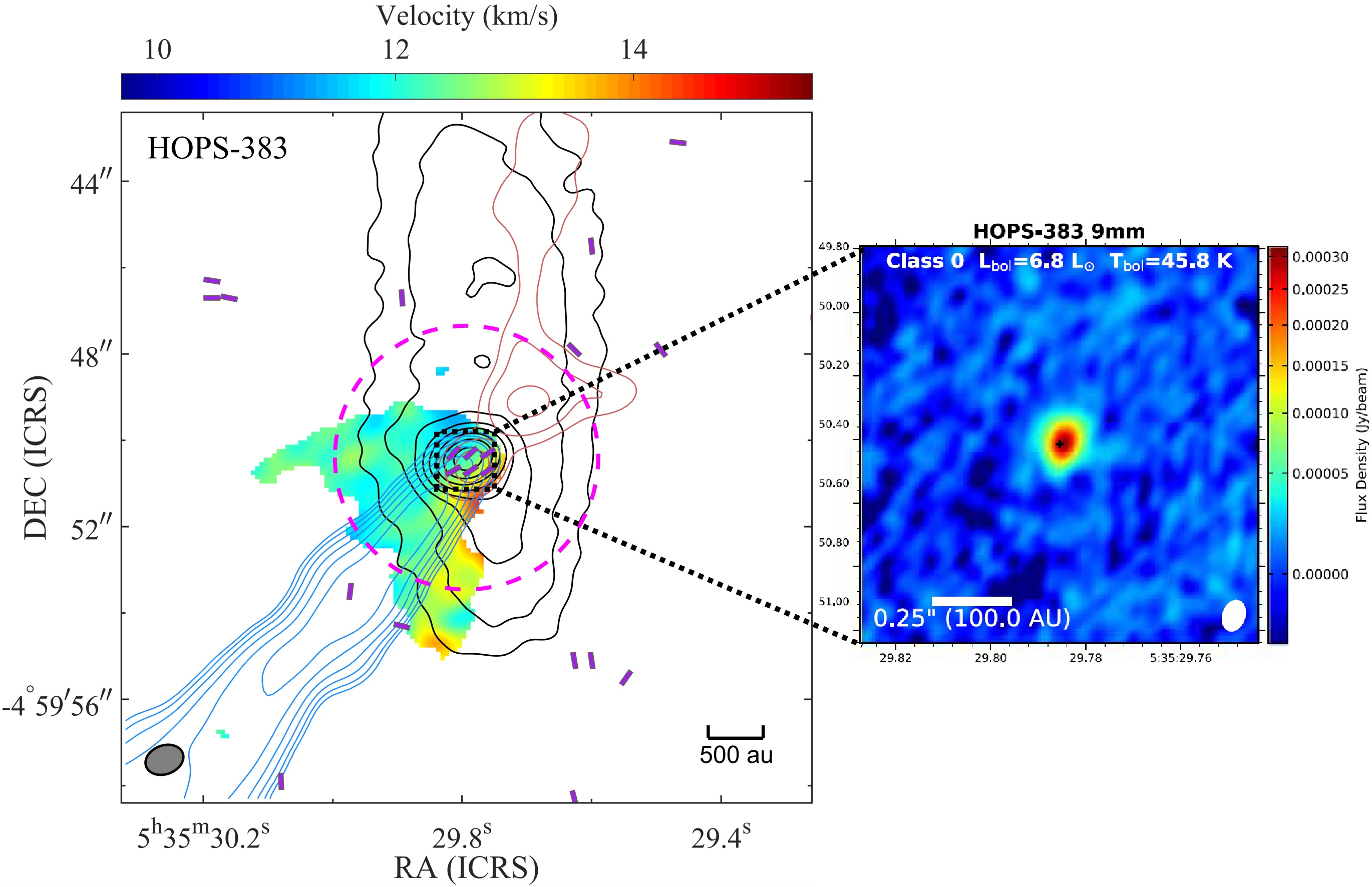}
\end{figure*}

~\\
~\\
~\\

\subsection{Aligned dust grain dominated protostars}
\noindent
Left panels: the velocity field in color scale (obtained from the C$^{17}$O (3--2) line) overlaid with the 870~$\mu$m dust polarization segments and Stokes {\em I} contours.
Blue/red contours indicate the blue-/red-shifted outflow, with counter levels set at 5 times the outflow {\em rms} × (1, 2, 4, 8, 16, 32).
The black contour levels for the Stokes {\textit I} image are 10 times the {\em rms} × (1, 2, 4, 8, 16, 32, 64, 128, 256, 512).
For the velocity field, regions with an S/N less than 4 have been flagged.
The white segments represent the magnetic fields.
The dashed magenta circle corresponds to scales with a radius of 1200 au.
The purple curves in categories of standard hourglass, rotated hourglass, and spiral indicate the {\em B}-field patterns.
Right panels: 9 mm dust maps obtained from the VANDAM survey \citep{tobin2020vla}.

\begin{figure*}[!ht]
\centering
\textbf{Standard hourglass}
\includegraphics[clip=true,trim=0cm 0cm 0cm 0cm,width=0.32 \textwidth]{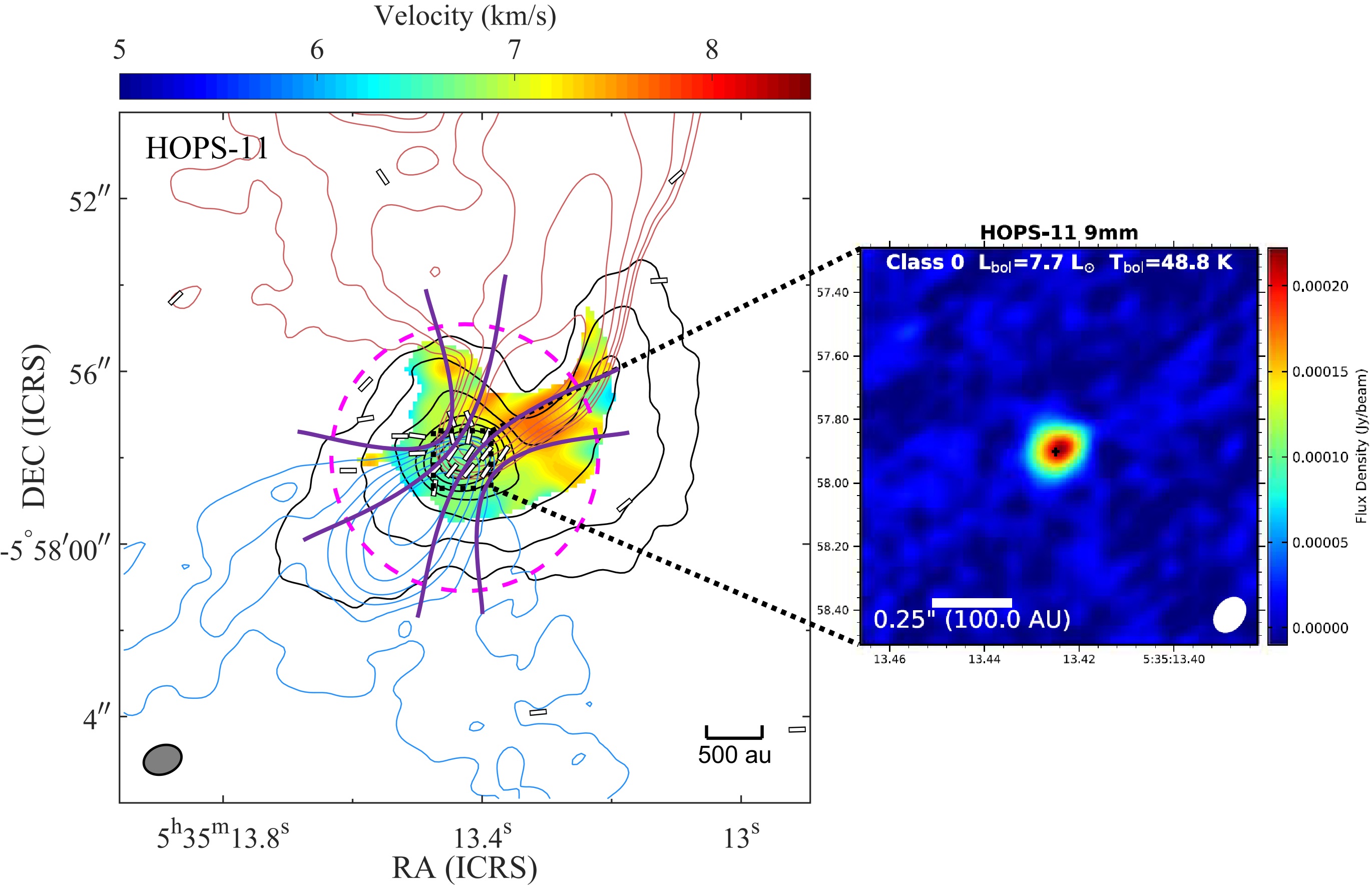}~~~~~
\includegraphics[clip=true,trim=0cm 0cm 0cm 0cm,width=0.32 \textwidth]{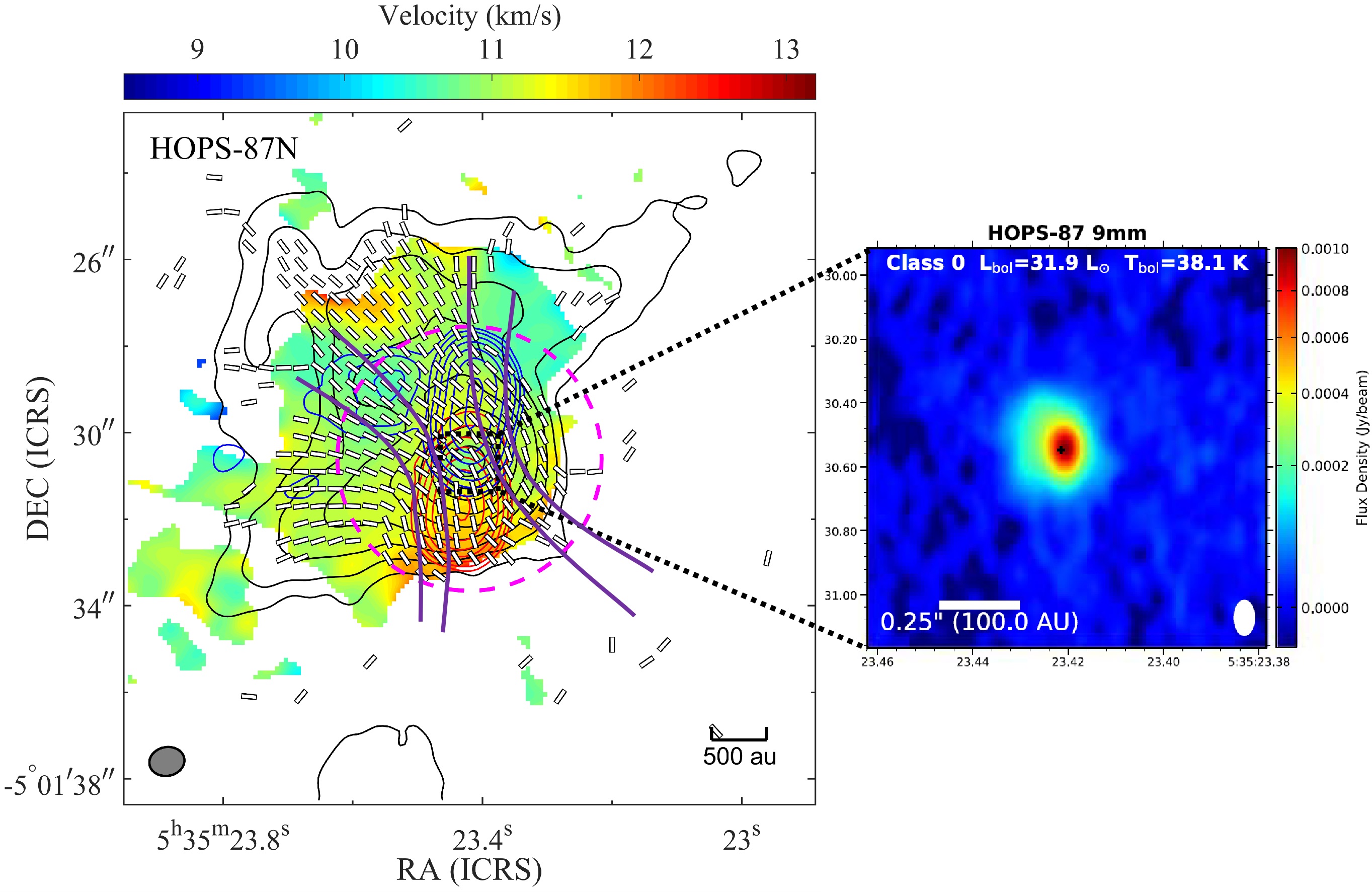}~~~~~
\includegraphics[clip=true,trim=0cm 0cm 0cm 0cm,width=0.32 \textwidth]{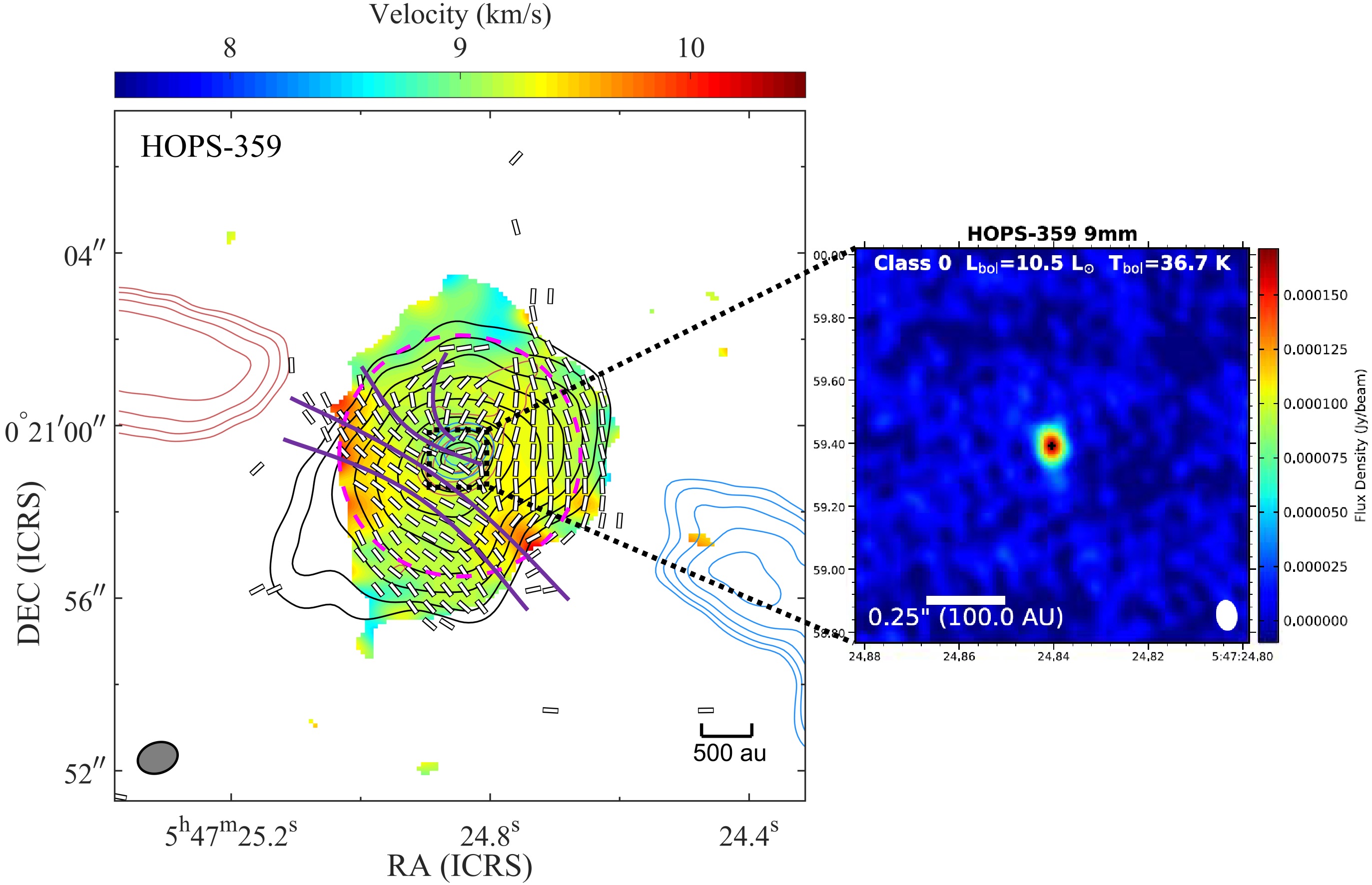}
~\\
~\\
\includegraphics[clip=true,trim=0cm 0cm 0cm 0cm,width=0.32 \textwidth]{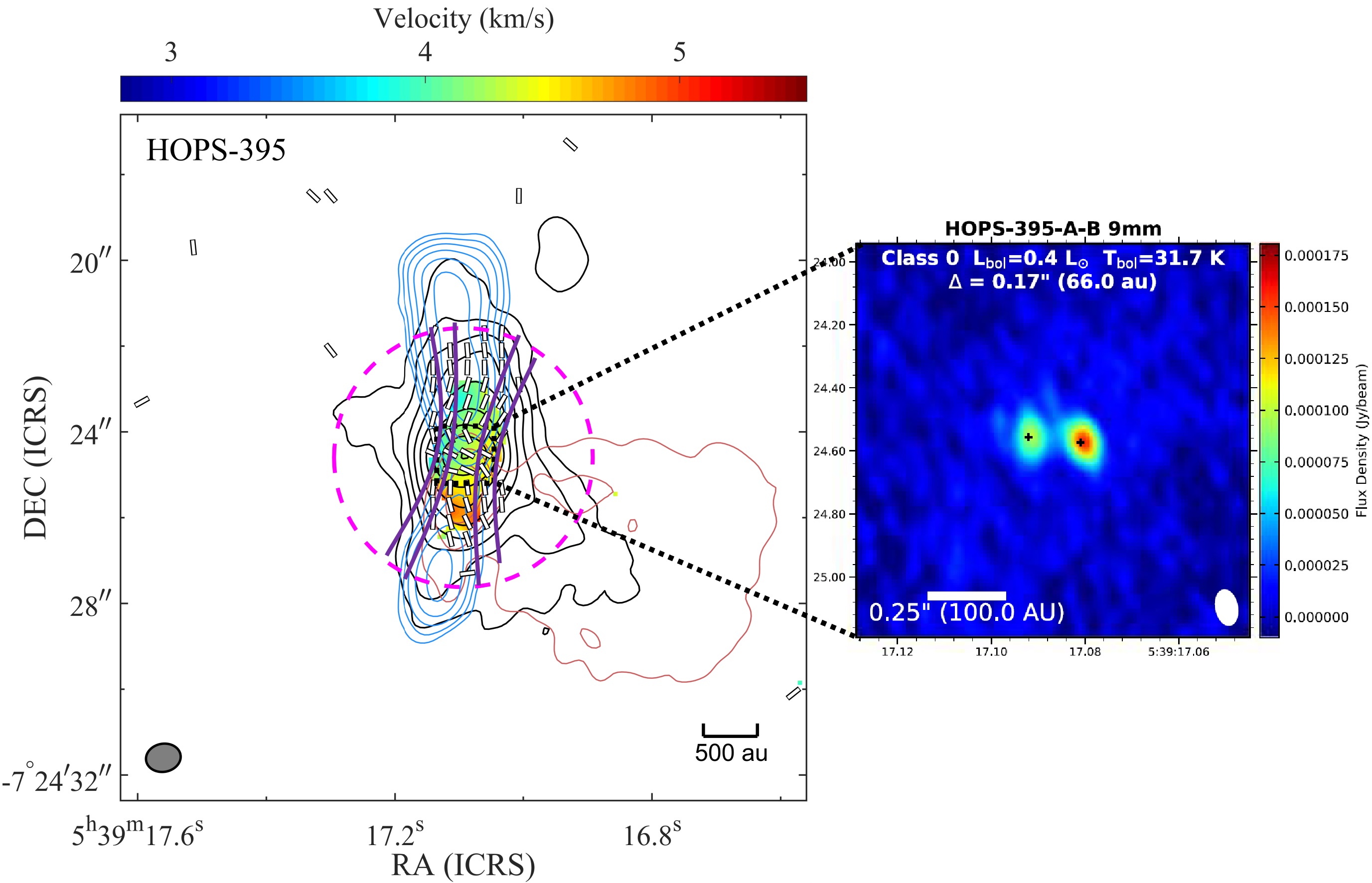}~~~~~
\includegraphics[clip=true,trim=0cm 0cm 0cm 0cm,width=0.32 \textwidth]{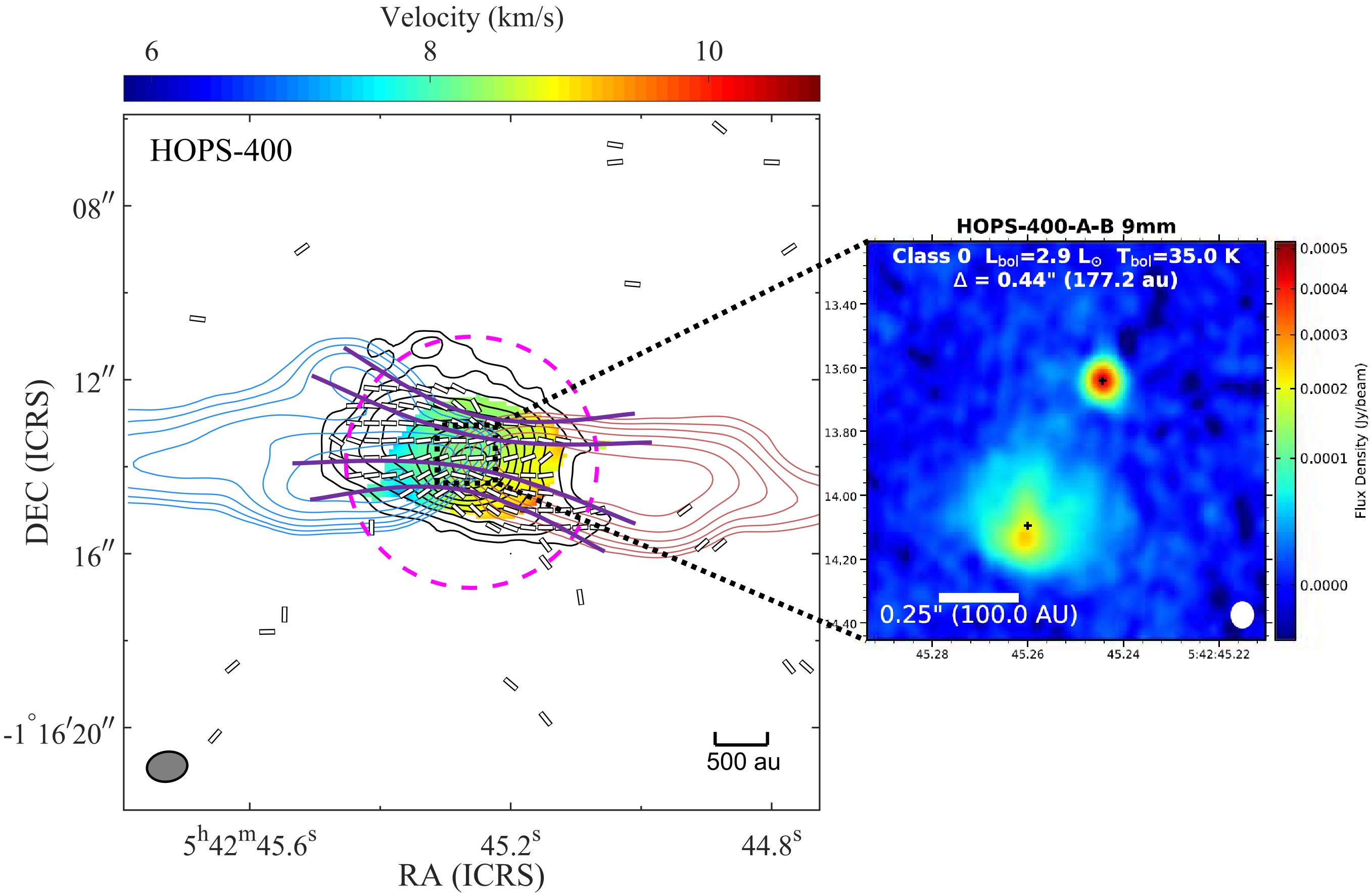}~~~~~
\includegraphics[clip=true,trim=0cm 0cm 0cm 0cm,width=0.32 \textwidth]{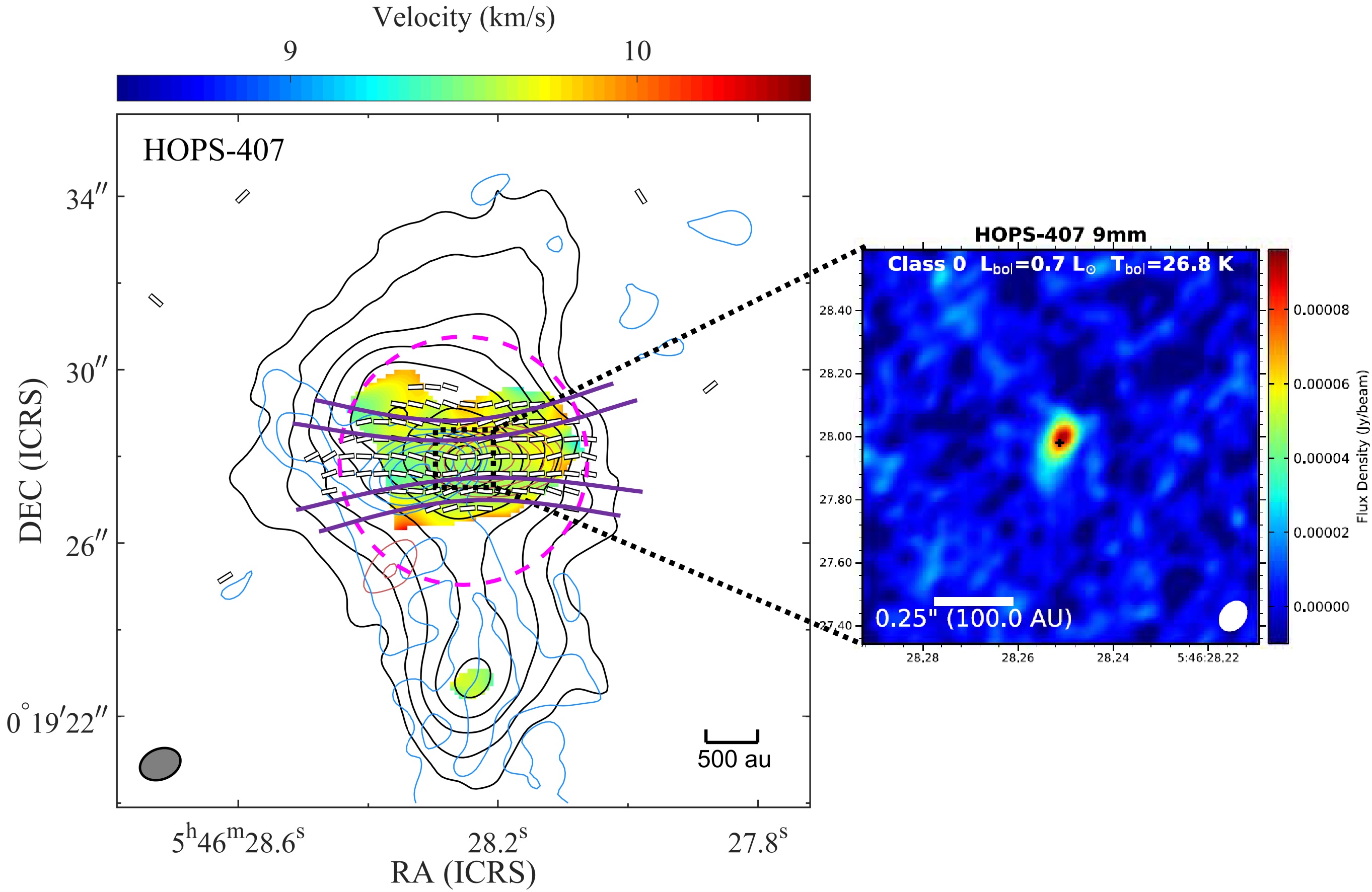}
~\\
~\\
\includegraphics[clip=true,trim=0cm 0cm 0cm 0cm,width=0.32 \textwidth]{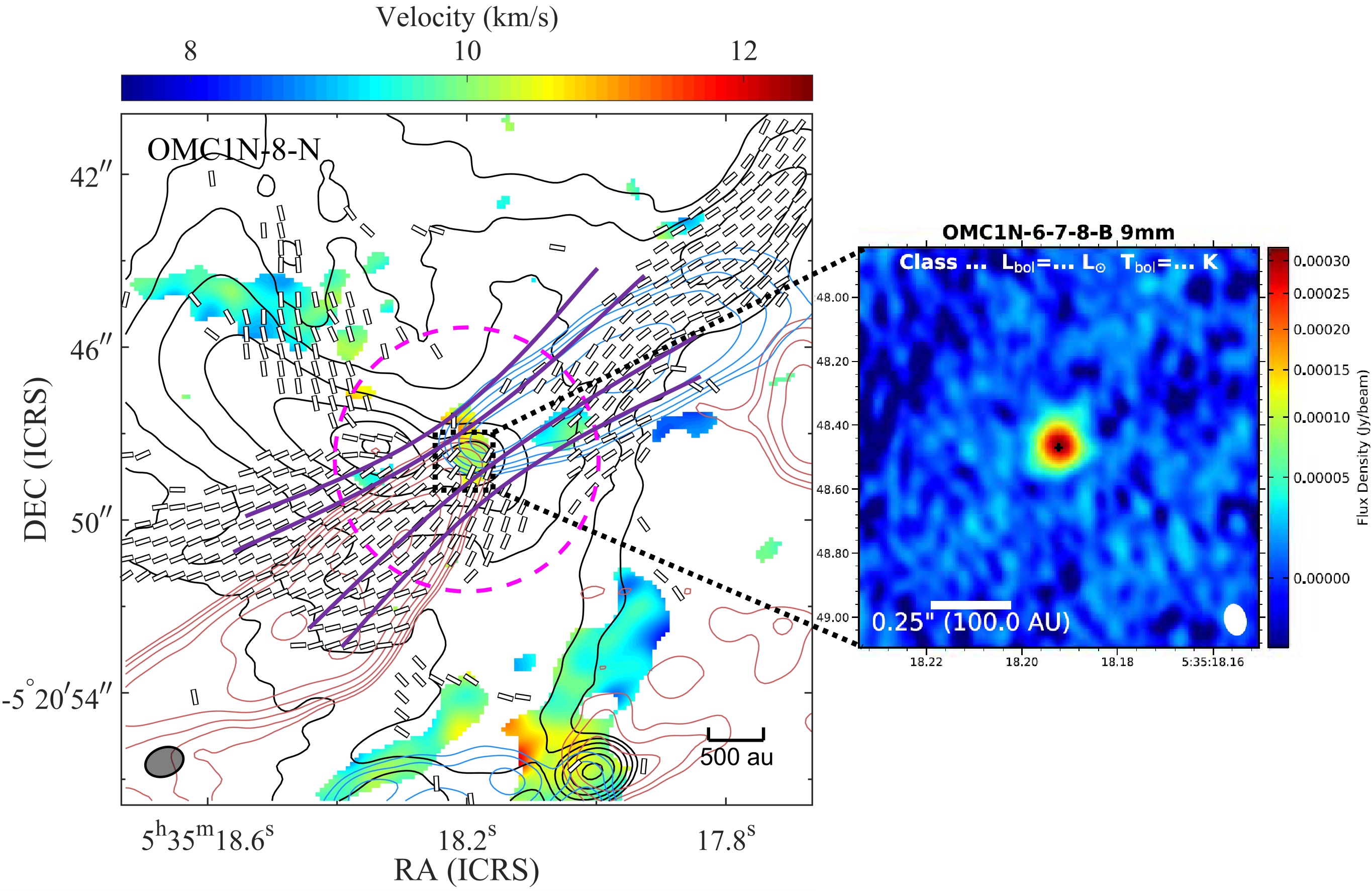}
\end{figure*}

\newpage
\begin{figure*}[!ht]
\centering
\textbf{Rotated hourglass}
\includegraphics[clip=true,trim=0cm 0cm 0cm 0cm,width=0.32 \textwidth]{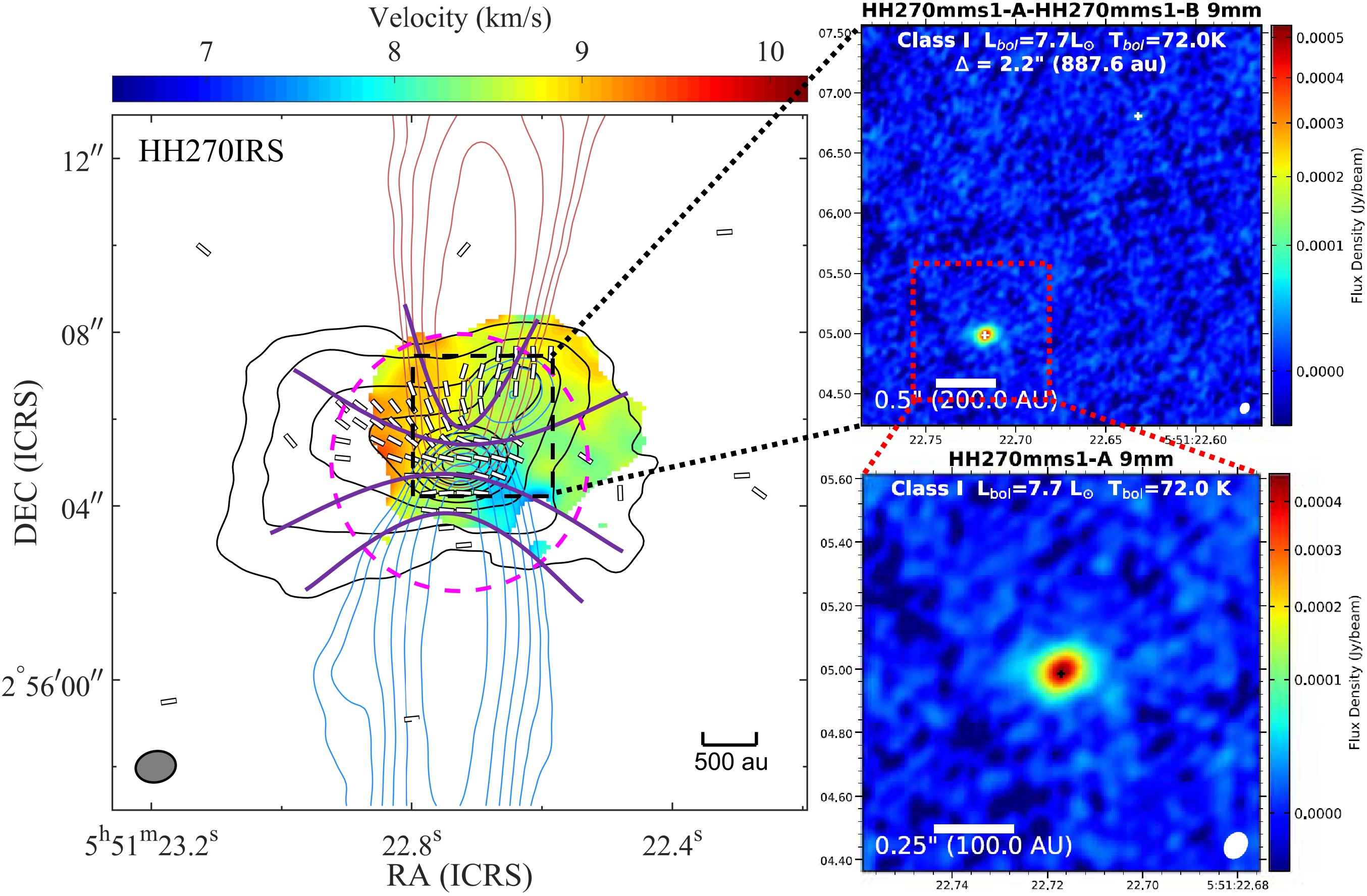}~~~~~
\includegraphics[clip=true,trim=0cm 0cm 0cm 0cm,width=0.32 \textwidth]{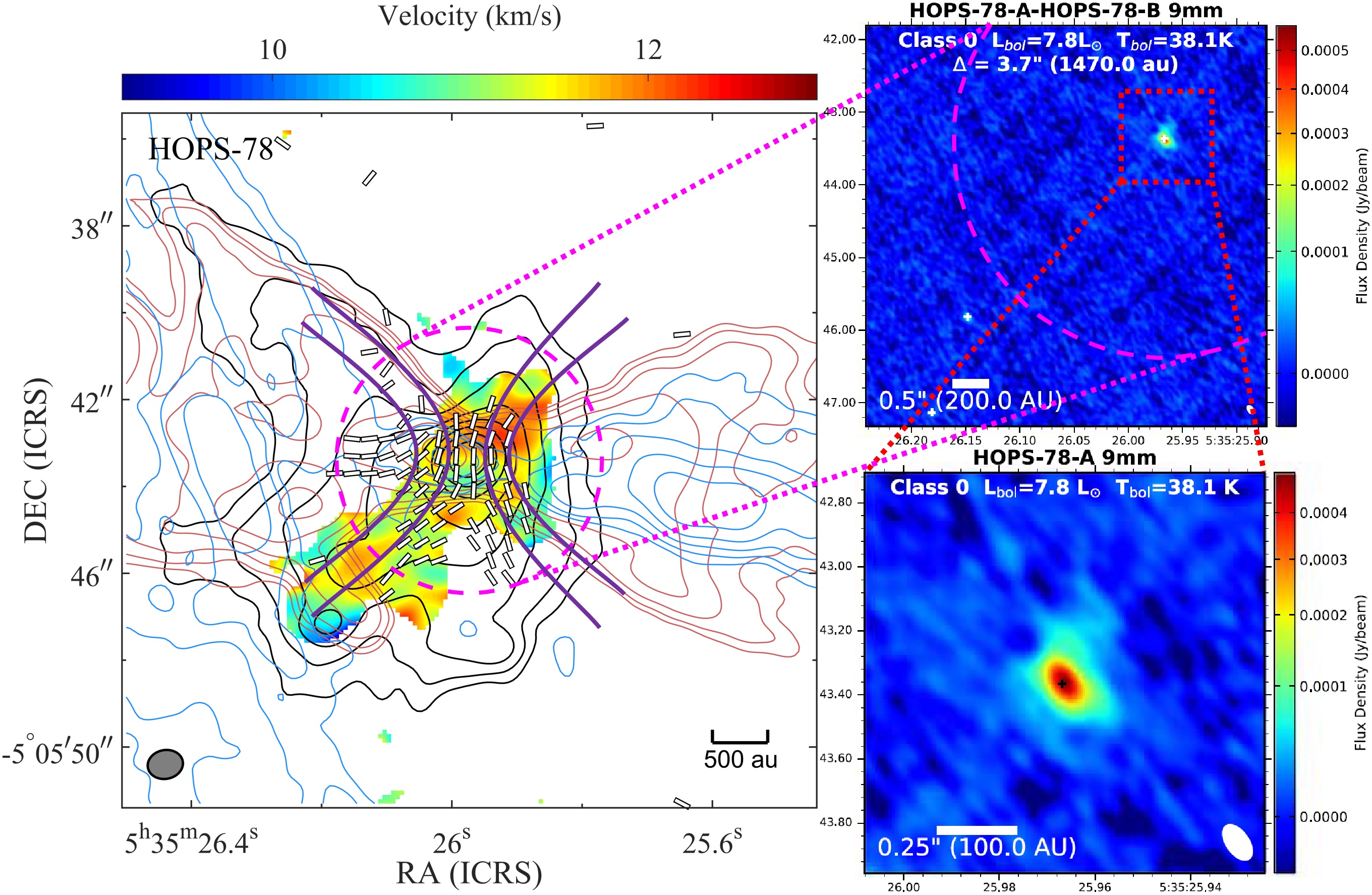}~~~~~
\includegraphics[clip=true,trim=0cm 0cm 0cm 0cm,width=0.327 \textwidth]{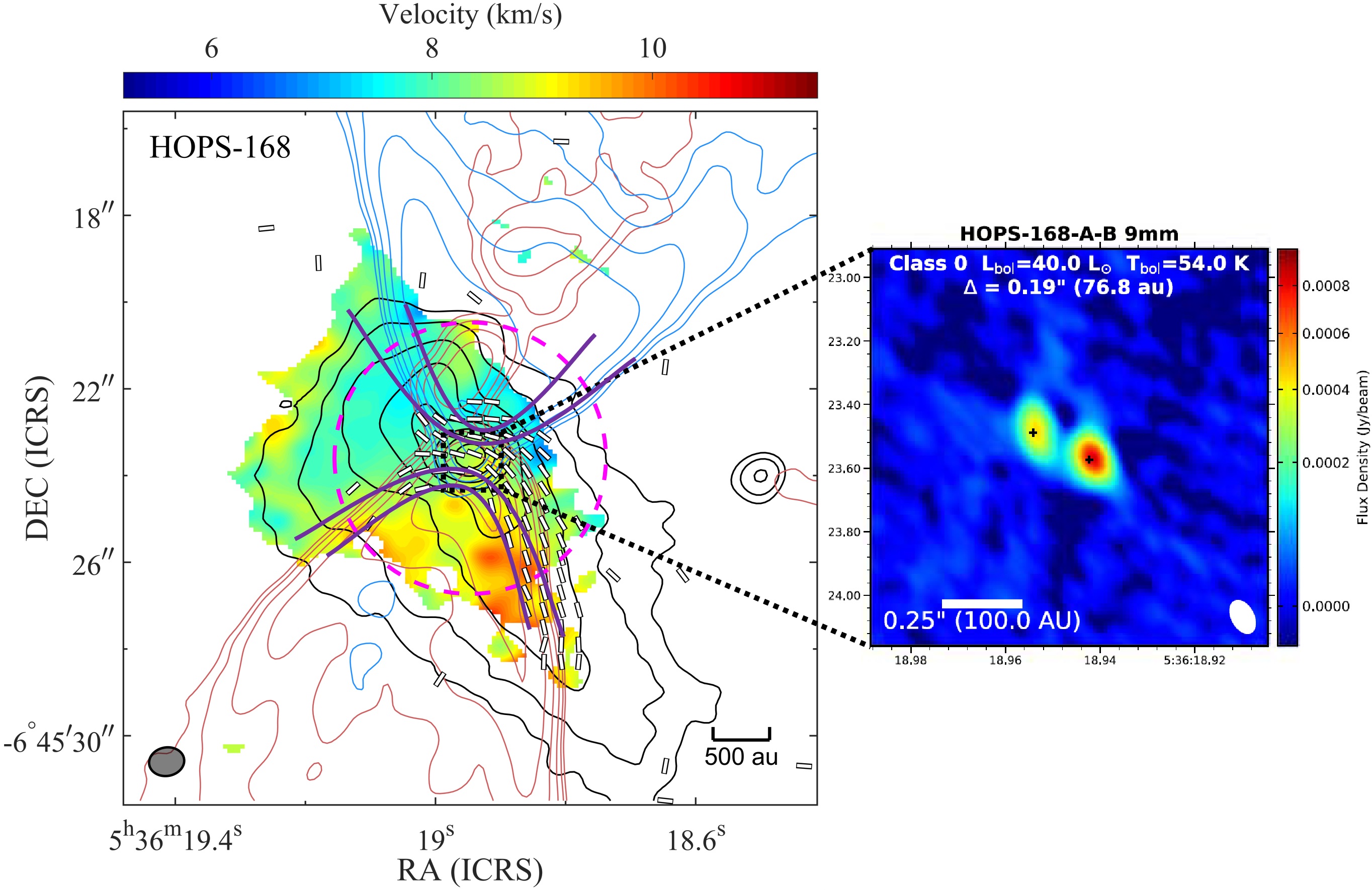}
~\\
~\\
\includegraphics[clip=true,trim=0cm 0cm 0cm 0cm,width=0.32 \textwidth]{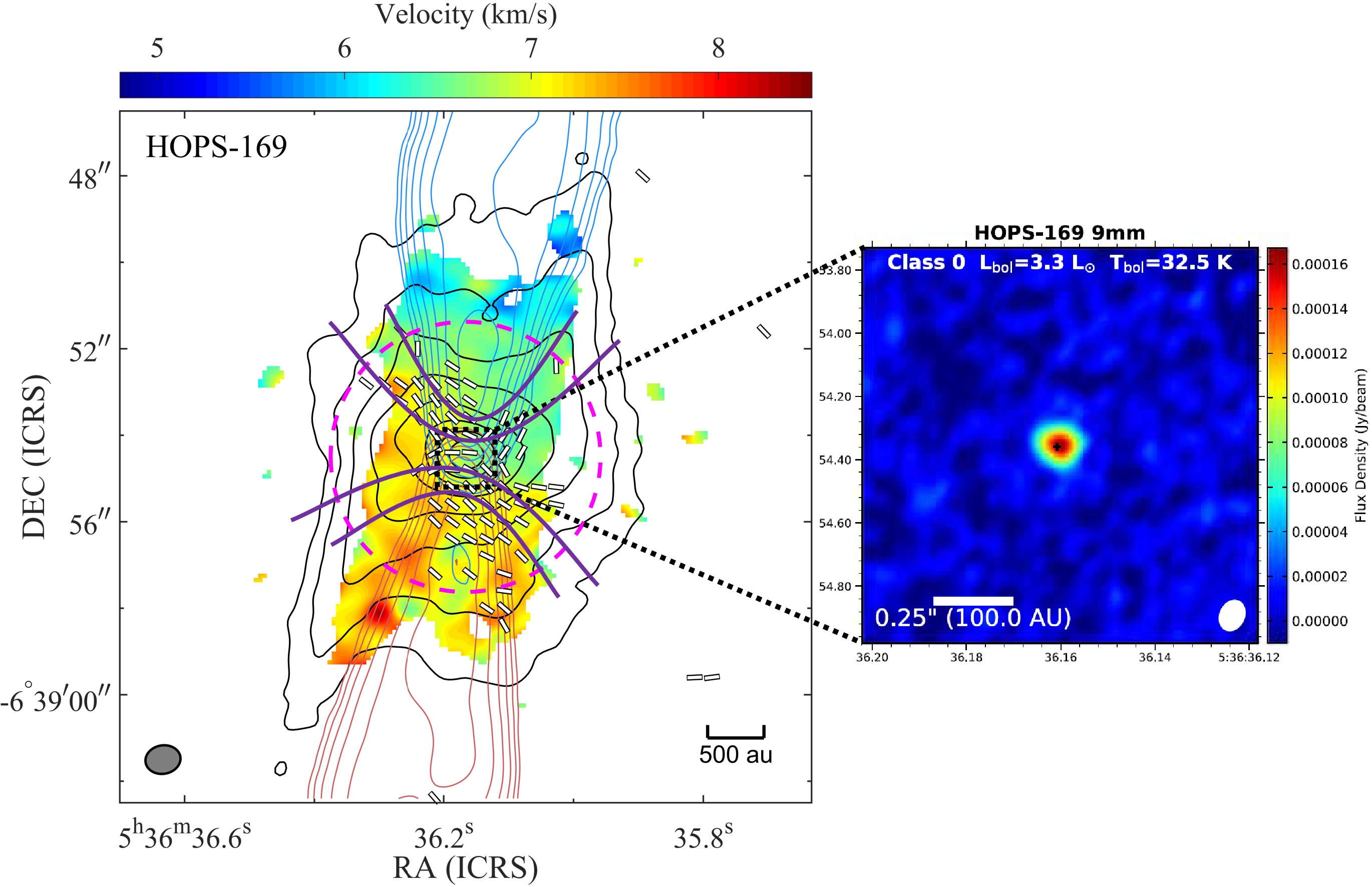}~~~~~
\includegraphics[clip=true,trim=0cm 0cm 0cm 0cm,width=0.32 \textwidth]{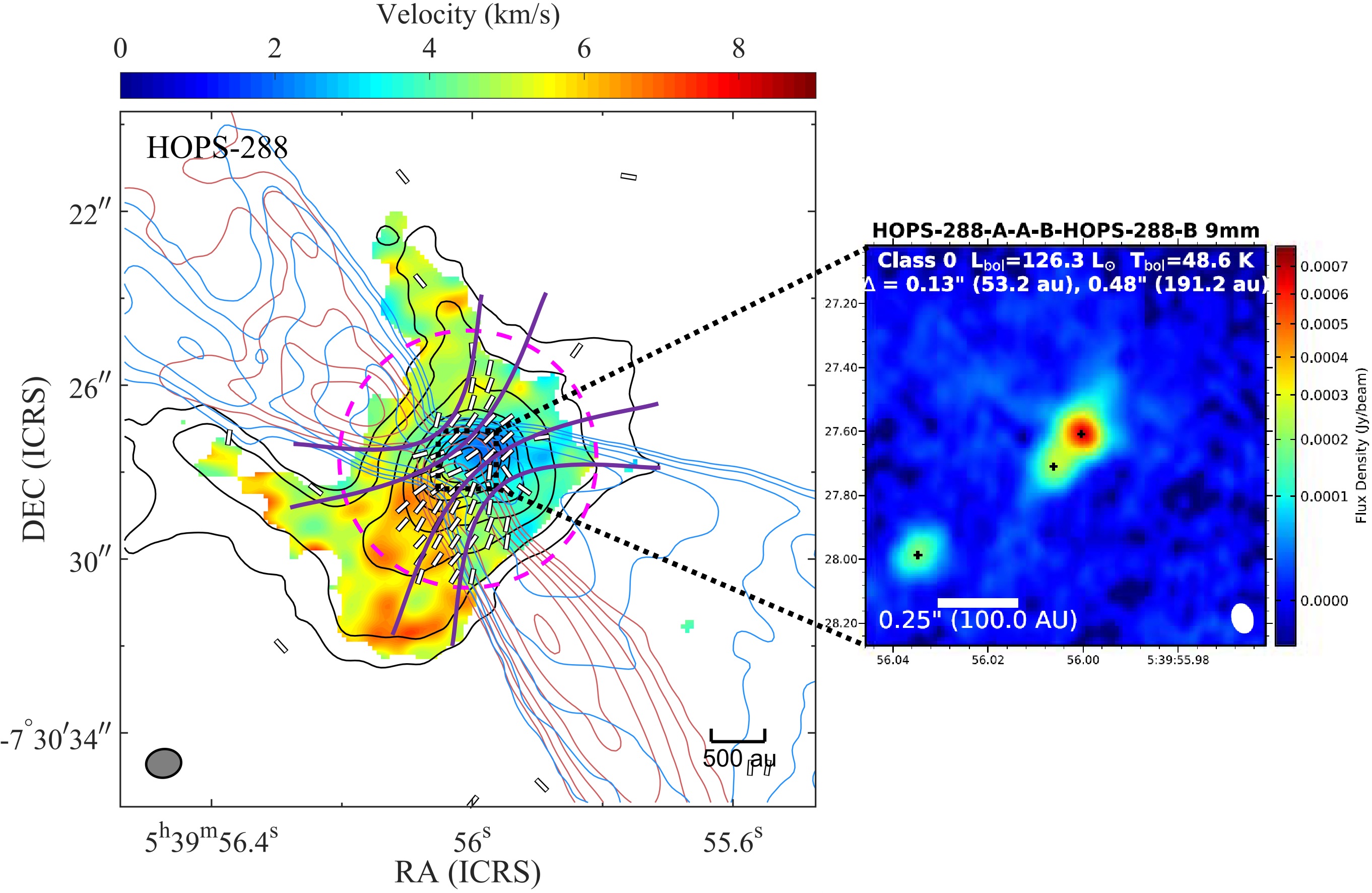}~~~~~
\includegraphics[clip=true,trim=0cm 0cm 0cm 0cm,width=0.32 \textwidth]{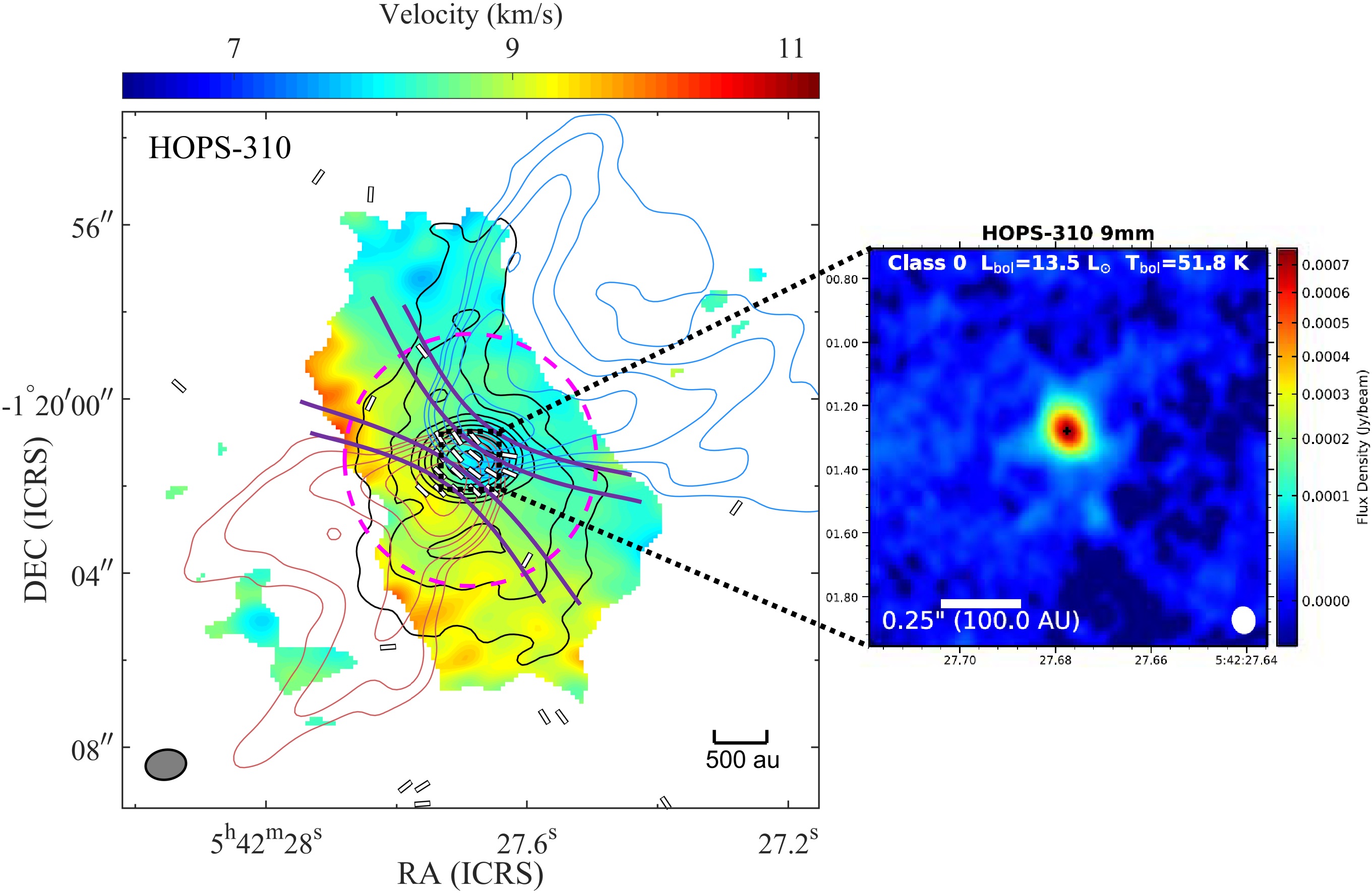}
~\\
~\\
\includegraphics[clip=true,trim=0cm 0cm 0cm 0cm,width=0.32 \textwidth]{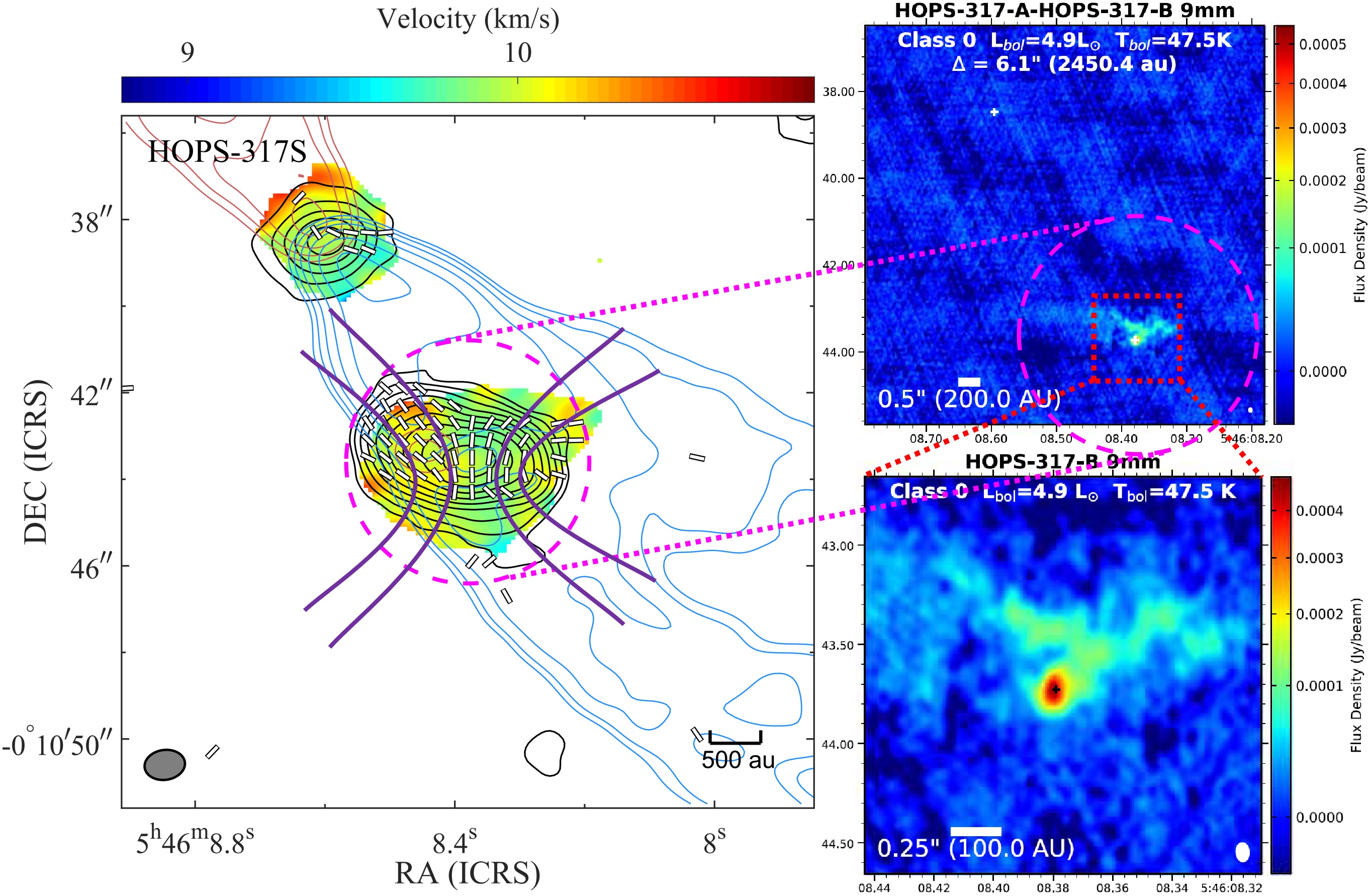}~~~~~
\includegraphics[clip=true,trim=0cm 0cm 0cm 0cm,width=0.32 \textwidth]{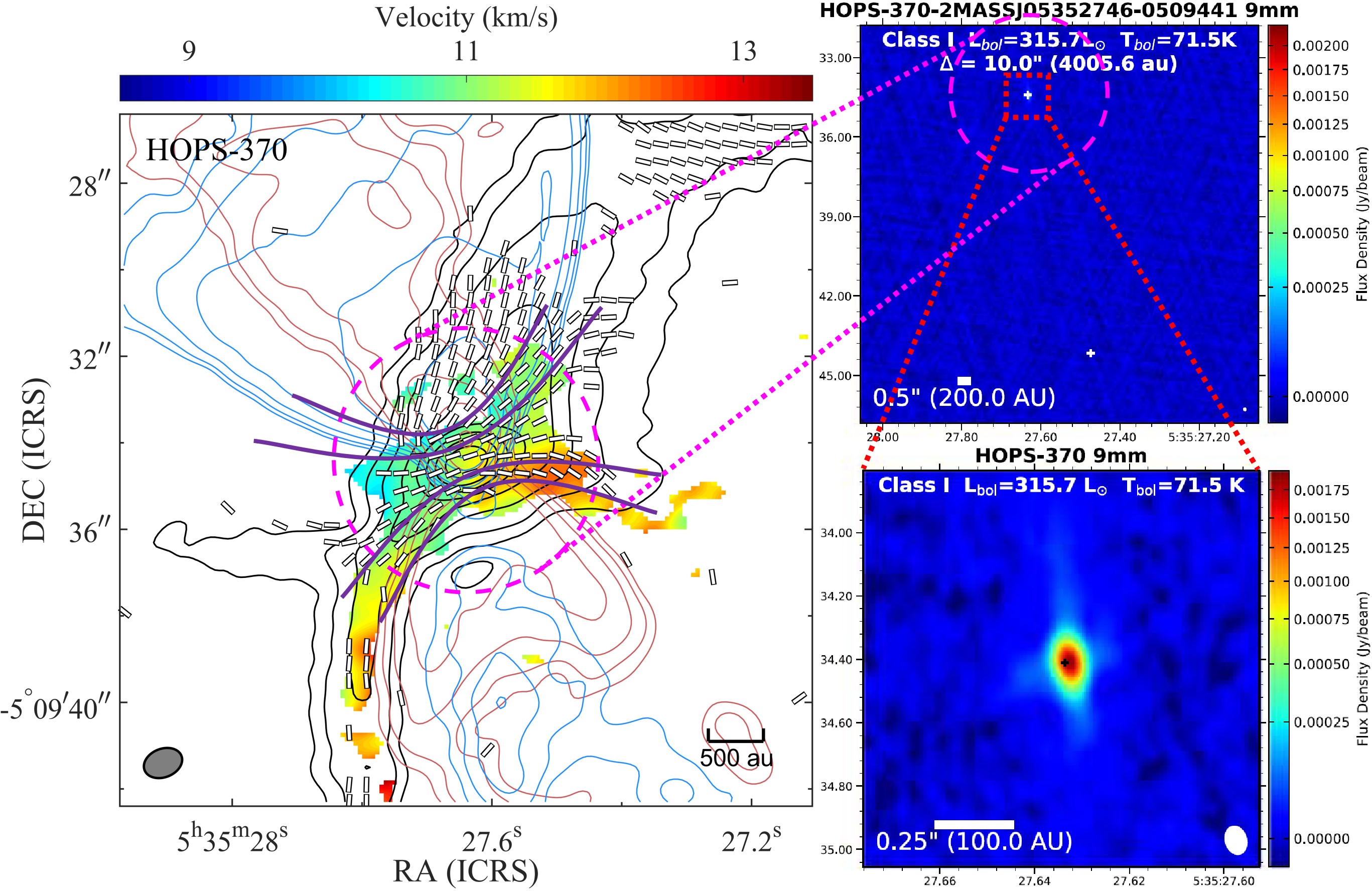}~~~~~
\includegraphics[clip=true,trim=0cm 0cm 0cm 0cm,width=0.32 \textwidth]{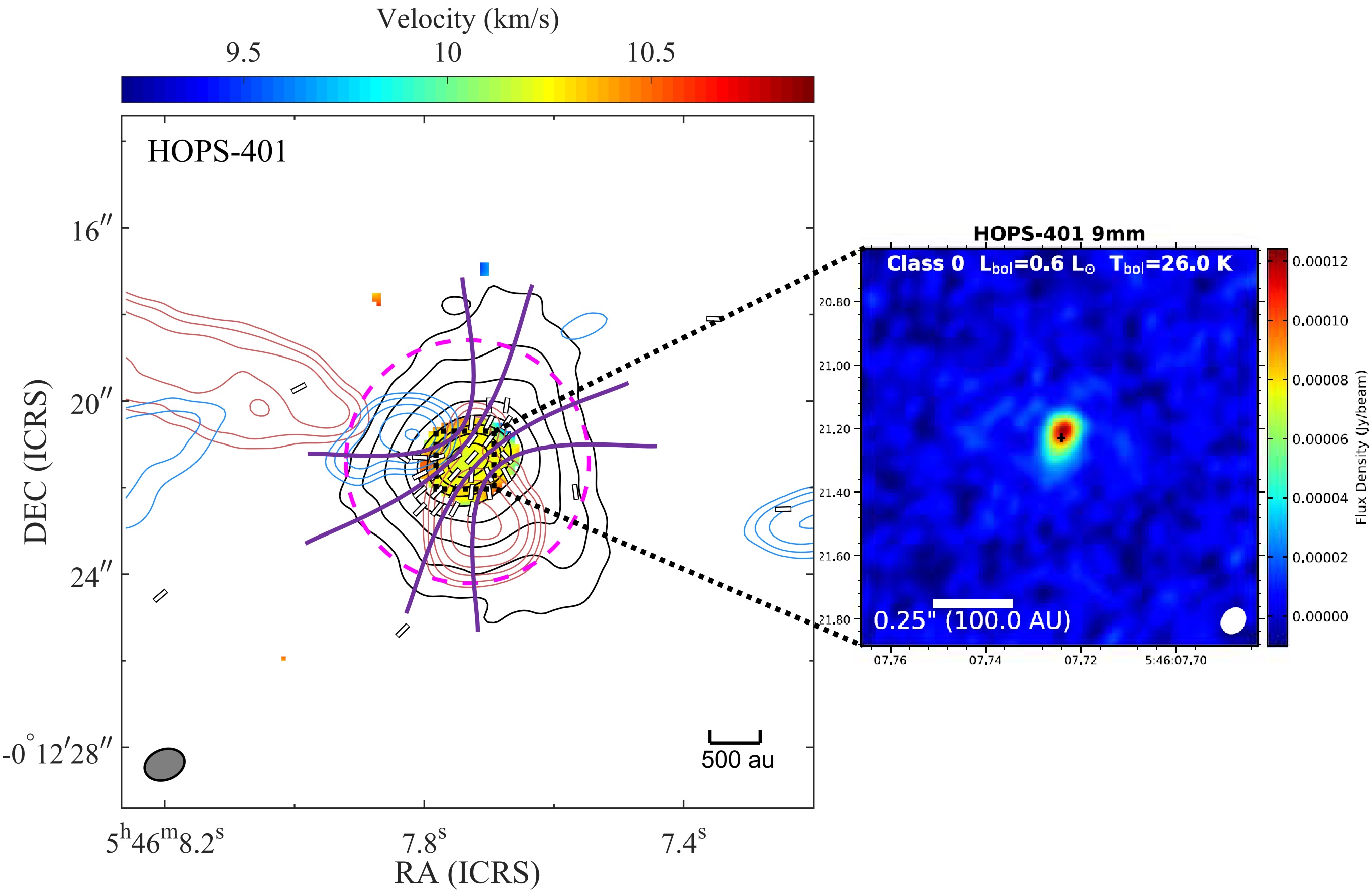}
~\\
~\\
\includegraphics[clip=true,trim=0cm 0cm 0cm 0cm,width=0.32 \textwidth]{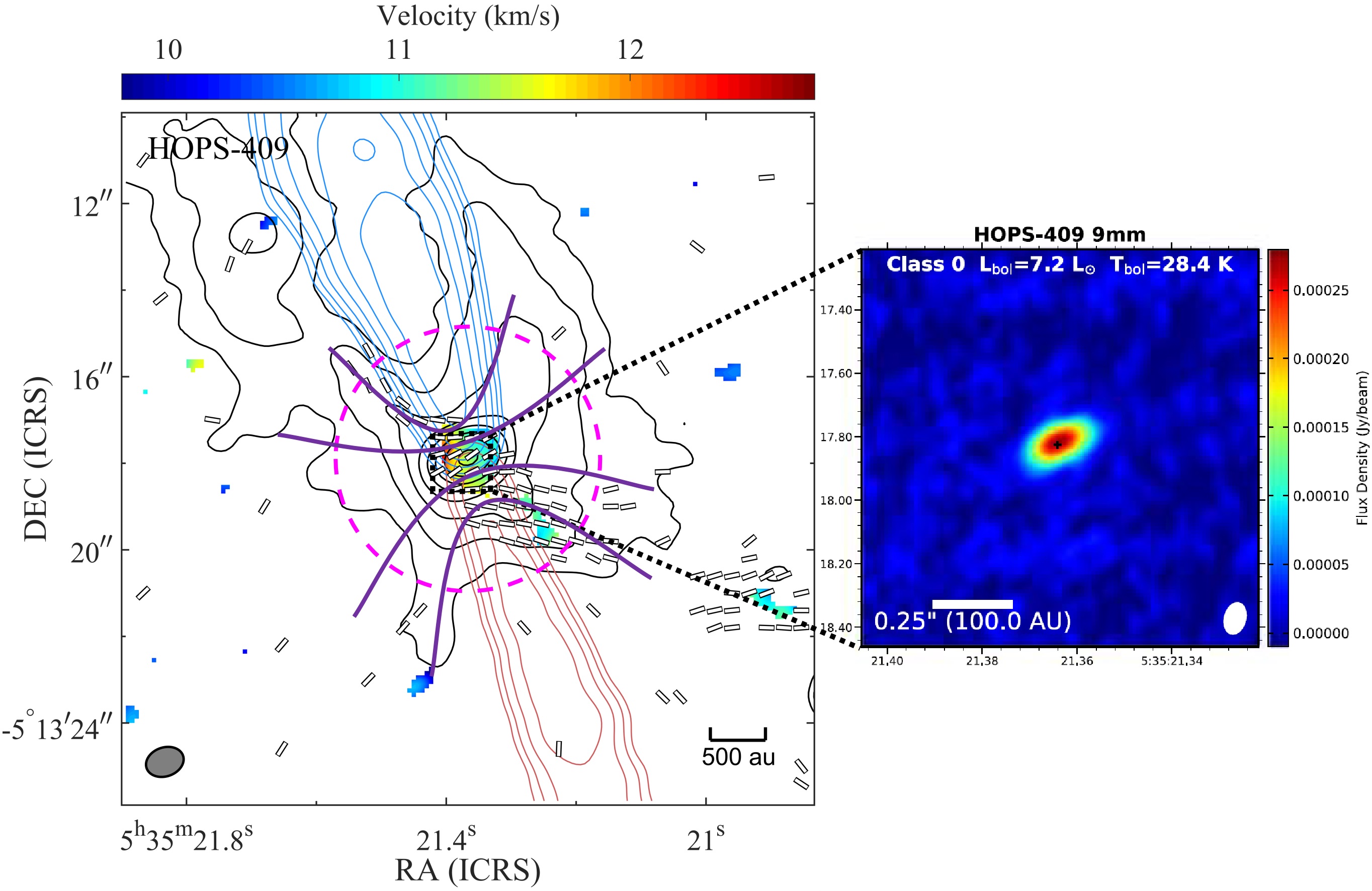}~~~~~
\includegraphics[clip=true,trim=0cm 0cm 0cm 0cm,width=0.32 \textwidth]{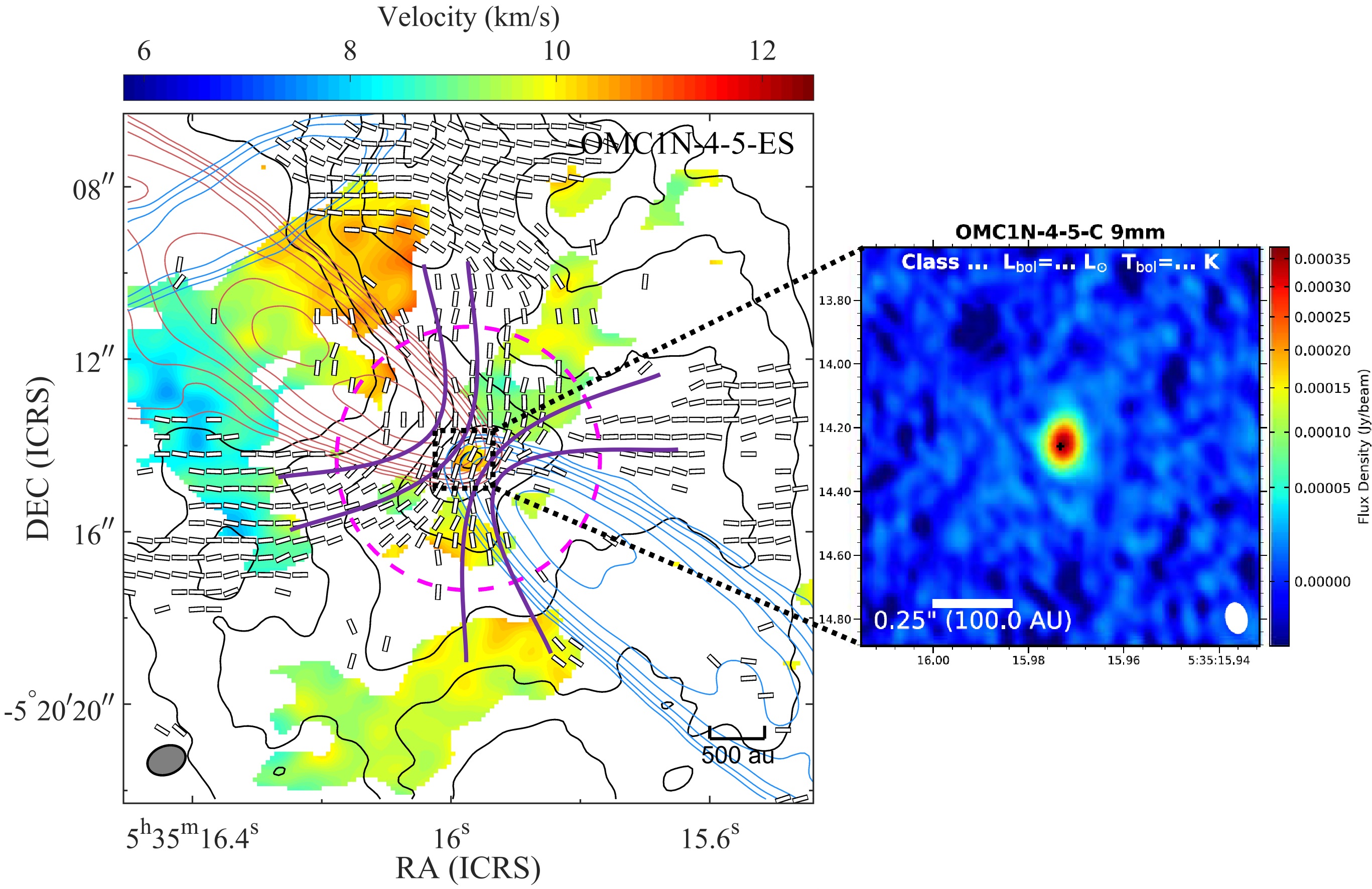}
\end{figure*}

\newpage
\begin{figure*}[!ht]
\centering
\textbf{Spiral}
\includegraphics[clip=true,trim=0cm 0cm 0cm 0cm,width=0.32 \textwidth]{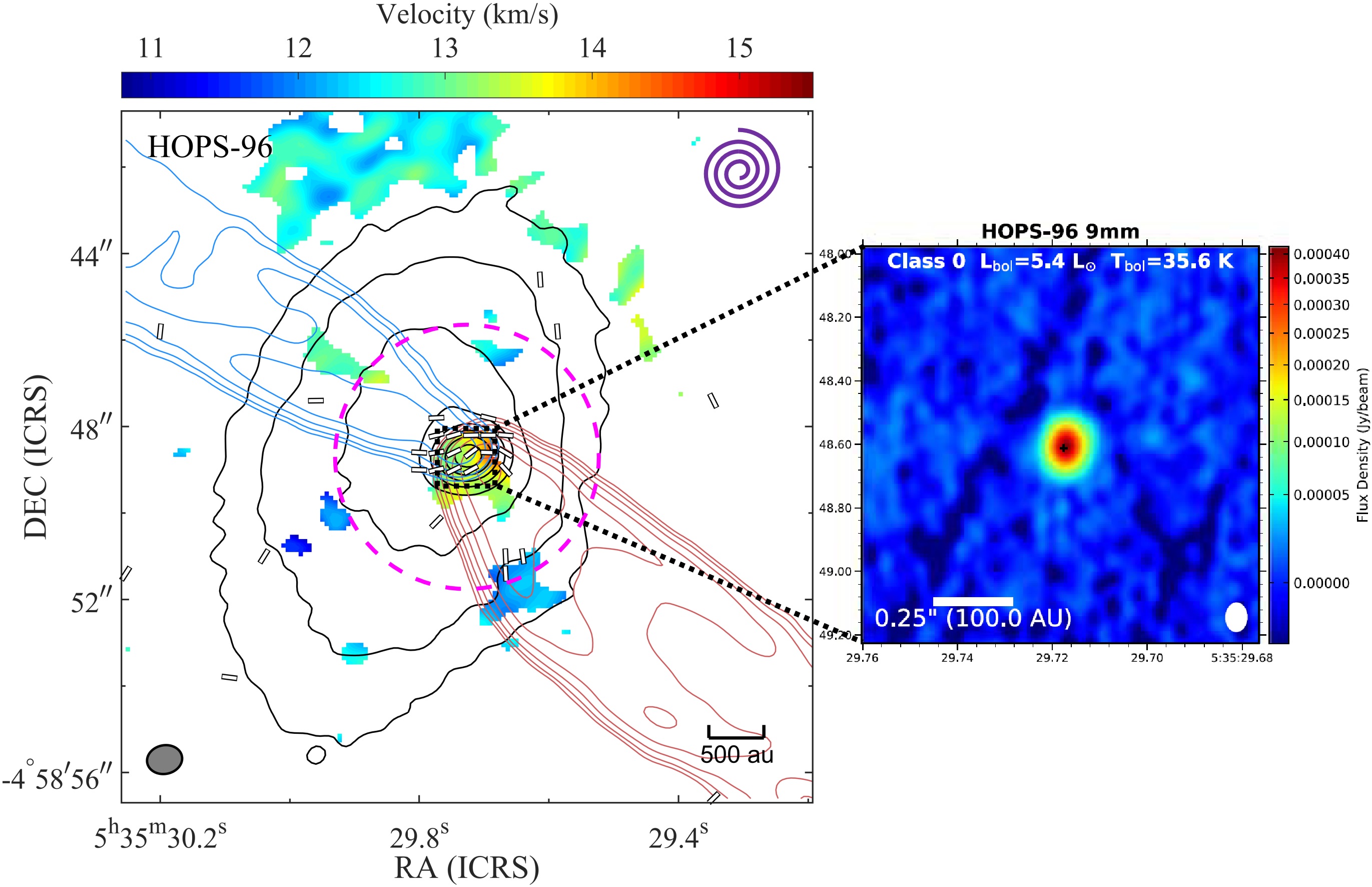}~~~~~
\includegraphics[clip=true,trim=0cm 0cm 0cm 0cm,width=0.32 \textwidth]{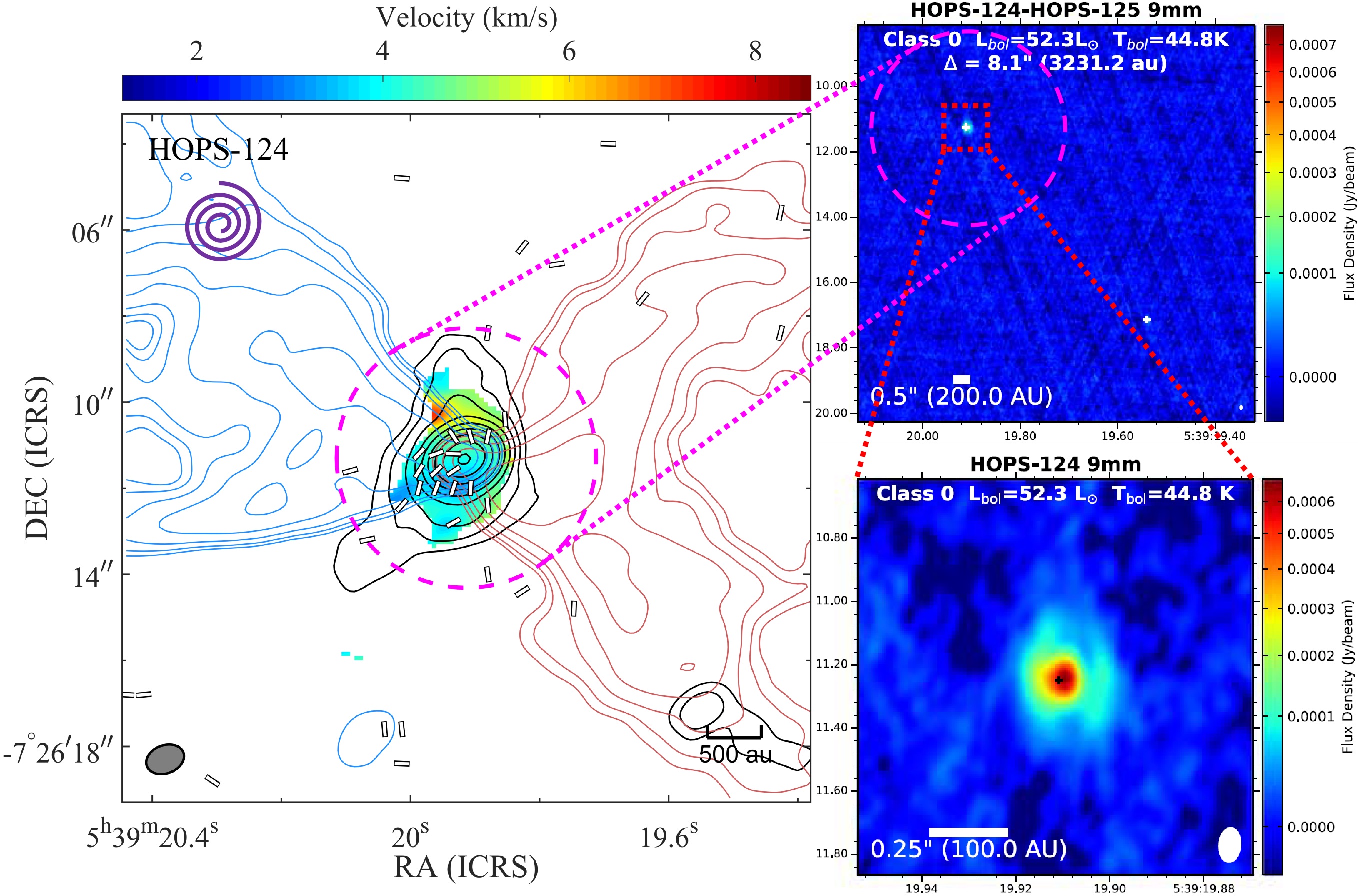}~~~~~
\includegraphics[clip=true,trim=0cm 0cm 0cm 0cm,width=0.327 \textwidth]{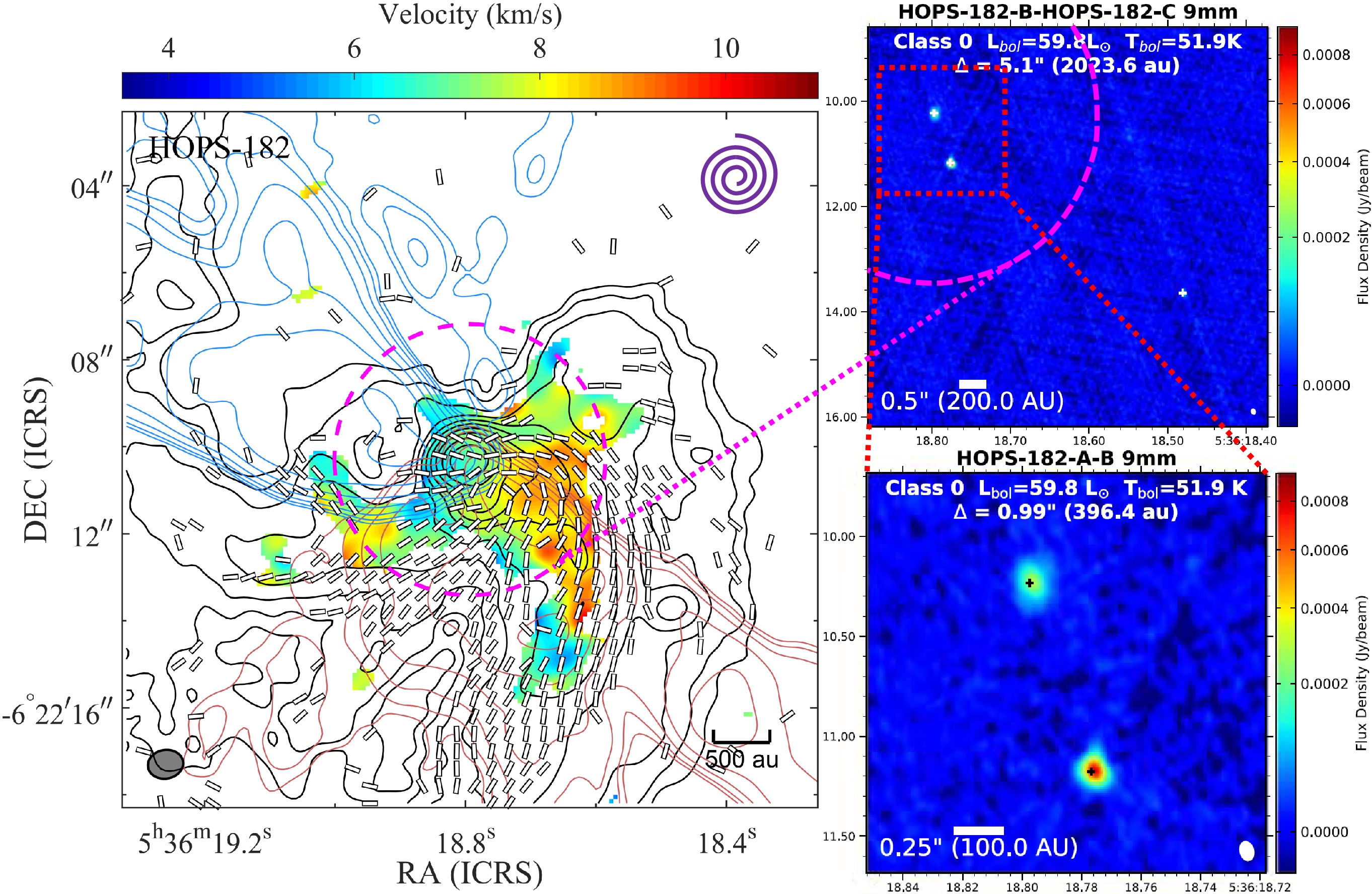}
~\\
~\\
\includegraphics[clip=true,trim=0cm 0cm 0cm 0cm,width=0.32 \textwidth]{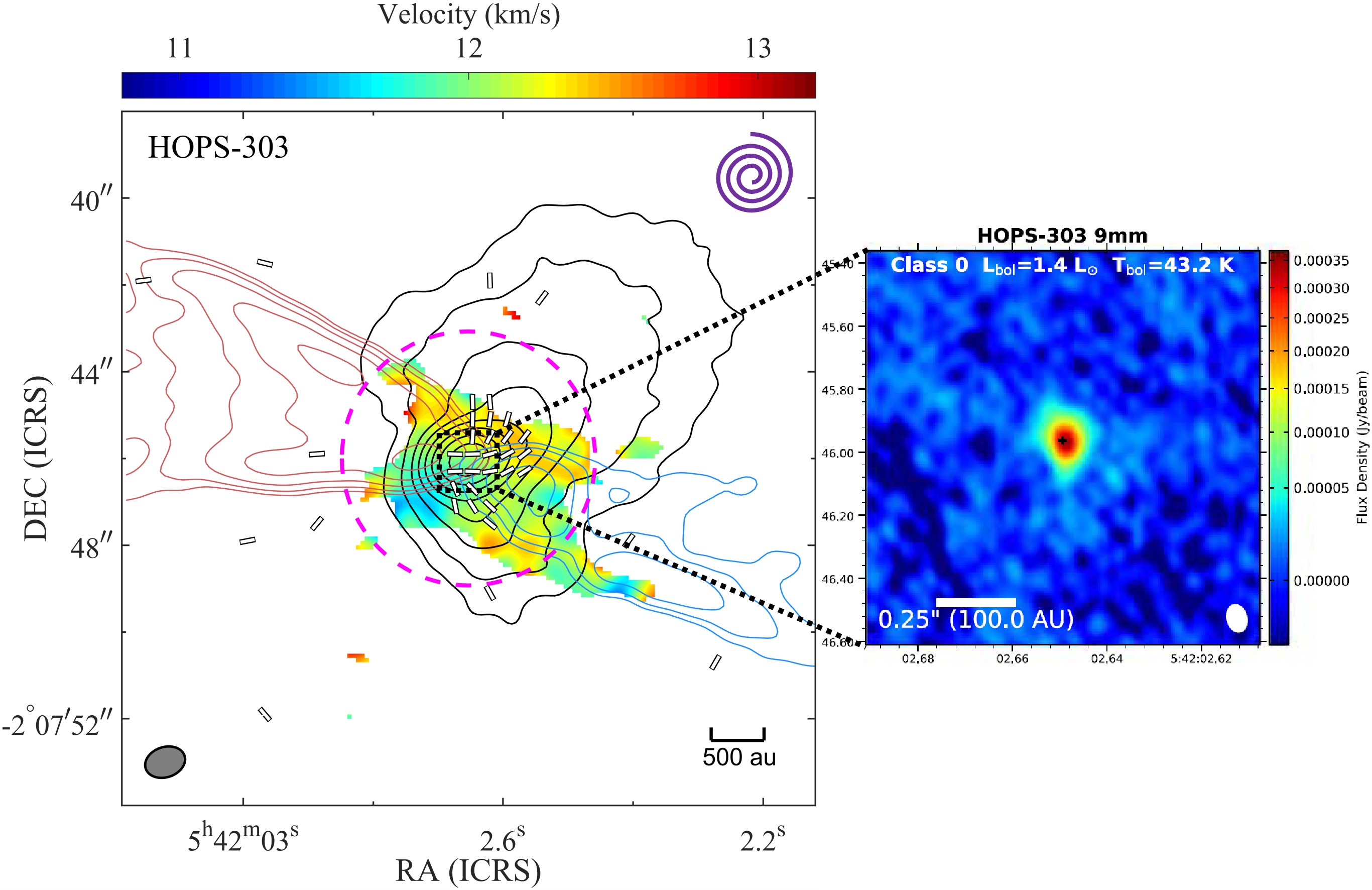}~~~~~
\includegraphics[clip=true,trim=0cm 0cm 0cm 0cm,width=0.32 \textwidth]{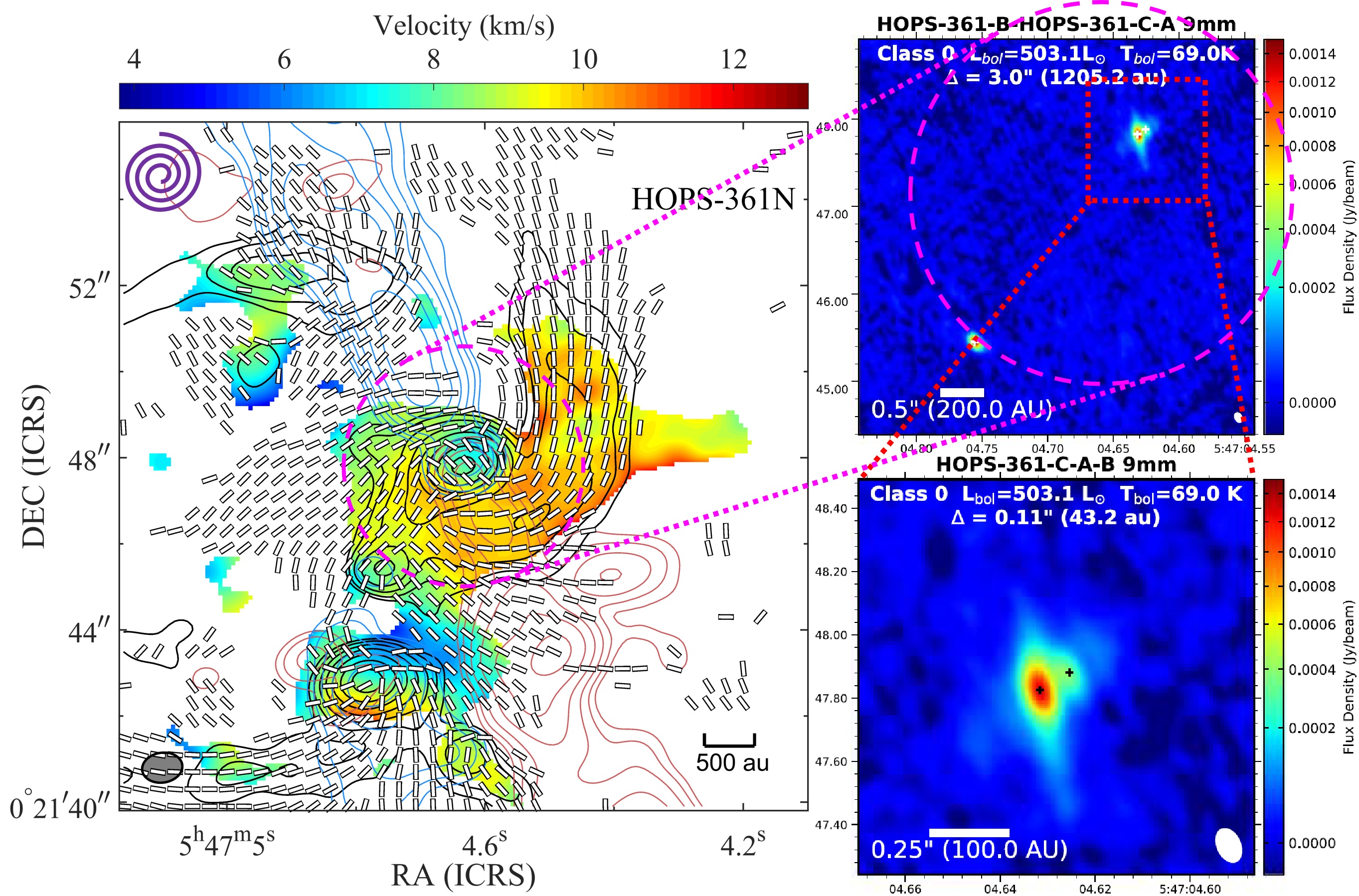}~~~~~
\includegraphics[clip=true,trim=0cm 0cm 0cm 0cm,width=0.32 \textwidth]{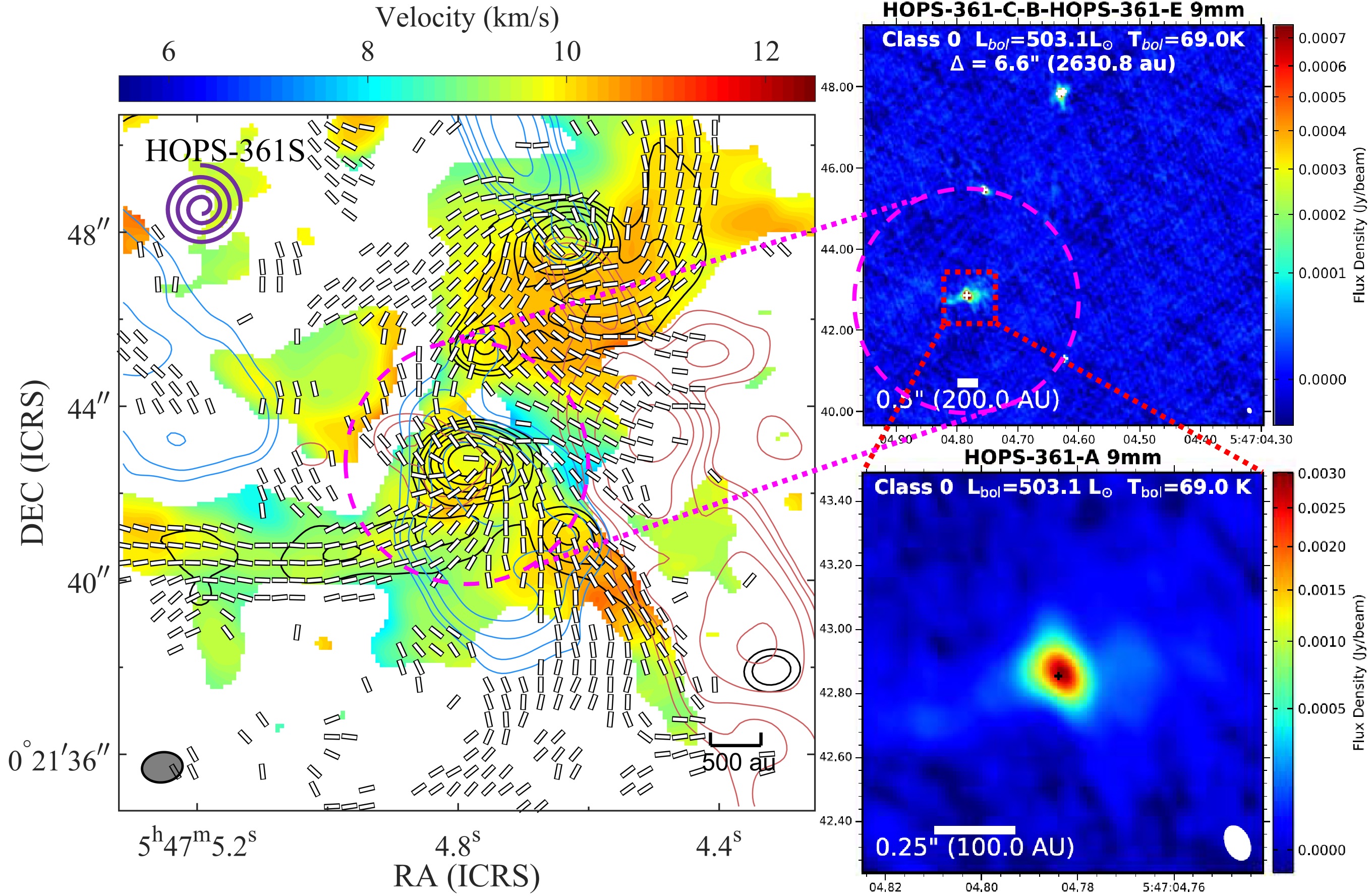}
~\\
~\\
\includegraphics[clip=true,trim=0cm 0cm 0cm 0cm,width=0.32 \textwidth]{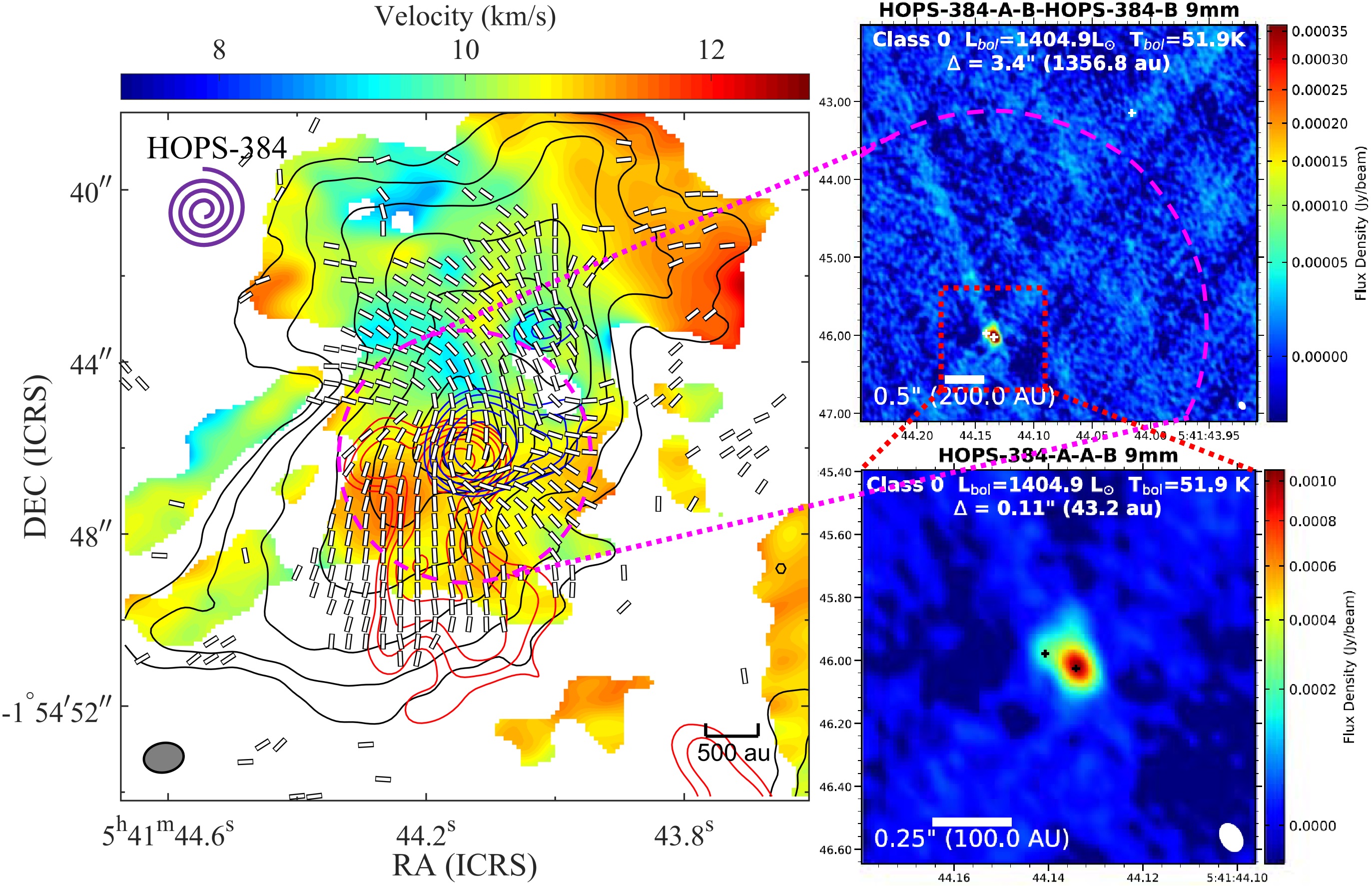}~~~~~
\includegraphics[clip=true,trim=0cm 0cm 0cm 0cm,width=0.32 \textwidth]{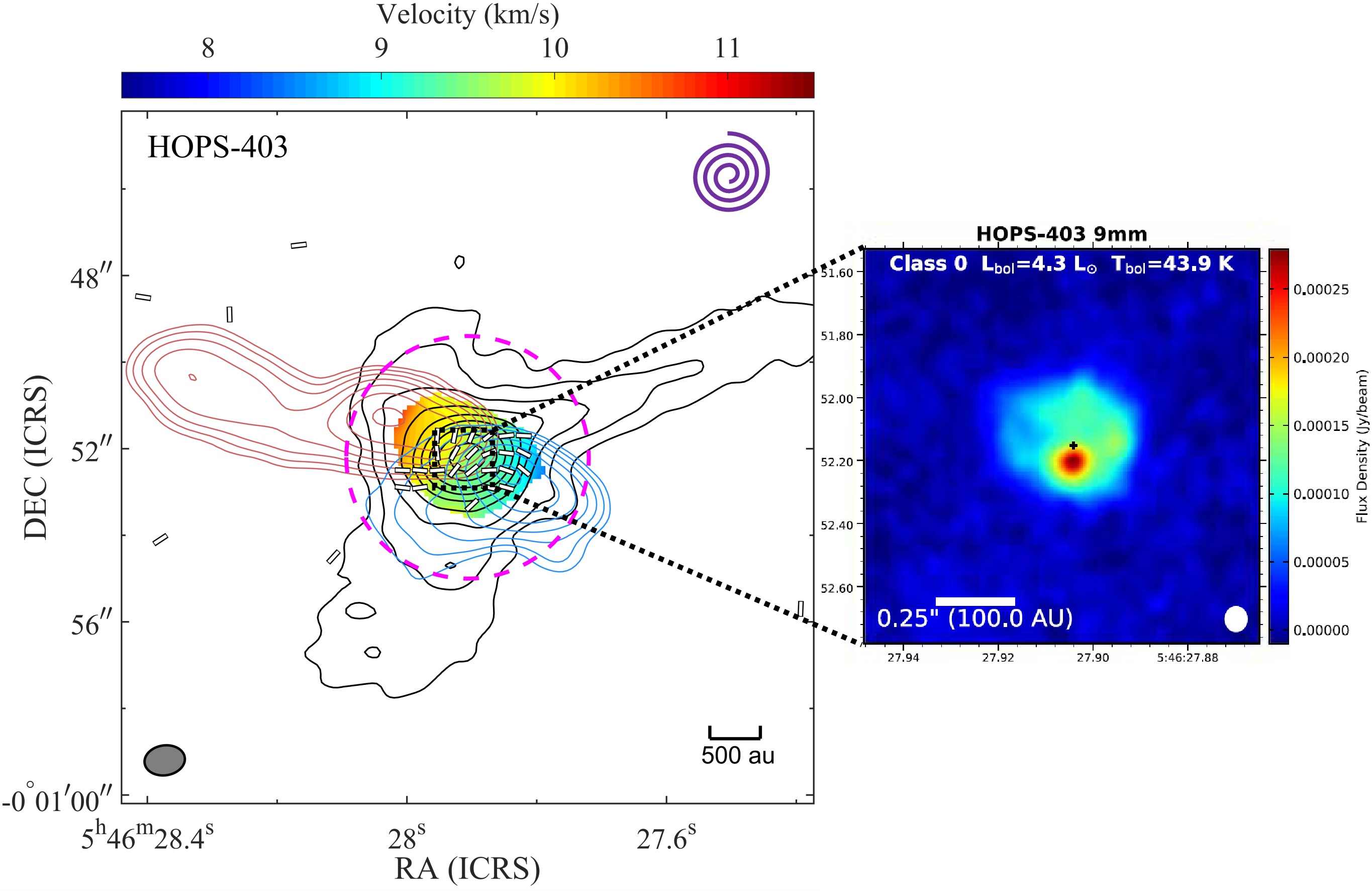}~~~~~
\includegraphics[clip=true,trim=0cm 0cm 0cm 0cm,width=0.32 \textwidth]{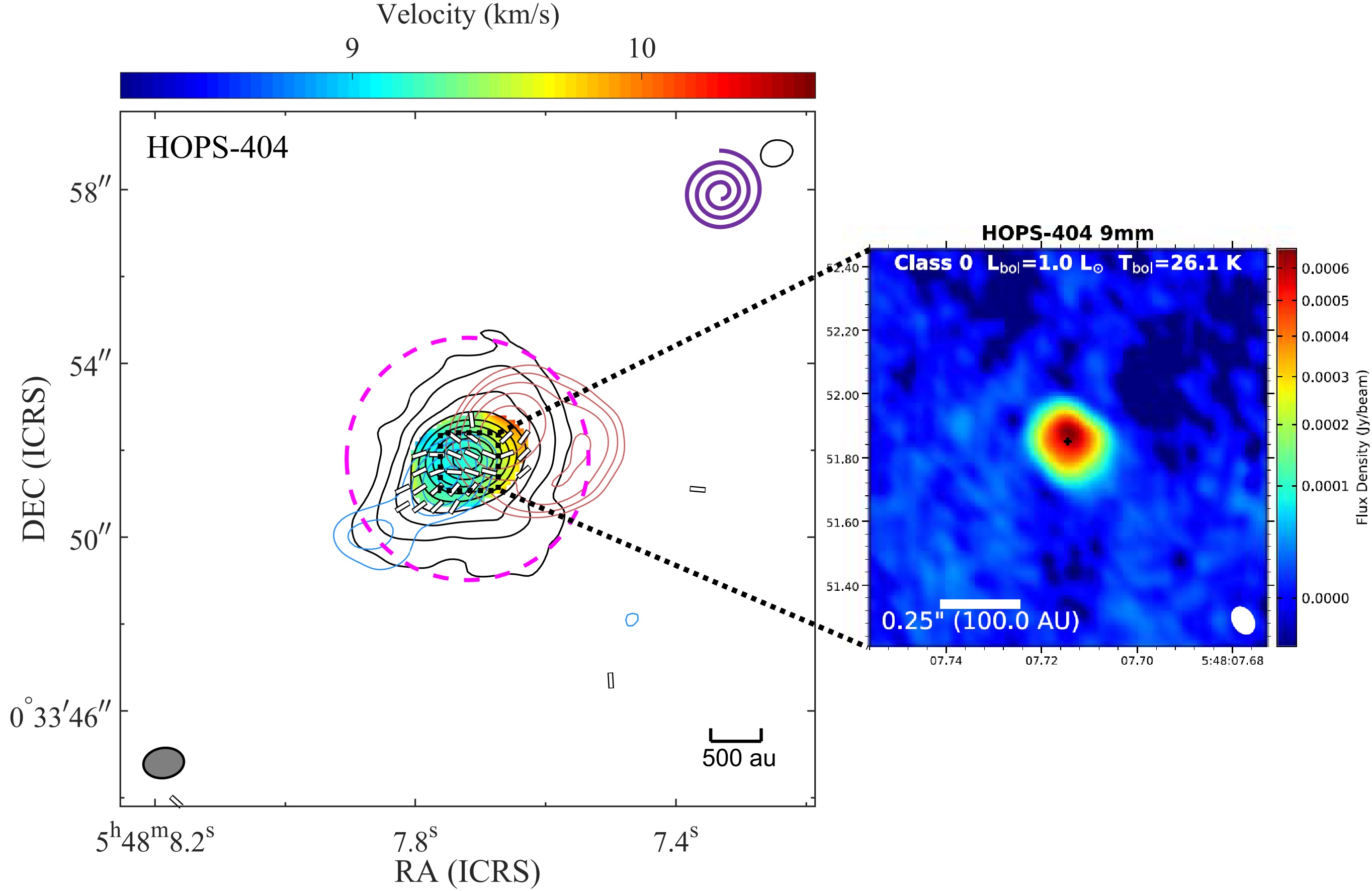}
\end{figure*}

~\\
~\\
~\\
~\\

\begin{figure*}[!ht]
\centering
\textbf{Others}
~\\
~\\
\includegraphics[clip=true,trim=0cm 0cm 0cm 0cm,width=0.32 \textwidth]{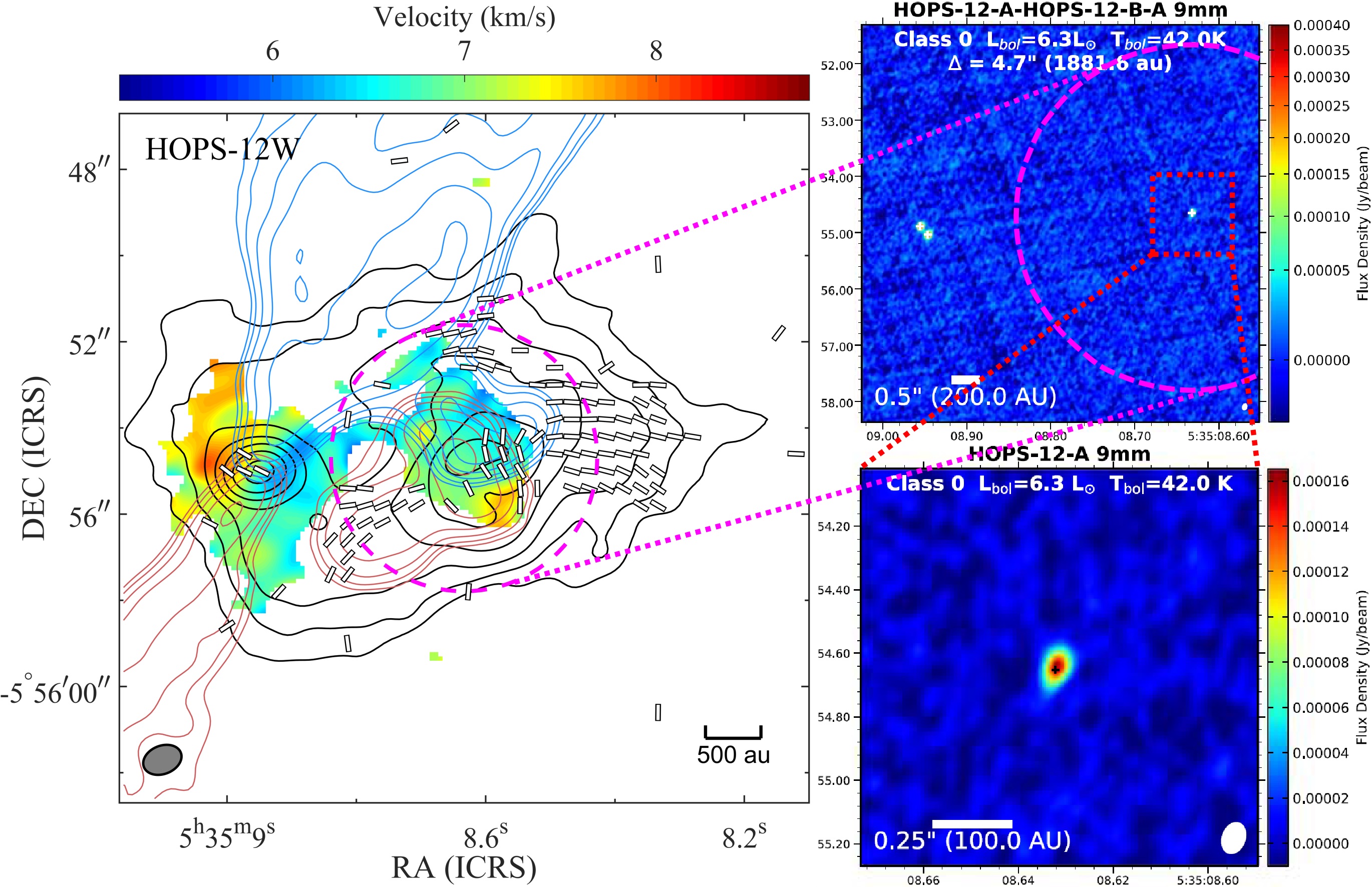}~~~~~~~~~~~~~
\includegraphics[clip=true,trim=0cm 0cm 0cm 0cm,width=0.2 \textwidth]{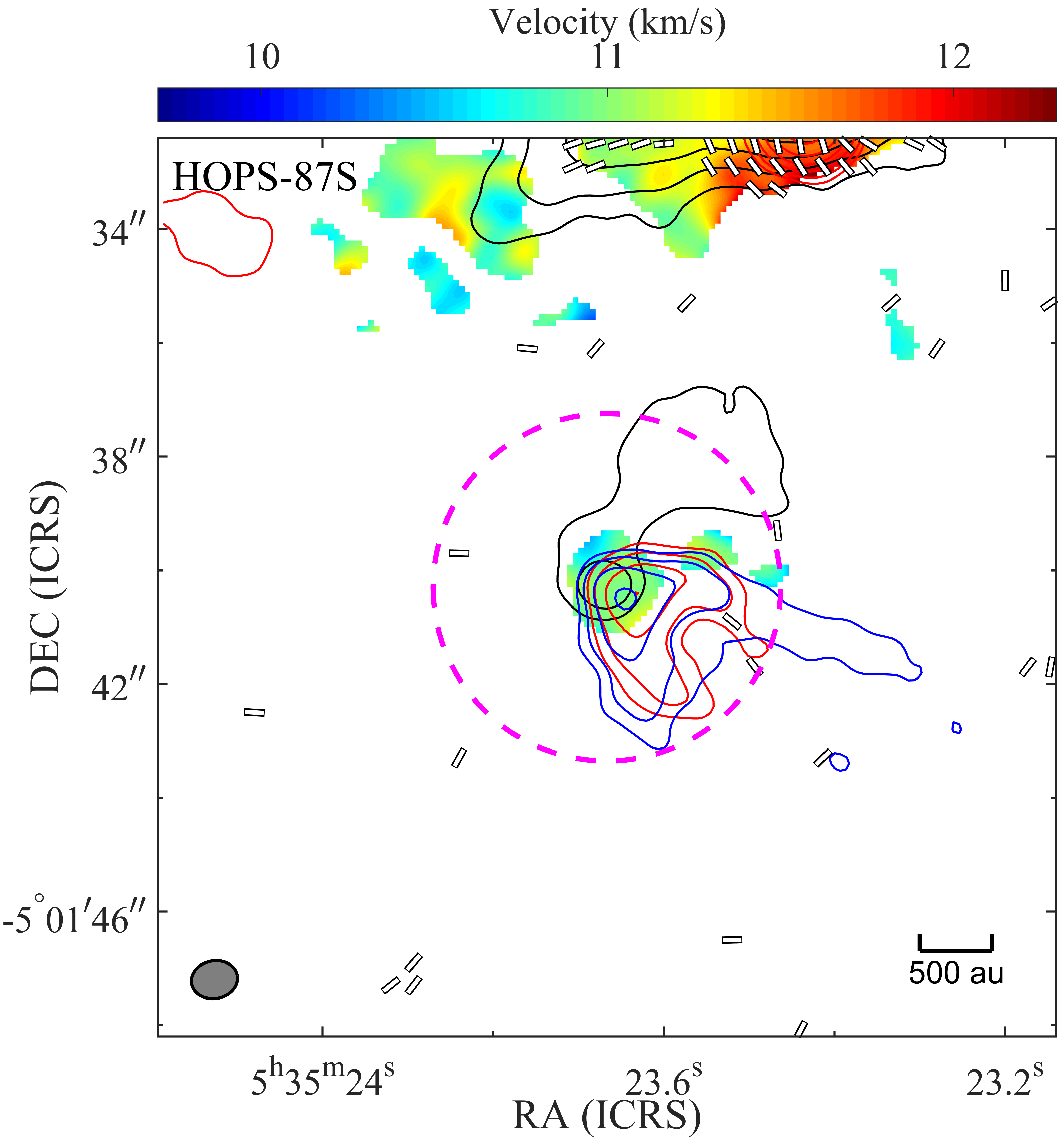}~~~~~~~~~~~~~
\includegraphics[clip=true,trim=0cm 0cm 0cm 0cm,width=0.327 \textwidth]{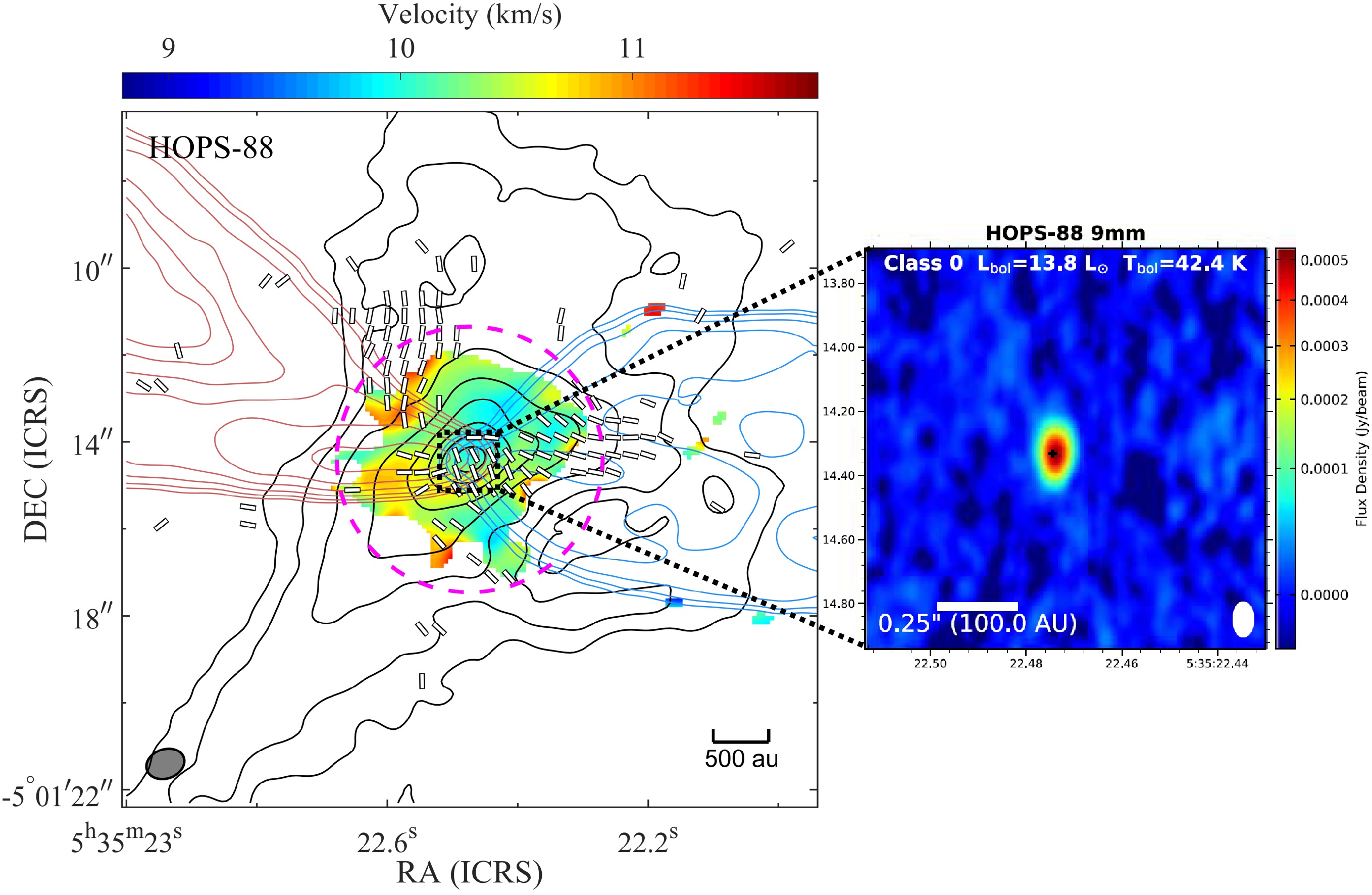}
\end{figure*}

\newpage
\begin{figure*}[!ht]
\centering
\textbf{Others (continue)}
\includegraphics[clip=true,trim=0cm 0cm 0cm 0cm,width=0.32 \textwidth]{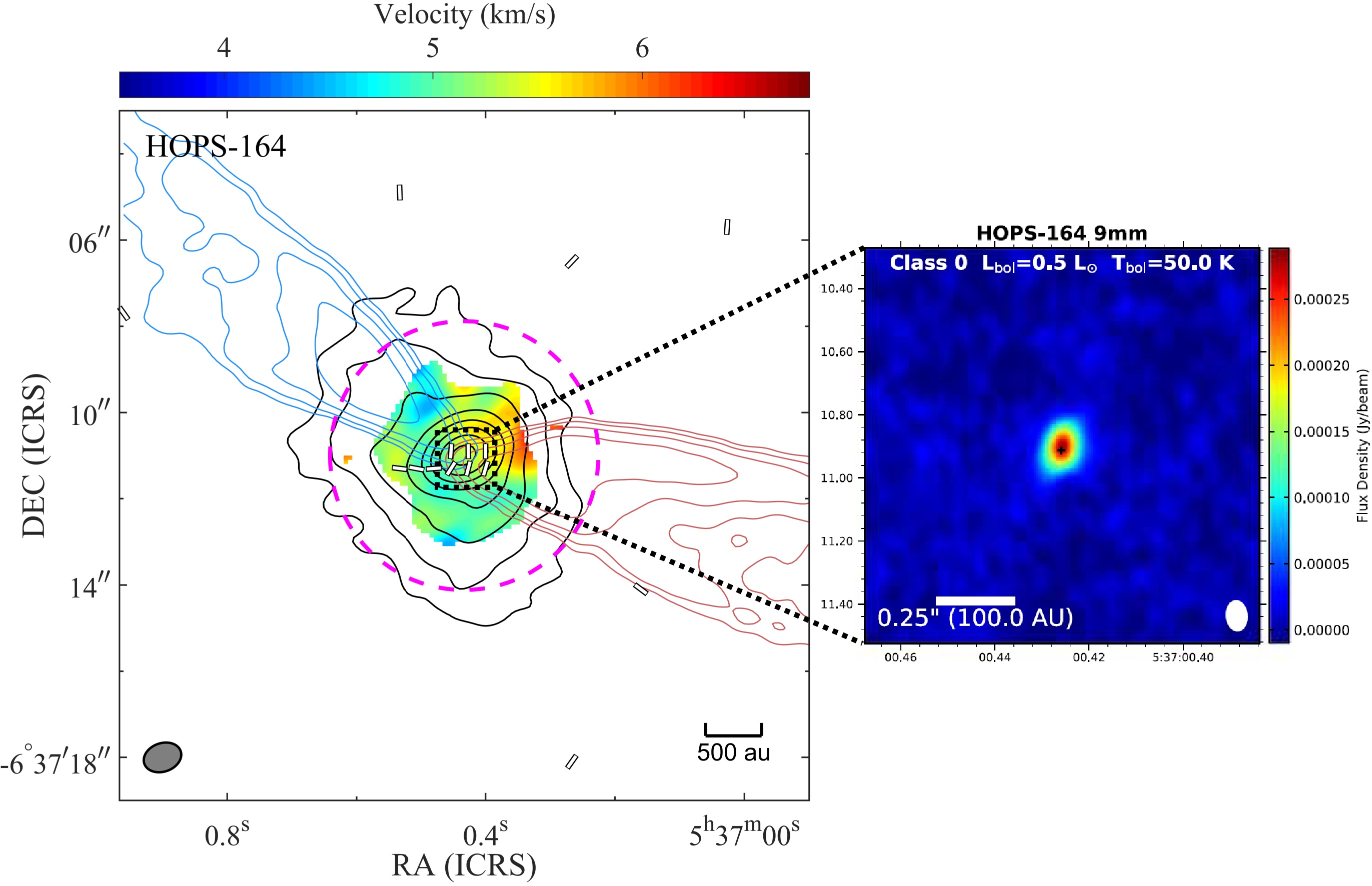}~~~~~
\includegraphics[clip=true,trim=0cm 0cm 0cm 0cm,width=0.32 \textwidth]{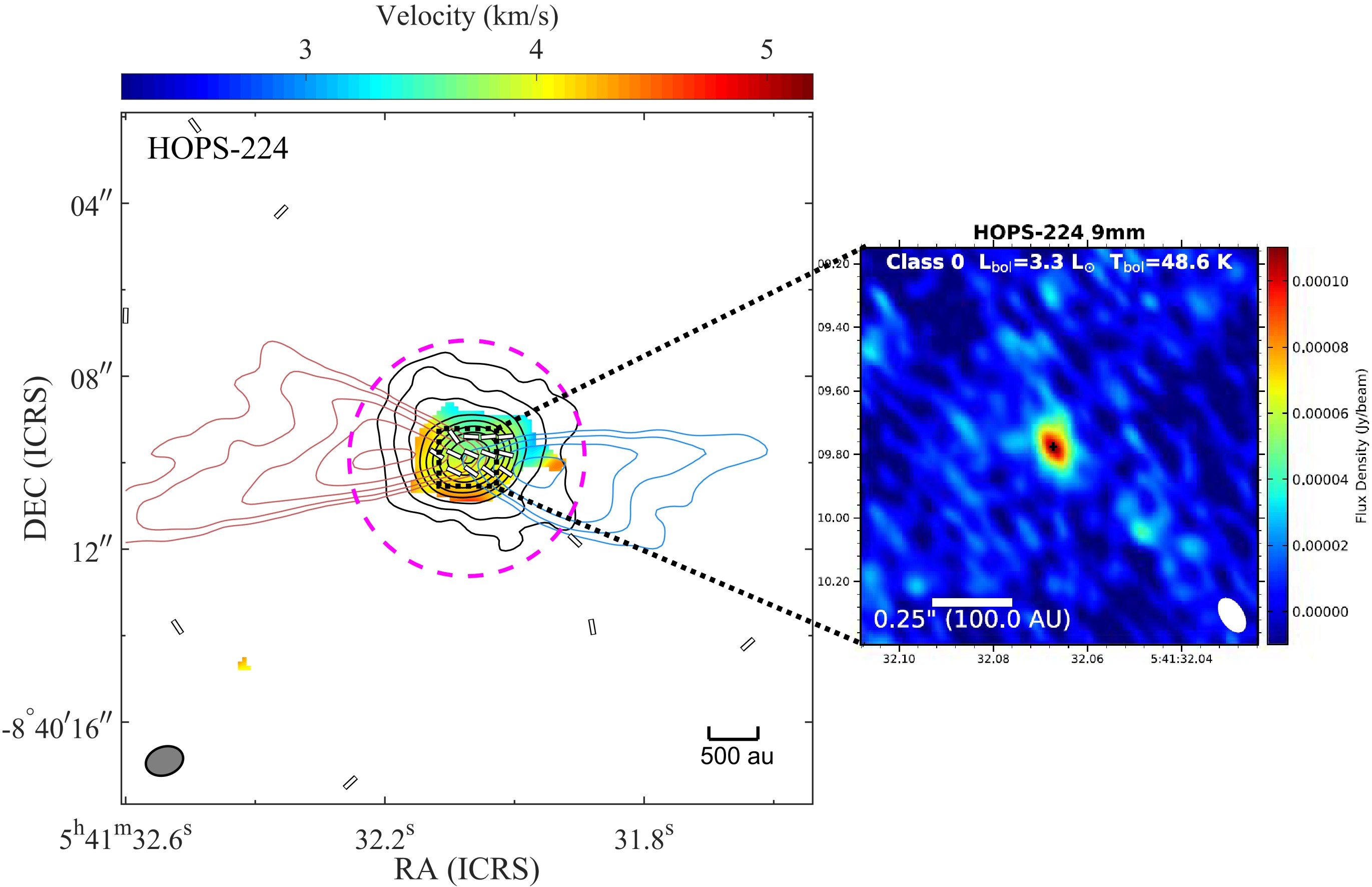}~~~~~
\includegraphics[clip=true,trim=0cm 0cm 0cm 0cm,width=0.32 \textwidth]{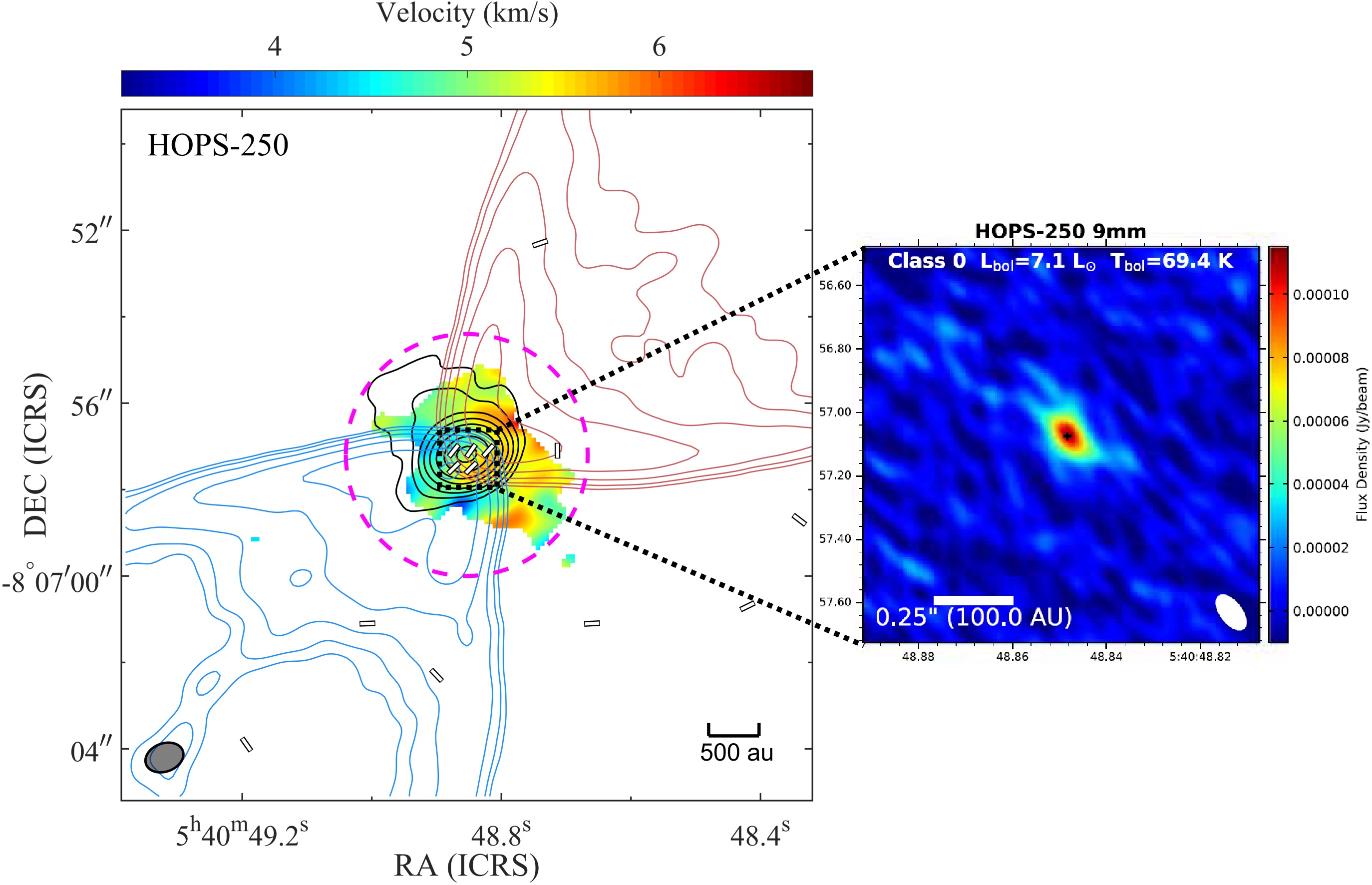}
~\\
~\\
\includegraphics[clip=true,trim=0cm 0cm 0cm 0cm,width=0.32 \textwidth]{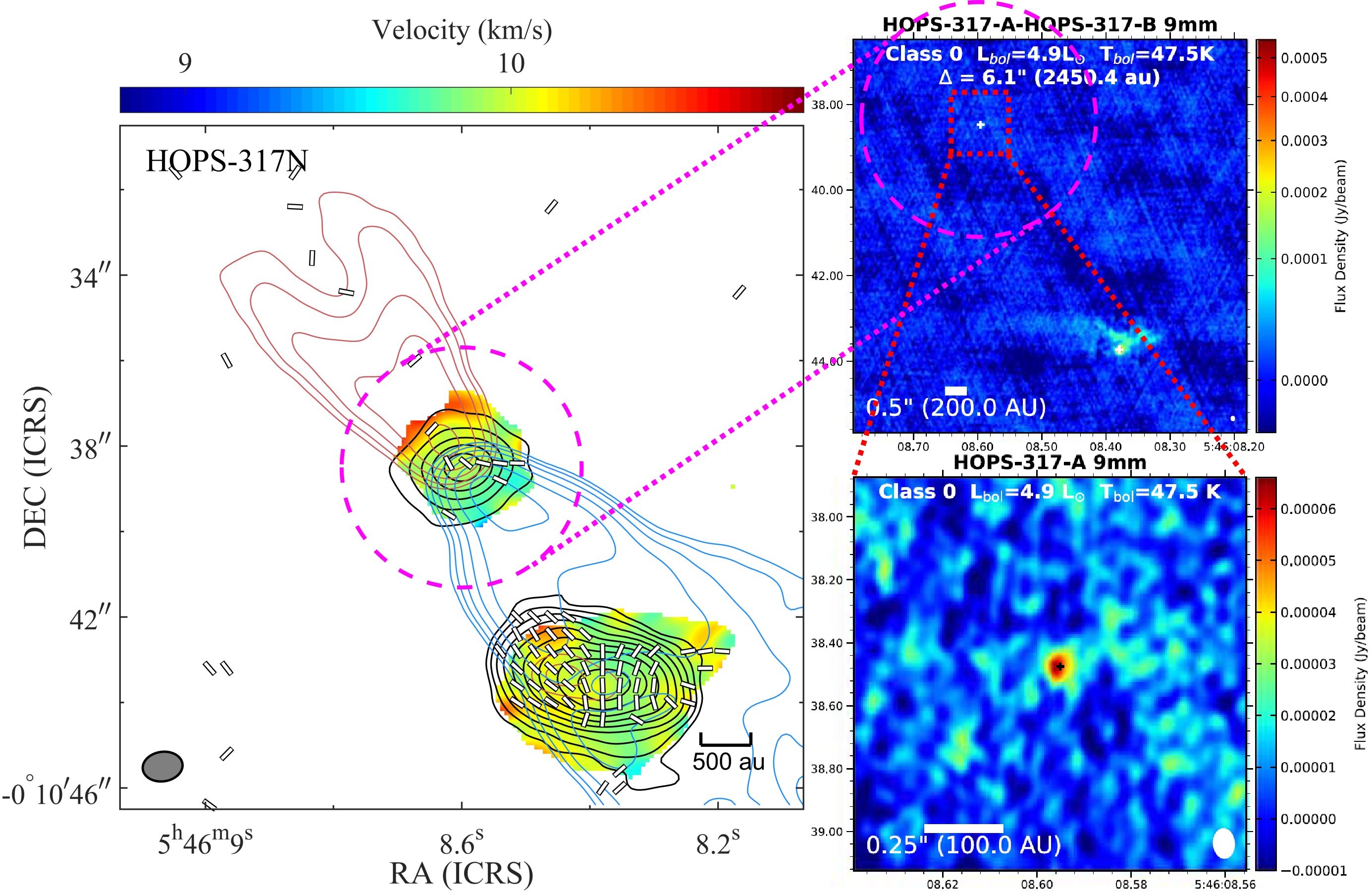}~~~~~
\includegraphics[clip=true,trim=0cm 0cm 0cm 0cm,width=0.32 \textwidth]{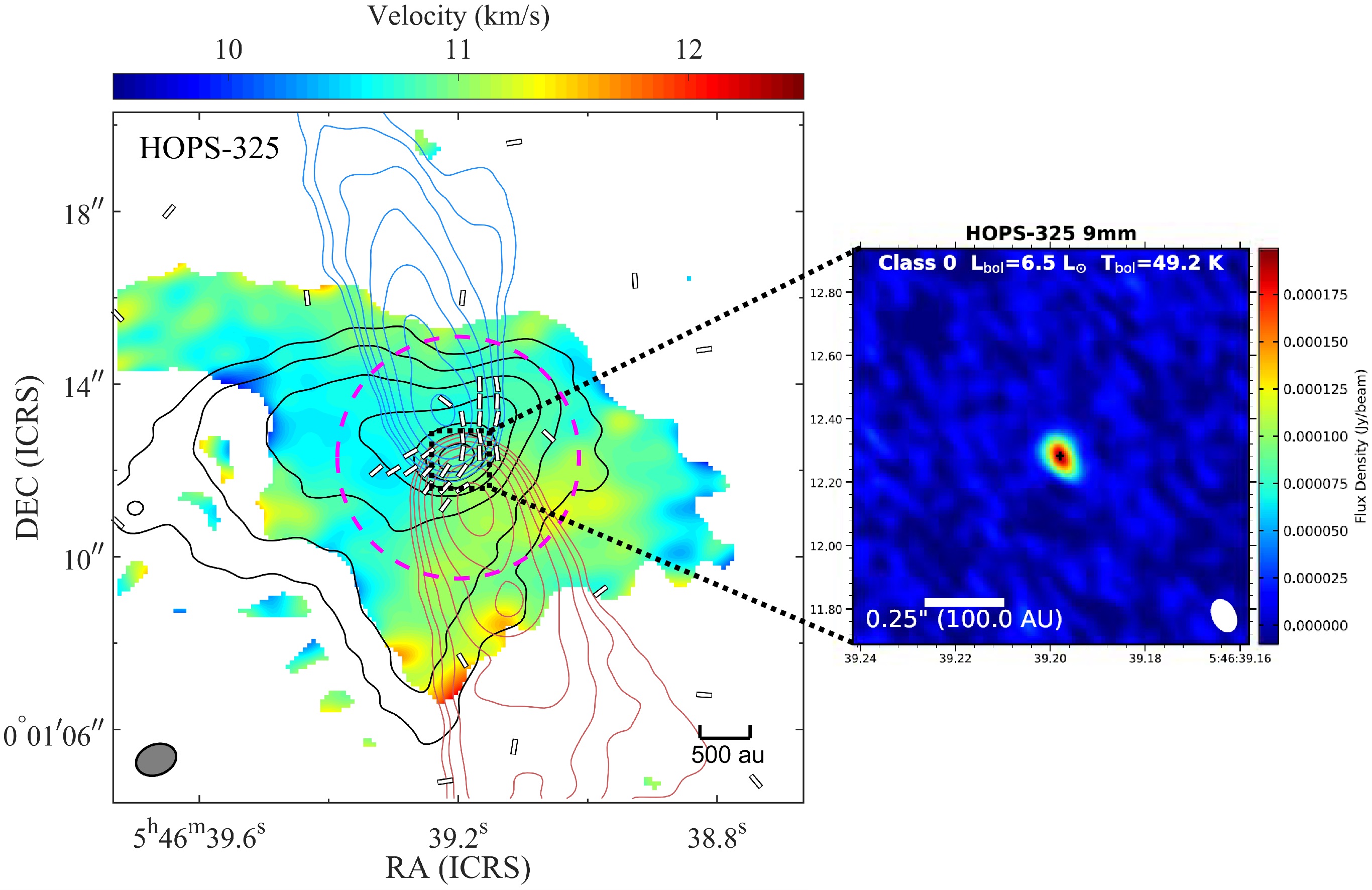}~~~~~
\includegraphics[clip=true,trim=0cm 0cm 0cm 0cm,width=0.32 \textwidth]{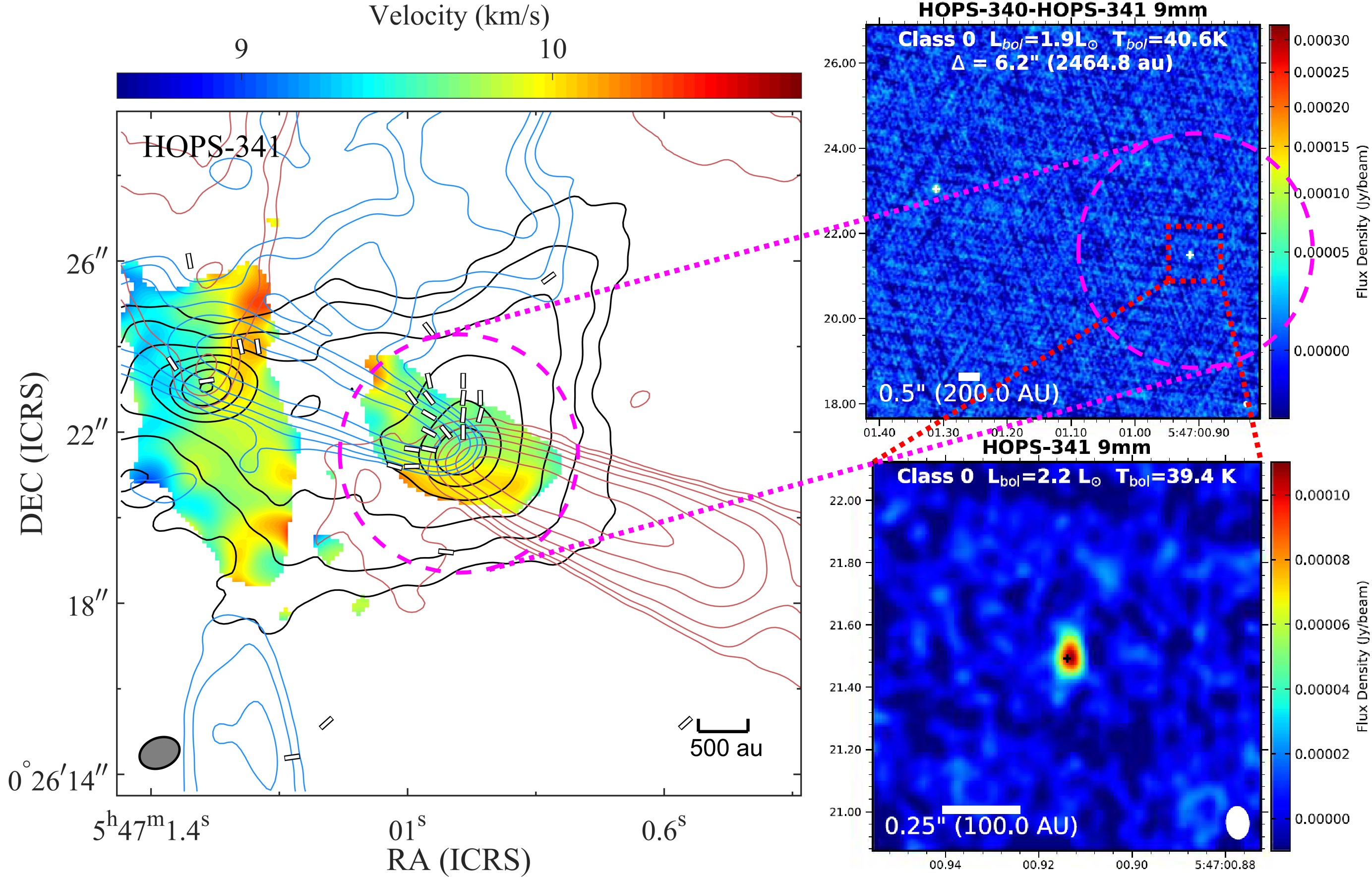}
~\\
~\\
\includegraphics[clip=true,trim=0cm 0cm 0cm 0cm,width=0.32 \textwidth]{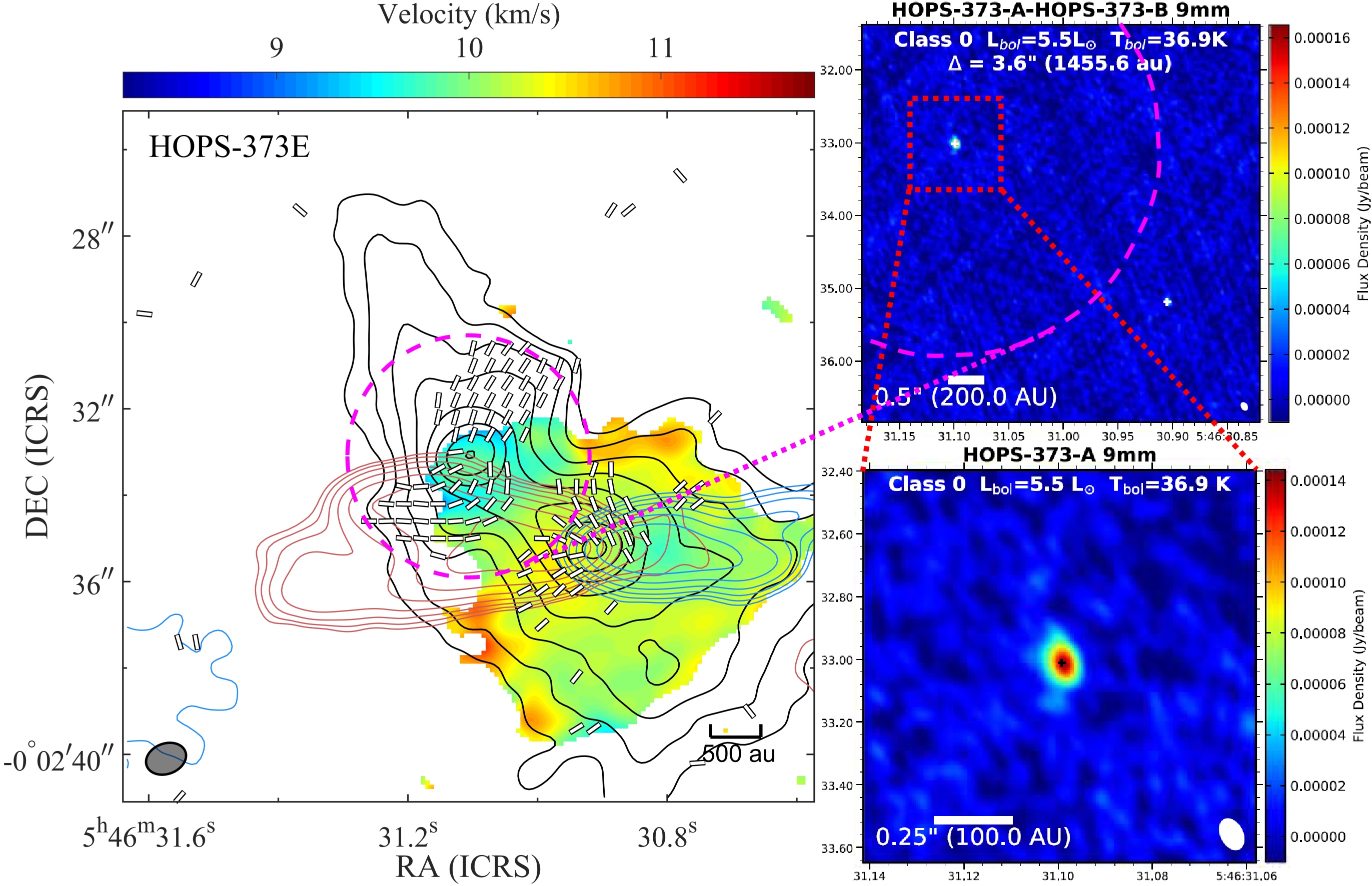}~~~~~
\includegraphics[clip=true,trim=0cm 0cm 0cm 0cm,width=0.32 \textwidth]{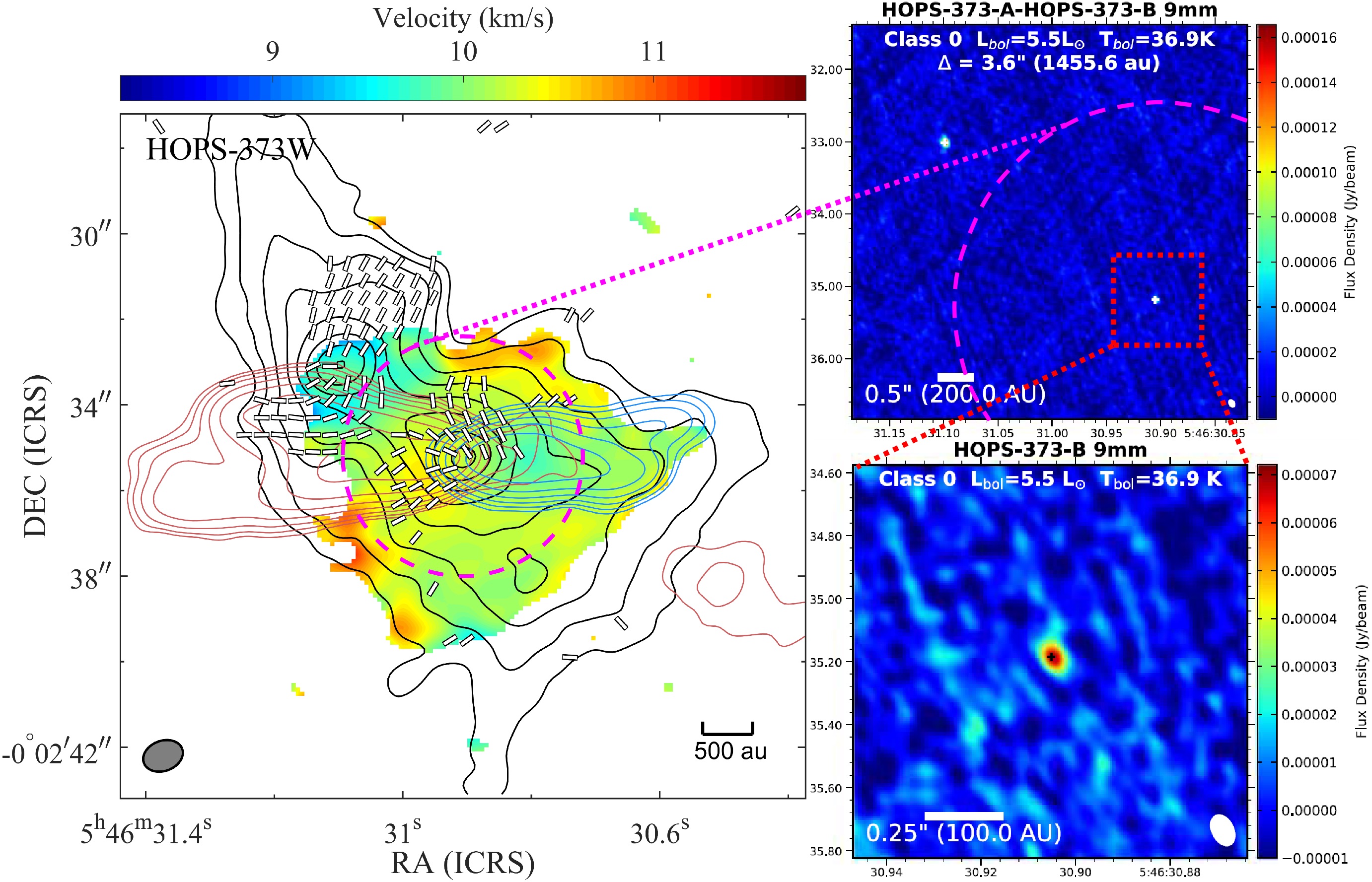}~~~~~
\includegraphics[clip=true,trim=0cm 0cm 0cm 0cm,width=0.32 \textwidth]{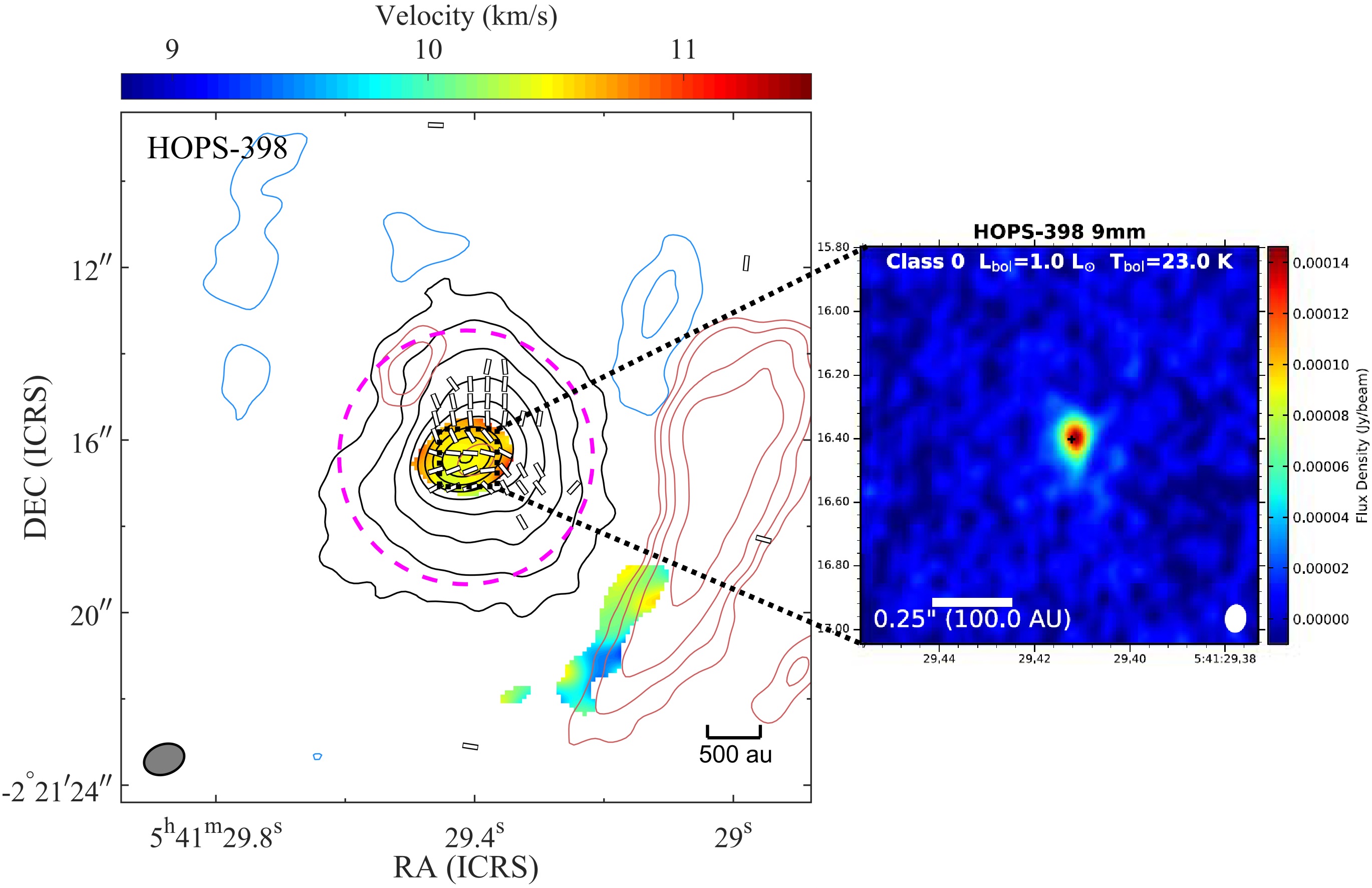}
~\\
~\\
\includegraphics[clip=true,trim=0cm 0cm 0cm 0cm,width=0.32 \textwidth]{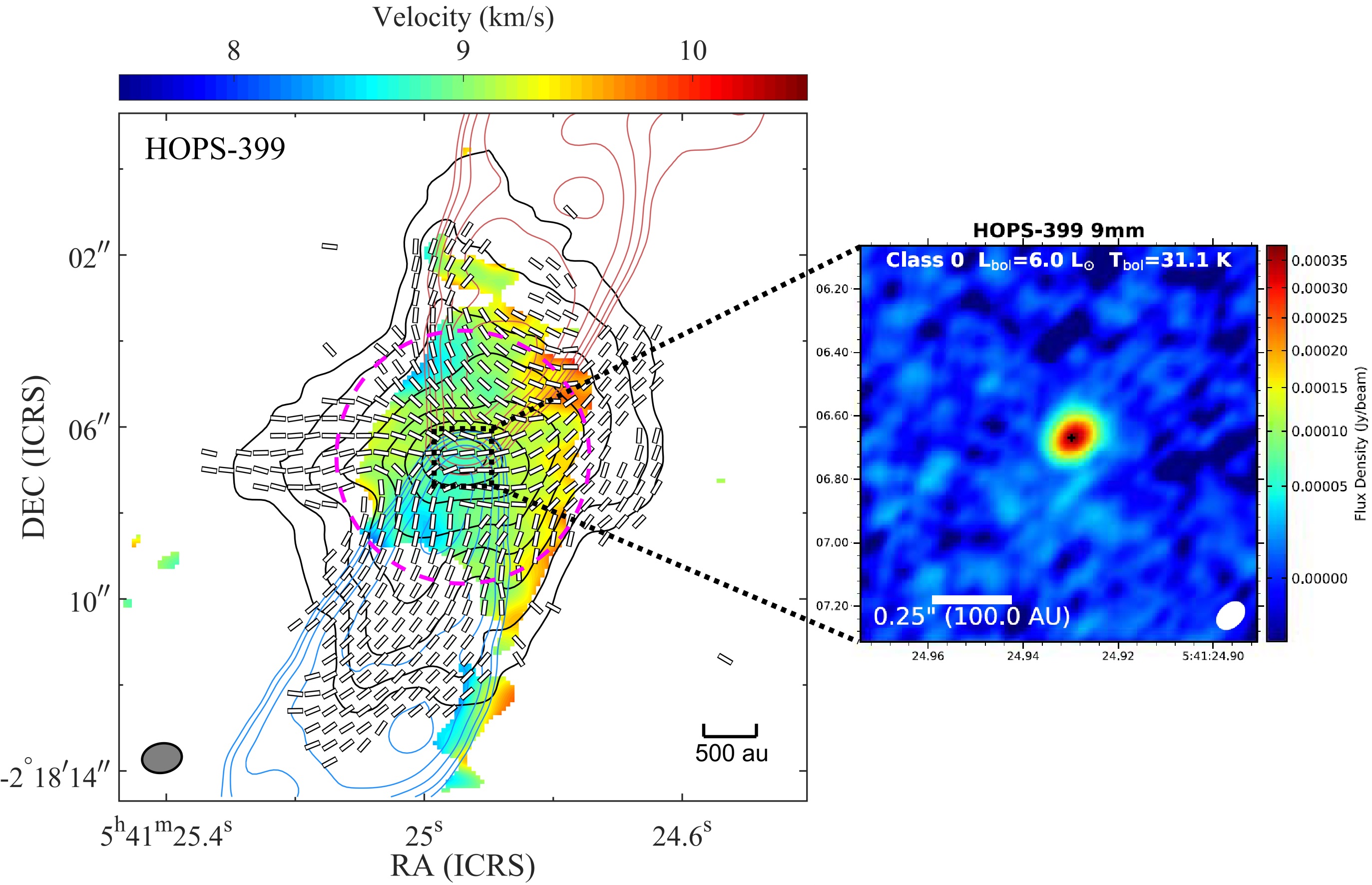}~~~~~
\includegraphics[clip=true,trim=0cm 0cm 0cm 0cm,width=0.32 \textwidth]{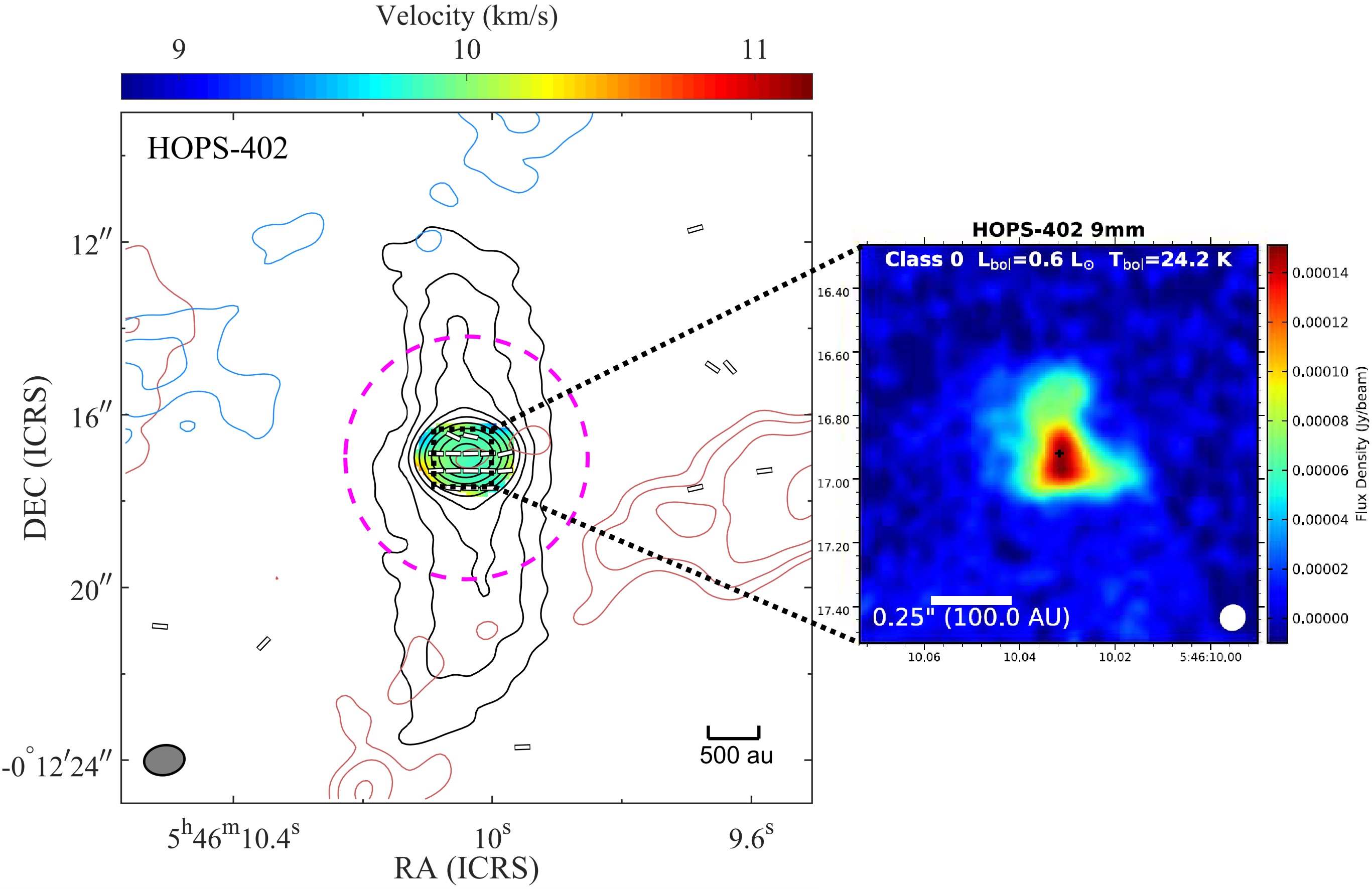}~~~~~
\includegraphics[clip=true,trim=0cm 0cm 0cm 0cm,width=0.32 \textwidth]{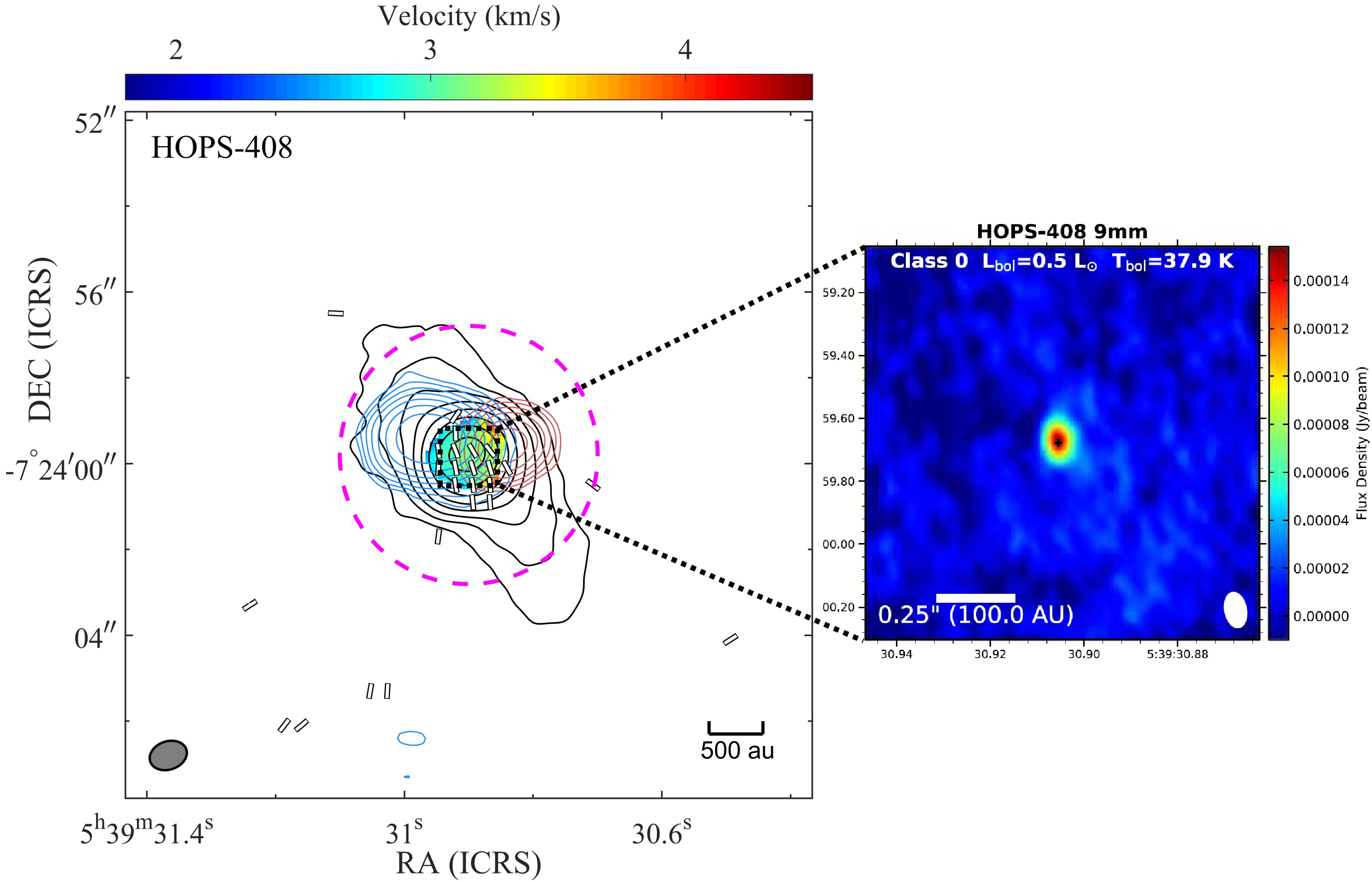}
~\\
~\\
\includegraphics[clip=true,trim=0cm 0cm 0cm 0cm,width=0.32 \textwidth]{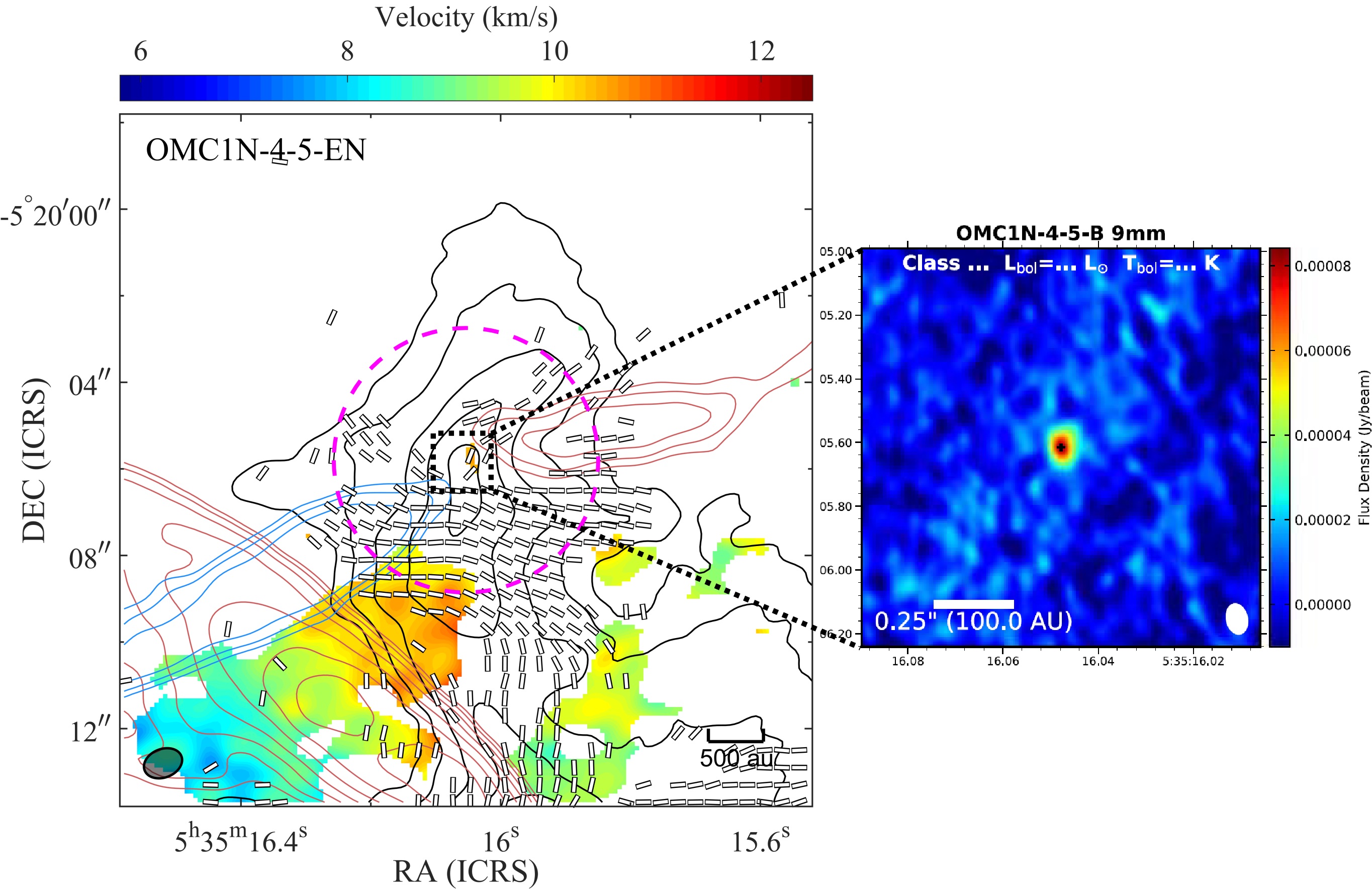}~~~~~
\includegraphics[clip=true,trim=0cm 0cm 0cm 0cm,width=0.32 \textwidth]{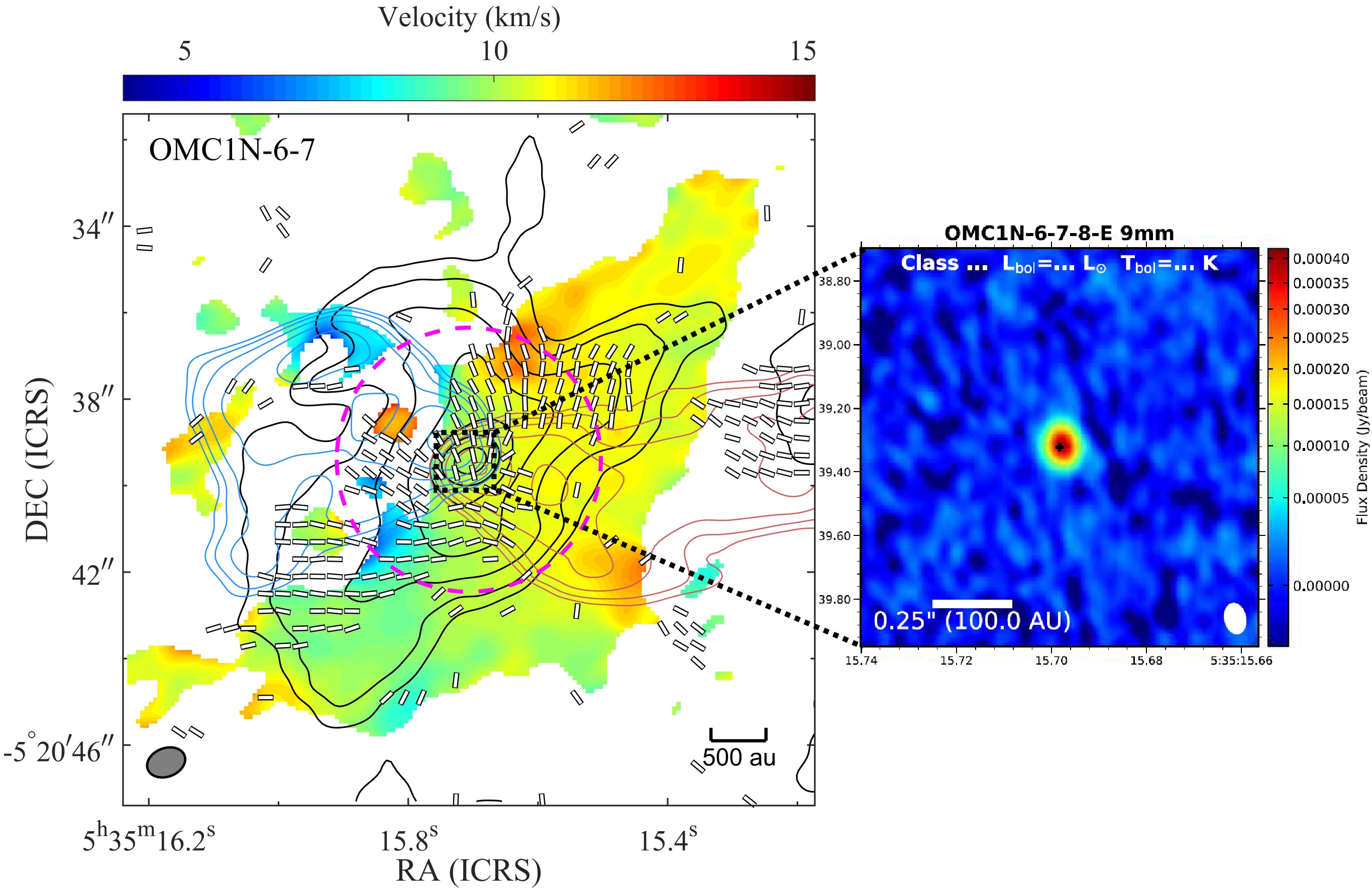}~~~~~
\includegraphics[clip=true,trim=0cm 0cm 0cm 0cm,width=0.32 \textwidth]{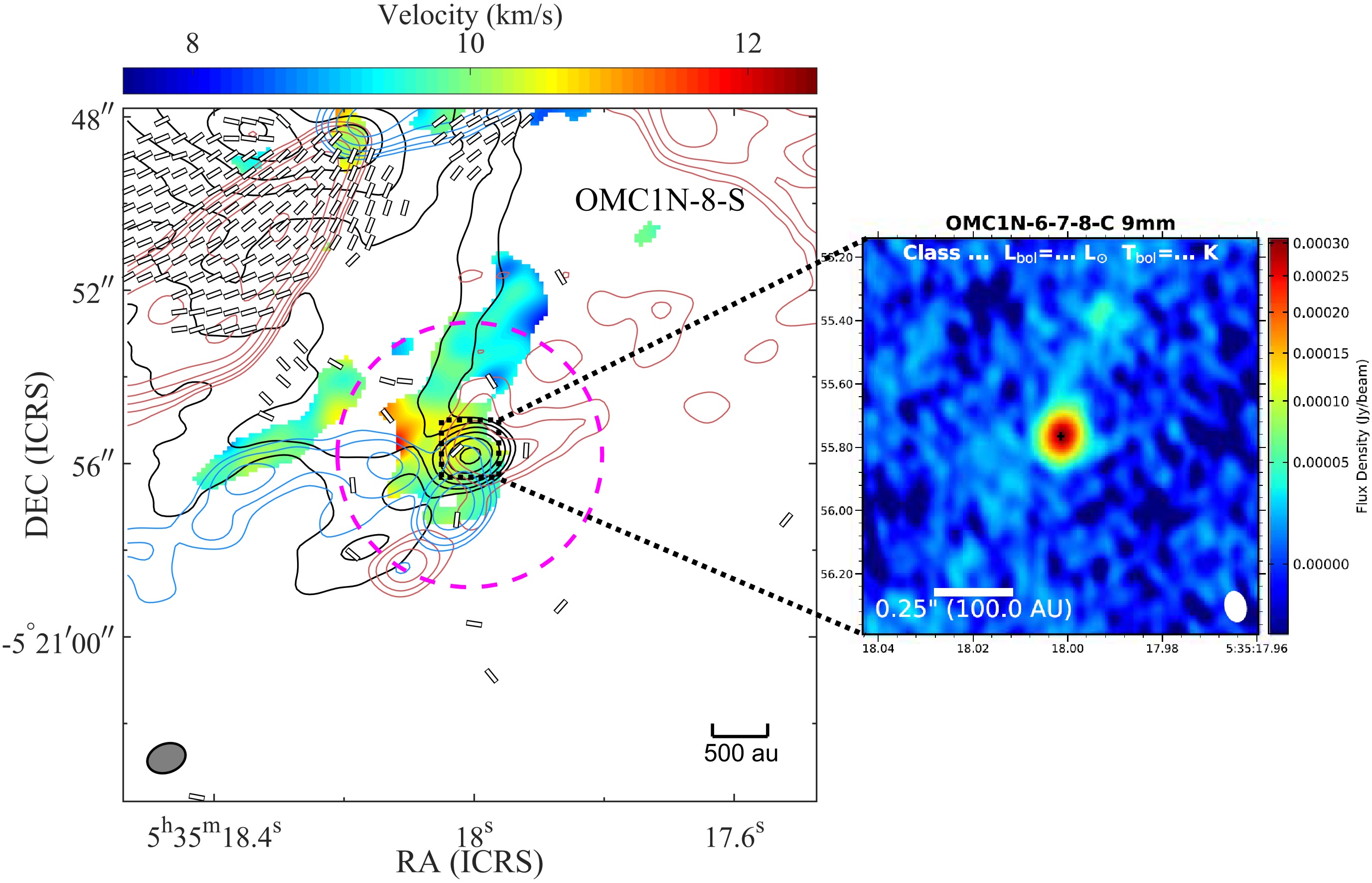}
\end{figure*}


\begin{thebibliography}{}
\expandafter\ifx\csname natexlab\endcsname\relax\def\natexlab#1{#1}\fi
\providecommand{\url}[1]{\href{#1}{#1}}
\providecommand{\dodoi}[1]{doi:~\href{http://doi.org/#1}{\nolinkurl{#1}}}
\providecommand{\doeprint}[1]{\href{http://ascl.net/#1}{\nolinkurl{http://ascl.net/#1}}}
\providecommand{\doarXiv}[1]{\href{https://arxiv.org/abs/#1}{\nolinkurl{https://arxiv.org/abs/#1}}}

\bibitem[{A{\~n}ez-L{\'o}pez {et~al.}(2020)A{\~n}ez-L{\'o}pez, {Busquet}, Koch, {et~al.}}]{nacho2020frag}
A{\~n}ez-L{\'o}pez, N., Busquet, G., Koch, P.~M., {et~al.} 2020, \aap, 644, A52

\bibitem[{Allen {et~al.}(2003)Allen, {Li}, \& Shu}]{allen2003collapse}
Allen, A., Li, Z.-Y., \& Shu, F.~H. 2003, \apj, 599, 363

\bibitem[{Andersson {et~al.}(2015)}]{andersson2015interstellar}
Andersson, B.~G., Lazarian, A., \& Vaillancourt, J.~E. 2015, \araa, 53, 501

\bibitem[{Beltr{\'a}n {et~al.}(2019)}]{beltran2019alma}
Beltr{\'a}n, M.~T., Padovani, M., Girart, J.~M., {et~al.} 2019, \aap, 630, A54

\bibitem[{Beuther {et~al.}(2015)}]{beuther2015frag}
Beuther, H., Henning, T., Linz, H., {et~al.} 2015, \aap, 581, A119

\bibitem[{Bik {et~al.}(2012)}]{bik2012fra}
Bik, A., Henning, T., Stolte, A., {et~al.} 2012, \apj, 744, 87

\bibitem[{Busquet {et~al.}(2016)}]{busquet2016frag}
Busquet, G., Estalella, R., Palau, A., {et~al.} 2016, \apj, 819, 139

\bibitem[{CASA {et~al.}(2022)}]{casa2022}
CASA Team, Bean, B., Bhatnagar, S., {et~al.} 2022, \pasp, 134, 114501

\bibitem[{Chandrasekhar {et~al.}(1953)}]{chandrasekhar1953dcf}
Chandrasekhar, S. \& Fermi, E. 1953, \apj, 118, 113

\bibitem[{Chini {et~al.}(2013)}]{chini2013multi}
Chini, R., Barr, A., Buda, L.~S., {et~al.} 2013, Central European Astrophysical Bulletin, 37, 295

\bibitem[{Chini {et~al.}(2012)}]{chini2012multi}
Chini, R., Hoffmeister, V.~H., Nasseri, A., {et~al.} 2012, \mnras, 424, 1925

\bibitem[{Commer{\c{c}}on {et~al.}(2011)}]{commercon2011collapse}
Commer{\c{c}}on, B., Hennebelle, P., \& Henning, T. 2011, \apjl, 742, L9

\bibitem[{Cort{\'e}s {et~al.}(2021)}]{cortes2021magnetic}
Cort{\'e}s, P.~C., Sanhueza, P., Houde, M., {et~al.} 2021, \apj, 923, 204

\bibitem[{Cort{\'e}s {et~al.}(2024)}]{cortes2024magmar}
Cort{\'e}s, P.~C., Girart, J.~M., Sanhueza, P., {et~al.} 2024, arXiv preprint arXiv:2406.14663

\bibitem[{Cox {et~al.}(2018)}]{cox2018alma}
Cox, E.~G., Harris, R.~J., Looney, L.~W., {et~al.} 2018, \apj, 855, 92

\bibitem[{Crutcher (1999)}]{crutcher1999ism}
Crutcher, R.~M. 1999, \apj, 520, 706

\bibitem[{Cunningham {et~al.}(2018)}]{cunningham2018mag}
Cunningham, A.~J., Krumholz, M.~R., McKee, C.~F., {et~al.} 2018, \mnras, 476, 771

\bibitem[{Davis (1999)}]{davis1951dcf}
Davis, R.~M. 1951, Phys. Rev., 81, 890

%\bibitem[{Federman {et~al.}(2023)}]{federman2023300}
%Federman, S., Megeath, S.~T., Tobin, J.~J., {et~al.} 2023, \apj, 944, 49

\bibitem[{Federrath (2015)}]{federrath2015ism}
Federrath, C. 2015, \mnras, 450, 4035

\bibitem[{Federrath {et~al.}(2008)}]{federrath2008frag}
Federrath, C., Klessen, R.~S., \& Schmidt, W. 2008, \apjl, 688, L79

\bibitem[{Federrath {et~al.}(2010)}]{federrath2010frag}
Federrath, C., Roman-Duval, J., Klessen, R.~S., {et~al.} 2010, \aap, 512, A81

\bibitem[{Fischer {et~al.}(2017)}]{fischer2017hops}
Fischer, W.~J., Megeath, S.~T., Furlan, E., {et~al.} 2017, \apj, 840, 69

\bibitem[{Frau {et~al.}(2017)}]{frau2011model}
Frau, P., Galli, D., \& Girart, J.~M. 2011, \aap, 535, A44

\bibitem[{Furlan {et~al.}(2017)}]{furlan2016herschel}
Furlan, E., Fischer, W.~J., Ali, B., {et~al.} 2016, \apjs, 224, 5

\bibitem[{Galametz {et~al.}(2018)}]{galametz2018sma}
Galametz, M., Maury, A., Girart, J.~M., {et~al.} 2018, \aap, 616, A139

\bibitem[{Galli {et~al.}(2006)}]{galli2006mag}
Galli, D., Lizano, S., Shu, F.~H., {et~al.} 2006, \apj, 647, 374

\bibitem[{Girart {et~al.}(2009)}]{girart2009magnetic}
Girart, J.~M., Beltr{\'a}n, M.~T., Zhang, Q., {et~al.} 2009, Sci, 24, 1408

\bibitem[{Girart {et~al.}(1999)}]{girart1999detection}
Girart, J.~M., Crutcher, R.~M., \& Rao, R. 1999, \apjl, 525, L109

\bibitem[{Girart {et~al.}(2013)}]{girart2013dr}
Girart, J.~M., Frau, P., Zhang, Q., {et~al.} 2013, \apj, 772, 69

\bibitem[{Girart {et~al.}(2009)}]{girart2009ism}
Girart, J.~M., Rao, R., \& Estalella, R. 2009, \apj, 694, 56

\bibitem[{Girart {et~al.}(2006)}]{girart2006magnetic}
Girart, J.~M., Rao, R., \& Marrone, D.~P. 2006, Sci, 313, 812

\bibitem[{Girichidis {et~al.}(2011)}]{girichidis2011frag}
Girichidis, P., Federrath, C., Banerjee, R., {et~al.} 2011, \mnras, 413, 2741

\bibitem[{Guszejnov {et~al.}(2021)}]{guszejnov2021starforge}
Guszejnov, D., Grudi{\'c}, M.~Y., Hopkins, P.~F., {et~al.} 2021, \mnras, 502, 3646

\bibitem[{Heitsch {et~al.}(2001)}]{heitsch2001magnetic}
Heitsch, F., Zweibel, E.~G., Mac Low, M.-M., {et~al.} 2001, \apj, 561, 800

\bibitem[{Hennebelle {et~al.}(2008)}]{hennebelle2008mag}
Hennebelle, P., \& Teyssier, R. 2008, \aap, 477, 25

\bibitem[{Hildebrand {et~al.}(2009)}]{hildebrand2009dispersion}
Hildebrand, R.~H., Kirby, L., Dotson, J.~L., {et~al.} 2009, \apj, 696, 567

\bibitem[{Hirano {et~al.}(2019)}]{hirano2019Bmisalign}
Hirano, S., \& Machida, M.~N. 2019, \mnras, 485, 4667

\bibitem[{Hoang {et~al.}(2009)}]{hoang2009grain}
Hoang, T., \& Lazarian, A. 2009, \apj, 697, 1316

\bibitem[{Houde {et~al.}(2009)}]{houde2009dispersion}
Houde, M., Vaillancourt, J.~E., Hildebrand, R.~H., {et~al.} 2009, \apj, 706, 1504

\bibitem[{Huang {et~al.}(2023)}]{huang2023clump}
Huang, B., Wang, K., Girart, J.~M., {et~al.} 2023, \apj, 949, 46

\bibitem[{Huang {et~al.}(2024)}]{huang2024magnetic}
Huang, B., Girart, J.~M., Stephens, I.~W., {et~al.} 2024, \apjl, 963, L31

\bibitem[{Hull {et~al.}(2020)}]{hull2020understanding}
Hull, C.~L.~H., Le Gouellec, V.~J.~M., Girart, J.~M., {et~al.} 2020, \apj, 892, 152

\bibitem[{Hull {et~al.}(2019)}]{hull2019interferometric}
Hull, C.~L.~H., \& Zhang, Q. 2019, FrASS, 6, 3

\bibitem[{Joos {et~al.}(2012)}]{joos2012protostellar}
Joos, M., Hennebelle, P., \& Ciardi, A. 2012, \aap, 543, A128

\bibitem[{J{\o}rgensen {et~al.}(2015)}]{jorgensen2015molecule}
J{\o}rgensen, J.~K., Visser, R., Williams, J.~P., {et~al.} 2015, \aap, 579, A23

%\bibitem[{Kataoka {et~al.}(2017)}]{kataoka2017evidence}
%Kataoka, A., Tsukagoshi, T., Pohl, A., {et~al.} 2017, \apjl, 844, L5

%\bibitem[{Kataoka {et~al.}(2015)}]{kataoka2015millimeter}
%Kataoka, A., Muto, T., Momose, M., {et~al.} 2015, \apj, 809, 78

\bibitem[{Kauffmann {et~al.}(2008)}]{kauffmann2008mambo}
Kauffmann, J., Bertoldi, F., Bourke, T.~L., {et~al.} 2008, \aap, 487, 993

\bibitem[{Kenyon {et~al.}(1993)}]{kenyon1993model}
Kenyon, S.~J., Calvet, N., \& Hartmann, L. 1993, \apj, 414, 676

%\bibitem[{Kim {et~al.}(2020)}]{kim2020molecular}
%Kim, G., Tatematsu, K., Liu, T., {et~al.} 2020, \apjs, 249, 33

\bibitem[{Krumholz {et~al.}(2005)}]{krumholz2005ism}
Krumholz, M.~R., \& McKee, C.~F. 2005, \apj, 630, 250

\bibitem[{Kwon {et~al.}(2009)}]{kwon2009opacity}
Kwon, W., Looney, L.~W., Mundy, L.~G., {et~al.} 2009, \apj, 696, 841

\bibitem[{Kwon {et~al.}(2019)}]{kwon2019mag}
Kwon, W., Stephens, I.~W., Tobin, J.~J., {et~al.} 2019, \apj, 879, 25

\bibitem[{Le Gouellec {et~al.}(2019)}]{le2019characterizing}
Le Gouellec, V.~J.~M., Hull, C.~L.~H., Maury, A.~J., {et~al.} 2019, \apj, 885, 106

\bibitem[{Le Gouellec {et~al.}(2020)}]{le2020IMS}
Le Gouellec, V.~J.~M., Maury, A.~J., Guillet, V., {et~al.} 2020, \aap, 644, A11

\bibitem[{Li {et~al.}(2013)}]{li2013misalignment}
Li, Z.-Y., Krasnopolsky, R., \& Shang, H. 2013, \apj, 774, 82

\bibitem[{Lin {et~al.}(2023)}]{lin2023disk}
Lin, Z.-Y.~D., Li, Z.-Y., Tobin, J.~J., {et~al.} 2023, \apj, 951, 9

\bibitem[{Lombardi {et~al.}(2014)}]{lombardi2014temperature}
Lombardi, M., Bouy, H., Alves, J., {et~al.} 2014, \aap, 566, A45

\bibitem[{Mac Low {et~al.}(2004)}]{mac2004tur}
Mac Low, M.-M., \& Klessen, R.~S. 2004, Reviews of Modern Physics, 76, 125

\bibitem[{Machida {et~al.}(2020)}]{machida2020misalignment}
Machida, M.~N., Hirano, S., \& Kitta, H. 2020, \mnras, 491, 2180

\bibitem[{Machida {et~al.}(2008)}]{machida2008formation}
Machida, M.~N., Tomisaka, K., Matsumoto, T., {et~al.} 2008, \apj, 677, 327

\bibitem[{Matsumoto {et~al.}(2003)}]{matsumoto2003fragmentation}
Matsumoto, T., \& Hanawa, T. 2003, \apj, 595, 913

\bibitem[{Maury {et~al.}(2022)}]{maury2022}
Maury, A., Hennebelle, P., \& Girart, J.~M. 2022, FrASS, 9, 949223

\bibitem[{McKee {et~al.}(2007)}]{mckee2007star}
McKee, C.~F., \& Ostriker, E.~C. 2007, \araa, 45, 565

\bibitem[{Mellon {et~al.}(2008)}]{mellon2008mag}
Mellon, R., Richard, R., \& Li, Z.-Y. 2008, \apj, 681, 1356

\bibitem[{Mignon-Risse {et~al.}(2021)}]{mignon2021collapse}
Mignon-Risse, R., Gonz{\'a}lez, M., Commer{\c{c}}on, B., {et~al.} 2021, \aap, 652, A69

\bibitem[{Myers {et~al.}(2014)}]{myers2014star}
Myers, A.~T., Klein, R.~I., Krumholz, M.~R., {et~al.} 2014, \mnras, 439, 3420

\bibitem[{Myers (2011)}]{myers2011ism}
Myers, P.~C. 2011, \apj, 743, 98

\bibitem[{Myers {et~al.}(2000)}]{myers2000obs}
Myers, P.~C., Evans, N.~J., \& Ohashi, N. 2000, Protostars and Planets IV, ed. V. Mannings, A.~P., Boss, \& S.~S., Russell, 217

\bibitem[{Narang {et~al.}(2024)}]{narang2024lumi}
Narang, M., Manoj, P., Tyagi, H. {et~al.}, 2024, \apjl, 962, L16

\bibitem[{Offner {et~al.}(2023)}]{offner2023ppvii}
Offner, S.~S.~R., Moe, M., Kratter, K.~M. 2023, Protostars and Planets VII, ed. S. Inutsuka, Y., Aikawa, T., Muto, K., Tomida, \& M., Tamura, 534, 275

\bibitem[{Ossenkopf {et~al.}(1994)}]{ossenkopf1994dust}
Ossenkopf, V., \& Henning, T. 1994, \aap, 291, 943

\bibitem[{Padoan {et~al.}(2001)}]{padoan2001turbulence}
Padoan, P., Juvela, M., Goodman, A.~A., {et~al.} 2001, \apj, 553, 227

\bibitem[{Padoan {et~al.}(2011)}]{padoan2011ism}
Padoan, P., \& Nordlund, \r{A}. 2011, \apj, 730, 40

\bibitem[{Palau {et~al.}(2013)}]{palau2013frag}
Palau, A., Fuente, A., Girart, J.~M., {et~al.} 2013, \apj, 762, 120

\bibitem[{Palau {et~al.}(2021)}]{palau2021frag}
Palau, A., Zhang, Q., Girart, J.~M., {et~al.} 2021, \apj, 912, 159

\bibitem[{Price {et~al.}(2009)}]{price2009magnetic}
Price, D.~J., \& Bate, M.~R. 2009, \mnras, 398, 33

\bibitem[{Qiu {et~al.}(2014)}]{qiu2014submillimeter}
Qiu, K., Zhang, Q., Menten, K.~M., {et~al.} 2014, \apjl, 794, L18

\bibitem[{Sheehan {et~al.}(2022)}]{sheehan2022vandam}
Sheehan, P.~D., Tobin, J.~J., Looney, L.~W., {et~al.} 2022, \apj, 851, 55

\bibitem[{Stephens {et~al.}(2013)}]{stephens2013hourglass}
Stephens, I.~W., Looney, L.~W., Kwon, W., {et~al.} 2013, \apjl, 769, L15

%\bibitem[{Stephens {et~al.}(2017)}]{stephens2017alma}
%Stephens, I.~W., Yang, H., Li, Z.-Y., {et~al.} 2017, \apj, 851, 55

\bibitem[{Tobin {et~al.}(2013)}]{tobin2013resolved}
Tobin, J.~J., Bergin, E.~A., Hartmann, L., {et~al.} 2013, \apj, 765, 18

\bibitem[{Tobin {et~al.}(2020)}]{tobin2020vla}
Tobin, J.~J., Sheehan, P.~D., Megeath, S.~T., {et~al.} 2020, \apj, 890, 130

\bibitem[{Tobin {et~al.}(2022)}]{tobin2022vla}
Tobin, J.~J., Offner, S.~S.~R., Kratter, K.~M., {et~al.} 2022, \apj, 925, 39

\bibitem[{Vaillancourt (2006)}]{vaillancourt2006placing}
Vaillancourt, J.~E. 2006, \pasp, 118, 1340

\bibitem[{van't Hoff {et~al.}(2023)}]{van2023disk}
van't Hoff, M.~L.~R., Tobin, J.~J., Li, Z.-Y., {et~al.} 2023, \apj, 951, 10

\bibitem[{V{\'a}zquez-Semadeni {et~al.}(2007)}]{vazquez2007frag}
V{\'a}zquez-Semadeni, E., G{\'o}mez, G.~C., Jappsen, A.~K., {et~al.} 2007, \apj, 657, 870

\bibitem[{Whitney {et~al.}(2003)}]{whitney2003radiative}
Whitney, B.~A., Wood, K., Bjorkman, J.~E., {et~al.} 2003, \apj, 591, 1049

\bibitem[{Wurster {et~al.}(2019)}]{wurster2019nonmhd}
Wurster, J., Bate, M.~R., \& Price, D.~J. 2019, \mnras, 489, 1719

\bibitem[{Yen {et~al.}(2021)}]{yen2021mag}
Yen, H.-W., Koch, P.~M., Hull, C.~L.~H., {et~al.} 2021, \apj, 907, 33

\bibitem[{Zakri {et~al.}(2022)}]{zakri2022orion}
Zakri, W., Megeath, S.~T., Fischer, W.~J., {et~al.} 2022, \apjl, 924, L23

\end{thebibliography}
\end{document}